%% file: main.tex
\documentclass[11pt, a4paper, oneside]{Thesis} 

\graphicspath{{Pictures/}} 

\usepackage[square,sort,compress,numbers]{natbib}
\usepackage{cite}
\usepackage{amsmath,latexsym} 
\usepackage{float}

\title{\ttitle} 
\DeclareUnicodeCharacter{2212}{-}
\DeclareUnicodeCharacter{2212}{+}
\begin{document}

\setstretch{1.3} 

\fancyhead{} 
\rhead{\thepage} 
\lhead{} 

\input{variables}

\maketitle

\addtocontents{toc}{\vspace{2em}} 
\frontmatter 
\Certificate

\Declaration

\begin{acknowledgements}

I would like to express my deepest gratitude to my supervisor, \textbf{Prof. Pradyumn Kumar Sahoo}, Professor, Department of Mathematics, BITS-Pilani, Hyderabad Campus, Hyderabad, Telangana. His dedication towards research, expertise in executing research projects, attentiveness towards new research and developments, and last but not least, his contribution to the development of cosmology and to attract young mathematicians in the field of cosmology is the root motivation of my work, as well as many others.

I sincerely thank my Doctoral Advisory Committee (DAC) members, \textbf{Prof. Bivudutta Mishra} and \textbf{Dr. Prasant Samantray}, for their valuable suggestions and constant encouragement to improve my research works.

It is my privilege to thank HoD, the DRC convener, faculty members, my colleagues, and the Department of Mathematics staff for supporting this amazing journey of my Ph.D. career.

I owe a deep sense of gratitude to my co-authors, \textbf{Dr. S. K. J. Pacif}, \textbf{Dr. Avik De}, \textbf{Dr. Tee-How Loo},  \textbf{Dr. P. H. R. S Moraes}, and my \textbf{colleagues} for their valuable suggestions, discussions, encouragement, and collaborations. 

I acknowledge \textbf{BITS-Pilani, Hyderabad Campus}, for providing me with the necessary facilities, and the \textbf{University Grant Commission (UGC), India}, for providing Research Fellowship (UGC Reference No. 201920-191620096030) to carry out my research works.

I also acknowledge \textbf{The MTTS Programm-2017} to shape my mathematical thinking, especially \textbf{Prof. S. Kumaresan} and \textbf{Prof. A. J. Jayanthan}. I also express my deepest thanks to my mentors, \textbf{Mr. Akash Shekatkar} and \textbf{Mr. Sumit Yadav}.

I would like to express my heartiest thanks to my parents \textbf{Mrs. Rekha Solanki} and \textbf{Mr. Narendra Solanki}, and my friends \textbf{Rajendra Singh Pawar}, \textbf{Chetan Likhar}, \textbf{Uday Sharma}, \textbf{Vishal Verma}. Last but not least, I thank \textbf{Dr. G. Parthasarathy, KIMS Hospital} for saving my life.

\vspace{1.2 cm}
Raja Solanki,\\
ID: 2020PHXF0003H.

\end{acknowledgements}

\begin{abstract}
In the last century, theoretical and experimental developments have established the General Relativity theory as the most successful theory for describing the gravitational phenomenon. On the other hand, in the last two decades, multiple observational probes have strongly favored the discovery of the acceleration of cosmic expansion. The observational enhancement and development in precision cosmology indicate a requirement to go beyond General Relativity and to search for an alternate description that can resolve the persistent issues.

In Chapter \ref{Chapter1}, we highlight some important elements of observational cosmology and attempt to shed light on the structure, evolution history, and constituents of the Universe. Further, we briefly discuss the physics of Newtonian, Special, and General Relativity. We also describe the mathematical foundations of General Relativity based on Riemannian geometry. Moreover, we discuss some alternative formulations of General Relativity based on the Non-Reimannian geometry, such as teleparallel and symmetric teleparallel gravity and its modifications. 

In Chapters \ref{Chapter2} and \ref{Chapter3}, we investigate the $f(Q)$ gravity in the presence of viscosity in the cosmic fluid. We explore a linear $f(Q)$ and a non-linear power-law $f(Q)$ model, and then we find the corresponding analytical solutions. Further, we employ the observational dataset to constrain the free parameters of the obtained solution. Then, the constrained solution predicts the evolution phase of the expansion. We find that the proposed model can effectively describe the acceleration of the cosmic expansion. However, these solutions cannot describe the early phases of the Universe.

In Chapters \ref{Chapter4} and \ref{Chapter5}, we explore the constraints on the various classes of non-linear $f(Q)$ gravity models in both coincident and non-coincident formalism, respectively. In the coincident formalism, we find the set of constraints on $f(Q)$ gravity models that can reproduce the quintessence, phantom, and cosmological constant type dark energy effects. Further, in a non-coincident formalism, we propose a model-independent parameterization form of the Hubble function that can efficiently describe the late-time de-Sitter limit, present deceleration parameter value, and the transition redshift. We explore constraints on several classes of $f(Q)$ gravity models utilizing this Hubble function and the energy conditions.

In Chapter \ref{Chapter6}, we present a covariant formulation and energy balance equation for the $f(Q,\mathcal{T})$ gravity, which is an extension of  $f(Q)$ gravity. Further, we investigate the physical capabilities of various classes of $f(Q,\mathcal{T})$ gravity models to describe the different cosmological epochs through the phase-space analysis. Finally, in Chapter \ref{Chapter7}, we briefly summarize the outcomes of the present thesis and the future scope.

\end{abstract} 

\clearpage
\setstretch{1.3} 

\pagestyle{empty} 
\pagenumbering{gobble}
\Dedicatory{ \begin{LARGE}
\textbf{Dedicated to}
\end{LARGE} 
\\
\vspace{3cm}
My Guru Late \textbf{Prof. K. Santaram}\\
        \textbf{\& }\\
My Source of Inspiration\\
My Wife \textbf{Purnima Solanki}\\}



\lhead{\emph{Contents}} 
\tableofcontents 
\addtocontents{toc}{\vspace{1em}}
\lhead{\emph{List of tables}}
\listoftables 
\addtocontents{toc}{\vspace{1em}}
\lhead{\emph{List of figures}}
\listoffigures 
\addtocontents{toc}{\vspace{1em}}




\lhead{\emph{List of symbols}}
\listofsymbols{ll}{

$g_{\mu\nu}$: \,\,\,\,\,\, Metric Tensor\\
$g$: \,\,\,\,\,\,\,\,\,\,\, Determinant of $g_{\mu\nu}$\\
$\hat{\Gamma}^{\lambda}_{\mu\nu}$: \,\,\,\,\, General Affine Connection\\
$\Gamma^{\lambda}_{\mu\nu}$: \,\,\,\,\, Levi-Civita Connection\\
$\tilde{\Gamma}^{\lambda}_{\mu\nu}$: \,\,\,\,\, Teleparallel Connection\\
$\mathring{\Gamma}^{\lambda}_{\mu\nu}$: \,\,\,\,\, Symmetric Teleparallel Connection\\
$\nabla_{\mu}$: \,\,\,\,\,\, Covariant derivative w.r.t. Levi-Civita Connection\\
$\hat{\nabla}_{\mu}$: \,\,\,\,\,\, Covariant derivative w.r.t. General Affine Connection\\
$\tilde{\nabla}_{\mu}$: \,\,\,\,\,\, Covariant derivative w.r.t. Teleparallel Connection\\
$\mathring{\nabla}_{\mu}$: \,\,\,\,\,\,\,  Covariant derivative w.r.t. Symmetric Teleparallel Connection\\
$(\mu\nu)$: \,\,\,\,\, Symmetrization over the indices $\mu$ and $\nu$\\
$[\mu\nu]$: \,\,\,\,\,\, Anti-Symmetrization over the indices $\mu$ and $\nu$\\
GR: \,\,\,\,\,\,\,\,\, General Relativity\\
$\Lambda$CDM:\, $\Lambda$ Cold Dark Matter\\
ECs: \,\,\,\,\,\, Energy Conditions\\
EoS:\,\,\,\,\,\,\,\,\,\,   Equation of State \\
SN Ia:\,\,\,\,\,\,\,   Type Ia Supernovae\\
CMB: \,\,\,\,\,\,Cosmic Microwave Background\\
BAO: \,\,\,\,\,  Baryon Acoustic Oscillations\\
DE:\,\,\,\,\,\,\,\,\,\,\,  Dark Energy\\
MCMC:  Markov Chain Monte Carlo}

\addtocontents{toc}{\vspace{2em}}

%
%


\clearpage 





\mainmatter 

\pagestyle{fancy} 

\input{Chapters/Introduction}

\input{Chapters/Chapter2}

\input{Chapters/Chapter3}

\input{Chapters/Chapter4}

\input{Chapters/Chapter5}

\input{Chapters/Chapter6}
\input{Chapters/Conclusion}






\addtocontents{toc}{\vspace{2em}} 

\backmatter


\label{References}
\lhead{\emph{References}}
\input{References.tex}
\cleardoublepage
\pagestyle{fancy}

\label{Publications}
\lhead{\emph{List of publications}}

\chapter{List of publications}
\section*{Thesis publications}
\begin{enumerate}

\item \textbf{Raja Solanki}, S. K. J. Pacif, A. Parida, and  P.K. Sahoo, \textit{Cosmic acceleration with bulk viscosity in modified f(Q) gravity}, \textcolor{blue}{Physics of the Dark Universe} \textbf{32}, 100820 (2021).

\item \textbf{Raja Solanki},  Simran Arora, P. K. Sahoo, and P. H. R. S. Moraes, \textit{Bulk viscous fluid in symmetric teleparallel cosmology: theory versus experiment}, \textcolor{blue}{Universe} \textbf{9(1)}, 12 (2023).

\item \textbf{Raja Solanki}, A. De, and P. K. Sahoo, \textit{Complete dark energy scenario in f(Q) gravity}, \textcolor{blue}{Physics of the Dark Universe} \textbf{36}, 100996 (2022).

\item \textbf{Raja Solanki}, A. De, S. Mandal, P. K. Sahoo, \textit{Accelerating expansion of the Universe in modified symmetric teleparallel gravity}, \textcolor{blue}{Physics of the Dark Universe} \textbf{36}, 101053 (2022).

\item S. Pradhan, \textbf{Raja Solanki}, P. K. Sahoo, \textit{Cosmological constraints on $f(Q)$ gravity models in the non-coincident formalism}, \textcolor{blue}{Journal of High Energy Astrophysics} \textbf{43}, 258-267 (2024).

\item T. H. Loo, \textbf{Raja Solanki}, A. De, and P. K. Sahoo, \textit{$f(Q,\mathcal{T})$ gravity, its covariant formulation, energy conservation and phase-space analysis}, \textcolor{blue}{The European Physical Journal C} \textbf{83(3)}, 261 (2023).
\end{enumerate}
\section*{Other publications}
\begin{enumerate}

\item A. Bhat, \textbf{Raja Solanki}, P. K. Sahoo, \textit{Extended Bose-Einstein condensate dark matter in $f(Q)$ gravity}, \textcolor{blue}{General Relativity and Gravitation} \textbf{56}, 63 (2024).

\item L. V. Jaybhaye, \textbf{Raja Solanki}, P. K. Sahoo, \textit{Bouncing cosmological models in $f(R,L_m)$ gravity}, \textcolor{blue}{Physica Scripta} \textbf{99}, 065031 (2024). 

\item S. Ghosh, \textbf{Raja Solanki}, P. K. Sahoo, \textit{Dynamical system analysis of Dirac-Born-Infeld scalar field cosmology in coincident f(Q) gravity}, \textcolor{blue}{Chinese Physics C} \textbf{48(9)}, 095102 (2024).

\item D. S. Rana, \textbf{Raja Solanki}, P. K. Sahoo, \textit{Phase-space analysis of the viscous fluid cosmological models in the coincident $f(Q)$ gravity}, \textcolor{blue}{Physics of the Dark Universe} \textbf{43}, 101421 (2024). 

\item S. Ghosh, \textbf{Raja Solanki}, P. K. Sahoo, \textit{Dynamical system analysis of scalar field cosmology in coincident $f(Q)$ gravity}, \textcolor{blue}{Physica Scripta} \textbf{99}, 055021 (2024).

\item L. V. Jaybhaye, \textbf{Raja Solanki}, P. K. Sahoo, \textit{Late time cosmic acceleration through parametrization of Hubble parameter in $f(R, L_m)$ gravity}, \textcolor{blue}{Physics of the Dark Universe} \textbf{46}, 101639 (2024).

\item \textbf{Raja Solanki}, A. Bhat, P. K. Sahoo, \textit{Bulk viscous cosmological model in $f(T,\mathcal{T})$ modified gravity}, \textcolor{blue}{Astroparticle Physics} \textbf{163}, 103013 (2024).

\item \textbf{Raja Solanki}, Z. Hassan, P. K. Sahoo, \textit{Wormhole solutions in $f(R,L_m)$ gravity}, \textcolor{blue}{Chinese Journal of Physics} \textbf{85}, 74-88 (2023). 

\item G. Leon, S. Chakraborty, S. Ghosh, \textbf{Raja Solanki}, P. K. Sahoo, E. Gonzalez, \textit{Scalar field evolution at background and perturbation levels for a broad class of potentials}, \textcolor{blue}{Fortschritte der Physik} \textbf{71}, 2300006 (2023).

\item \textbf{Raja Solanki}, D. S. Rana, S. Mandal, P. K. Sahoo, \textit{Viscous fluid cosmology in symmetric teleparallel gravity}, \textcolor{blue}{Fortschritte der Physik} \textbf{71}, 2200202 (2023).

\item \textbf{Raja Solanki}, B. Patel, L. V. Jaybhaye, \textbf{Raja Solanki}, \textit{Cosmic acceleration with bulk viscosity in an anisotropic $f(R,L_m)$ background}, \textcolor{blue}{Communications in Theoretical Physics} \textbf{75(7)}, 075401 (2023).

\item L. V. Jaybhaye, \textbf{Raja Solanki}, S. Mandal, P.K. Sahoo, \textit{Cosmology in $f(R,L_m)$ gravity}, \textcolor{blue}{Physics Letters B} \textbf{831}, 137148 (2022).

\item \textbf{Raja Solanki}, P. K. Sahoo, \textit{Statefinder Analysis of Symmetric Teleparallel Cosmology}, \textcolor{blue}{Annalen der Physik} \textbf{534(6)}, 2200076 (2022). 

\item A. De, T. H. Loo, \textbf{Raja Solanki}, and P. K. Sahoo, \textit{How a conformally flat (GR)4 impacts Gauss-Bonnet gravity $?$}, \textcolor{blue}{Fortschritte der Physik} \textbf{69(10)}, 2100088 (2021).

\item A. De, T. H. Loo, \textbf{Raja Solanki}, and P. K. Sahoo, \textit{A Conformally Flat Generalized Ricci Recurrent Spacetime in $F(R)$ Gravity}, \textcolor{blue}{Physica Scripta} \textbf{96(8)}, 085001 (2021).

\end{enumerate}
\cleardoublepage
\pagestyle{fancy}

\lhead{\emph{Paper presented at conferences}}

\chapter{Paper presented at conferences}
\label{Paper presented at conferences}

\begin{enumerate}
\item Presented research paper entitled “\textit{Extended Bose-Einstein condensate dark matter in $f(Q)$ gravity}" at the conference “\textbf{17th International Conference on Interconnections between Particle Physics and Cosmology}” organized by the IIT Hyderabad, during the period $14^{th}-18^{th}$ October, $2024$.
\item Best paper award for the research paper entitled “\textit{Extended Bose-Einstein condensate dark matter in $f(Q)$ gravity}” at the conference “\textbf{International Conference on Gravitation, Astrophysics and Cosmology}” jointly organized by the Tensor Society of India and GLA University, Mathura, during the period $14^{th}-16^{th}$ June, $2024$.
\item Presented research paper entitled “\textit{Viscous fluid cosmology in symmetric teleparallel gravity}” at the conference “\textbf{Differential Geometry and Relativity}” jointly organized by the Tensor Society of India and the Department of Mathematics, SSJ University, Almora, Uttarakhand, during the period $18^{th}-20^{th}$ May, $2023$.
\item Presented research paper entitled “\textit{Statefinder analysis of symmetric teleparallel cosmology}” at the conference “\textbf{International Conference on Mathematical Sciences and its Applications}” organized by the Department of Mathematics, SRTM University, Nanded, during the period $28^{th}-30^{th}$ July, $2022$.
\item Presented research paper entitled “\textit{Bulk viscous fluid in symmetric teleparallel cosmology: theory versus experiment}” at the conference “\textbf{Differential Geometry and its Applications}” jointly organized by the Tensor Society of India and Department of Mathematics, Kuvempu University, Shivamogga, Karnataka, during the period $4^{th}-5^{th}$ March, $2022$.
\item Presented research paper entitled “\textit{Cosmic acceleration with bulk viscosity in modified f(Q) gravity}" at the conference “\textbf{International Academy of Physical Sciences on Advances in Relativity and Cosmology}” organized by Department of Mathematics, BITS-Pilani, Hyderabad Campus, during the period $26^{th}-28^{th}$ October, $2021$.

\end{enumerate}

\cleardoublepage
\pagestyle{fancy}
\lhead{\emph{Biography}}

\chapter{Biography}

\section*{Brief biography of the candidate:}
\textbf{Mr. Raja Solanki} obtained his Bachelor's degree in 2016 and Master's degree in Mathematics from the School of Mathematics, Devi Ahilya University, Indore, in 2018. During his Masters, he was selected for the Mathematical Training and Talent Search (MTTS) Program held in RIE, Mysuru. He qualified for the Council of Scientific and Industrial Research (CSIR), the National Eligibility Test (NET) for Lectureship (LS) in 2018, the Junior Research Fellowship (JRF) in 2019 with $97.7$ percentile, and the Graduate Aptitude Test in Engineering (GATE) in 2019 with $92$ percentile. He has experience as a visiting faculty member at the School of Mathematics, Devi Ahilya University. He was also selected as a Senate Member (Ph.D. representative) in the BITS Pilani, Hyderabad Campus, during the period 2023-2024. He is also a member of the European Cooperation in Science and Technology (COST) (CA21136) working group, which works on addressing observational tensions in cosmology with systematics and fundamental physics. He has published $21$ research articles in renowned international journals during his Ph.D. research career. He has presented his research at several National and International conferences. He also received the Best Paper award at the International Conference on Gravitation, Astrophysics, and Cosmology, jointly organized by the Tensor Society of India and GLA University, Mathura.

\section*{Brief biography of the supervisor:}

\textbf{Prof. P.K. Sahoo} received his Ph.D. degree from Sambalpur University, Odisha, India, in 2004. He is currently serving as a Professor in the Department of Mathematics, Birla Institute of Technology and Science-Pilani, Hyderabad Campus. He also holds the position of Head of the Department during the period from October 2020 to September 2024. He has conducted several academic and scientific events in the department, including the 89th Annual Conference of The Indian Mathematical Society in 2023. He contributed to BITS through five sponsored research projects from University Grants Commission (UGC 2012-2014),  DAAD-Research Internships in Science and Engineering (RISE) Worldwide (2018, 2019, 2023 and 2024), Council of Scientific and Industrial Research (CSIR 2019-2022), National Board for Higher Mathematics (NBHM 2022-2025), Science and Engineering Research Board (SERB), Department of Science and Technology (DST 2023-2026). He is an expert reviewer of Physical Science Projects, SERB, DST, Govt. of India, and University Grants Commission (UGC) research schemes. 

\textbf{Prof. Sahoo} has published more than 250 research articles in various renowned national and international journals. He has participated in many international and national conferences, most of which he has presented his work as an invited speaker. He has various research collaborations at both the national and international levels. He has been placed among the top $2\%$ scientists of the world according to the survey by researchers from Stanford University, USA, in Nuclear and Particle Physics for five consecutive years.   He serves as an expert reviewer and editorial member for a number of reputed scientific journals and is also a Ph.D. examiner at several universities. He has been awarded with visiting professor fellowship at Transilvania University of Brasov, Romania. He is also the recipient of the Science Academics Summer Research Fellowship, a UGC Visiting Fellow, a Fellow of the Institute of Mathematics and its Applications (FIMA), London, a Fellow of the Royal Astronomical Society (FRAS), London, elected as a foreign member of the Russian Gravitational Society. Prof. Sahoo is also a COST (CA21136) member: Addressing observational tensions in cosmology with systematics and fundamental physics. As a visiting scientist, he had the opportunity to visit the European Organization for Nuclear Research (CERN), Geneva, Switzerland, a well-known research center for scientific research.

\end{document}

%% file: variables.tex
%

\thesistitle{Accelerating Expansion of the Universe in Modified Symmetric Teleparallel Gravity}
\documenttype{\bf{THESIS}}
\supervisor{\bf{Prof. Pradyumn Kumar Sahoo}}
\supervisorposition{Professor}
\supervisorinstitute{BITS-Pilani, Hyderabad Campus}
\examiner{}
\degree{Ph.D. Research Scholar}
\coursecode{\textbf{DOCTOR OF PHILOSOPHY}}
\coursename{THESIS}
\authors{\textbf{RAJA SOLANKI}}
\IDNumber{2020PHXF0003H}
\addresses{}
\subject{}
\keywords{}
\university{\texorpdfstring{\href{http://www.bits-pilani.ac.in/} 
                {Birla Institute of Technology and Science, Pilani}} 
                {Birla Institute of Technology and Science, Pilani}}
\UNIVERSITY{\texorpdfstring{\href{http://www.bits-pilani.ac.in/} 
                {\textbf{BIRLA INSTITUTE OF TECHNOLOGY AND SCIENCE, PILANI}}} 
                {BIRLA INSTITUTE OF TECHNOLOGY AND SCIENCE, PILANI}}



\department{\texorpdfstring{\href{http://www.bits-pilani.ac.in/pilani/Mathematics/Mathematics} 
                {Mathematics}} 
                {Mathematics}}
\DEPARTMENT{\texorpdfstring{\href{http://www.bits-pilani.ac.in/pilani/Mathematics/Mathematics} 
                {Mathematics}} 
                {Mathematics}}
\group{\texorpdfstring{\href{Research Group Web Site URL Here (include http://)}
                {Research Group Name}} 
                {Research Group Name}}
\GROUP{\texorpdfstring{\href{Research Group Web Site URL Here (include http://)}
                {RESEARCH GROUP NAME (IN BLOCK CAPITALS)}}
                {RESEARCH GROUP NAME (IN BLOCK CAPITALS)}}
\faculty{\texorpdfstring{\href{Faculty Web Site URL Here (include http://)}
                {Faculty Name}}
                {Faculty Name}}
\FACULTY{\texorpdfstring{\href{Faculty Web Site URL Here (include http://)}
                {FACULTY NAME (IN BLOCK CAPITALS)}}
                {FACULTY NAME (IN BLOCK CAPITALS)}}

%% file: Chapters/Introduction.tex

\chapter{Preliminaries} 
\label{Chapter1}

\lhead{Chapter 1. \emph{Preliminaries}} 

\clearpage
\pagebreak

\section{The cosmos observed}
We live in an expanding Universe filled with billions of stars, galaxies, and several mysterious objects; all the galaxies are now rushing away from each other. Physics, the study of energy, matter, and their mutual interactions, serves as a fundamental force driving humanity's intellectual advancement. Cosmology, on the other hand, explores the Universe as a whole, seeking to understand its origins, composition, and future. According to the cosmological principle, the Universe appears uniform in every direction and has no unique regions when observed on a sufficiently large scale. On the cosmological scales, the galaxies are the constituent elements of the whole Universe, and the only relevant interaction among them is gravitation. Gravity is the weakest of the four fundamental interactions acting between individual elementary particles. Gravity, according to Newtonian theory, acts between all masses, and in the context of relativistic gravity, it interacts with all forms of energy. It is an unscreened force and has a long-range interaction. These facts are key to understand the influence of gravity in physical phenomena, particularly in governing the organization of the Universe on the largest scales observed in cosmology and astrophysics.

\subsection{The cosmological scale}

In astronomy, the standard unit of distance is the parsec (pc), and it is the measure of the distance at which a star exhibits a parallax of one arcsecond for a base line equal to an astronomical unit distance i.e. the distance between the Sun and the Earth. One parsec is roughly around 3.26 light-years. The nearest star to our solar system is about 1.2 pc away. The Milky Way galaxy, our home galaxy, contains nearly $\mathcal{O}(10^{11})$ stars within a disk of thickness of about 2 kpc and a diameter of 30 kpc. The Milky Way galaxy further belongs to a small cluster of around 30 galaxies in a volume roughly 1 Mpc across, called the Local Group. For example, the distance to our neighboring Andromeda Galaxy is 0.7 Mpc. The Virgo cluster, consisting of nearly 2000 galaxies within a distance scale of 5 Mpc radius, belongs to the Virgo Supercluster that spans a volume of about 50 Mpc, along with the Local group. This and other clusters of galaxies appear to reside near vast empty regions known as voids. These galaxies in the Universe are not randomly scattered around us but rather form clustered structures in a coherent way that spread out up to 100 Mpc. Such distribution is characterized by a network of large filamentary structures consisting of huge voids forming a cosmic web. Beyond this cosmological scale, the Universe looks almost uniform. The cosmic microwave background (CMB) radiation, which is the largest observable entity, also appears to be relatively uniform.

\subsection{The Hubble's discovery: the expanding Universe}

In the period 1910-1930, the cosmic distance ladder extended beyond 100 kpc. When observing a galaxy in a visible wavelength, its spectrum usually displays absorption lines. Note that the wavelength of light obtained from a receding object is elongated, leading to what is known as redshift that measures this stretching factor or wavelength shift as follows,
\begin{equation}\label{e1}
 z \equiv \frac{\lambda_{obs}-\lambda_{emit}}{\lambda_{emit}}\text{.}
\end{equation}
For non-relativistic motion i.e. low redshift case, the standard Doppler's effect gives $z \approx \frac{v}{c}$, where $v$ is the velocity of the object receding from us. Over a decade of observations, of around 40 galaxies, Vesto Slipher at Arizona's Lowell Observatory found that nearly all galaxies except for a few in the Local Group, exhibited redshift. Edwin Hubble at Mt. Wilson Observatory in California then attempted to find the distances to these galaxies by correlating these observed redshifts. He discovered that this observed redshift was directly proportional to the distance $d$ of the galaxy emitting the light. In the year 1929, Hubble presented his following groundbreaking results \cite{R0},
\begin{equation}\label{e2}
z=\frac{H_0}{c}d\text{.}
\end{equation}
Utilizing the Doppler's relation $z \approx \frac{v}{c}$, we have 
\begin{equation}\label{e3}
v=H_0 d\text{.}
\end{equation}
This proportionality constant $H_0$ is well known as the Hubble constant, and it represents the recession speed per unit separation.

\subsection{The $H_0$ tension}

Determining the precise value of $H_0$ has been a significant challenge as it was obtained via measuring large cosmic distances. At the beginning, Hubble estimated a value of $H_0$ as $ 500 \: km/s/Mpc$ \cite{R1}, which was overestimated due to inaccuracies in distance calibrations. After the launch of the Hubble Space Telescope (HST), the range of $H_0$ value approximated between $50-100 \: km/s/Mpc$. Over the last decade, the Planck collaboration obtained $H_0$ value as $67.4 \pm 0.5 \: km/s/Mpc $ \cite{R2} by analyzing the CMB data. 
Another measurements of the $H_0$ value that align with the Planck results i.e. favoring the lower values of $H_0$ are the ground-based CMB telescope, the Wilkinson Microwave Anisotropy Probe (WMAP) + Atacama Cosmology Telescope (ACT-DR4) \cite{R3} estimates a value of $H_0$ as $67.6 \pm 1.1 \: km/s/Mpc$ at $68$\% confidence level (CL) in $\Lambda$CDM, whereas the South Pole Telescope (SPT-3G) reports $H_0 = 68.8 \pm 1.5 \: km/s/Mpc $ at the same CL \cite{R4}. In addition, the Big Bang Nucleosynthesis (BBN) + Baryon Acoustic Oscillations (BAO) from eBOSS and BOSS survey \cite{R5} estimates $H_0 = 67.35 \pm 0.97 \: km/s/Mpc $ at $68$\% CL. On the other hand, the HST survey involves measuring the local distance ladder by combining the photometry of Cepheids with other local distance anchors such as Milky Way parallaxes and calibration distances to Cepheids in nearby galaxies that host SN Ia. Particularly, the HST survey includes measurements of $70$ Cepheids in the Large Magellanic Cloud. One of the most prominent tensions of the observational as well as theoretical cosmology is the $H_0$ tension, which is the disagreement in $H_0$ values obtained from the local measurements via the SH0ES collaboration, for instance, R19 \cite{R6} report $H_0 = 74.03 \pm 1.42 \: km/s/Mpc $ at a $68$\% CL, or R20 \cite{R7} when incorporating Gaia EDR3 parallax measurements, which give $H_0 = 73.2 \pm 1.3 \: km/s/Mpc $ at the same CL, and that of the Planck \cite{R2} estimate $H_0 = 67.27 \pm 0.60 \: km/s/Mpc$ at $68$\% CL, incorporating the $\Lambda$CDM as the base cosmological model.

\subsection{Type Ia Supernovae (SN Ia): the discovery of accelerating Universe}

The key method for measuring the distance to any stellar object is to approximate its intrinsic brightness and then compare it with the light flux. Such stellar objects with known luminosity are known as standard candles. The intrinsic luminosity $L$ of a nearby emitting object can be estimated by utilizing the apparent brightness $f$ along with the luminosity distance $d_L$ of the stellar object obtained through triangulation via the inverse square law technique as follows,
\begin{equation}\label{e4}
f=\frac{L}{4 \pi d_L^2}\text{.}
\end{equation}
Astronomers invoke logarithmic scales to estimate the luminosity and flux, that are known as absolute magnitude $M$ and apparent magnitude $m$ defined as follows,
\begin{equation}\label{e5}
M=-2.5 log_{10}\left( \frac{L}{L_0} \right)    
\end{equation}
and 
\begin{equation}\label{e6}
m=-2.5 log_{10}\left( \frac{f}{f_0} \right)\text{.}     
\end{equation}
Here, the reference flux is taken to be $f_0 = 2.52 \times 10^{-8} \: Wm^{-2}$, and $L_0$ denotes the reference luminosity of the stellar object when it is kept away at a distance of $10 \: pc$. We define the distance modulus as follows,
\begin{equation}\label{e7}
\mu = m-M = 5 log_{10}\left( \frac{d_L}{10 \: pc} \right)\text{.}   
\end{equation}
This distance modulus is analogous to the inverse square law in a flat space. 

Cepheid variables are luminous, yellow, and massive stars. When found in relatively close clusters, their luminosities can be calculated via distances obtained from earlier steps in the distance ladder, making them ideal candidates for use as standard candles. Supernovae are dramatic stellar explosions whose brightness is comparable to that of an entire galaxy, making them detectable at large cosmic distances. SN Ia occur when a normal star in a binary system transfers mass to a white dwarf, leading to a thermonuclear explosion \cite{R8}. By calibrating the relationship between peak brightness and the time at which this brightness decays, along with the Cepheid variables, one can achieve the cosmic distance scale beyond the range detected by Cepheid variables alone. In 1998, Reiss et al. \cite{R9} first identified the accelerating behavior of the expansion phase of the Universe by investigating $34$ nearby and $16$ distant SN Ia utilizing the Hubble Space Telescope. Shortly thereafter, Perlmutter et al. \cite{R10} analyzed $42$ high-redshift and $18$ nearby supernovae and confirmed the discovery of an accelerating Universe.

After the groundbreaking discovery of the accelerating Universe, surveys of SN Ia have been paid much attention by the research community in the last two decades. Several observational projects initiated in this area, such as the Nearby Supernova Factory (NSF) \cite{R11}, the Sloan Digital Sky Survey (SDSS) \cite{R12}, the Lick Observatory Supernova Search (LOSS) \cite{R13}, the Supernova Legacy Survey (SNLS) \cite{R14}, and the Supernova Cosmology Project (SCP). In 2008, SCP published the Union dataset with $307$ SN Ia samples \cite{R15}. The Union dataset was updated in 2010 to the Union2 \cite{R16}, which included $557$ sample points, and then further revised to the Union2.1 dataset consisting of $580$ SN Ia samples \cite{R17}. Recently, in the year 2022, Brout et al. published the Pantheon+SH0ES dataset, an updation of Pantheon 2018 version of $1048$ data points \cite{R18}, that features $1,701$ light curves from $1,550$ SN Ia in the redshift range $ 0.001 \leq z \leq 2.26$ \cite{R19}.

\subsection{The Cosmic Microwave Background (CMB): an evidence of big bang}

In 1927, George Lemaitre proposed that our Universe emerged from a single massive primeval atom that exploded, leading to the creation of the smaller atoms we see today \cite{R20}. However, he did not use the term Big Bang. Despite Lemaître's proposal, most cosmologists at the time rejected the idea of a dynamic, expanding Universe, favoring the concept of a static and eternal Universe instead. In 1948, George Gamow, Robert Herman, and Ralph Alpher predicted that the Big Bang model would result in a relic background radiation with a tiny temperature. Penzias and Wilson conducted a precise measurement of some excess noise during a trial satellite communication experiment, discovering that it was consistent regardless of time, direction, or season \cite{R21}. As they tried to determine the source of this radiation, a colleague informed them about the Princeton group's interest in detecting cosmic background radiation. This led to the simultaneous publication of two papers in 1965: one by Penzias and Wilson announcing their discovery of the CMB and another by the Princeton group discussing its cosmological importance \cite{R22}.

In the early Universe, radiation and matter were coupled as extremely hot plasma. The matter coalesced into neutral atoms, and the photons decoupled from the cosmic plasma as the Universe cooled along with the cosmic expansion. These free thermal photons persisted as the CMB radiation we observe today. Observations from the Cosmic Background Explorer (COBE) satellite revealed that the CMB radiation follows a perfect blackbody spectrum with a temperature $T_0= 2.725 \pm 0.002 \: K$, confirming the uniform thermal nature of cosmic background radiation \cite{R23}. The discovery of CMB radiation marked a significant milestone in modern science. For many years, a strong debate existed between supporters of the steady-state Universe, according to which the Universe was eternal and unchanging, and proponents of the Big Bang theory of the Universe. This debate was ended when Arno Penzias and Robert Wilson provided evidence of cosmic background radiation, the remnants of the Big Bang that occurred billions of years ago.

\subsection{The galaxy rotation curves: existence of dark matter}

The critical density $\rho_c$ serves as the benchmark to define density scale of the Universe, it is given by
\begin{equation}\label{e8}
\rho_c = \frac{3H^2}{8 \pi G}\text{,}    
\end{equation}
where G denotes the Newton's constant. The current value of critical density is roughly obtained as $\rho_c = 1.88h^2 \times 10^{-26} \:\: kg/m^3$, where $H_0= 100h \:\: km/s/Mpc$. Further, we define the dimensionless density parameter as the following ratio,
\begin{equation}\label{e9}
\Omega = \frac{\rho}{\rho_c}\text{.}
\end{equation}
The present density parameter value $\Omega_m$ is used to quantify the total matter density in the Universe. It's important to understand how this quantity is distributed among the various types of matter content in the Universe. The most straightforward approach is to examine all the stars within a sufficiently large region. Stellar structure theory provides a reliable estimate of the star density parameter, based on its temperature and luminosity, as follows,
\begin{equation}\label{e10}
 \Omega_{star} = \frac{\rho_{star}}{\rho_c} \approx 0.005\text{.} 
\end{equation}
The nucleosynthesis theory provides a lower limit on baryonic matter density $\Omega_B$, indicating that our Universe likely contains significantly more baryonic content other than stars, potentially amounting to more than $4$ percent of the total critical density.

There is substantial evidence suggesting the presence of more matter other than the visible one. In year 1932, Oort identified the excessive amount of hidden matter in our galaxy, and a year later, Zwicky detected the same within the galaxy clusters. The strongest evidence for the existence of unseen dark matter comes from the observed rotation curves of spiral galaxies. These galaxies consist of a disk of dust and stars orbiting around a central core. A galaxy's rotation curve illustrates how the velocity of its matter content changes with distance from the center. The galaxy of mass $M(r)$ related to the velocity $v(r)$ at the radius $r$ via following relation,
\begin{equation}\label{e11}
v(r) = \sqrt{\frac{GM(r)}{r}}\text{.}
\end{equation}
It is evident from the above relation that at a sufficiently large radius, encompassing most of the visible region inside a galaxy, the mass $M(r)$ is expected to remain nearly constant, which implies the rotational velocity $v(r) \longrightarrow 0$ as the square root of 
$r$. But this is not the case. In fact, the rotational velocity $v(r)$ remains constant at a large radius, implying that $M(r)$ is proportional to $r$ for a large radius. Hence, one can easily infer the existence of a halo of unseen dark matter around each galaxy, exhibiting approximately $10$ times mass as compared to that of the visible one. An estimate of this dark matter halo reads as,
\begin{equation}\label{e12}
\Omega_{halo} \approx 0.1  \text{.}
\end{equation}

\subsection{The constituents}

Galaxy clusters, being the most massive gravitationally-bound structures in the Universe, serve as an excellent tool for studying various types of matter. These clusters primarily consist of two visible entities: the stars within the galaxies and the surrounding hot diffuse gas. Interestingly, the proportion of stars to hot gas in these clusters aligns well with the observed star density when compared to the total baryon density predicted by nucleosynthesis. Additionally, the hot gas provides a means to predict the dark matter density in the clusters, as the gravitational pull of the gas by itself alone is insufficient, since the estimated mass of the gas is nearly a tenth fraction of the entire cluster's mass. This suggests that the additional gravitational force is likely provided by dark matter. If this is the case, the density of dark matter would need to be approximately ten times higher than the density of baryonic content predicted by nucleosynthesis. For instance, observations from the Chandra X-ray satellite estimated the following,
\begin{equation}\label{e13}
\frac{\Omega_B}{\Omega_m} = 0.065 h^{-{\frac{3}{2}}}  \approx 0.1 \text{.}
\end{equation}
On utilizing the constrain $0.021 < \Omega_B h^2 < 0.025$ estimated by nucleosynthesis, we can have
\begin{equation}\label{e14}
\Omega_m \approx 0.35 h^{-{\frac{1}{2}}} \approx 0.4 \text{.}
\end{equation}
The analysis clearly shows that dark matter dominates the majority of the matter density. However, the density of dark matter is much less than that of the critical density. The Planck 2018 results \cite{R2}, estimated $\Omega_B \approx 0.049$, whereas $\Omega_m = 0.315 \pm 0.007$.  Several key observational results indicate that the baryonic matter makes up around 5\% of the entire matter-energy content of the Universe, whereas the remaining 95\% is composed of two unseen dark components, first one is the dark matter that makes up around 27\%, while the second one is known as the unknown mysterious force that causes the acceleration of the Universe called dark energy, that is the highest contributing constituent roughly 68\% of the total matter-energy content.

\section{The general relativity}
In the earlier discussion, we ended up with a plethora of observational results along with several challenges emerging in the field of astrophysics and cosmology. We seek a theoretical framework that can effectively describe the observed cosmos and provide consistent solutions to the existing problems. 

\subsection{Newtonian relativity}
Inertial frames refer to coordinate systems where a particle will persistently follow its state of rest or motion at a constant velocity if no outer forces are applied to it, as per Newton's first law. Within such a frame, an observer can identify a parameter $t$, denoted as time, in relation to which the position of a freely moving particle changes at a constant rate. Specifically, the motion of any freely moving particle can be expressed by its coordinates as a function of time $t$ i.e. $(x_1(t), x_2(t), x_3(t))$, having no acceleration. A key insight provided by Newton and Galileo is that the most straightforward way to describe the physical world (i.e. laws of physics) is through the use of inertial reference frames. The coordinate transformation from one inertial frame to another is achieved by the Galilean transformation. In the case of the Galilean transformation, our focus is primarily on transforming coordinates between inertial frames that share the same orientation. This type of transformation is referred to as a (Galilean) boost, defined as follows,
\begin{equation}\label{e15}
 x_j \longrightarrow   x'_j = x_j - v_j t  \:\:\: \text{and} \:\:\:  t \longrightarrow t' = t \text{,}
\end{equation}
where $v_j$ are the constant velocity components between two inertial frames. Moreover, we obtain the following addition rule for velocity,
\begin{equation}\label{e16}
u_j  \longrightarrow   u'_j = u_j - v_j \text{,}
\end{equation}
where $u_j= \frac{dx_j}{dt}$ and $u'_j = \frac{dx'_j}{dt}$.

In classical mechanics, when the relative velocity between inertial frames is negligible to that of the speed of light $c$, these frames are connected through Galilean transformations. Consequently, by Newtonian relativity, we interpret that the Galilean transformations make the laws of Newtonian mechanics covariant.

\subsection{Special relativity}
It can be demonstrated that Maxwell's equations of electromagnetism are not covariant under Galilean transformations. This is because the constant nature of the speed of light $c$ does not follow the velocity addition rule of the Galilean transformations. Its equations are valid in the rest frame of the ether (the absolute rest frame), and this fact was widely accepted by many physicists. However, the second interpretation was ultimately proven correct. The Michelson-Morley experiment \cite{R24} revealed that the speed of light $c$ is identical in all moving coordinate systems. This effectively brought back the concept of absolute space (the ether frame) into physics. 

Utilizing the previous analysis by Lorentz, Poincare independently discovered that Maxwell's equations of electromagnetism were covariant in a new set of boost transformations called the Lorentz transformation. These equations retain their form when both the position and time variables are transformed between the two moving frames with a velocity $\Vec{v} = v \hat{x}$ and have the following transformation rule,
\begin{equation}\label{e17}
x'= \tau (x-vt), \:\: y'=y, \:\: z'=z, \:\: \text{and} \:\: t'= \tau \left( t - \frac{v}{c^2} x \right)\text{,}
\end{equation}
where $\tau = \frac{1}{\sqrt{1-(\frac{v}{c^2})^2}}$. In the non-relativistic limit i.e. $v <<< c$, the Lorentz transformation reduces to the Galilean transformation with a factor of $\tau$. We obtain the velocity addition rule for the Lorentz transformation as follows,
\begin{equation}\label{e18}
u'_x = \frac{u_x - v}{1-\frac{u_x v}{c^2}}\text{,} 
\end{equation}
where $u'_x = \frac{dx'}{dt'}$ and $u_x =\frac{dx}{dt}$. In particular, for the case $u_x=c$, one can obtain $u'_x =c$ using the above velocity addition rule. This shows that the Lorentz transformation is consistent with the constant nature of $c$. The Poincare and Lorentz independently identified the invariance of Maxwell's equations under Lorentz transformations. However, it was Einstein who proposed the key principles and emphasized the foundation of this symmetry, called principles of special relativity \cite{R25}. The first principle is the principle of relativity, which states that the laws of physics are identical in all inertial reference frames, meaning no experiment can detect the absolute motion of an inertial frame i.e. there is no ether. The second principle asserts that the speed of light remains constant in every reference frame, regardless of the observer's motion. Einstein indicated that a new approach to kinematics was necessary to fully accommodate the new symmetry. Specifically, distinct time labels were needed for different inertial frames when signal transmission speed was finite. Before this, the absolute nature of time was widely accepted. However, Einstein stressed that time itself is fundamentally defined by the concept of simultaneity. In order to agree on the simultaneity of two events i.e. $\Delta t = 0$ or any time interval $\Delta t$, for different observers in two different frames, their clocks must be synchronized.

The kinematics of special relativity were effectively described by a geometric formalism, where the time coordinate was treated equally alongside spatial coordinates. In this context, the notion of flat spacetime was introduced, which was suitable for special relativity physics and referred to as Minkowski spacetime since it was originally developed by Hermann Minkowski. An event is described as a specific point in both space and time, uniquely identified by the coordinates $(ct, x, y, z)$. We then define the spacetime interval that separates two events as follows,
\begin{equation}\label{e19}
s^2 = -(c \Delta t)^2 + (\Delta x)^2 + (\Delta y)^2 + (\Delta z)^2\text{.}  
\end{equation}
In the matrix form, we can have,
\begin{equation}\label{e20}
s^2 = \eta_{\mu\nu} \Delta x^\mu \Delta x^\nu \text{,}
\end{equation}
where $\eta_{\mu\nu} = diag(-1,1,1,1)$. It is important to note that we apply the summation convention, where indices that appear as both subscripts and superscripts are automatically summed over. Note that the translation $x^{\mu'} = x^\mu + a^\mu$ keeps the spacetime interval to be invariant. Now, consider the transformation 
\begin{equation}\label{e21}
x'= Px \text{,}
\end{equation}
where $P$ represents a matrix with spacetime independent entries. Then the invariance of spacetime interval gives,
\begin{equation}\label{e22}
 s^2 = (\Delta x')^T\eta (\Delta x')  =  (P \Delta x)^T\eta (P \Delta x) = (\Delta x)^T P^T \eta P (\Delta x) =  (\Delta x)^T\eta (\Delta x) \implies P^T \eta P  = \eta \text{.}
\end{equation}
Such matrices $P$ transform spacetime interval in an invariant way, called Lorentz transformation. A set of all such matrices forms a nonabelian group with respect to operation matrix multiplication, called the Lorentz group. Moreover, the set of Lorentz transformations, along with the translations, also forms a non-abelian group, known as the Poincare group.

Inspired from the invariance property of the spacetime interval, we define its infinitesimal version describing the geometry of the four-dimensional flat spacetime, known as Minkowski metric, as follows,
\begin{equation}\label{e23}
ds^2 = -(cdt)^2 + dx^2 + dy^2 + dz^2 \text{.}
\end{equation}
The negative sign in the line element or in the spacetime interval is a novel feature of this geometry. Now, we describe the physical essence of the assumed line element in the following discussion,
\begin{itemize}
    \item \textbf{Spacelike separated:} We say two events are spacelike separated when $(\Delta s)^2 > 0$. For instance, consider the case $\Delta t =0$ with atleast one of the spatial interval is non-zero, say $\Delta x \neq 0$, then it is evident that two events are separated in space but occur at the same time.
    \item \textbf{Timelike separated:} We say two events are timelike separated when $(\Delta s)^2 < 0$. For instance, consider the case $\Delta t \neq 0$ but $\Delta x = \Delta y = \Delta z = 0$, then two events are occur at the same position but at different times.
    \item \textbf{Lightlike separated:} We say two events are lightlike or null separated when $(\Delta s)^2 = 0$. For instance, consider the case $\Delta x = c \Delta t$ but $\Delta y = \Delta z = 0$, then two events are connected via light rays with speed $c$.
\end{itemize}

\subsection{Curved spacetime: gravity is geometry}
Gravity is a manifestation of the spacetime curvature. When we describe a physical phenomenon in geometric terms, it means that the results of physical measurements are directly related to the spacetime geometry. For instance, the distances measured on the surface of a sphere vary with direction, such as the north-south direction exhibits a constant distance measurement corresponding to a fixed angle $\phi$, whereas the east-west distances between two points having the azimuthal angle $\Delta \phi$ decrease as the points move further from the equator. Note that one can obtain the following relation between the rate of signals emitted from a point $A$ and received at point $B$ and having the gravitational potential $\Phi_A$ and $\Phi_B$ respectively \cite{B1},
\begin{equation}\label{e24}
\text{rate of signals observed at B}\:\: = \left( 1+\frac{\Phi_A-\Phi_B}{c^2} \right) \times \:\: \text{rate of signals emitted at A} \text{.}
\end{equation}
Thus, from the above relation, it is evident that the rate at which the signals are observed at $B$ is lesser than the rate at the emitter $A$ when we kept $B$ at higher gravitational potential as compared to that of $A$. The description of the variation in rates of signals emitted and observed can be attributed to the impact of gravity on how clocks run. Without a gravitational field, two stationary clocks in an inertial frame within flat spacetime will both measure time in sync with that frame. However, when a gravitational field is present, the clocks tick at a rate that differs by a factor of $\left( 1+\frac{\Phi_A-\Phi_B}{c^2} \right)$ compared to their rates in the absence of gravity, keeping the spacetime remains flat. Clocks run slower if $(\Phi_A-\Phi_B) < 0$, whereas it run faster if $(\Phi_A-\Phi_B) > 0$.
The description of time intervals measured by different clocks in a gravitational field within a flat spacetime reveals an underlying geometry that cannot be directly observed. It is admissible that clocks accurately measure timelike distances within spacetime, but actually, the spacetime geometry is curved, as similar to the case we obtain differences in east-west distances between two points having the azimuthal angle $\Delta \phi$ away from the equator, are actually attributed to the curved geometry of the spherical surface of the earth.

GR can be understood as a gravity theory in a curved spacetime. A famous saying, ``Spacetime tells matter how to move and matter tells spacetime how to curve" by John A. Wheeler briefly describes gravity as a geometric theory of curved spacetime. The principle of GR follows the general covariance, which states that the laws of physics are invariant under general coordinate transformations, including non-inertial frames i.e. accelerating reference frames. This general principle implies a democracy for choosing different coordinate systems, along with the most insightful perspective, is to treat them as local Lorentz transformations, and in the limit of low velocities, these reduce to the Galilean transformations.

\subsection{Metric: the line element in spacetime geometry}
To represent the geometry of curved spacetime, one can utilize a four component coordinate system, $x^\mu$, to identify points. The following line element $ds^2$ defines the distance between two points situated apart with the coordinate intervals $dx^\mu$, 
\begin{equation}\label{e25}
 ds^2 = g_{\mu\nu}(x) dx^\mu dx^\nu \text{,}
\end{equation}
where $g_{\mu\nu}(x)$ is a coordinate dependent, $4 \times 4$ symmetric matrix, known as the metric or a metric tensor. When the metric tensor is positive definite, the corresponding line element is referred to as an Euclidean or Riemannian metric. However, if the time component of the metric tensor contains a negative sign, whereas the space components exhibit a positive sign, it is called a pseudo-Riemannian or Lorentzian metric. In theoretical physics, spacetime geometry is typically represented using a Lorentzian metric. It is important to note that no coordinate transformation can reduce $g_{\mu\nu}(x)$ to $\eta_{\mu\nu}$ across the entire curved spacetime, meaning $g_{\mu\nu}(x)= \eta_{\mu\nu}$  cannot hold globally. However, corresponding to each point $P$ of the spacetime geometry, one can identify the coordinates $x^{\mu'}$ such that,
\begin{equation}\label{e26}
g_{\mu\nu}(x^{\mu'}_P)= \eta_{\mu\nu} = diag(-1,1,1,1) \:\: \text{and} \:\: \left( \frac{\partial g_{\mu\nu}}{\partial x^{\mu'}} \right)_{x=x_P}= 0 \text{.}
\end{equation}
Such a coordinate system is recognized as the local inertial frame at the point $P$. As $g_{\mu\nu}(x)$ is a symmetric matrix, one can always find a transformation $x^{\mu}_P \longrightarrow x^{\mu'}_P$ such that $g_{\mu\nu}(x^{\mu'}_P)$ is a diagonal matrix and that can further transform to the Minkowski's metric. This is an inertial frame within flat spacetime, but only in the vicinity of the point $P$, which is why it is referred to as a local inertial frame. Einstein’s crucial insight is that the equivalence principle enables us to eliminate the effects of gravity within a sufficiently small region. In this specific region, due to the absence of gravity, GR mirrors the special relativity and follows the same lightcone structure locally at each point, such as, $ds^2 > 0$ being spacelike, $ds^2 < 0$ timelike, and  $ds^2 = 0$ representing lightlike paths.

\subsection{The Levi-Civita connection}
Recall that a manifold $M$ represents a geometry of curved spacetime having a complex topology, but within local regions, it resembles $\mathbb{R}^n$. Consider a subset $S \subset M$ and a function $\phi: S \longrightarrow \mathbb{R}^n$ such that $\phi$ is one-one and $\phi(S)$ is an open subset of $\mathbb{R}^n$. Such a pair of $(S, \phi)$ is called a coordinate system or a chart. A collection $\{ (S_i,\phi_i)\}_{i\in I}$ of coordinate systems (charts) is known as a smooth atlas if the collection $\{S_i\}_{i\in I}$ forms an open cover for $M$ and all charts are smoothly sewn together i.e. whenever $S_i \cap S_j \neq \emptyset$ then the function $\phi_i \circ \phi_j^{-1}: \phi_j(S_i \cap S_j) \longrightarrow \phi_i(S_i \cap S_j)$ must be onto, provided all these functions are infinitely differentiable i.e. $C^\infty$ maps. Such a set $M$ along with a maximal smooth atlas is known as a smooth $n$-dimensional manifold. This definition can efficiently express the idea of a set that resembles $\mathbb{R}^n$ locally \cite{B2}.

Consider a vector $A^\mu$ placed at a point $P(x^\mu)$. Now, we shift the vector $A^\mu$ to a nearby point $Q(x^\mu+\delta x^\mu)$ in a way that it is translated parallel to itself that is well known as parallel transport. Let us assume that $dA^\mu$ denotes the coordinate difference and $\delta A^\mu$ represents the changes in the component of $A^\mu$ in such a parallel transport, then we have
\begin{equation}\label{e27}
dA^\mu = A^\mu(x^\mu+\delta x^\mu) - A^\mu(x^\mu) \approx  A^\mu(x^\mu) + \delta x^\eta \frac{\partial A^\mu}{\partial x^\eta}  - A^\mu(x^\mu) = \delta x^\eta \frac{\partial A^\mu}{\partial x^\eta}
\end{equation}
and 
\begin{equation}\label{e28}
\delta A^\mu = - \Gamma^\mu_{\sigma \eta} A^\sigma \delta x^\eta \text{.} 
\end{equation}
The actual change in vector $A^\mu$ after a parallel transport is $dA^\mu - \delta A^\mu $. We define the covariant derivative of $A^\mu$ to be the rate of change of $A^\mu$ in the parallel transport when $\delta x^\eta \longrightarrow 0$ as follows,
\begin{equation}\label{e29}
 \nabla_\eta A^\mu = \frac{\partial A^\mu}{\partial x^\eta} + \Gamma^\mu_{\sigma \eta} A^\sigma = \partial_\eta A^\mu + \Gamma^\mu_{\sigma \eta} A^\sigma  \text{.}
\end{equation}
The quantity $\Gamma^\mu_{\sigma \eta}$ is  known as the Christoffel symbols or Levi-Civita connection or an affine connection of the curved spacetime geometry. Note that an affine connection satisfying the symmetric condition i.e. $\Gamma^\mu_{\sigma \eta} = \Gamma^\mu_{\eta \sigma }$ and the metricity (or metric compatibility) condition i.e. $\nabla_\eta g_{\mu\nu} = 0$, is called a Riemannian connection and the underlying geometry is known as the Reimannain geometry. On utilizing these two conditions, one can obtain the Levi-Civita connection in terms of a metric tensor as follows,
\begin{equation}\label{e30}
 \Gamma_{\mu \sigma \eta} = \frac{1}{2} \left[\partial_\mu g_{\sigma \eta} + \partial_\sigma g_{\eta \mu} - \partial_\eta g_{\mu \sigma} \right]  \text{.}
\end{equation}

\subsection{Field equation of the GR}
Using the previously discussed affine connection, one can measure the curvature of the spacetime geometry with the help of the following quantity known as the Riemann curvature tensor,
\begin{equation}\label{e31}
R^{\rho}_{\sigma\mu\nu}=\partial_{\mu}\Gamma^{\rho}_{\nu\sigma}-\partial_{\nu}\Gamma^{\rho}_{\mu\sigma}+\Gamma^{\rho}_{\mu\lambda} \Gamma^{\lambda}_{\nu\sigma}-\Gamma^{\rho}_{\nu\lambda} \Gamma^{\lambda}_{\mu\sigma} \text{.}
\end{equation}
Note that, the Reimann curvature tensor satisfy the following Bianchi identity,
\begin{equation}\label{e32}
\nabla_\alpha R^{\rho}_{\sigma\mu\nu} + \nabla_\mu R^{\rho}_{\sigma\nu\alpha} + \nabla_\nu R^{\rho}_{\sigma\alpha\mu} = 0 \text{.}
\end{equation}
A Riemannian space is said to be a flat space or curvature free if the Riemann curvature tensor vanishes identically i.e. $R^{\rho}_{\sigma\mu\nu}=0$. A contraction of the Riemann curvature tensor is known as the Ricci tensor given as follows,
\begin{equation}\label{e33}
R_{\sigma\mu} = R^{\nu}_{\sigma\mu\nu} \text{.}
\end{equation}
Its further contraction give rise to the Ricci curvature scalar as follows,
\begin{equation}\label{e34}
R = g^{\sigma\mu} R_{\sigma\mu} \text{.}
\end{equation}
Now we define an important tensor that encodes the geometry of the spacetime known as the Einstein tensor, reads as,
\begin{equation}\label{e35}
G_{\mu\nu} \equiv R_{\mu\nu}-\frac{1}{2}g_{\mu\nu}R \text{.} 
\end{equation}
Utilizing the Bianchi identity, one can obtain the following vanishing covariant divergence of the Einstein tensor,
\begin{equation}\label{e36}
\nabla^\mu G_{\mu\nu}  = 0 \text{.}
\end{equation}

From the perspective of relativity, mass is simply a form of energy (known as rest energy). As energy and momentum are interchangeable, they can be converted into one another depending on the observer's reference frame. In the context of relativity, we express the mass density $\rho$ in a more generalized fashion known as the energy–momentum tensor, denoted by $\mathcal{T_{\mu\nu}}$. It is a symmetric tensor, and hence having only $10$ independent components, out of which the component $T_{00}$ represents the matter-energy density $\rho c^2$, whereas the components $T_{ii}$ denotes the pressure. The remaining off-diagonal components represent the energy and momentum flux, momentum density, and shear stress.

Note that, the gravity field equation for the Newtonian case is given by,
\begin{equation}\label{e37}
\nabla^2 \Phi = 4\pi G_N \rho \text{.}  
\end{equation}
In other words, the second derivative $\nabla^2$ of the gravitational potential $\Phi$ is directly proportional to the mass density $\rho$ i.e. $\nabla^2 \Phi \propto \rho$. We have seen that the Levi-Civita connection $\Gamma$ consists of the first derivation of the metric, and hence the Reimann curvature tensor, which is equivalent to the quantity $d\Gamma + \Gamma \Gamma$, exhibits a second derivative of the metric tensor. This suggests the relativistic extension of the Newtonian field equation having the same form i.e. the left-hand side being the Einstein tensor $G_{\mu\nu}$ that includes the second derivative of the metric $g_{\mu\nu}$, a relativistic gravitational potential, and the right hand side is the energy–momentum tensor $\mathcal{T_{\mu\nu}}$, which generalizing the matter field. Thus, we obtained the Einstein's field equation of the GR as follows \cite{R26},
\begin{equation}\label{e38}
G_{\mu\nu} \propto \mathcal{T}_{\mu\nu} \:\: \text{i.e.} \:\: G_{\mu\nu} = \kappa \mathcal{T}_{\mu\nu} \text{,}
\end{equation}
where $\kappa = \frac{8\pi G}{c^4}$ is a proportionality constant, also known as coupling constant. One can also formulate the above Einstein's field equation of the GR by varying the following Einstein-Hilbert action through variation principle,
\begin{equation}\label{e39}
S=\frac{1}{2\kappa}\int R \sqrt{-g} d^4x +\int L_m \sqrt{-g} d^4x \text{.}
\end{equation}
Here, $\kappa = \frac{8\pi G}{c^4}$, $L_m$ is the matter Lagrangian density, and $d^4x\sqrt{-g}$ is the coordinate invariant volume element in a four dimensional spacetime manifold with $g=det(g_{\mu\nu})$.

\section{$\Lambda$CDM cosmology}
In the previous section, we describe the physics of GR and its mathematical foundations. Now, we will see that how these formulations can describe the various cosmological observations.

\subsection{Comoving distances}
The cosmological principle presents the Universe as a cosmic fluid, wherein galaxies act as the fundamental particles. A fluid element represents a volume that encompasses numerous galaxies but remains very small in comparison to the entire Universe. Therefore, the movement of a cosmic fluid element reflects the smear motion of the galaxies within it. This motivates us to use a special coordinate system known as the comoving coordinate system, which evolves along with the Universe's expansion. The relation between comoving distance $x$ and the actual physical distance $r$ is given by,
\begin{equation}\label{e40}
 r = a(t) x \text{,}  
\end{equation}
where $a(t)$ is the cosmic expansion factor known as the scale factor. Now, consider a galaxy with recessional velocity $v$, then 
\begin{equation}\label{e41}
v= \frac{dr}{dt}  = \dot{r} = \dot{a} x = \frac{\dot{a}}{a} a x = \frac{\dot{a}}{a} r \text{.}
\end{equation}
According to the Hubble's law of expansion, we have $v= H r$, hence on comparing we obtain the following,
\begin{equation}\label{e42}
 H = \frac{\dot{a}}{a} \text{.}   
\end{equation}
This expression we call the Hubble parameter. 

\subsection{FLRW Universe}
The terms open, closed, and flat are traditionally used to differentiate between the three possible isotropic and homogeneous spatial geometries. In the case of flat space, spatial curvature is zero at every point. The closed scenario is characterized by a positive constant spatial curvature, whereas the open Universe exhibits a constant negative spatial curvature. The following line element represents these three different geometries of an isotropic and homogeneous spatial background,
\begin{equation}\label{e43}
ds^2=-c^2dt^2+a^2(t)\left[d\xi^2+V(\xi)^2(d\theta^2+sin^2\theta d\phi^2)\right] \text{,}
\end{equation}
where 
\begin{equation}\label{e44}
V(\xi)=
\begin{dcases}
sin\xi ; &  closed \\
\xi ;&  flat\\
sinh\xi ; &  open 
\end{dcases} \text{.}
\end{equation}
One can rewrite the above metric in a more unified form as follows,
\begin{equation}\label{e45}
ds^2=-dt^2+a^2(t)\left[\frac{dr^2}{1-k r^2}+r^2(d\theta^2+sin^2\theta d\phi^2)\right]\text{.}
\end{equation}
The above line element characterizes the spatially isotropic and homogeneous Universe's evolution, as the scale factor $a(t)$ varies. This line element is commonly known as the FLRW (Friedmann-Lemaitre-Robertson-Walker) metric \cite{R27,R28}. Note that, from now on, we will work in units for which $c=1$. The quantity $k$ represents the spatial curvature, such as the case $k = -1$ represents the open Universe, the case $k = +1$ represents the closed Universe, whereas the case $k = 0$ represents the spatially flat Universe.

\subsection{History of $\Lambda$: an evolution of cosmological constant}
Now, on using the line element \eqref{e45}, we obtain the components of the Einstein's equation \eqref{e38} as follows,
\begin{equation}\label{e46}
\left(\frac{\dot{a}}{a}\right)^2 + \frac{k}{a^2}=\frac{8\pi G}{3}\rho 
\end{equation}
and
\begin{equation}\label{e47}
2\frac{\ddot{a}}{a} + \left(\frac{\dot{a}}{a}\right)^2 + \frac{k}{a^2} = - 8\pi G p \text{.}
\end{equation}
On combining above two equations, we obtain the following acceleration equation,
\begin{equation}\label{e48}
\frac{\ddot{a}}{a} = - \frac{4\pi G}{3}(\rho + 3p) \text{.} 
\end{equation}
At the time of the formulation of GR, the scientific community strongly believed in the static Universe i.e. the scale factor $a(t)$ must be constant. As a consequence of constant scale factor $a(t)$ and the above set of field equations, one can obtain the following results,
\begin{equation}\label{e49}
\rho=-3p= \frac{3k}{8\pi G a^2} \text{.}   
\end{equation}
Note that the positivity of energy density $\rho$ implies that the pressure component exhibits negative value, or if we assume $p=0$, then we obtain $\rho=0$. In both cases, the obtained results are absurd. To bypass this issue, later, Einstein modified his field equation by adding a constant term $\Lambda$ called the cosmological constant. Thus, the new set of equations that are compatible with the static Universe read as follows,
\begin{equation}\label{e50}
\left(\frac{\dot{a}}{a}\right)^2 + \frac{k}{a^2}=\frac{8\pi G}{3}\rho + \frac{\Lambda}{3}
\end{equation}
and
\begin{equation}\label{e51}
2\frac{\ddot{a}}{a} + \left(\frac{\dot{a}}{a}\right)^2 + \frac{k}{a^2} = - 8\pi G p + \Lambda \text{.}
\end{equation}
Note that, in the year 1917, Einstein incorporated this cosmological constant $\Lambda$ into his field equation to achieve a static Universe model \cite{R29}. However, following the discovery of cosmic expansion, he removed $\Lambda$ in the year 1931 \cite{R30}. Later, in 1967, Zel'dovich revived the idea of the cosmological constant $\Lambda$, considering vacuum fluctuations \cite{R31}. In the year 1987, Weinberg depicted a tiny non-vanishing cosmological constant $\Lambda$ \cite{R32}. Finally, in the year 1998, the discovery of the accelerating expansion of the Universe brought back the cosmological constant $\Lambda$ into focus as a potential candidate for dark energy, which is driving this accelerated expansion.

\subsection{The standard cosmological model}
Recall that, an isotropic and homogeneous Universe having cosmic matter as perfect fluid characterize by the following energy-momentum tensor,
\begin{equation}\label{e52}
\mathcal{T}_{\mu\nu}=(\rho+p)u_{\mu}u_{\nu}+ p g_{\mu\nu} \text{,}
\end{equation}
where $u_\mu = (1,0,0,0)$ represents the four velocity vector of the cosmic fluid in a comoving coordinate. Note that, the vanishing covariant divergence of the energy-momentum tensor i.e. $\nabla^\mu \mathcal{T}_{\mu\nu} = 0 $ implies
\begin{equation}\label{e53}
\dot{\rho}+3 \frac{\dot{a}}{a}(\rho+p)=0 \text{.}  
\end{equation}
On employing the barotropic equation of state $p=\omega \rho$ in the above equation, we can have
\begin{equation}\label{e54}
\rho \propto a^{-3(1+\omega)} \text{.} 
\end{equation}
Note that, for different values of equation of state (EoS) parameter $\omega$, one can obtained the various expressions for the energy densities that corresponds to different cosmological epochs, as follows,
\begin{itemize}
\item $\omega=1/3 \implies \rho_r = \rho_{r_0}a^{-4}$ represents the radiation phase,
\item $\omega=0 \implies \rho_m = \rho_{m_0}a^{-3}$ represents the matter phase,
\item $\omega=-1 \implies \rho_\Lambda = constant$ representing cosmological constant case as dark energy.
\end{itemize}
Now on utilizing the equation \eqref{e50}, along with the assumption $\rho = \rho_m + \rho_r$ and $ \rho_\Lambda = \frac{\Lambda}{8 \pi G}$, we obtain the following expression,
\begin{equation}\label{e55}
 H^2 = \frac{8 \pi G}{3} \left[ \rho_{m_0}a^{-3} + \rho_{r_0}a^{-4}  + \rho_{\Lambda} \right] - \frac{k}{a^2} \text{.} 
\end{equation}
Here, $\rho_m = \rho_b + \rho_{cdm}$, where $\rho_b$ denotes the density of ordinary (baryonic) matter and $\rho_{cdm}$ represents the cold (non-relativistic) dark matter density. We define the dimensionless density parameter for the various component as follows,
\begin{equation}\label{e56}
\Omega_{m_0} = \frac{\rho_{m_0}}{\rho_{crit_0}}, \:\: \Omega_{r_0} = \frac{\rho_{r_0}}{\rho_{crit_0}},  \:\: \Omega_{\Lambda} =\Omega_{\Lambda_0} = \frac{\rho_{\Lambda}}{\rho_{crit_0}}=\frac{\Lambda}{3H_0^2}, \:\: \text{and} \:\: \Omega_{k_0} = -\frac{k}{H_0^2} \text{,} 
\end{equation}
where $a_0=1$ (conventional assumption) and $\rho_{crit_0}=\frac{3H_0^2}{8 \pi G}$. Now, on utilizing the above setting along with the scale factor-redshift relation $a^{-1} = 1+z$, the equation \eqref{e55} becomes,
\begin{equation}\label{e57}
H^2 = H_0^2  \left[ \Omega_{m_0}(1+z)^{3} + \Omega_{r_0}(1+z)^{4}  + \Omega_{\Lambda_0} + \Omega_{k_0}(1+z)^2 \right] \text{.} 
\end{equation}
The model discussed above is widely known as the standard cosmological model or the $\Lambda$CDM model. The Planck 2018 results \cite{R2} estimated the free parameter constraints of this standard model and predicted the values $H_0 = 67.4 \pm 0.5 \: km/s/Mpc$, $\Omega_{m_0}=0.315 \pm 0.007$, and $\Omega_{k_0}= 0.001 \pm 0.002$ favoring the spatial curvature of the Universe to be flat.

\subsection{Problems persistent with the $\Lambda$CDM}
In cosmology, the vacuum energy and the cosmological constant are often used interchangeably. However, within the realm of quantum field theory, the term vacuum energy carries more significant implications. One can acquire the zero-point (vacuum) energy density utilizing the quantum theory as $\rho_\Lambda(QFT) = 3.7873 \times 10^{73} \:\: GeV^{-4}$, while utilizing the cosmological observations, we have $\rho_\Lambda = \Omega_{\Lambda_0} \rho_{crit_0} = 3.7151  \times 10^{-47} \:\: GeV^{4}$, and hence we end up with $\frac{\rho_\Lambda(QFT)}{\rho_\Lambda}= 1.01194 \times 10^{120}$ a huge discrepancy. This problem is widely known as the cosmological constant problem. Also note that the observational findings of the Planck's base-$\Lambda$CDM results align well with findings from BAO, SN Ia, and certain galaxy lensing studies. However, there is a slight discrepancy when compared to the Dark Energy Survey’s results, which involves galaxy clustering. Additionally, there is a notable $3.6 \: \sigma$ tension with local measurements of the Hubble constant, which tends to favor a higher value \cite{R2}.

The $\Lambda$CDM model is considered the standard model of cosmology due to its excellent agreement with most observational findings and its computational simplicity. The $\Lambda$CDM model based on three pillar assumption as follows \cite{R33},
\begin{itemize}
    \item A slow-rolling and minimally coupled scalar field describe the inflationary scenario.
    \item The dark matter is a pressureless fluid composed of cold particles, characterized by non-colliding low momentum particles.
    \item The cosmological constant $\Lambda$ is the dark energy candidate.
\end{itemize}
The $\Lambda$CDM model lacks the robust theoretical foundations needed to fully engage with fundamental physics laws. Therefore, it is important to avoid becoming overly attached to the model. As observational study becomes more precise and abundant, departures from the $\Lambda$CDM model might expected. For instance, several theoretical unanswered questions motivate us to go beyond the standard cosmological model, such as,
\begin{itemize}
    \item Why are the densities of dark matter and dark energy observed to be of same order$?$
    \item Are inflation and dark energy linked together$?$
    \item What is the solution of big bang singularity$?$
    \item Is quantum gravity or a unified theory that combines GR and quantum field theory necessary to fully complete the standard model$?$
    \item What is the origin of $H_0$, $S_8$, and $f_{\sigma_8}$ tension$?$
\end{itemize}
Given the above persistent issues with the standard $\Lambda$CDM model, it is not reasonable to consider the $\Lambda$CDM cosmology as the ultimate explanation of the Universe that can resolve all long standing issues. With the increasing observational sensitivity, it is crucial to find a suitable theoretical framework that can address all issues.

\section{Modified gravity: dark energy without dark energy}
The key motivation for the modified gravity scenario is to find suitable alternatives for the main three missing pieces of the standard model of cosmology, particularly inflation, dark matter, and dark energy. There are various methods to modify the GR, which can be categorized into two ways. One approach involves introducing new fields, while the other focuses on modification of the geometric framework. In this investigation, we will focus on the geometrical modification to GR, and hence, from now on, by the term modified gravity, we mean the geometrical modification. Notably, some modified gravity models have shown promise in addressing the $H_0$ tension, especially through late-time solutions. Modified gravity models are effective in studying late time epochs, particularly in producing late time acceleration without the need for a dark energy component \cite{R34}.

\subsection{Geodesics and autoparallel}
Einstein's GR is based on a four-dimensional Lorentzian manifold, characterized by a metric potential $g_{\mu\nu}$ and a connection $\Gamma$ that defines the covariant derivative $\nabla$. The connection $\Gamma$ is torsion free and satisfies the metricity (metric-compatiblity) condition, leading to the unique determination of the connection coefficients as the components of the Christoffel symbols $\Gamma^\lambda_{\mu\nu}$. A natural extension of this framework involves relaxing these conditions, allowing the affine connection $\hat{\Gamma}$ to be neither torsion-free nor metric-compatible. A manifold endowed with an affine connection $\hat{\Gamma}$ and a metric $g_{\mu\nu}$ produces two different curves, first one is geodesics, and the second is autoparallels. Geodesics are the curves with the possible shortest path between two fixed points on the manifold, while autoparallels describe the possible straightest curves connecting two points. Consider a curve $C$ that describe by the local coordinates $x^\mu(\tau)$ having the tangent vectors $T^\mu = \frac{dx^\mu}{d\tau}$ and $\tau$ is the affine parameter. Now consider the following functional,
\begin{equation}\label{e58}
s= \int^{\tau_2}_{\tau_1} \sqrt{g_{\mu\nu}\frac{dx^\mu}{d\tau}\frac{dx^\nu}{d\tau}} d\tau \text{.} 
\end{equation}
One can obtain the following well known geodesic equation by extremising the above functional,
\begin{equation}\label{e59}
\frac{d^2x^\mu}{d\tau^2} + \Gamma^\mu_{\alpha\beta} \frac{dx^\alpha}{d\tau}\frac{dx^\beta}{d\tau} = 0 \:\: \text{or} \:\: \frac{dT^\mu}{d\tau} + \Gamma^\mu_{\alpha\beta} T^\alpha T^\beta = 0 \text{.} 
\end{equation}
Note that, in the special relativistic limit, the above geodesic equation becomes,
\begin{equation}\label{e60}
\frac{d^2x^\mu}{d\tau^2} = 0 \text{.} 
\end{equation}
Thus, the test particle exhibits a constant velocity when there is no external force applied or in the absence of gravity. We know that, in a parallel transport the tangent vectors along the curve are transported in such a way that they remain parallel to themselves, and hence the vector remain parallel as possible along the curve, making it autoparallel. The autoparallel equation of a vector $V^\alpha$ that is parallely transported along the curve $C$ with tangent $T^\alpha$, is given by
\begin{equation}\label{e61}
\frac{dT^\mu}{d\tau} + \hat{\Gamma}^\mu_{\alpha\beta} T^\alpha T^\beta = 0 \text{.}   
\end{equation}
Observe that, the equations \eqref{e59} and \eqref{e61} have the same form but differ in the connection appeared. Also, it is interesting to note that on exchanging the tangent vectors $T^\alpha$ and $T^\beta$, the equation \eqref{e61} depends on the symmetric part of connection i.e. $\hat{\Gamma}^\mu_{(\alpha\beta)}$. One can also observe that
\begin{equation}\label{e62}
\hat{\Gamma}^\mu_{(\alpha\beta)} \neq \Gamma^\mu_{\alpha\beta} \text{.} 
\end{equation}
Thus, the symmetric part of the general affine connection $\hat{\Gamma}^\mu_{(\alpha\beta)}$ is not identical to the Levi-Civita connection $\Gamma$. Moreover, it contains the Levi-Civita components along with torsion and non-metricity contributions. The geodesic and autoparallel will coincide when the affine connection $\hat{\Gamma}$ and Christoffel symbol $\Gamma$ differs by a totally skew-symmetric piece. Hence, GR is a special scenario where the autoparallels and geodesics coincide since $\hat{\Gamma}=\Gamma$, but in the extended framework, where torsion and non-metricity contribute in $\hat{\Gamma}$, this is not the case.

\subsection{Geometric trinity}
It is well known that an affine connection $\hat{\Gamma}$ and a metric structure $g$ are the building blocks for a spacetime manifold. Such a manifold is represented by $(M,g, \hat{\Gamma})$. The metric component describes distances, inner products, and the mappings between contravariant and covariant tensor fields, whereas an affine connection characterizes the parallel transport of tensor fields with the help of covariant differentiation, enabling the comparison of vectors located in different vector spaces. In a general affine connection $\hat{\Gamma}$, both non-metricity and torsion components contribute, and hence it is neither metric-compatible nor symmetric. This spacetime manifold structure equipped with a metric and a non-metric and non-symmetric affine connection is referred to as non-Riemannian geometry \cite{R35}. We define the curvature tensor (Reimann tensor) with respect to the generic affine connection $\hat{\Gamma}$ as follows,
\begin{equation}\label{e63}
\hat{R}^\lambda_{\gamma\alpha\beta} =  2\partial_{[\alpha}\hat{\Gamma}^\lambda_{|\gamma|\beta]} + 2\hat{\Gamma}^\lambda_{\rho[\alpha}\hat{\Gamma}^\rho_{|\gamma|\beta]} \text{,} 
\end{equation}
where the vertical bar denotes the left out index in the anti-symmetrization. Now, we define the torsion tensor which is anti-symmetric part of the generic affine connection $\hat{\Gamma}$, given as follows,
\begin{equation}\label{e64}
\hat{T}^\lambda_{\alpha\beta} = -2\hat{\Gamma}^\lambda_{[\alpha\beta]} \text{.} 
\end{equation}
As we have assumed that the affine connection $\hat{\Gamma}$ is not metric compatible, hence we define the non-metricity tensor that captures this deviation from the metricity condition, as follows,
\begin{equation}\label{e65}
\hat{Q}_{\lambda\alpha\beta} = \hat{\nabla}_\lambda g_{\alpha\beta} \text{.} 
\end{equation}
One can write the affine connection $\hat{\Gamma}$ as the following decomposition \cite{R36,R37},
\begin{equation}\label{e66}
\hat{\Gamma}^{\lambda}_{\mu\nu}= \frac{1}{2}g^{\lambda \beta}\left(\partial_{\mu}g_{\beta \nu}+\partial_{\nu}g_{\beta \mu}-\partial_{\beta}g_{\mu \nu}\right)  + \frac{1}{2}g^{\lambda \beta}\left(  \hat{T}_{\nu\beta\mu} + \hat{T}_{\mu\beta\nu} - \hat{T}_{\beta\mu\nu} \right) + \frac{1}{2}g^{\lambda \beta}\left(  -\hat{Q}_{\mu\nu\beta} - \hat{Q}_{\nu\beta\mu} + \hat{Q}_{\beta\mu\nu} \right) \text{.} 
\end{equation}
In a more compact way, one can re-write the above decomposition as,
\begin{equation}\label{e67}
\hat{\Gamma}^{\lambda}_{\mu\nu}= \Gamma^{\lambda}_{\mu\nu} + K^{\lambda}_{\mu\nu} + L^{\lambda}_{\mu\nu} \text{,} 
\end{equation}
where, the first one is usual Christoffel symbols,
\begin{equation}\label{e68}
 \Gamma^{\lambda}_{\mu\nu} =   \frac{1}{2}g^{\lambda \beta}\left(\partial_{\mu}g_{\beta \nu}+\partial_{\nu}g_{\beta \mu}-\partial_{\beta}g_{\mu \nu}\right) \text{,} 
\end{equation}
second one is the contortion tensor,
\begin{equation}\label{e69}
 K^{\lambda}_{\mu\nu} = \frac{1}{2}g^{\lambda \beta}\left(  \hat{T}_{\nu\beta\mu} + \hat{T}_{\mu\beta\nu} - \hat{T}_{\beta\mu\nu} \right) \text{,} 
\end{equation}
and the last one is disformation tensor, given as follows,
\begin{equation}\label{e70}
 L^{\lambda}_{\mu\nu}  =  \frac{1}{2}g^{\lambda \beta}\left(  -\hat{Q}_{\mu\nu\beta} - \hat{Q}_{\nu\beta\mu} + \hat{Q}_{\beta\mu\nu} \right) \text{.}  
\end{equation}
This decomposition is well known as the geometric trinity of gravity that allows one to decompose the generic affine connection into the Riemannian part having Levi-Civita components and the non-Riemannian part having non-metricity and torsional contributions.

\subsection{Geometric interpretation of torsion}
Consider two curves $C$ and $\bar{C}$ describe by $x^\mu(\tau)$ and $\bar{x}^\mu(\tau)$, along with the tangent vectors $u^\mu = \frac{dx^\mu}{d\tau}$ and $\bar{u}^\mu = \frac{d\bar{x}^\mu}{d\tau}$ respectively. Let us assume that in a parallel transport of the vector $u^\sigma$ along the curve $\bar{C}$ with the displacement $d\bar{x}^\mu$, we obtain the components of the displaced vector $u'^\sigma$ (upto first order) as follows,
\begin{equation}\label{e71}
 u'^\sigma = u^\sigma + (\partial_\mu u^\sigma)d\bar{x}^\mu \text{.} 
\end{equation}
Since $u^\sigma$ is parallel transported along $\bar{C}$, we have
\begin{equation}\label{e72}
 \frac{d\bar{x}^\mu}{d\tau} (\hat{\nabla}_\mu u^\sigma) = 0 \implies \frac{d\bar{x}^\mu}{d\tau} (\partial_\mu u^\sigma + \hat{\Gamma}^\sigma_{\nu\mu} u^\nu) = 0 \implies (\partial_\mu u^\sigma)d\bar{x}^\mu = - \hat{\Gamma}^\sigma_{\nu\mu} u^\nu \bar{u}^\mu d\lambda \text{.} 
\end{equation}
From equations \eqref{e71} and \eqref{e72}, we have
\begin{equation}\label{e73}
u'^\sigma = u^\sigma - \hat{\Gamma}^\sigma_{\nu\mu} u^\nu \bar{u}^\mu d\lambda \text{.} 
\end{equation}
Similarly, in the parallel transport of the vector $u'^\sigma$ along the curve $C$ with the displacement $dx^\mu$, we obtain
\begin{equation}\label{e74}
\bar{u}'^\sigma = \bar{u}^\sigma - \hat{\Gamma}^\sigma_{\mu\nu} \bar{u}^\mu u^\nu d\lambda \text{.} 
\end{equation}
On subtracting equation \eqref{e74} from \eqref{e73}, we obtain
\begin{equation}\label{e75}
 (u'^\sigma + \bar{u}^\sigma) - (u^\sigma + \bar{u}'^\sigma)  =  - ( \hat{\Gamma}^\sigma_{\nu\mu} - \hat{\Gamma}^\sigma_{\mu\nu} ) u^\nu \bar{u}^\mu d\lambda = - \hat{T}^\sigma_{\mu\nu} u^\nu \bar{u}^\mu d\lambda \text{.} 
\end{equation}
Observe that, in the case of the Levi-Civita connection of GR, i.e., $\hat{\Gamma} = \Gamma$, we obtain the right hand side of the above equation to be zero, and thus an infinitesimal parallelogram exists, since this connection is symmetric. But for the general affine connection $\hat{\Gamma}$ in \eqref{e67}, this is not the case; this parallelogram has been cracked into a pentagon due to the presence of torsion $\hat{T}^\sigma_{\mu\nu}$ in the spacetime geometry. One can define the vector $V^\sigma =  - \hat{T}^\sigma_{\mu\nu} u^\nu \bar{u}^\mu$ that measures this deviation of cracked parallelogram \cite{R38}.

\subsection{Geometric interpretation of non-metricity}
Consider a curve $C$ describe by $x^\mu(\tau)$. Let us assume that two vectors $a^\mu$ and $b^\mu$ are parallely transported along this curve $C$ and having the inner product $a.b = a^\mu b^\nu g_{\mu\nu}$, then the total covariant derivative, denoted by $\hat{D}$, of the inner product $a.b$ along the curve $C$ is given by
\begin{equation}\label{e76}
\frac{\hat{D}}{d\tau}(a.b) = \frac{\hat{D}}{d\tau}(a^\mu b^\nu g_{\mu\nu}) = \frac{dx^\sigma}{d\tau} (\hat{\nabla}_\sigma a^\mu) b_\mu + \frac{dx^\sigma}{d\tau} (\hat{\nabla}_\sigma b^\nu) a_\nu  + \frac{dx^\sigma}{d\tau} (\hat{\nabla}_\sigma g_{\mu\nu} ) a^\mu b^\nu \text{.} 
\end{equation}
As the vectors $a^\mu$ and $b^\mu$ are parallely transported along this curve $C$, we have
\begin{equation}\label{e77}
 \frac{dx^\sigma}{d\tau} (\hat{\nabla}_\sigma a^\mu) = 0 \:\: \text{and}  \:\:  \frac{dx^\sigma}{d\tau} (\hat{\nabla}_\sigma b^\nu) = 0 \text{.} 
\end{equation}
On utilizing the above fact and the relation $\hat{Q}_{\sigma\mu\nu}=\hat{\nabla}_\sigma g_{\mu\nu} $, the equation \eqref{e76} becomes,
\begin{equation}\label{e78}
\frac{\hat{D}}{d\tau}(a.b) = \hat{Q}_{\sigma\mu\nu}  \frac{dx^\sigma}{d\tau} a^\mu b^\nu \text{.} 
\end{equation}
On taking $b^\mu=a^\mu$, we can have
\begin{equation}\label{e79}
\frac{\hat{D}}{d\tau}(|a|^2) =  \hat{Q}_{\sigma\mu\nu}  \frac{dx^\sigma}{d\tau} a^\mu a^\nu \text{.} 
\end{equation}
From the equations \eqref{e78} and \eqref{e79}, one can easily observe that both the inner product and the length of a vector change in a parallel transport along a given curve. Thus, in the case of the Levi-Civita connection of GR i.e. $\hat{\Gamma} = \Gamma$, we obtain the right hand side of the equations \eqref{e78} and \eqref{e79} to be zero, and hence the length of the vector is invariant under a parallel transport. But for the general affine connection $\hat{\Gamma}$ in \eqref{e67}, this is not the case; the length of a vector changes due to the non-metricity condition (metric incompatibility) or presence of non-metricity tensor $\hat{Q}_{\sigma\mu\nu}$ \cite{R38}.

\subsection{Teleparallel gravity}
Over the past few decades, teleparallel gravity theories have gained increasing interest. Teleparallel gravity presents an alternative geometric framework that involves a flat, curvature-free, and metric-compatible connection entirely based on torsion, denoted by $\tilde{\Gamma}$. The flatness of the teleparallel connection $\tilde{\Gamma}$ ensures that the parallel transport is independent of the path taken, preserving the concept of parallelism over long distances. This characteristic is the basis for the term teleparallel \cite{R39}. As the teleparallel connection $\tilde{\Gamma}$ is curvature free and metric compatible, one can have,
\begin{equation}\label{e80}
\tilde{R}^{\rho}_{\sigma\mu\nu}=\partial_{\mu}\tilde{\Gamma}^{\rho}_{\nu\sigma}-\partial_{\nu}\tilde{\Gamma}^{\rho}_{\mu\sigma}+\tilde{\Gamma}^{\rho}_{\mu\lambda} \tilde{\Gamma}^{\lambda}_{\nu\sigma}-\tilde{\Gamma}^{\rho}_{\nu\lambda} \tilde{\Gamma}^{\lambda}_{\mu\sigma} = 0 
\end{equation}
and 
\begin{equation}\label{e81}
Q_{\sigma\mu\nu} =  \tilde{\nabla}_\sigma g_{\mu\nu}  = 0 \text{.} 
\end{equation}
Moreover, the torsion tensor which is anti-symmetric part of the teleparallel connection $\tilde{\Gamma}$, given as follows,
\begin{equation}\label{e82}
T^\lambda_{\alpha\beta} = -2\tilde{\Gamma}^\lambda_{[\alpha\beta]} = \tilde{\Gamma}^\lambda_{\beta\alpha}  - \tilde{\Gamma}^\lambda_{\alpha\beta} \neq 0 \text{.} 
\end{equation}
Note that, in the Einstein's original framework, the tetrad is the only fundamental variable, given as $e^i = e^i_\mu dx^\mu$. This tetrad field characterizes both the metric and teleparallel connection components as follows,
\begin{equation}\label{e83}
g_{\mu\nu}=\eta_{ij}e^i_{\mu}e^j_{\nu}
\end{equation}
and 
\begin{equation}\label{e84}
\tilde{\Gamma}^{\sigma}_{\mu\nu}\equiv e^{\sigma}_i\partial_{\nu}e^i_{\mu} \text{,} 
\end{equation}
where $e_i = e_i^\mu \partial_\mu$ represents the inverse tetrad obeying the condition $e^i_\mu e_i^\nu = \delta^\nu_\mu$ and $e^i_\mu e_j^\mu = \delta^i_j$. This teleparallel connection, whose connection coefficients can be obtained via the above tetrad field, is well known as the Weitzenb$\ddot{o}$ck connection \cite{R40}. Now, we will derive the relation between the usual Ricci scalar of the Levi-Civita connection and the torsion of the teleparallel connection. In order to probe this, we define the contortion tensor, which is nothing but the difference between the teleparallel Weitzenb$\ddot{o}$ck connection and the Levi-Civita one, as follows,
\begin{equation}\label{e85}
K^\sigma_{\mu\nu} =  \tilde{\Gamma}^\sigma_{\mu\nu} - \Gamma^\sigma_{\mu\nu} = \frac{1}{2}
\left( T_\mu^\sigma{}_\nu + T_\nu^\sigma{}_\mu - T^\sigma_{\mu\nu} \right) \text{.} 
\end{equation}
On utilizing the above relation, we obtain the expression of Reimann tensor for the Levi-Civita connection as follows,
\begin{equation}\label{e86}
R^{\rho}_{\sigma\mu\nu}  =  \tilde{R}^{\rho}_{\sigma\mu\nu} - \nabla_\mu K^\rho_{\sigma\nu} + \nabla_\nu K^\rho_{\sigma\mu} -  K^\rho_{\alpha\mu} K^\alpha_{\sigma\nu} + K^\rho_{\alpha\nu} K^\alpha_{\sigma\mu} \text{.} 
\end{equation}
Now using the constraint $\tilde{R}^{\rho}_{\sigma\mu\nu}=0$ from the equation \eqref{e80} and the antisymmetry $K^{\rho\sigma\mu} = 0$, we obtain the following expression of the Ricci scalar for the Levi-Civita connection,
\begin{equation}\label{e87}
R = -2\nabla_\rho K^{\rho\sigma}{}_\sigma + K^{\mu\rho}{}_\rho  K_{\mu\sigma}{}^\sigma - K^{\mu\rho\sigma} K_{\mu\sigma\rho} \text{.}  
\end{equation}
We define the another important quantity called torsion scalar that is obtained from the contraction of the torsion tensor as follows,
\begin{equation}\label{e88}
T =  T^\mu{}_{\rho\sigma} S_\mu{}^{\rho\sigma} \text{,}  
\end{equation}
where 
\begin{equation}\label{e89}
S_\mu{}^{\rho\sigma} = \frac{1}{2} \left( K^{\rho\sigma}{}_\mu -\delta^\rho_\mu T_\alpha{}^{\alpha\sigma} + \delta^\sigma_\mu T_\alpha{}^{\alpha\rho}  \right) \text{.}   
\end{equation}
is the superpotential tensor. Now, on utilizing the above definitions in the equation \eqref{e87}, we obtained the following important relation,
\begin{equation}\label{e90}
 R = -T + 2\nabla_\rho T_\sigma{}^{\sigma\rho} = -T + B \text{.} 
\end{equation}
Thus it is evident that the Ricci scalar $R$, corresponding to the Levi-Civita connection, differ the torsion scalar $T$, corresponding to the teleparallel Weitzenb$\ddot{o}$ck connection, by a total divergence called boundary term $B$. As this boundary term does not make any contribution to the field equations, we neglect it. Thus, one can define an equivalent formulation to the GR by utilizing the torsion scalar instead of Ricci scalar via following action \cite{R41},
\begin{equation}\label{e91}
S=-\frac{1}{2\kappa}\int T e d^4x + \int e L_m d^4x \text{,} 
\end{equation}
where $e=\sqrt{-g}$ and $L_m$ is the usual matter lagrangian. This equivalent formulation to the GR in terms of torsion is widely known as the Teleparallel Equivalent to the GR (TEGR).

\subsection{Modified teleparallel gravity: the $f(T)$ gravity}
In the context of curvature, one of the simplest and most direct modifications is the theory known as $f(R)$ gravity \cite{R42,R43}. This approach generalizes the Einstein-Hilbert action by extending it to an arbitrary function of the Ricci scalar, thereby providing a broader range of phenomena to explore. A decade ago, following the same spirit, the $f(T)$ theory was introduced \cite{R44}, which is a generalization of the TEGR case involving non-linear functions of the torsion scalar. Note that the teleparallel equivalent to the GR formulation is equivalent to GR at the level of field equations, however, the modification of these theories i.e. $f(R)$ and $f(T)$ are not equivalent. The action of the $f(T)$ theory reads as follows,
\begin{equation}\label{e92}
S= \frac{1}{2\kappa}\int f(T) \sqrt{-g} d^4x + \int  L_m\sqrt{-g} d^4x \text{.} 
\end{equation}
On varying the above gravity action for the tetrad, we obtain the following field equation for the $f(T)$ theory,
\begin{equation}\label{e93}
 e^{-1}\partial_{\mu}(ee^{\gamma}_i S_{\gamma}{}^{\mu
\nu})f_T-f_Te^{\gamma}_i T^{\gamma}_{\mu\lambda}S_{\gamma}{}^{\lambda
\mu} +e^{\gamma}_i S_{\gamma}{}^{\mu\nu}\partial_{\mu}(T)f_{TT}+
\frac{1}{4}e^{\nu}_i f(T)= 4\pi G e^{\gamma}_i\mathcal{T}^\nu_\gamma \text{,} 
\end{equation}
where $f_T=df(T)/dT$ and $f_{TT}=d^2f(T)/dT^2$. Note that the term $R=-T+B$ is a unique Lorentz scalar, but the single torsion scalar $T$ or the boundary term $B$ are not Lorentz scalars. Thus, GR and its curvature extension $f(R)$ modified gravity are both locally Lorentz invariant, but this standard formulation of the $f(T)$ is not locally Lorentz invariant. Therefore, the selection of tetrads is essential in $f(T)$ cosmological models, as various tetrads result in different field equations, which subsequently lead to distinct solutions. A tetrad can be considered as a good tetrad if it does not place any limitations on the $f(T)$ functional form \cite{R45}. However, this issue was eradicated in another invariant formulation of $f(T)$ theory that is widely known as the covariant formulation of the $f(T)$ gravity \cite{R46}. For a detailed view of $f(T)$ gravity and its cosmological implications, one can check the references \cite{R47,R47aa,R47ab,R47ac}.

\subsection{Symmetric teleparallel gravity}
The symmetric teleparallel gravity has recently gained the attention of cosmologists. Similar to teleparallel gravity, its formulation involves a flat curvature-free connection $\mathring{\Gamma}$. However, this connection $\mathring{\Gamma}$ is metric incompatible and has vanishing torsion. The flatness of the connection $\mathring{\Gamma}$ ensures that the parallel transport is independent of the path taken, preserving the concept of parallelism over long distances, and hence this formulation is also referred to as the teleparallel gravity. Moreover, the vanishing torsion makes the connection $\mathring{\Gamma}$ symmetric in its last two indices. Thus, the connection $\mathring{\Gamma}$ is known as the symmetric teleparallel connection and the corresponding gravity formulation is called symmetric teleparallel gravity \cite{R48}. As the symmetric teleparallel connection $\mathring{\Gamma}$ is curvature free, torsion free, and satisfies non-metricity condition, one can have,
\begin{equation}\label{e94}
\mathring{R}^{\rho}_{\sigma\mu\nu}=\partial_{\mu}\mathring{\Gamma}^{\rho}_{\nu\sigma}-\partial_{\nu}\mathring{\Gamma}^{\rho}_{\mu\sigma}+\mathring{\Gamma}^{\rho}_{\mu\lambda} \mathring{\Gamma}^{\lambda}_{\nu\sigma}-\mathring{\Gamma}^{\rho}_{\nu\lambda} \mathring{\Gamma}^{\lambda}_{\mu\sigma} = 0
\end{equation}
and 
\begin{equation}\label{e95}
T^\lambda_{\alpha\beta} = -2\mathring{\Gamma}^\lambda_{[\alpha\beta]} = \mathring{\Gamma}^\lambda_{\beta\alpha}  - \mathring{\Gamma}^\lambda_{\alpha\beta} = 0 \implies \mathring{\Gamma}^\lambda_{\alpha\beta} = \mathring{\Gamma}^\lambda_{\beta\alpha} \text{.} 
\end{equation}
Moreover, we define the following non-metricity tensor arises due to the metric incompatibility of the symmetric teleparallel connection $\mathring{\Gamma}$,
\begin{equation}\label{e96}
Q_{\sigma\mu\nu} =  \mathring{\nabla}_\sigma g_{\mu\nu} = \partial_\sigma g_{\mu\nu} - \mathring{\Gamma}^\rho_{\sigma\mu} g_{\rho\nu} - \mathring{\Gamma}^\rho_{\sigma\nu} g_{\rho\mu} \neq 0 \text{.} 
\end{equation}
Note that, one can express the symmetric teleparallel connection through the general element of $GL(4, \mathbb{R})$, as follows \cite{R49}, 
\begin{equation}\label{e97}
\mathring{\Gamma}^\sigma_{\mu\nu}  = A_\lambda{}^\sigma \partial_\mu A^\lambda{}_\nu \text{,} 
\end{equation}
where $A^\mu{}_\nu$ is a $4 \times 4$ matrix having inverse $A_\nu{}^\mu$. As the connection $\mathring{\Gamma}$ follows a torsion free constraint, the matrix $A^\sigma{}_\mu$ must be reduce to Jacobian of the coordinate transformation $A^\sigma{}_\mu = \partial_\mu \xi^\sigma$, for some arbitrary vector field $\xi^\sigma$. Thus we obtain,
\begin{equation}\label{e98}
\mathring{\Gamma}^\sigma_{\mu\nu}  =  \frac{\partial x^\sigma}{\partial \xi^\rho} \frac{\partial^2 \xi^\rho}{\partial x^\mu \partial x^\nu} \text{.} 
\end{equation}
Thus from the above expression, it is evident that one can always find a coordinate transformation such that the general symmetric teleparallel connection $\mathring{\Gamma}$ completely vanishes. Such a transformation choice is known as the coincident guage \cite{R49}. For this particular, vanishing $\mathring{\Gamma}$ case, the non-metricity tensor becomes $Q_{\sigma\mu\nu} = \partial_\sigma g_{\mu\nu}$ which also implies that the covariant derivative of the connection reduces to partial one and hence the corresponding field equations becomes much simplified. Now, for the general symmetric teleparallel connection, one can obtain the relation between the usual Ricci scalar of the Levi-Civita connection and the non-metricity of the symmetric teleparallel connection. In order to probe this, we define the disformation tensor, which is nothing but the difference between the symmetric teleparallel connection and the Levi-Civita one, as follows,
\begin{equation}\label{e99}
L^\sigma_{\mu\nu} =  \mathring{\Gamma}^\sigma_{\mu\nu} - \Gamma^\sigma_{\mu\nu} = \frac{1}{2}
\left( Q^\sigma{}_{\mu\nu} - Q_\mu{}^\sigma{}_\nu - Q_\nu{}^\sigma{}_\mu \right) \text{.} 
\end{equation}
On utilizing the above relation, we obtain the expression of Reimann tensor for the Levi-Civita connection as follows,
\begin{equation}\label{e100}
R^{\rho}_{\sigma\mu\nu}  =  \mathring{R}^{\rho}_{\sigma\mu\nu} - \nabla_\mu L^\rho_{\sigma\nu} + \nabla_\nu L^\rho_{\sigma\mu} -  L^\rho_{\alpha\mu} L^\alpha_{\sigma\nu} + L^\rho_{\alpha\nu} L^\alpha_{\sigma\mu} \text{.} 
\end{equation}
We define the another important quantity called non-metricity scalar that is obtained from the contraction of the non-metricity tensor as follows,
\begin{equation}\label{e101}
Q =  - Q_{\sigma\mu\nu} P^{\sigma\mu\nu} \text{,} 
\end{equation}
where 
\begin{equation}\label{e102}
P^\sigma_{\mu\nu} = \frac{1}{4} \left( -Q^\sigma\:_{\mu\nu} + 2Q_{(\mu}\:^\sigma\:_{\nu)} + (Q^\sigma - \bar{Q}^\sigma) g_{\mu\nu} - \delta^\sigma_{(\mu}Q_{\nu)} \right) 
\end{equation}
is the superpotential tensor. Here, $Q_\alpha = Q_\alpha\:^\mu\:_\mu $ and $ \bar{Q}_\alpha = Q^\mu\:_{\alpha\mu} $ are two non-metricity vectors. Now, on utilizing the above definitions and the constraint $\mathring{R}^{\rho}_{\sigma\mu\nu}=0$ from the equation \eqref{e94} in the equation \eqref{e100}, we obtained the following important relation,
\begin{equation}\label{e103}
 R = Q - 2\nabla_\rho \left( Q^\rho - \bar{Q}^\rho \right) = Q - B \text{.}  
\end{equation}
Thus, it is evident that the Ricci scalar $R$, corresponding to the Levi-Civita connection, differs from the non-metricity scalar $Q$, corresponding to the symmetric teleparallel connection, by a total derivative term called boundary term $B$. As this boundary term does not make any contribution to the field equations, we neglect it. Thus, one can define an equivalent formulation to the GR by utilizing the non-metricity scalar instead of the Ricci scalar via the following action,
\begin{equation}\label{e104}
S=-\frac{1}{2\kappa}\int Q  \sqrt{-g}d^4x + \int  L_m \sqrt{-g}d^4x \text{,} 
\end{equation}
where $L_m$ is the usual matter lagrangian. This equivalent formulation to the GR in terms of non-metricity is widely known as the Symmetric Teleparallel Equivalent to the GR (STEGR) \cite{R50}.

\subsection{Modified symmetric teleparallel gravity: the $f(Q)$ gravity}
In the year 2018, the $f(Q)$ theory was introduced \cite{R51}, which is a generalization of the STEGR case involving non-linear functions of non-metricity scalar. Note that, the STEGR formulation is equivalent to GR at the level of field equations, however, the modification of this theory i.e. modified symmetric teleparallel gravity or $f(Q)$ gravity is neither equivalent to $f(R)$ gravity nor to $f(T)$ gravity. The action of the $f(Q)$ theory reads as follows,
\begin{equation}\label{e105}
S= \frac{1}{2\kappa}\int f(Q) \sqrt{-g} d^4x + \int  L_m\sqrt{-g} d^4x \text{.} 
\end{equation}
On varying the generic action \eqref{e105} with respect to metric, we obtained the following metric field equation of the $f(Q)$ gravity as follows,
\begin{equation}\label{e106}
\frac{2}{\sqrt{-g}}\nabla_{\sigma}\left( \sqrt{-g}f_Q {P^{\sigma}}_{\mu\nu}\right)+\frac{1}{2}g_{\mu\nu}f
+f_Q\left(P_{\mu\sigma\rho}{Q_{\nu}}^{\sigma\rho}-2Q_{\sigma\rho\mu}{P^{\sigma\rho}}_{\nu} \right)=-\mathcal{T}_{\mu\nu} \text{,} 
\end{equation} 
where $f_Q=\frac{df}{dQ}$. Again, on varying the generic action \eqref{e105} with respect to symmetyric teleparallel connection, we obtained the following connection field equation of the $f(Q)$ gravity as follows, 
\begin{equation}\label{e107}
\nabla_{\mu}\nabla_\nu \left( \sqrt{-g}f_Q {P^{\mu\nu}}_{\sigma} \right) = 0 \text{.} 
\end{equation}

\section{Modelling and statistics}
In the previous section, we have discussed several cosmological observations and their persistent issues. Further, we have studied the different geometrical structures of the spacetime manifold. Now, in this section, we will see how these geometrical structures can describe the observable Universe and make statistically significant predictions.

\subsection{Modelling the Universe}
\begin{figure}[H]
    \centering
    \includegraphics[width=17 cm, height= 9 cm]{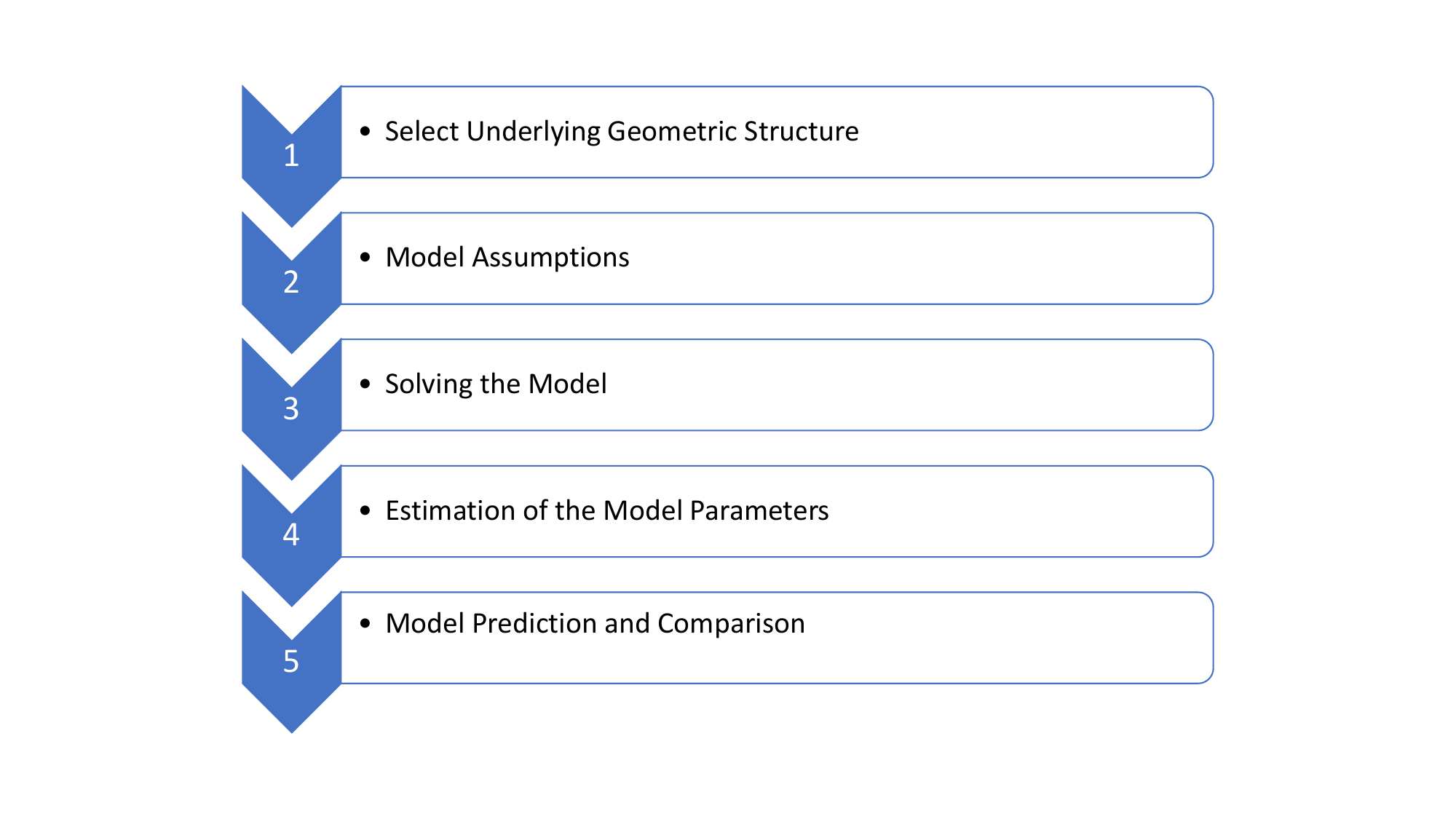}
    \caption{Flow chart representing the steps to construct a cosmological model.}
    \label{fig:int-1}
\end{figure}

To construct a cosmological model that can describe the physical Universe we follow some key steps described as below,
\begin{itemize}
    \item \textbf{Specifying the underlying geometry:} We begin with a spacetime manifold structure endowed with a suitable connection and a metric that describes the notion of distances and covariant differentiation.
    \item \textbf{Model assumptions:} After specifying the underlying geometric structure, we consider some realistic assumptions on the model, such as fixing the functional form of the considered modified gravity, considering a realistic equation of state, etc. These assumptions make the model realistic to describe some specific cosmological epochs.
    \item \textbf{Solving the model:} The model assumptions are also helpful in obtaining an analytical solution. Thus, we find the analytical solution of the assumed cosmological model by utilizing differential equation techniques.
    \item \textbf{Parameter’s estimation:} Once we obtain the exact solution of the model, we are left with some free parameters in the obtained solution. Hence, in the next step, we obtain the free parameter values by employing various observational dataset along with some standard statistical methods.
    \item \textbf{Model prediction and comparison:} One can study the various cosmological parameters such as deceleration, statefinder, equation of state parameter, etc, using the constrained solution from the observational data. The evolution of these cosmological parameters predicts the evolutionary behavior of the assumed cosmological model, and further, these predictions can be compared to the predictions of established models of cosmology.
\end{itemize}
One can easily understand the methodology and framework to study any cosmological model by following flow chart, presented in Figure \eqref{fig:int-1}.

\subsection{$\chi^2$ minimization}
Consider a function $f_{model}(x,\theta)$ in the independent variable $x$ and the set of free parameters $\theta$. If the set $\{f_{k,obs}(x_{k,obs})\}_{k=1}^n$ represents $n$ independent observation along with the standard deviation $\sigma_{k,obs}$, then we define the $\chi^2$ as follows \cite{R52},
\begin{equation}\label{e108}
\chi^2(\theta) = \sum_{k=1}^n  \frac{\left( f_{model}(x_{k,obs},\theta) - f_{k,obs} \right)^2}{\sigma_{k,obs}^2} \text{.} 
\end{equation}
Moreover, if these $n$ observations are not independent, then in this case a covariance matrix $C$ is utilized instead of the standard deviation. In such a case the $\chi^2$ function becomes,
\begin{equation}\label{e109}
\chi^2(\theta) = \sum_{i,j=1}^n X_i C^{-1}_{ij} X_j \text{,} 
\end{equation}
where $X_i = f_{i,obs} - f_{model}(x_{i,obs},\theta) $. One can define another important statistical quantity called likelihood function as follows,
\begin{equation}\label{e110}
\mathcal{L}(\theta) = P(\theta| D = Data) \text{.} 
\end{equation}
One can obtained the following relation between the above two statistical parameters as follows,
\begin{equation}\label{e111}
\mathcal{L}(\theta) \propto exp\left( -\frac{1}{2} \chi^2(\theta) \right) \text{.}  
\end{equation}
Our ultimate aim is to find the best fit values of parameter set $\theta$ so that the function $f_{model}(x,\theta)$ agrees with the observational data. In order to probe this, one has to minimize the $\chi^2(\theta)$ function (that is equivalent to maximize the likelihood function), and the parameter corresponds to the $\chi^2_{min}$ value, say $\theta_{min}$ is the required best fit parameter value. Several approaches exist in the literature to solve this optimization issue, such as the Gradient Descent algorithm, Newton's method, and the Random Walk algorithm \cite{R53,R54}. The disadvantage of these algorithms is that whenever the parameter space exhibits too many local minima, the best fit value obtained by these methods is local instead of global. However, by utilizing the features of these algorithms, a new efficient algorithm can be constructed. One such algorithm is widely adopted in the computational field of science known as the Markov Chain Monte Carlo (MCMC) algorithm. 

\subsection{The MCMC approach }
Over the past decade, probabilistic data analysis, particularly Bayesian statistical inference, has revolutionized scientific research. This approach involves utilizing either the posterior probability density function (PDF) or the likelihood function. Generally, one can find the optimum value of these functions easily using some algorithms, but a more detailed understanding of the posterior PDF is often required. Markov Chain Monte Carlo methods are specifically developed to efficiently sample from the posterior PDF, offering sampling even in high dimensional parameter spaces \cite{R55}. Mathematically, a chain with several arrays of parameter's values, given a probability distribution, obtained by a random process and that stores the memory of  current and its preceding step only is known as the Markov Chain. Further, the Monte Carlo process helps to explore the parameter space in the entire domain. The core idea of MCMC is to create a Markov chain that samples a model's parameter space based on a specified probability distribution. This chain consists of a sequence of parameter values, where each value is derived from the previous one through a defined set of transition rules associated with a proposal distribution. The proposal distribution suggests a new parameter value, and its acceptance depends on its posterior probability, considering both the observational data and the prior probability function. After the chain converges, the posterior distribution for the parameters can be estimated by calculating the frequency of parameter values within the chain. This posterior distribution then enables the estimation of optimal parameter values and their associated uncertainties, aiding in predictions for various observables. The Metropolis Hasting algorithm based on MCMC and the random walk approach is one of the most popular methods. Another more advanced algorithm is recently proposed by Mackey called the emcee: The MCMC Hammer \cite{R56}. The emcee algorithm performs better than the traditional MCMC sampling techniques. When using an emcee, one can go with a large number of walkers, often in the hundreds. Theoretically, there is no downside to increasing the number of walkers unless you encounter performance limitations. The detailed discussion on the different MCMC algorithms compared with the emcee sampling method is beyond the scope of the present thesis.

\section{Conclusions}

In this chapter, we have discussed the several cosmological observations and its implications. Then we have recalled the fundamentals of relativity. Further, we have discussed the modified gravity scenario as an alternative to the dark energy candidate. In the present thesis, we present the accelerating cosmological models in the non-metricity based modified symmetric teleparallel gravity. This thesis aims to reproduce the dark energy effects originating from the modification of spacetime geometry, bypassing the need of controversial cosmological constant. Further, for the statistical assessment and model parameter estimation, the MCMC approach, along with the emcee sampler and Bayesian statistical inference, is utilized. The following chapters enclose the detailed investigation and the corresponding outcomes.

%% file: Chapters/Chapter2.tex

\chapter{Cosmic acceleration with bulk viscous matter in linear $f(Q)$ cosmology} 

\label{Chapter2} 

\lhead{Chapter 2. \emph{Cosmic acceleration with bulk viscous matter in linear $f(Q)$ cosmology}} 

\vspace{10 cm}
* The work, in this chapter, is covered by the following publications: \\
 
\textit{Cosmic acceleration with bulk viscosity in modified $f(Q)$ gravity}, Physics of the Dark Universe {\bf 32}, 100820 (2021).

\clearpage
\pagebreak

The current chapter presents the role of bulk viscosity to study the dynamics of late-time expansion phase of the Universe within the context of linear $f(Q)$ gravity. We consider a bulk viscous matter-dominated cosmological model with the bulk viscosity coefficient of the form $\xi =\xi _{0}+\xi_{1}H+\xi _{2}\left( \frac{\dot{H}}{H}+H\right) $ which is proportional to the velocity and acceleration of the expanding Universe. We obtain two sets of limiting conditions on the bulk viscous parameters $\xi _{0},$ $\xi _{1},$ $\xi _{2}$ and model parameter $\alpha$, out of which one condition favours the present scenario of cosmic acceleration with a phase transition and corresponds to the Universe with a Big Bang origin. Moreover, we discuss the cosmological behaviour of some geometrical parameters. In addition, we obtain the suitable values of the arbitrary parameters $\xi_{0},$ $\xi _{1},$ $\xi _{2}$ and $\alpha $ by constraining our model with updated Hubble datasets having $57$ data points and Pantheon datasets having $1048$ supernovae points which show that our model has good compatibility with observations. Further, we also include the BAO datasets having six data points with the Hubble \& Pantheon datasets. Finally, we analyze our model with the statefinder diagnostic analysis and we find some interesting results that are discussed below in details.


\section{Introduction}\label{sec1z}

Over the
last two decades, a convergence of several cosmological observations
indicate that our Universe is going through a period of accelerated
expansion. The
observational evidence such as SN Ia \cite{R9,R10}, BAO \cite{D.J.,W.J.}, large scale structure \cite{T.Koivisto,S.F.}, galaxy redshift
survey \cite{C.Fedeli} and CMB
\cite{R.R.,Z.Y.} strongly support this accelerating expansion. Plenty of models have been proposed in
the literature to describe this recent acceleration. Basically, there are
two approaches to interpret this recent acceleration of the Universe. The
first approach is the assumption of the existence of mysterious force with
high negative pressure so-called dark energy as responsible for the
current acceleration of the Universe. The simplest candidate for dark energy
is the cosmological constant $\Lambda$ (or vacuum energy) i.e. the fluid
responsible for such an effective negative pressure with constant energy
density \cite{S.Weinberg,Carroll}. Another time-varying dark energy models have been proposed in the
literature like quintessence \cite{Carroll-2,Y.Fujii}, k-essence \cite%
{T.Chiba,C.Arm.} and perfect fluid models (like the Chaplygin gas model) 
\cite{M.C.,A.Y.}. The second route to characterize the current expansion phase is to modify spacetime's geometry. Recently, modified theories of gravity have
attracted the interest of cosmologists for understanding the role of dark
energy. In modified gravity, the origin of dark energy is recognized as a
modification of gravity. There are several modified theories have
been proposed in the literature like $\:f(R)\:$ theory \cite{R42,A.A.,S.Nojiri}, $\:f(T)\:$ theory \cite{R44,E.V.,K.B.}, $f(T,B)$
theory \cite{Sebastian}, $\:f(R,T)\:$ theory \cite{Harko,Hamid}, $f(Q,T)$
theory \cite{Yixin}, $\:f(G)\:$ theory \cite{S.Nojiri-2}, $f(R,G)$
theory \cite{E.E.,K.B.-2}, etc. Nowadays, $f(Q)$ theories of gravity have
been extremely investigated. Recently, there are several studies done in $f(Q)$
gravity. T. Harko et al. studied the extension of symmetric teleparallel gravity 
\cite{Harko-2}. S. Mandal et al. studied energy conditions in $f(Q)$ gravity and 
also did a comparative study between $f(Q)$ gravity and $\Lambda$CDM 
\cite{Sanjay}. Moreover, they used the cosmographic idea to constrain the
Lagrangian function $f(Q)$ using the latest pantheon data \cite{Sanjay2}. An
interesting investigation on $f(Q)$ gravity was done by Noemi, where he
explored the signatures of non-metricity gravity in its' fundamental level 
\cite{Noemi/2021}.

Earlier, to study inflationary epoch in the early Universe bulk viscosity
has been proposed in the literature without any requirement of dark energy 
\cite{T.P.,I.W.}. Hence, it is very natural to expect that the bulk
viscosity can be responsible for the current accelerated expansion of the
Universe. Nowadays, several authors are attempted to explain the late-time
acceleration via bulk viscosity without any dark energy constituent or
cosmological constant \cite{Athira,Mohan,G.C.,J.C.,A.Av.}.
Theoretically, deviations that occur from the local thermodynamic stability
can originate the bulk viscosity but a detailed mechanism for the formation
of bulk viscosity is still not achievable \cite{W.Z.}. In cosmology, when
the matter content of the Universe expands or contract too fast as a
cosmological fluid then the effective pressure is generated to bring back
the system to its thermal stability. The bulk viscosity is the manifestation
of such an effective pressure \cite{J.R.,H.O.}.

In cosmology, there are two main formalism for the description of bulk
viscosity. The first one is the non-casual theory, where the deviation of
only first-order is considered and one can find that the heat flow and
viscosity propagate with infinite speed while in the second one i.e. the
casual theory it propagates with finite speed. In the year 1940, Eckart
proposed the non-casual theory \cite{C.E.}. Later, Lifshitz and Landau gave
a similar theory \cite{L.D.}. The casual theory was developed by Israel,
Hiscock and Stewart. In this theory, second-order deviation from equilibrium
is considered \cite{W.I.,W.I.-2,W.I.-3,W.A.,W.A.-2}. Moreover, Eckart theory
can be acquired from it as a first-order approximation. Hence, Eckart's
theory is a good approximation to the Israel theory in the limit of
vanishing relaxation time. To analyze the late acceleration of the Universe,
the casual theory of bulk viscosity has been used. Cataldo et al. have
investigated the late time acceleration using the casual theory \cite%
{M.Cataldo}. Basically, they used an ansatz for the Hubble parameter
(inspired by the Eckart theory) and they have shown the transition of the
Universe from the big rip singularity to the phantom behavior.

The expansion process of an accelerating Universe is a collection of states
that lose their thermal stability in a small fragment of time \cite{A.Av.-2}%
. Hence, it is quite natural to consider the existence of bulk viscosity
coefficient to describe the expansion of the Universe. The accelerated
expansion scenario of the Universe (the mean stage of low redshift) can be
justified by the geometrical modification in Einstein's equation. Also,
without any requirement of cosmological constant bulk viscosity can generate
an acceleration. It contributes to the pressure term and applies additional
pressure to drive the acceleration \cite{S.Od.}. C. P. Singh and Pankaj
Kumar has investigated the role of bulk viscosity in modified $f(R,T)$
theories of gravity \cite{Singh}. S. Davood has investigated the effect of
bulk viscous matter in modified $f(T)$ theories of gravity \cite{S.Davood}.

In this chapter, we have focused on studying the cosmic acceleration within $f(Q)$ gravity background in the presence of bulk viscous fluid. The
motivation of working in the non-metricity $f(Q)$ gravity is that in this
framework, the motion equations are in the second-order, which is easy to
solve. In $f(R)$ gravity, an extra scalar mode appears because the model is
the higher derivative theory as the Ricci scalar includes the second-order
derivatives of the metric tensor. This scalar mode generates additional
force, and it is often inconsistent with the Newton law observations and
also for a density of a canonical scalar field $\phi$; the non-minimal
coupling between geometry and the matter Lagrangian produces an additional
kinetic term which is not an agreement with the stable Horndeski class \cite%
{Olmo/2015}. Nevertheless, the non-metricity formalism overcomes the above
problems, which are induced by the higher-order theory. 

The outline of the
present chapter is as follows. In Sec. \ref{sec3z} we describe the
FLRW Universe dominated with bulk viscous matter and also we derive the
expression for the Hubble parameter. In Sec. \ref{sec4z} we derive the scale
factor and found two sets of limiting conditions on the coefficients of bulk
viscosity which corresponds to the Universe which begins with a Big Bang and
then making a transition from deceleration phase to the accelerating expansion phase.
In Sec. \ref{sec5z} we show the evolution of deceleration parameter $q$. In
Sec. \ref{sec6z} we have constrained the free parameters by utilizing the Hubble
data and Pantheon data sets. In Sec. \ref{sec7z} we adopt the statefinder
diagnostic pair to differentiate present bulk viscous model with other
models of dark energy. Finally, in the last section Sec. \ref{sec8z}, we
briefly discuss our conclusions.

\section{FLRW Universe dominated with bulk viscous matter}\label{sec3z}

We consider that the Universe is described by the spatially flat
FLRW line element 
\begin{equation}
ds^{2}=-dt^{2}+a^{2}(t)[dx^{2}+dy^{2}+dz^{2}]\text{.}  \label{3az}
\end{equation}%
Here, $a(t)$ denotes the expansion factor of the Universe dominated with bulk viscous
matter. The trace of non-metricity tensor with respect to line element \eqref{3az} is given as,
\begin{equation}
Q=6H^{2}\text{.}  \label{3bz}
\end{equation}%
For a bulk viscous fluid, described by its effective pressure $\bar{p}$ and
the energy density $\rho $, the energy-momentum tensor takes the form 
\begin{equation}
\mathcal{T}_{\mu \nu }=(\rho +\bar{p})u_{\mu }u_{\nu }+\bar{p}g_{\mu \nu }\text{,}
\label{3c}
\end{equation}%
where $\bar{p}=p-3\xi H$.
Here, $\xi $ is the coefficient of bulk viscosity which can be a function of
Hubble parameter and its derivative and the components of four-velocity $%
u^{\mu }$ are $u^{\mu }=(1,0)$ and $p$ is the normal pressure which is 0 for
non-relativistic matter.

The Friedmann equations describing the Universe dominated with bulk viscous
matter are 
\begin{equation}
3H^{2}=\frac{1}{2f_{Q}}\left( -\rho +\frac{f}{2}\right)  \label{3dz}
\end{equation}%
and 
\begin{equation}
\dot{H}+3H^{2}+\frac{\dot{f_{Q}}}{f_{Q}}H=\frac{1}{2f_{Q}}\left( \bar{p}+%
\frac{f}{2}\right) \text{.}  \label{3ez}
\end{equation}%
In an accelerated expanding Universe, the coefficient of viscosity should
depend on velocity and acceleration. In this paper, we consider a time
dependent bulk viscosity of the form \cite{J.Ren} 
\begin{equation}
\xi =\xi _{0}+\xi _{1}\left( \frac{\dot{a}}{a}\right) +\xi _{2}\left( \frac{{%
\ddot{a}}}{\dot{a}}\right) =\xi _{0}+\xi _{1}H+\xi _{2}\left( \frac{\dot{H}}{%
H}+H\right) \text{.}  \label{3fz}
\end{equation}%
It is a linear combination of three terms, first one is a constant, second
one is proportional to the Hubble parameter, which indicates the dependence
of the viscosity on speed, and the third one is proportional to the $\frac{%
\ddot{a}}{\dot{a}}$, indicating the dependence of the bulk viscosity on
acceleration.

In this paper, we consider the following functional form of $f(Q)$, 
\begin{equation}
f(Q)=\alpha Q,\ \ \ \alpha \neq 0\text{.}  \label{3gz}
\end{equation}%
Then, for this particular choice of the function, the field equation becomes, 
\begin{equation}
\rho =-3\alpha H^{2}  \label{3hz}
\end{equation}%
and 
\begin{equation}
\bar{p}=2\alpha \dot{H}+3\alpha H^{2}\text{.}  \label{3iz}
\end{equation}%
As we are concerned with late-time acceleration, we have considered the
non-relativistic matter dominates the Universe. From the Friedmann equations
\eqref{3iz} and equation \eqref{3fz}, we have first-order differential
equation for the Hubble parameter by replacing $\frac{d}{dt}$ with $\frac{d}{%
dln(a)}$ via $\frac{d}{dt}=H\frac{d}{dln(a)}$,

\begin{equation}
\frac{dH}{dln(a)}+\left( \frac{3\alpha +3\xi _{1}+3\xi _{2}}{2\alpha +3\xi
_{2}}\right) H+\left( \frac{3\xi _{0}}{2\alpha +3\xi _{2}}\right) =0\text{.}
\label{3jz}
\end{equation}

Now, we set 
\begin{equation}
3\xi _{0}=\bar{\xi _{0}}H_{0}, \ \ 3\xi _{1}=\bar{\xi _{1}}, \ \ 3\xi _{2}=%
\bar{\xi _{2}} \ \ \text{and} \ \ \bar{\xi}_{12}=\bar{\xi _{1}}+\bar{\xi _{2}%
}\text{,}  \label{3kz}
\end{equation}%
where $H_{0}$ is present value of the Hubble parameter and $\bar{\xi}_{0},\
\ $ $\bar{\xi}_{1},\ \ $ $\bar{\xi}_{2}$ are the dimensionless bulk viscous
parameters, then by using above equation \eqref{3jz} becomes,

\begin{equation}
\frac{dH}{dln(a)}+\left( \frac{3\alpha +\bar{\xi}_{12}}{2\alpha +\bar{\xi
_{2}}}\right) H+\left( \frac{\bar{\xi _{0}}}{2\alpha +\bar{\xi _{2}}}\right)
H_{0}=0\text{.}  \label{3lz}
\end{equation}

After integrating above equation we obtain the Hubble parameter as, 
\begin{equation}
H(a)=H_{0}\left[ a^{-\left( \frac{3\alpha +\bar{\xi}_{12}}{2\alpha +\bar{\xi
_{2}}}\right) }\left( 1+\frac{\bar{\xi _{0}}}{3\alpha +\bar{\xi}_{12}}%
\right) -\frac{\bar{\xi _{0}}}{3\alpha +\bar{\xi}_{12}}\right] \text{.}
\label{3mz}
\end{equation}

At $\bar{\xi}_{0}=\bar{\xi}_{1}=\bar{\xi}_{2}=0$, equation \eqref{3mz}
becomes, 
\begin{equation}
H=H_{0}a^{-\frac{3}{2}}\text{.}  \label{3nz}
\end{equation}%
The equation \eqref{3nz} gives the value of the Hubble parameter in case of
ordinary matter-dominated Universe i.e. when all the bulk viscous parameters
are $0$. \newline
Now, by applying the scale factor-redshift relation i.e. $a(t)=\frac{%
1}{1+z}$, the explicit expression of the Hubble function in the redshift terms is given as, 
\begin{equation}
H(z)=H_{0}\left[ (1+z)^{\left( \frac{3\alpha +\bar{\xi}_{12}}{2\alpha +\bar{%
\xi _{2}}}\right) }\left( 1+\frac{\bar{\xi _{0}}}{3\alpha +\bar{\xi}_{12}}%
\right) -\frac{\bar{\xi _{0}}}{3\alpha +\bar{\xi}_{12}}\right] \text{.}
\label{3oz}
\end{equation}

\section{Scale factor}\label{sec4z}

Now, using the definition of Hubble parameter, the equation \eqref{3mz}
becomes, 
\begin{equation}
\frac{1}{a}\frac{da}{dt}=H_{0}\left[ a^{-\left( \frac{3\alpha +\bar{\xi}_{12}%
}{2\alpha +\bar{\xi _{2}}}\right) }\left( 1+\frac{\bar{\xi _{0}}}{3\alpha +%
\bar{\xi}_{12}}\right) -\frac{\bar{\xi _{0}}}{3\alpha +\bar{\xi}_{12}}\right]
\text{.}  \label{4az}
\end{equation}%
On integrating the above equation we get the scale factor, 
\begin{equation}
a(t)=\left[ \frac{3\alpha +\bar{\xi}_{12}+\bar{\xi}_{0}}{\bar{\xi}_{0}}%
-\left( \frac{3\alpha +\bar{\xi}_{12}}{\bar{\xi}_{0}}\right)
e^{-H_{0}(t-t_{0})\frac{\bar{\xi}_{0}}{2\alpha +\bar{\xi}_{2}}}\right] ^{%
\frac{2\alpha +\bar{\xi}_{2}}{3\alpha +\bar{\xi}_{12}}}\text{,}  \label{4bz}
\end{equation}%
where $t_{0}$ is the present cosmic time.

Now, let $y=H_0(t-t_0)$, then the second order derivative of $a(t) $ with
respect to $y$ is

\begin{equation}
\frac{d^2a}{dy^2} = \frac{e^{\frac{- \bar{\xi}_0 y}{2\alpha+ \bar{\xi}_2}}}{2\alpha+\bar{\xi}_2} \left(-(\bar{\xi}_0+ \bar{\xi}_{12}+3\alpha) + (2\alpha + \bar{\xi}_2) e^{\frac{- \bar{\xi}_0 y}{2\alpha+ \bar{\xi}_2}} \right)  
 \times \left[\frac{\bar{\xi}_0 + \bar{\xi}_{12} + 3\alpha - (3\alpha + \bar{\xi}_{12}) e^{\frac{- \bar{\xi}_0 y}{2\alpha + \bar{\xi}_2}}}{\bar{\xi}_0} \right]^{\frac{-2(2\alpha + \bar{\xi}_1)- \bar{\xi}_2}{3\alpha + \bar{\xi}_{12}}}\text{.}
\end{equation}

From the above expression it is clear that we have two limiting conditions
based on the values of $\bar{\xi}_{0},$ $\bar{\xi}_{1}$ and $\bar{\xi}_{2}$.
By equation \eqref{3hz} and the fact that density of ordinary matter in the
Universe is always positive, so we must have $\alpha <0$. Assuming, $\alpha
=-\bar{\alpha}$ where $\bar{\alpha}>0$, these two limiting conditions are

\begin{equation}
\bar{\xi}_{0}>0,\text{ }\bar{\xi}_{12}<3\bar{\alpha},\text{ }\bar{\xi}_{2}<2%
\bar{\alpha},\text{ }\bar{\xi}_{0}+\bar{\xi}_{12}<3\bar{\alpha}  \label{4cz}
\end{equation}%
and 
\begin{equation}
\bar{\xi}_{0}<0,\text{ }\bar{\xi}_{12}>3\bar{\alpha},\text{ }\bar{\xi}_{2}>2%
\bar{\alpha},\text{ }\bar{\xi}_{0}+\bar{\xi}_{12}>3\bar{\alpha}\text{.}
\label{4dz}
\end{equation}
These two limiting condition implies that the Universe experienced a
deceleration phase at early times and then making a transition into the
accelerated phase in the latter times. If we place $\bar{\xi}_{0}+\bar{\xi}%
_{12}>3\bar{\alpha}$ and $\bar{\xi}_{0}+\bar{\xi}_{12}<3\bar{\alpha}$ in the
first and second limiting condition respectively, then the Universe will
experience an everlasting accelerated expansion.

\section{Deceleration parameter}\label{sec5z}

\begin{figure}[H]
\centering
\includegraphics[scale=0.65]{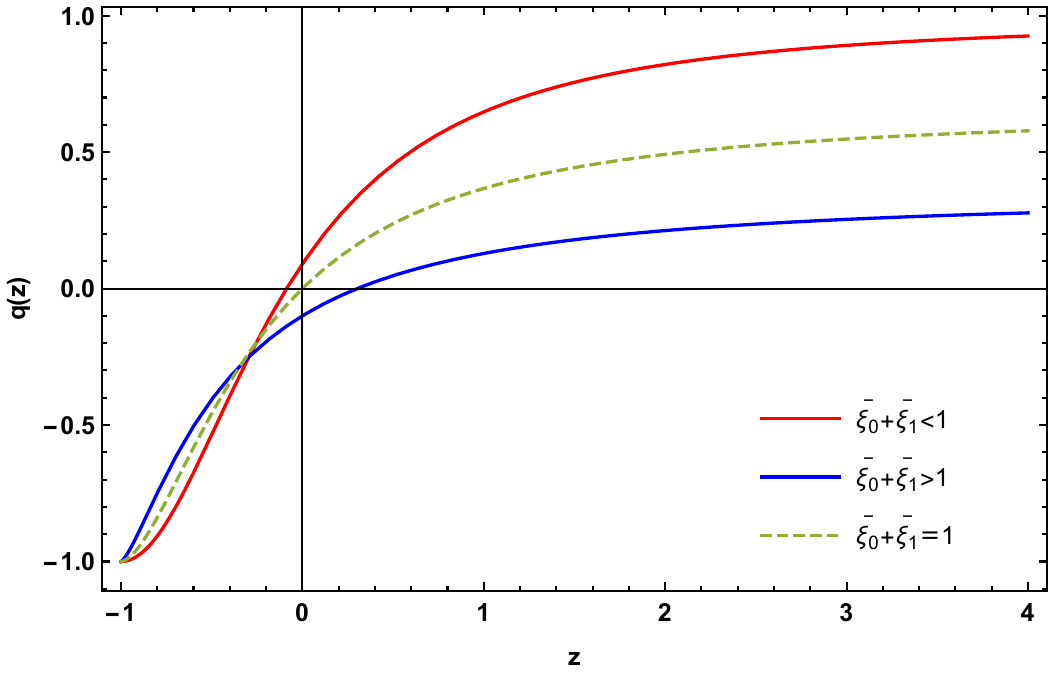}
\caption{Variation of the deceleration parameter with redshift $z$ for the
first limiting conditions $\bar{\protect\xi}_{0}>0, \bar{\protect\xi}%
_{12}<3, \bar{\protect\xi}_{2}<2, \bar{\protect\xi}_{0}+\bar{\protect\xi}%
_{12}<3$. Here, we took $\protect\alpha =-1$ i.e., $\bar{\protect\alpha}=1 $%
. $q$ enters the negative region in the recent past if $\bar{\protect\xi}%
_{0}+\bar{\protect\xi}_{1}>1$, at present if $\bar{\protect\xi}_{0}+\bar{%
\protect\xi}_{1}=1$ and in the future if $\bar{\protect\xi}_{0}+\bar{\protect%
\xi}_{1}<1$. For Red, Blue and Green plots the value of $(\bar{\protect\xi}%
_{0},\bar{\protect\xi}_{1},\bar{\protect\xi}_{2})$ are $%
(0.9,0.01,1),(0.45,0.65,1),(0.65,0.35,1)$ respectively. }
\label{f1z}
\end{figure}

\begin{figure}[H]
\centering
\includegraphics[scale=0.65]{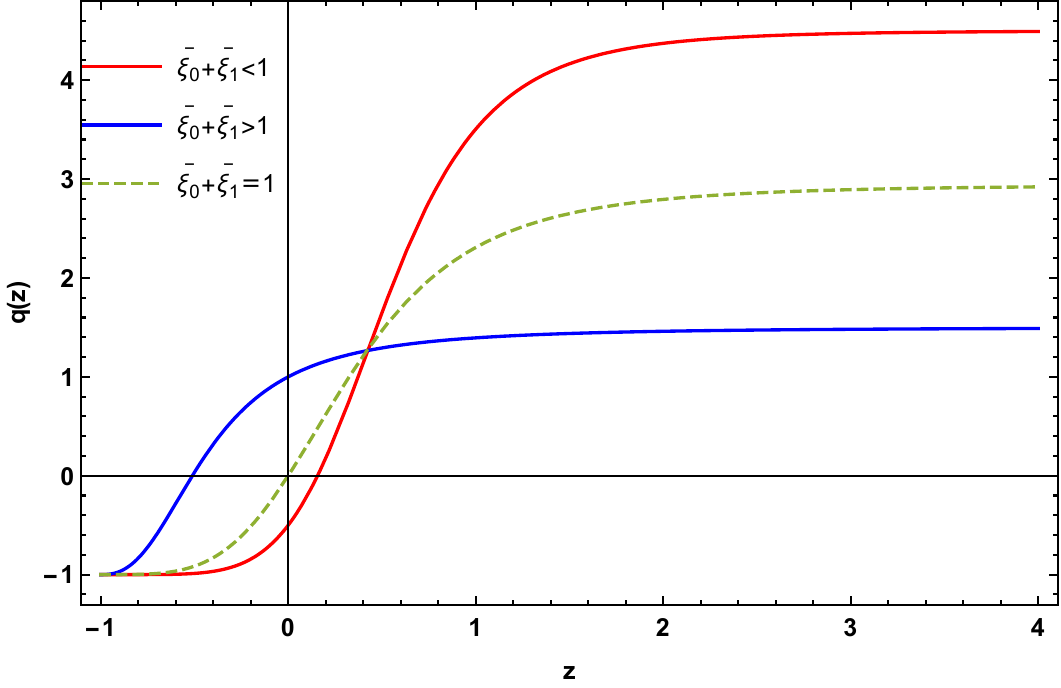}
\caption{Variation of the deceleration parameter with redshift $z$ for the
second limiting conditions $\bar{\protect\xi}_{0}<0, \bar{\protect\xi}%
_{12}>3, \bar{\protect\xi}_{2}>2, \bar{\protect\xi}_{0}+\bar{\protect\xi}%
_{12}>3$. Here, we took $\protect\alpha =-1$ i.e., $\bar{\protect\alpha}=1 $%
. $q$ enters the negative region in the recent past if $\bar{\protect\xi}%
_{0}+\bar{\protect\xi}_{1}<1$, at present if $\bar{\protect\xi}_{0}+\bar{%
\protect\xi}_{1}=1$ and in the future if $\bar{\protect\xi}_{0}+\bar{\protect%
\xi}_{1}>1$. For Red, Blue and Green plots the value of $(\bar{\protect\xi}%
_{0},\bar{\protect\xi}_{1},\bar{\protect\xi}_{2})$ are $%
(-0.5,1.45,2.1),(-0.5,2.5,3),(-0.5,1.5,2.17)$ respectively}
\label{f2z}
\end{figure}

The deceleration parameter is defined as, 
\begin{equation}
q=-\frac{a\ddot{a}}{\dot{a}^{2}}=-\frac{\ddot{a}}{a}\frac{1}{H^{2}}\text{.}
\label{5az}
\end{equation}%
From the Friedmann equation \eqref{3iz} one can obtain, 
\begin{equation}
\frac{\ddot{a}}{a}=-\frac{1}{2\alpha }\left[ \alpha H^{2}+3H\left( \xi
_{0}+\xi _{1}H+\xi _{2}\left( \frac{\dot{H}}{H}+H\right) \right) \right] 
\text{.}  \label{5bz}
\end{equation}%
Using the equation \eqref{3kz} the deceleration parameter becomes, 
\begin{equation}
q=\frac{1}{2\alpha }\left[ \bar{\xi}_{0}\frac{H_{0}}{H}+\left( \bar{\xi}%
_{12}+\alpha \right) +\bar{\xi}_{2}\frac{\dot{H}}{H^{2}}\right] \text{.}
\label{5cz}
\end{equation}
Now, using the value of Hubble parameter given by equation \eqref{3mz} and
equations \eqref{3iz}-\eqref{3lz}, the equation \eqref{5cz} becomes, 
\begin{equation}
q(a)=\frac{1}{2\alpha +\bar{\xi}_{2}}\left[ \bar{\xi}_{1}+\alpha +\frac{\bar{%
\xi}_{0}}{a^{-\left( \frac{3\alpha +\bar{\xi}_{12}}{2\alpha +\bar{\xi}_{2}}%
\right) }\left[ 1+\frac{\bar{\xi}_{0}}{3\alpha +\bar{\xi}_{12}}\right] -%
\frac{\bar{\xi}_{0}}{3\alpha +\bar{\xi}_{12}}}\right] \text{.}  \label{5dz}
\end{equation}
Now, by applying the scale factor-redshift relation i.e., $a(t)=\frac{1%
}{1+z}$, in terms of redshift deceleration parameter is given as,

\begin{equation}\label{5ez}
q(z)=\frac{1}{2\alpha + \bar{\xi}_2} \left[\bar{\xi}_1+\alpha + \frac{\bar{\xi}_0}{(1+z)^{\left(\frac{3\alpha+ \bar{\xi}_{12}}{2\alpha+ \bar{\xi}_2}\right)} \left[ 1+ \frac{\bar{\xi}_0}{3\alpha+ \bar{\xi}_{12}}\right] - \frac{\bar{\xi}_0}{3\alpha+\bar{\xi}_{12}}} \right].
\end{equation}
The present value of deceleration parameter i.e., the value of 
$q$ at $z=0$ or $a=1$ is, 
\begin{equation}
q_{0}=\frac{\alpha +\bar{\xi}_{0}+\bar{\xi}_{1}}{2\alpha +\bar{\xi}_{2}}%
\text{.}  \label{5fz}
\end{equation}%
If the value of all bulk viscous parameters are 0, then the deceleration
parameter becomes $q=\frac{1}{2}$ which correspond to a matter dominated
decelerating Universe with null bulk viscosity. For the two sets of limiting
conditions based on the dimensionless bulk viscous parameter, the variation
of deceleration parameter with respect to redshift $z$ can be plotted as
shown in figures \eqref{f1z} and \eqref{f2z}.


\begin{figure}[H]
\centering
\includegraphics[scale=0.65]{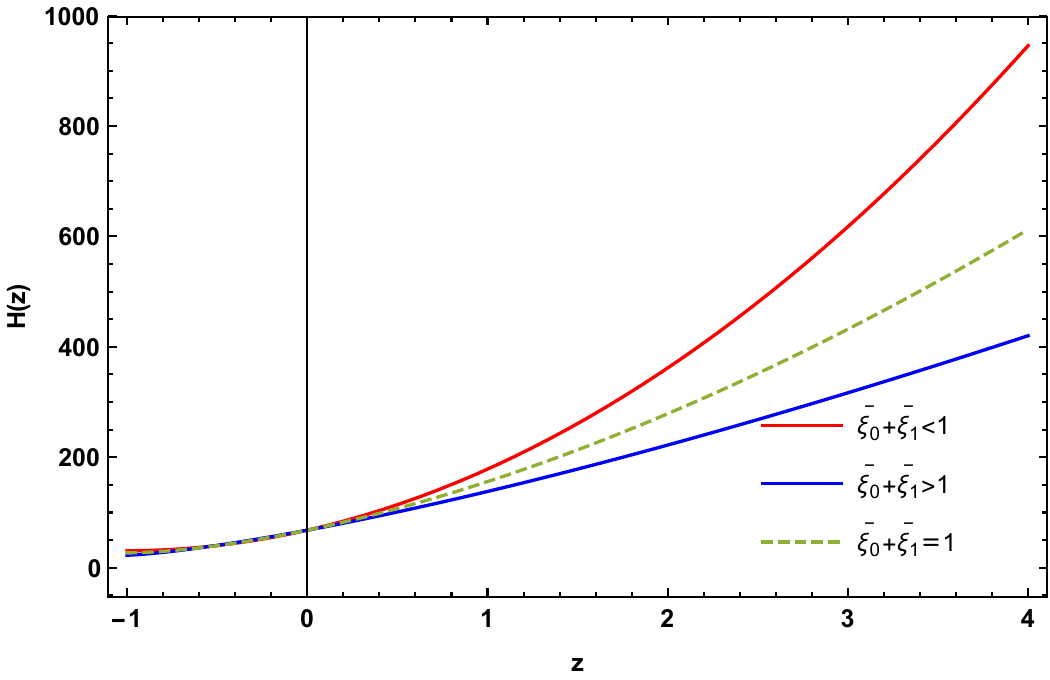}
\caption{Variation of the Hubble parameter with redshift $z$ for the first
limiting conditions $\bar{\protect\xi}_{0}>0, \bar{\protect\xi}_{12}<3, \bar{%
\protect\xi}_{2}<2, \bar{\protect\xi}_{0}+\bar{\protect\xi}_{12}<3$. Here we
took $\protect\alpha =-1$ i.e., $\bar{\protect\alpha}=1$. For Red, Blue and
Green plots the value of $(\bar{\protect\xi}_{0},\bar{\protect\xi}_{1},\bar{%
\protect\xi}_{2})$ are $(0.9,0.01,1),(0.45,0.65,1),(0.65,0.35,1)$
respectively.}
\label{f3z}
\end{figure}

\begin{figure}[H]
\centering
\includegraphics[scale=0.65]{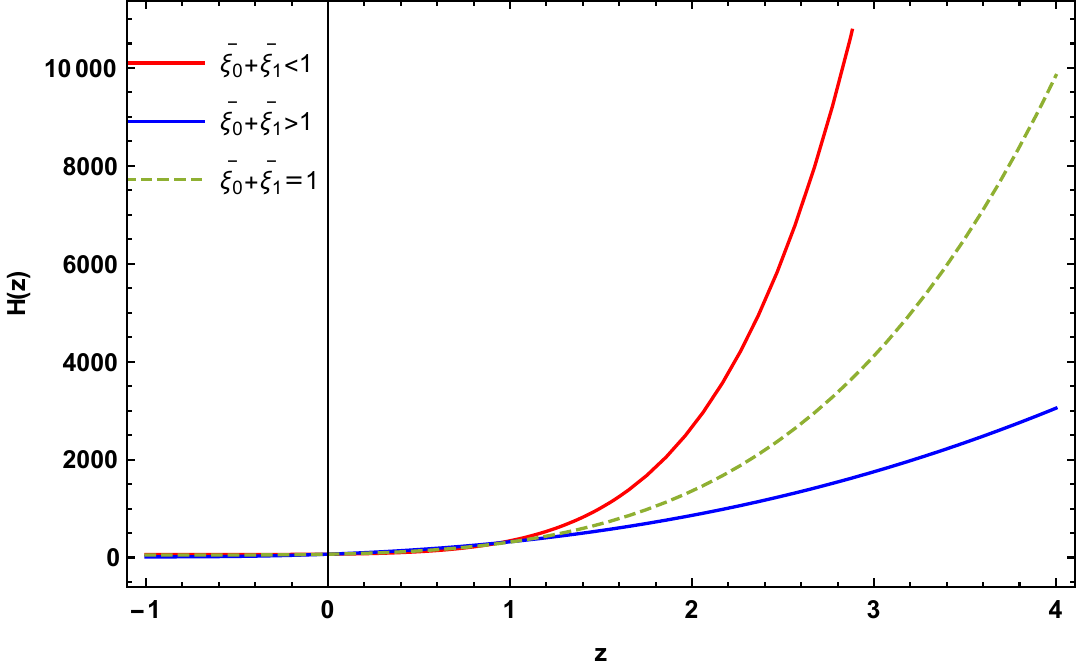}
\caption{Variation of the Hubble parameter with redshift $z$ for the second
limiting conditions $\bar{\protect\xi}_{0}<0, \bar{\protect\xi}_{12}>3, \bar{%
\protect\xi}_{2}>2, \bar{\protect\xi}_{0}+\bar{\protect\xi}_{12}>3$. Here we
took $\protect\alpha =-1$ i.e., $\bar{\protect\alpha}=1$. For Red, Blue and
Green plots the value of $(\bar{\protect\xi}_{0},\bar{\protect\xi}_{1},\bar{%
\protect\xi}_{2})$ are $(-0.5,1.45,2.1),(-0.5,2.5,3),(-0.5,1.5,2.17)$
respectively}
\label{f4z}
\end{figure}

From the above figures of $q(z)\sim z$, we can see that, only first limiting
condition with $\bar{\xi}_{0}+\bar{\xi}_{1}>1$ (blue line in figure \eqref{f1z}
shows a phase transition from early deceleration to present acceleration and
second limiting condition is not be suitable to discuss the present
observational scenario. Also the first limiting condition with $\bar{\xi}%
_{0}+\bar{\xi}_{1}=1$ and $\bar{\xi}_{0}+\bar{\xi}_{1}<1$, which can be
inferred from the following plots of Hubble parameter $H(z)\sim z$ as shown
in figures \eqref{f3z} and \eqref{f4z}.


With a discussion of the geometrical behavior of our obtained model, we
shall now try to find out suitable numerical values of the model parameters
consistent with the present observations. We only consider the first
limiting condition for our further analysis.

\section{Best fit values of model parameters from observations}\label{sec6z}

We have found an exact solution of the Einstein field equations with the
bulk viscous matter in $f(Q)$ gravity having four model parameters $\alpha $%
, $\xi _{0}$, $\xi _{1}$ and $\xi _{2}$. We must discuss the approximate
values of these model parameters describing well the present Universe
through some observational datasets. To constrain the arbitrary parameters of the model, we have used mainly two datasets namely, the Hubble
datasets containing $57$ data points and the Pantheon datasets containing $%
1048$ datasets. Furthermore, we have discussed our results together with the
BAO datasets. To constrain the model
parameters with the above discussed datasets, we have used the Python's
Scipy optimization technique. First, we have estimated the global minima corresponding to
the Hubble function in equation \eqref{3oz}. For the numerical analysis, we
employ the Python's emcee library and consider a Gaussian prior with above
estimates as means and a fixed $\sigma =1.0$ as dispersion. The idea behind
the analysis is to check the parameter space in the neighborhood of the
local minima. More about the Hubble datasets, Pantheon datasets and BAO
datasets and methodology are discussed below in some detail and finally, the
results are discussed as 2-dimensional contour plots with $1-\sigma $ \& $%
2-\sigma $ errors.

\subsection{H(z) datasets}

We are familiar with the well known cosmological principle which assumes
that on the large scale, our Universe is homogeneous and isotropic. This
principle is the backbone of modern cosmology. In the last few decades this
principle had been tested several times and have been supported by many
cosmological observations. In the study of observational cosmology, the
expansion scenario of the Universe be directly investigated by the Hubble
parameter i.e. $H=\frac{\dot{a}}{a}$ where $\dot{a}$ represents derivative
of cosmic scale factor $a$ vs cosmic time $t$. The Hubble
parameter as a function of redshift can be expressed as $H(z)$ $=-\frac{1}{%
1+z}\frac{dz}{dt}$, where $dz$ is acquired from the spectroscopic surveys
and therefore a measurement of $dt$ furnishes the model independent value of
the Hubble function. In general, there are two well known methods that are
used to measure the $H(z)$ values at some
definite redshift. The first one is the measurement of $H(z)$ from
line-of-sight BAO technique and the second one is the differential age method \cite%
{H1}-\cite{H19}. In this manuscript, we have taken an updated set of $57$
data points. In this set of $57$ Hubble data points, $31$ points measured
via the method of differential age (DA) and remaining $26$ points through
BAO technique and other methods in the redshift range given as $0.07\leqslant
z\leqslant 2.42$ \cite{sharov}. Furthermore, we have taken $H_{0}=69$ $%
Km/s/Mpc$ for the investigation. To find out the values of free
parameters $\alpha $, $\xi _{0}$, $\xi _{1}$ and $\xi _{2}$ of the model, we have taken the chi-square
function (which is equivalent to the maximum likelihood analysis) as,

\begin{equation}
\chi _{H}^{2}(\alpha ,\xi _{0},\xi _{1},\xi _{2})=\sum\limits_{i=1}^{57}%
\frac{[H_{th}(z_{i},\alpha ,\xi _{0},\xi _{1},\xi _{2})-H_{obs}(z_{i})]^{2}}{%
\sigma _{H(z_{i})}^{2}},  \label{chihzz}
\end{equation}%
where the theoretical value of Hubble parameter is represented by $H_{th}$
and the observed value by $H_{obs}$ and $\sigma _{H(z_{i})}$ represents the observed
standard error. The $57$ $H(z)$ measurements with the corresponding errors $\sigma _{H}$ from differential age ($31$
points) method and BAO and other ($26$ points) methods are tabulated in
Table \eqref{Table-1z} with references.

\begin{table}[H]
\begin{center}
\begin{tabular}{|c|c|c|c|c|c|c|c|}\hline
\multicolumn{8}{|c|}{57 points of $H(z)$ datasets} \\ \hline
\multicolumn{8}{|c|}{31 points from DA method}  \\ \hline
$z$ & $H(z)$ & $\sigma _{H}$ & Ref. & $z$ & $H(z)$ & $\sigma _{H}$ & Ref. \\ \hline
$0.070$ & $69$ & $19.6$ & \cite{H1} & $0.4783$ & $80$ & $99$ & \cite{H5} \\ \hline
$0.90$ & $69$ & $12$ & \cite{H2} & $0.480$ & $97$ & $62$ & \cite{H1} \\ \hline
$0.120$ & $68.6$ & $26.2$ & \cite{H1} & $0.593$ & $104$ & $13$ & \cite{H3} \\ \hline
$0.170$ & $83$ & $8$ & \cite{H2} & $0.6797$ & $92$ & $8$ & \cite{H3} \\ \hline
$0.1791$ & $75$ & $4$ & \cite{H3} & $0.7812$ & $105$ & $12$ & \cite{H3} \\ \hline
$0.1993$ & $75$ & $5$ & \cite{H3} & $0.8754$ & $125$ & $17$ & \cite{H3} \\ \hline
$0.200$ & $72.9$ & $29.6$ & \cite{H4} & $0.880$ & $90$ & $40$ & \cite{H1} \\ \hline
$0.270$ & $77$ & $14$ & \cite{H2} & $0.900$ & $117$ & $23$ & \cite{H2} \\ \hline 
$0.280$ & $88.8$ & $36.6$ & \cite{H4} & $1.037$ & $154$ & $20$ & \cite{H3} \\ \hline 
$0.3519$ & $83$ & $14$ & \cite{H3} & $1.300$ & $168$ & $17$ & \cite{H2} \\ \hline 
$0.3802$ & $83$ & $13.5$ & \cite{H5} & $1.363$ & $160$ & $33.6$ & \cite{H7} \\ \hline 
$0.400$ & $95$ & $17$ & \cite{H2} & $1.430$ & $177$ & $18$ & \cite{H2} \\ \hline 
$0.4004$ & $77$ & $10.2$ & \cite{H5} & $1.530$ & $140$ & $14$ & \cite{H2} \\ \hline
$0.4247$ & $87.1$ & $11.2$ & \cite{H5} & $1.750$ & $202$ & $40$ & \cite{H2} \\ \hline
$0.4497$ & $92.8$ & $12.9$ & \cite{H5} & $1.965$ & $186.5$ & $50.4$ & \cite{H7}  \\ \hline
$0.470$ & $89$ & $34$ & \cite{H6} &  &  &  &   \\ \hline
\multicolumn{8}{|c|}{26 points from BAO \& other method} \\ \hline
$z$ & $H(z)$ & $\sigma _{H}$ & Ref. & $z$ & $H(z)$ & $\sigma _{H}$ & Ref. \\ \hline
$0.24$ & $79.69$ & $2.99$ & \cite{H8} & $0.52$ & $94.35$ & $2.64$ & \cite{H10} \\ \hline
$0.30$& $81.7$ & $6.22$ & \cite{H9} & $0.56$ & $93.34$ & $2.3$ & \cite{H10} \\ \hline
$0.31$ & $78.18$ & $4.74$ & \cite{H10} & $0.57$ & $87.6$ & $7.8$ & \cite{H14} \\ \hline
$0.34$ & $83.8$ & $3.66$ & \cite{H8} & $0.57$ & $96.8$ & $3.4$ & \cite{H15} \\ \hline
$0.35$ & $82.7$ & $9.1$ & \cite{H11} & $0.59$ & $98.48$ & $3.18$ & \cite{H10} \\ \hline
$0.36$ & $79.94$ & $3.38$ & \cite{H10} & $0.60$ & $87.9$ & $6.1$ & \cite{H13} \\ \hline
$0.38$ & $81.5$ & $1.9$ & \cite{H12} & $0.61$ & $97.3$ & $2.1$ & \cite{H12} \\ \hline
$ 0.40$ & $82.04$ & $2.03$ & \cite{H10} & $0.64$ & $98.82$ & $2.98$ & \cite{H10}  \\ \hline
$0.43$ & $86.45$ & $3.97$ & \cite{H8} & $0.73$ & $97.3$ & $7.0$ & \cite{H13} \\ \hline
$0.44$ & $82.6$ & $7.8$ & \cite{H13} & $2.30$ & $224$ & $8.6$ & \cite{H16} \\ \hline
$0.44$ & $84.81$ & $1.83$ & \cite{H10} & $2.33$ & $224$ & $8$ & \cite{H17} \\ \hline
$0.48$ & $87.79$ & $2.03$ & \cite{H10} & $2.34$ & $222$ & $8.5$ & \cite{H18} \\ \hline
$0.51$ & $90.4$ & $1.9$ & \cite{H12} & $2.36$ & $226$ & $9.3$ & \cite{H19} \\ \hline
\end{tabular}
\end{center}
\caption{Table shows 57 points of $H(z)$ dataset.}\label{Table-1z}
\end{table}

\begin{figure}[H]
\centering
\includegraphics[scale=0.85]{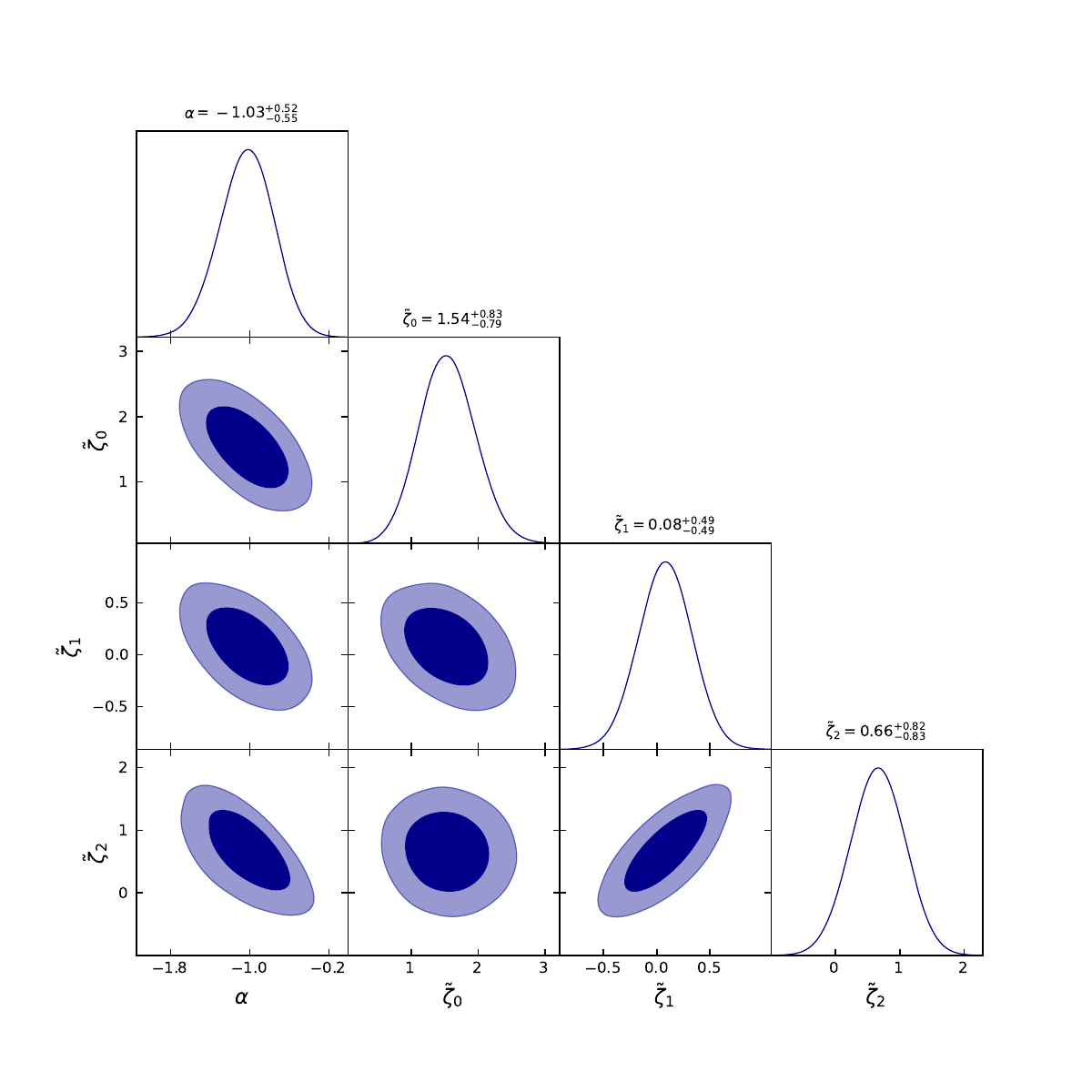}  
\caption{The plot shows the 2-d contour plots of the model parameters with $1-\protect%
\sigma $ and $2-\protect\sigma $ errors and also shows the best fit values of the model parameters $%
\protect\alpha $, $\protect\xi _{0}$, $\protect\xi _{1}$ and $\protect\xi %
_{2}$ obtained from the $57$ points of Hubble datasets.}
\label{fig:Hubblez}
\end{figure}

\begin{figure}[H]
\centering
\includegraphics[scale=0.6]{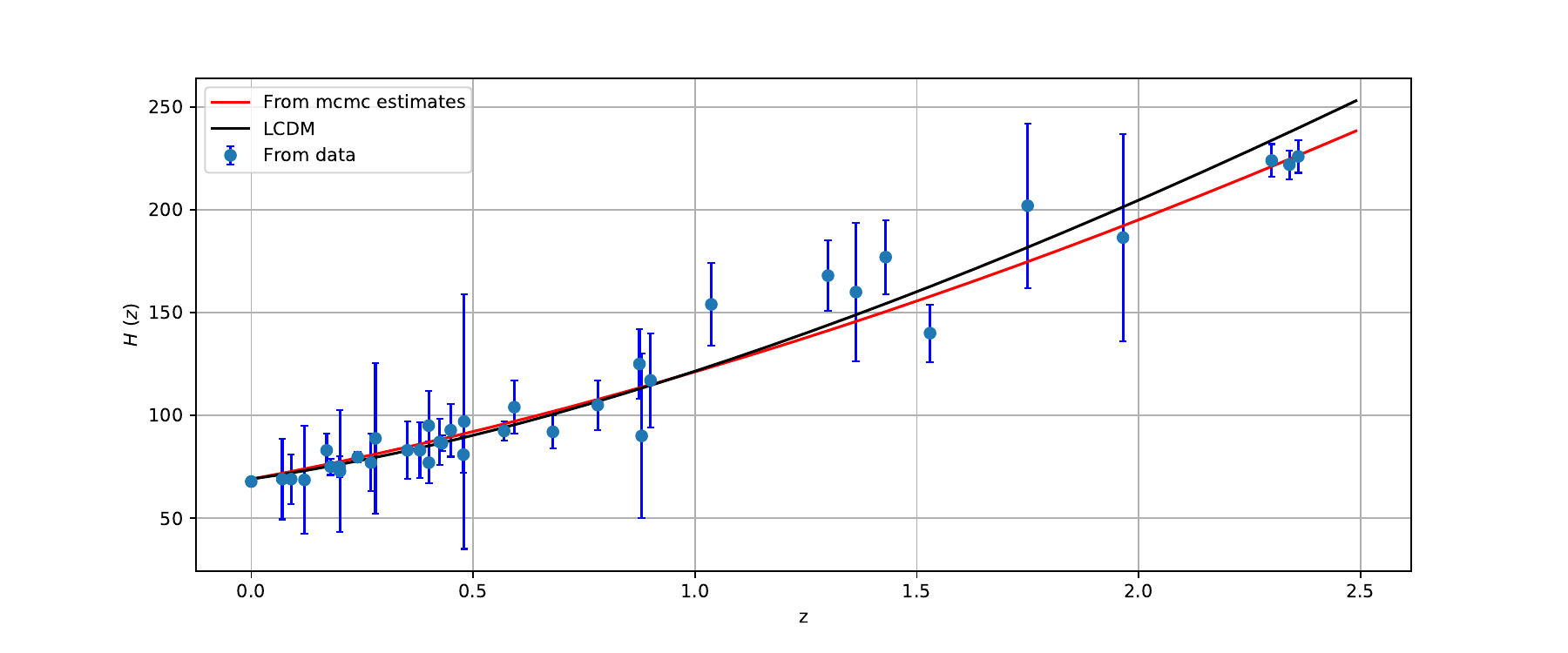}  
\caption{The plot shows the plot of Hubble function $H(z)$ vs.
redshift $z$ for our model shown in red line which shows nice fit to the $57$
points of the Hubble datasets shown in dots with it's error bars and also compared to the $\Lambda$CDM model shown in black solid line with $\Omega_{m0}=0.3$ \& $\Omega_{\Lambda 0}=0.7$.}
\label{fig:final-hzz}
\end{figure}

Using the above datasets, we have estimated the values of the free parameters $\alpha $, $\xi _{0}$, $\xi _{1}$ and $\xi
_{2}$ as and is shown in the following plot \eqref{fig:Hubblez} as 2-d contour
sub-plots with $1-\sigma $ \& $2-\sigma $ errors. The best fit values are
obtained as $\alpha =-1.03_{-0.55}^{+0.52}$, $\xi _{0}=1.54_{-0.79}^{+0.83}$%
, $\xi _{1}=0.08_{-0.49}^{+0.49}$ and $\xi _{2}=0.66_{-0.83}^{+0.82}$ with
the $57$ points of Hubble datasets as given in Table-1. 
Also, we have shown
the error bar plot for the discussed Hubble datasets and is shown in the
following plot figure \eqref{fig:final-hzz} together with our obtained model compared
with the $\Lambda $CDM model (with $\Omega _{m0}=0.3$ and $\Omega _{\Lambda
0}=0.7$). The plot shows nice fit of our model to the observational Hubble
datasets.

\subsection{Pantheon datasets}

Initially, the observational studies on supernovae of the golden sample of $%
50$ points of type \textit{Ia} suggested that our Universe is in an
accelerating phase of expansion. After the result, the studies on more and
more samples of supernovae datasets increased during the past two decades.
Recently, the latest sample of SN Ia datasets are
released containing $1048$ data points. In this article, we have used this
set of datasets known as Pantheon datasets \cite{R18} with $1048$ samples of
spectroscopically confirmed SN Ia covering the range in the
redshift range $0.01<z<2.26$. In the redshift range $0<z_{i}\leq 1.41$,
these data points gives the estimation of the distance moduli $\mu _{i}=\mu
_{i}^{obs}$. Here, we fit our model parameters of the obtained model,
comparing the theoretical $\mu _{i}^{th}$ value and the observed $\mu
_{i}^{obs}$ value of the distance modulus. The distance moduli which are the
logarithms given as $\mu _{i}^{th}=\mu (D_{L})=m-M=5\log _{10}(D_{L})+\mu
_{0}$, where $m$ and $M$ represents apparent and absolute magnitudes and $%
\mu _{0}=5\log \left( H_{0}^{-1}/Mpc\right) +25$ is the marginalized
nuisance parameter. The luminosity distance is taken to be, 
\begin{eqnarray*}
D_{l}(z) &=&\frac{c(1+z)}{H_{0}}S_{k}\left( H_{0}\int_{0}^{z}\frac{1}{%
H(z^{\ast })}dz^{\ast }\right) , \\
\text{where }S_{k}(x) &=&\left\{ 
\begin{array}{c}
\sinh (x\sqrt{\Omega _{k}})/\Omega _{k}\text{, }\Omega _{k}>0 \\ 
x\text{, \ \ \ \ \ \ \ \ \ \ \ \ \ \ \ \ \ \ \ \ \ \ \ }\Omega _{k}=0 \\ 
\sin x\sqrt{\left\vert \Omega _{k}\right\vert })/\left\vert \Omega
_{k}\right\vert \text{, }\Omega _{k}<0%
\end{array}%
\right. .
\end{eqnarray*}%
Here, $\Omega _{k}=0$ (flat spacetime). We have estimated distance $%
D_{L}(z)$ and corresponding chi square function that measures difference
between predictions of our model and the SN Ia observational data.
The $\chi _{SN}^{2}$ function for the Pantheon datasets is taken to be,

\begin{equation}
\chi _{SN}^{2}(\mu _{0},\alpha ,\xi _{0},\xi _{1},\xi
_{2})=\sum\limits_{i=1}^{1048}\frac{[\mu ^{th}(\mu _{0},z_{i},\alpha ,\xi
_{0},\xi _{1},\xi _{2})-\mu ^{obs}(z_{i})]^{2}}{\sigma _{\mu (z_{i})}^{2}},
\label{chisnz}
\end{equation}%
$\sigma _{\mu (z_{i})}^{2}$ is the observed standard error.
After marginalizing $\mu _{0}$, the chi square function is written as,

$\qquad \qquad \qquad \qquad  \chi
_{SN}^{2}(\alpha ,\xi _{0},\xi _{1},\xi _{2})=A(\alpha ,\xi _{0},\xi
_{1},\xi _{2})-[B(\alpha ,\xi _{0},\xi _{1},\xi _{2})]^{2}/C(\alpha ,\xi
_{0},\xi _{1},\xi _{2})$

where,

$\qquad \qquad \qquad \qquad  A(\alpha ,\xi
_{0},\xi _{1},\xi _{2})=\sum\limits_{i=1}^{1048}\frac{[\mu ^{th}(\mu
_{0}=0,z_{i},\alpha ,\xi _{0},\xi _{1},\xi _{2})-\mu ^{obs}(z_{i})]^{2}}{%
\sigma _{\mu (z_{i})}^{2}},$

$\qquad \qquad \qquad \qquad  B(\alpha ,\xi
_{0},\xi _{1},\xi _{2})=\sum\limits_{i=1}^{1048}\frac{[\mu ^{th}(\mu
_{0}=0,z_{i},\alpha ,\xi _{0},\xi _{1},\xi _{2})-\mu ^{obs}(z_{i})]^{2}}{%
\sigma _{\mu (z_{i})}^{2}},$

$\qquad \qquad \qquad \qquad  C(\alpha ,\xi
_{0},\xi _{1},\xi _{2})=\sum\limits_{i=1}^{1048}\frac{1}{\sigma _{\mu
(z_{i})}^{2}}$.

\begin{figure}[H]
\centering
\includegraphics[scale=0.87]{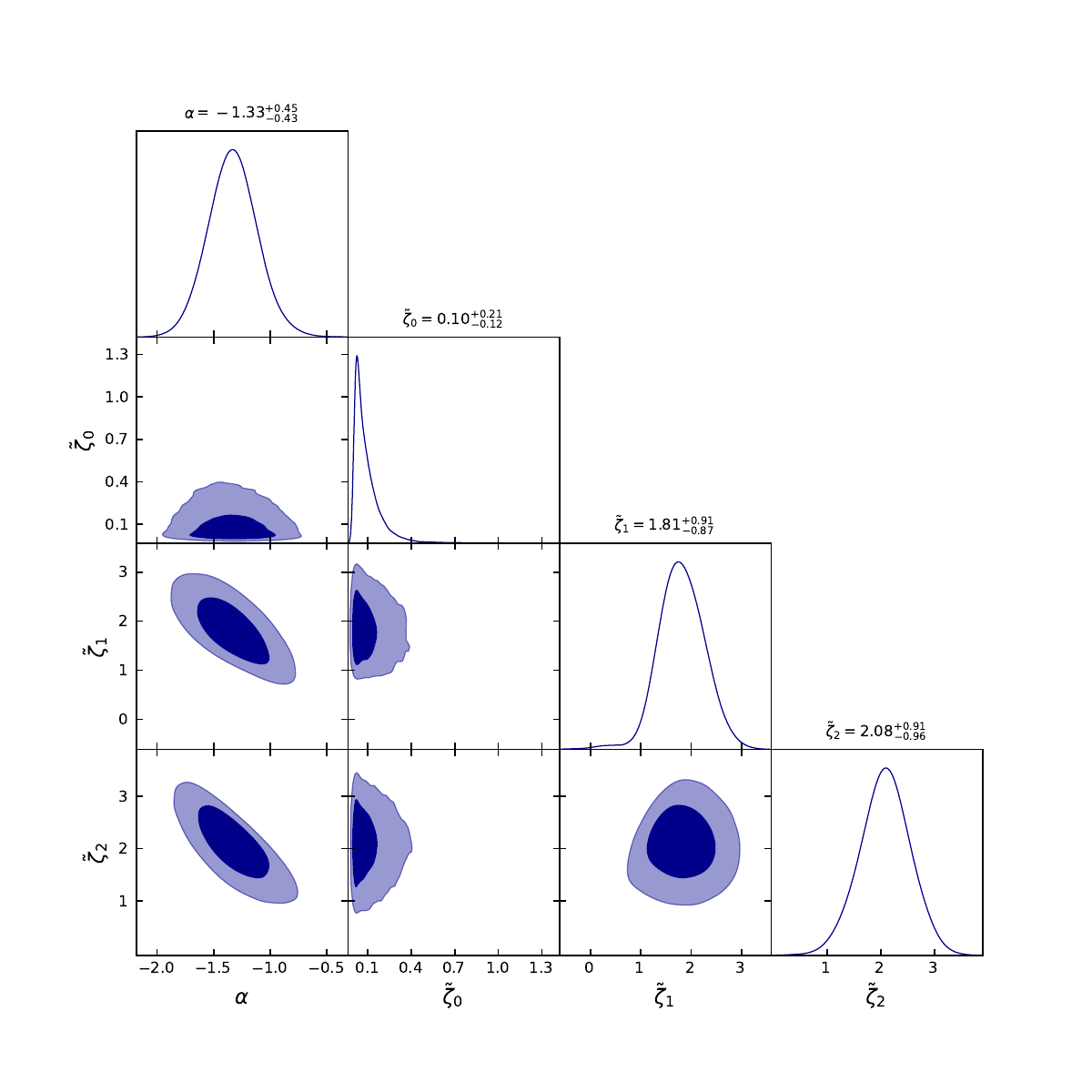}  
\caption{The plot shows the best fit values of the model parameters $%
\protect\alpha $, $\protect\xi _{0}$, $\protect\xi _{1}$ and $\protect\xi %
_{2}$ obtained w.r.t to the $1048$ points of Pantheon datasets at $1-\protect%
\sigma $ and $2-\protect\sigma $ confidence level.}
\label{fig:Pantheonz}
\end{figure}

\begin{figure}[H]
\centering
\includegraphics[scale=0.6]{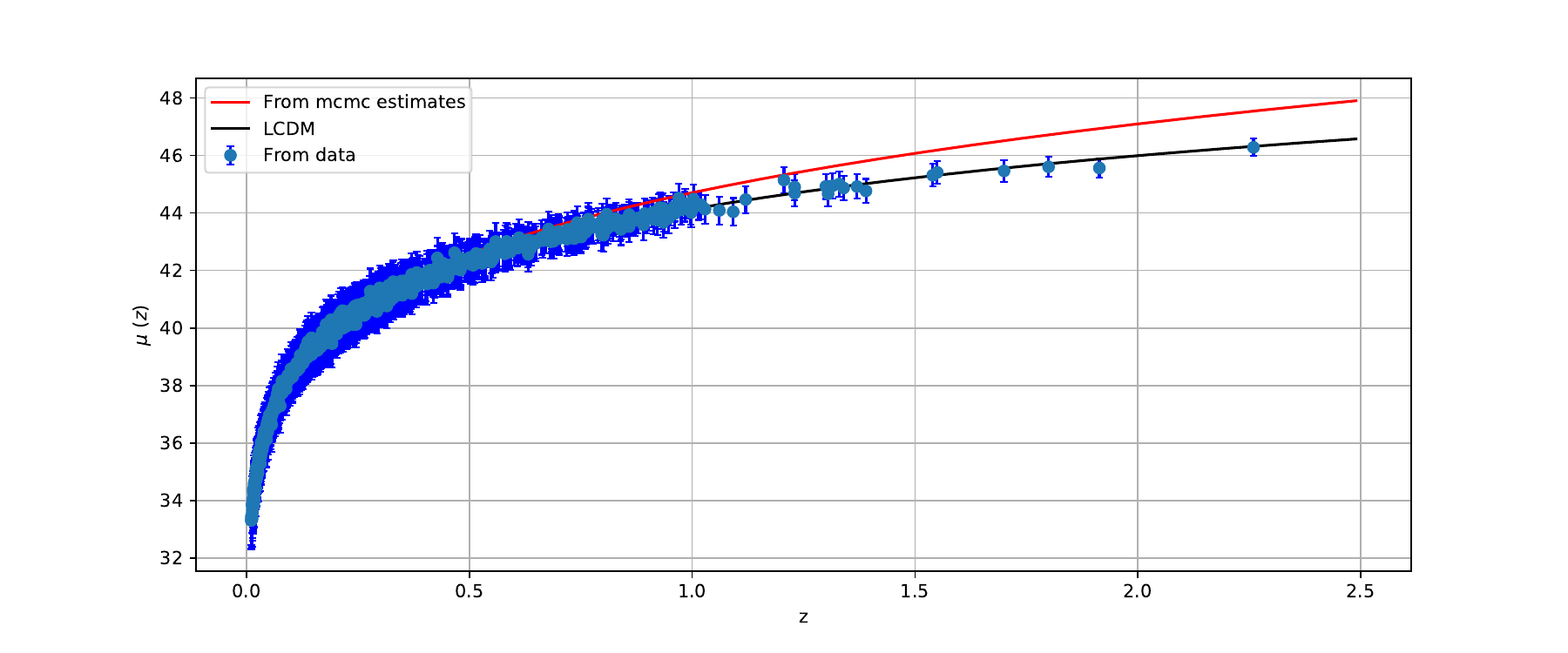}  
\caption{{}{}The plot shows the plot of distance modulus $\protect\mu (z)$
vs. redshift $z$ for our model shown in red line which shows nice fit to the 
$1048$ points of the Pantheon datasets shown in dots with it's error bars.}
\label{fig:final-muzz}
\end{figure}

Using the above Pantheon datasets, we have obtained the values of
free parameters $\alpha $, $\xi _{0}$, $\xi _{1}$ and $\xi _{2}$ and is
shown in the following plot figure \eqref{fig:Pantheonz} as 2-d contour sub-plots with 
$1-\sigma $ \& $2-\sigma $ errors. The best fit values are obtained as $%
\alpha =-1.33_{-0.43}^{+0.45}$, $\xi _{0}=0.10_{-0.12}^{+0.21}$, $\xi
_{1}=1.81_{-0.87}^{+0.91}$ and $\xi _{2}=2.08_{-0.96}^{+0.91}$ with $1048$
points of Pantheon datasets. Also, we have shown the error bar plot for the
discussed Pantheon datasets and is shown in the following plot figure \eqref{fig:final-muzz} together with our obtained model compared with the $\Lambda $%
CDM model (with $\Omega _{m0}=0.3$ and $\Omega _{\Lambda 0}=0.7$). The plot
shows nice fit of our model to the observational Pantheon datasets.

\subsection{BAO datasets}

In the study of the early Universe baryons, photons and dark
matter come into picture and act as a single fluid coupled tightly through
the Thompson scattering but do not collapse under gravity and oscillate due
to the large pressure of photons. BAO is an
analysis that discusses these oscillations in the early Universe. The
characteristic scale of BAO is governed by the sound horizon $r_{s}$ at the photon decoupling epoch $z_{\ast }$ and is given by the
following relation, 
\begin{equation*}
r_{s}(z_{\ast })=\frac{c}{\sqrt{3}}\int_{0}^{\frac{1}{1+z_{\ast }}}\frac{da}{%
a^{2}H(a)\sqrt{1+(3\Omega _{0b}/4\Omega _{0\gamma })a}},
\end{equation*}%
where $\Omega _{0b}$ and $\Omega _{0\gamma }$
are respectively referred to baryon density and photon density at present
time.

The angular diameter distance $D_{A}$ and the Hubble
expansion rate $H$ as a function of $z$ are derived
using the BAO sound horizon scale. If the measured angular separation of the
BAO feature is denoted by $\triangle \theta $ in the 2 point
correlation function of the galaxy distribution on the sky and if the
measured redshift separation of the BAO feature is denoted by $\triangle z$ in same the 2 point correlation function along the line of sight
then we have the relation, $\triangle \theta =\frac{r_{s}}{d_{A}(z)}$%
 where $d_{A}(z)=\int_{0}^{z}\frac{dz^{\prime }}{H(z^{\prime })}$%
 and $\triangle z=H(z)r_{s}$. Here, in this chapter, a simple
BAO datasets of six points for $d_{A}(z_{\ast })/D_{V}(z_{BAO})$
is considered from the references \cite{BAO1, W.J., BAO3, BAO4, D.J., BAO6},
where the photon decoupling redshift is taken as $z_{\ast }\approx 1091$%
 and $d_{A}(z)$ is the co-moving angular diameter
distance together with the dilation scale $D_{V}(z)=\left(
d_{A}(z)^{2}z/H(z)\right) ^{1/3}$. The following Table \eqref{Table-2z} shows the
six points of the BAO datasets,

\begin{table}[H]
\begin{center}
\begin{tabular}{|c|c|c|c|c|c|c|}
\hline
\multicolumn{7}{|c|}{Values of $d_{A}(z_{\ast })/D_{V}(z_{BAO})$
for distinct values of $z_{BAO}$} \\ \hline
$z_{BAO}$ & $0.106$ & $0.2$ & $0.35$ & $0.44$ & $0.6$ & $0.73$ \\ \hline
$\frac{d_{A}(z_{\ast })}{D_{V}(z_{BAO})}$ & $30.95\pm 1.46$ & $17.55\pm 0.60$
& $10.11\pm 0.37$ & $8.44\pm 0.67$ & $6.69\pm 0.33$ & $5.45\pm 0.31$ \\ 
\hline
\end{tabular}%
\end{center}  
\caption{Table shows the values of $d_{A}(z_{\ast })/D_{V}(z_{BAO})$
for distinct values of $z_{BAO}$.}
\label{Table-2z}
\end{table}

Also, the chi square function for BAO is given by \cite{BAO6}, 
\begin{equation}
\chi _{BAO}^{2}=X^{T}C^{-1}X\,,  \label{chibaoz}
\end{equation}%
where
\begin{equation*}
X=\left( 
\begin{array}{c}
\frac{d_{A}(z_{\star })}{D_{V}(0.106)}-30.95 \\ 
\frac{d_{A}(z_{\star })}{D_{V}(0.2)}-17.55 \\ 
\frac{d_{A}(z_{\star })}{D_{V}(0.35)}-10.11 \\ 
\frac{d_{A}(z_{\star })}{D_{V}(0.44)}-8.44 \\ 
\frac{d_{A}(z_{\star })}{D_{V}(0.6)}-6.69 \\ 
\frac{d_{A}(z_{\star })}{D_{V}(0.73)}-5.45%
\end{array}%
\right) \,,
\end{equation*}
and the inverse covariance matrix $C^{-1}$ is defined in 
\cite{BAO6}. 

\begin{equation*}
C^{-1}=\left( 
\begin{array}{cccccc}
0.48435 & -0.101383 & -0.164945 & -0.0305703 & -0.097874 & -0.106738 \\ 
-0.101383 & 3.2882 & -2.45497 & -0.0787898 & -0.252254 & -0.2751 \\ 
-0.164945 & -2.454987 & 9.55916 & -0.128187 & -0.410404 & -0.447574 \\ 
-0.0305703 & -0.0787898 & -0.128187 & 2.78728 & -2.75632 & 1.16437 \\ 
-0.097874 & -0.252254 & -0.410404 & -2.75632 & 14.9245 & -7.32441 \\ 
-0.106738 & -0.2751 & -0.447574 & 1.16437 & -7.32441 & 14.5022%
\end{array}%
\right) \,.
\end{equation*}

\begin{figure}[H]
\centering
\includegraphics[scale=0.85]{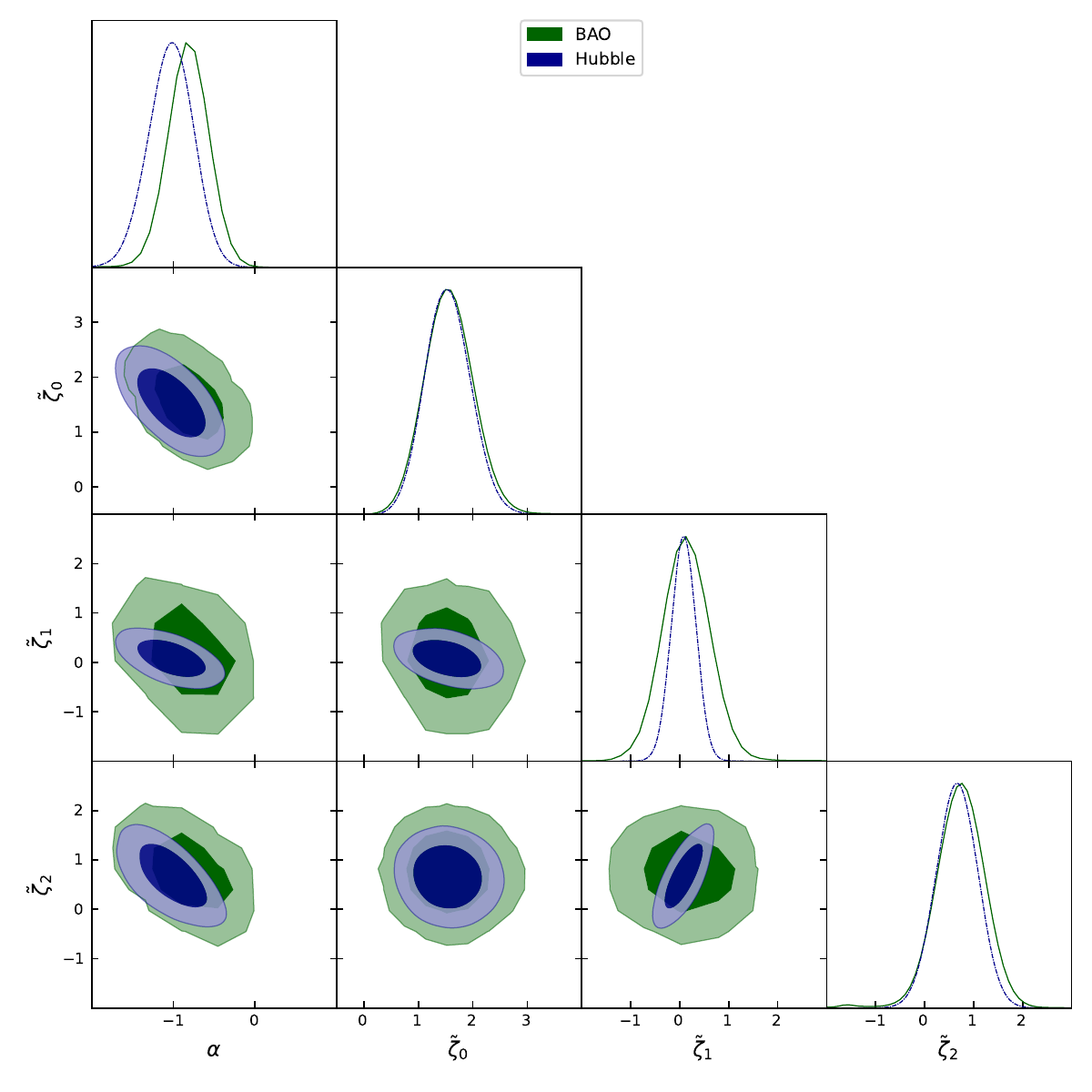}  
\caption{The plot shows the 2-d contour plots of the model parameters with $1-\protect%
\sigma $ and $2-\protect\sigma $ errors and also shows the best fit values of the model parameters $%
\protect\alpha $, $\protect\xi _{0}$, $\protect\xi _{1}$ and $\protect\xi %
_{2}$ obtained from the $57$ points of Hubble datasets together with six points of BAO datasets.}
\label{fig:Hubble_BAOz}
\end{figure}

\begin{figure}[H]
\centering
\includegraphics[scale=0.85]{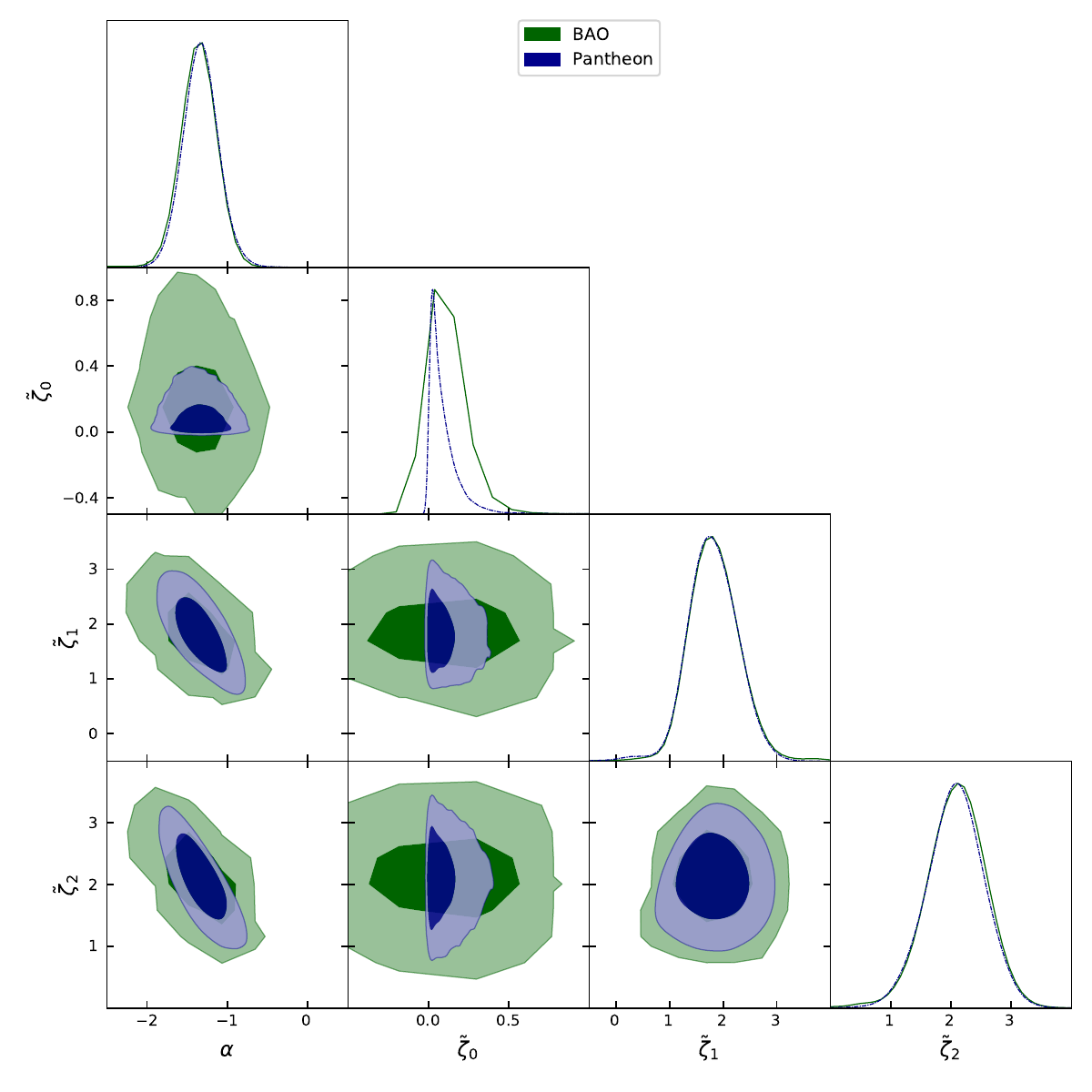}  
\caption{The plot shows the best fit values of the model parameters $%
\protect\alpha $, $\protect\xi _{0}$, $\protect\xi _{1}$ and $\protect\xi %
_{2}$ obtained w.r.t to the $1048$ points of Pantheon datasets together with six points of BAO datasets at $1-\protect%
\sigma $ and $2-\protect\sigma $ confidence level.}
\label{fig:Pantheon_BAOz}
\end{figure}

Including these six data points of the BAO datasets with the Hubble
datasets and combine the above results of Hubble constrained values, we
obtain the values of the model parameters as, $\alpha
=-1.06_{-0.82}^{+0.34} $, $\xi _{0}=2.25_{-1.7}^{+0.37}$, $%
\xi _{1}=-0.08_{-0.96}^{+1.1}$ and $\xi _{2}=0.7_{-1.1}^{+1.1}$%
. Similarly, including these six data points of the BAO datasets
with the Pantheon datasets and combine the above results of Pantheon
constrained values, we obtain the values of the model parameters as, $%
\alpha =-1.65_{-0.25}^{+0.85}$, $\xi _{0}=0.86_{-1.1}^{+0.29}$%
, $\xi _{1}=-1.90_{-0.99}^{+0.97}$ and $\xi
_{2}=1.93_{-0.91}^{+1.2}$. The combined results are shown as 2-d
contour sub-plots with $1-\sigma $ \& $2-\sigma $ errors
in the following plot figures \eqref{fig:Hubble_BAOz} and \eqref{fig:Pantheon_BAOz}.

\section{Statefinder diagnostic}\label{sec7z}

It is well-known that the deceleration parameter $q$ and the Hubble
parameter $H$ are the oldest geometric variable. During the last few
decades, plenty of DE (Dark Energy) models have been proposed and the
remarkable growth in the precision of observational data both motivate us to
go beyond these two parameters. In this direction, a new pair of geometrical
quantities have been proposed by V. Sahni et al. \cite{V.Sahni} called
statefinder diagnostic parameters $(r,s) $. The state finder parameters
investigate the expansion dynamics through second and third derivatives of
the cosmic scale factor which is a natural succeeding step beyond the
parameters $H $ and $q $. These parameters are defined as follows

\begin{equation}
r=\frac{\dddot{a}}{aH^{3}}
\end{equation}%
and 
\begin{equation}
s=\frac{r-1}{3(q-\frac{1}{2})}.
\end{equation}

\begin{figure}[H]
\centering
\includegraphics[scale=0.45]{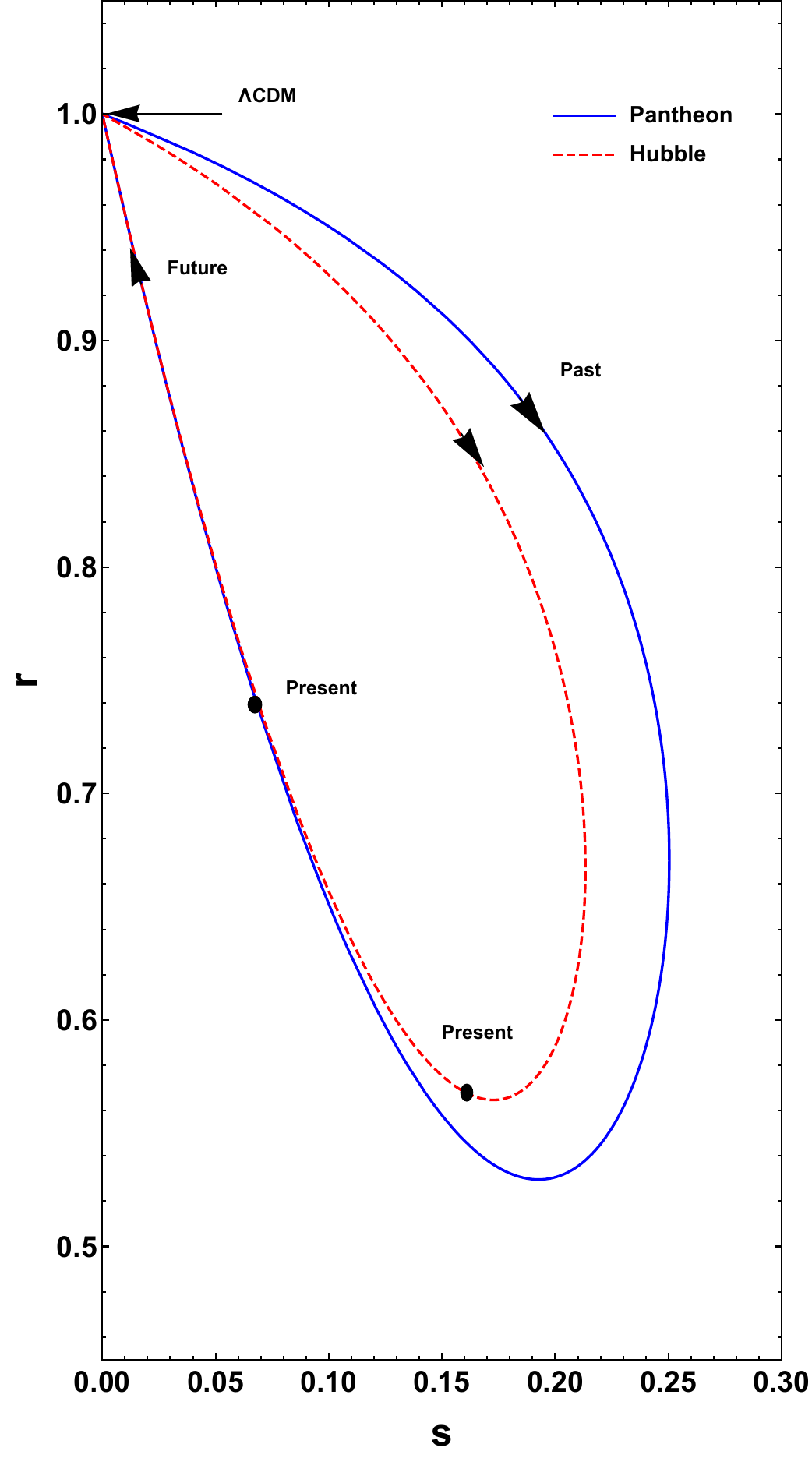}
\caption{The evolution trajectories of the given model in $s-r$ plane
for the free parameter values obtained by the Hubble
and Pantheon datasets. }
\label{r-sz}
\end{figure}

\begin{figure}[H]
\centering
\includegraphics[scale=0.55]{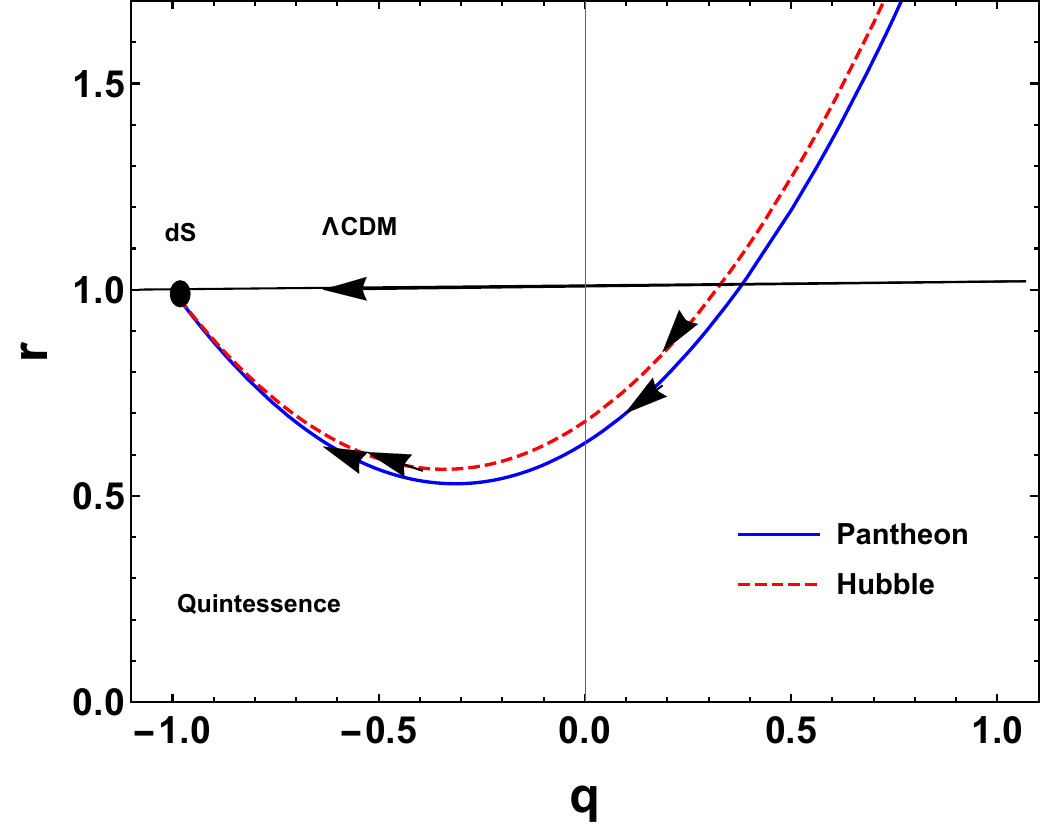}
\caption{The evolution trajectories of the given model in $q-r$ plane
for the free parameter values obtained by the Hubble
and Pantheon datasets.}
\label{r-qz}
\end{figure}

\begin{figure}[H]
\centering
\includegraphics[scale=0.45]{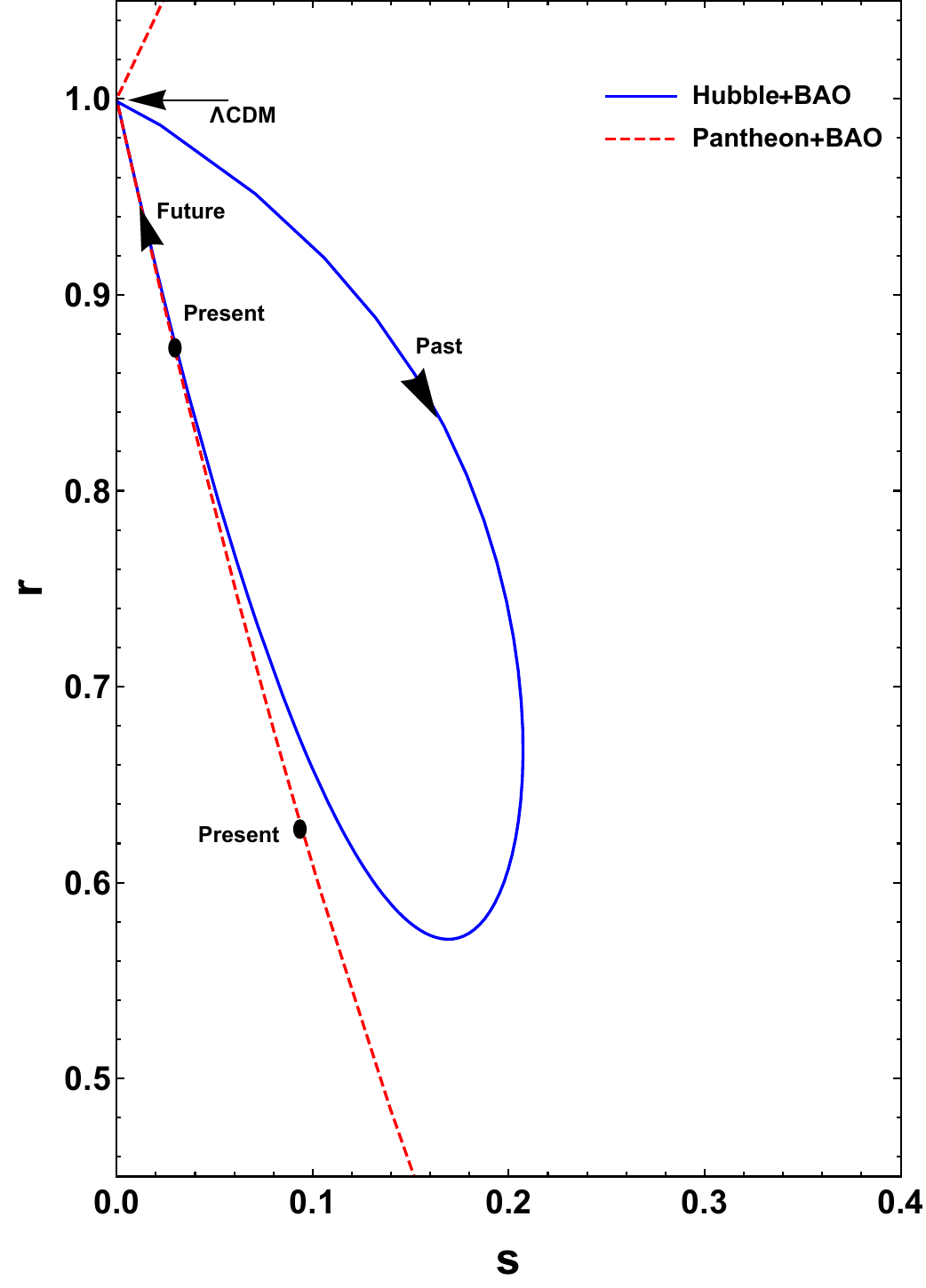}
\caption{The evolution trajectories of the given model in $s-r$ plane
for the free parameter values obtained by the Hubble
and Pantheon datasets together with BAO datasets. }
\label{r-s-BAOz}
\end{figure}

\begin{figure}[H]
\centering
\includegraphics[scale=0.38]{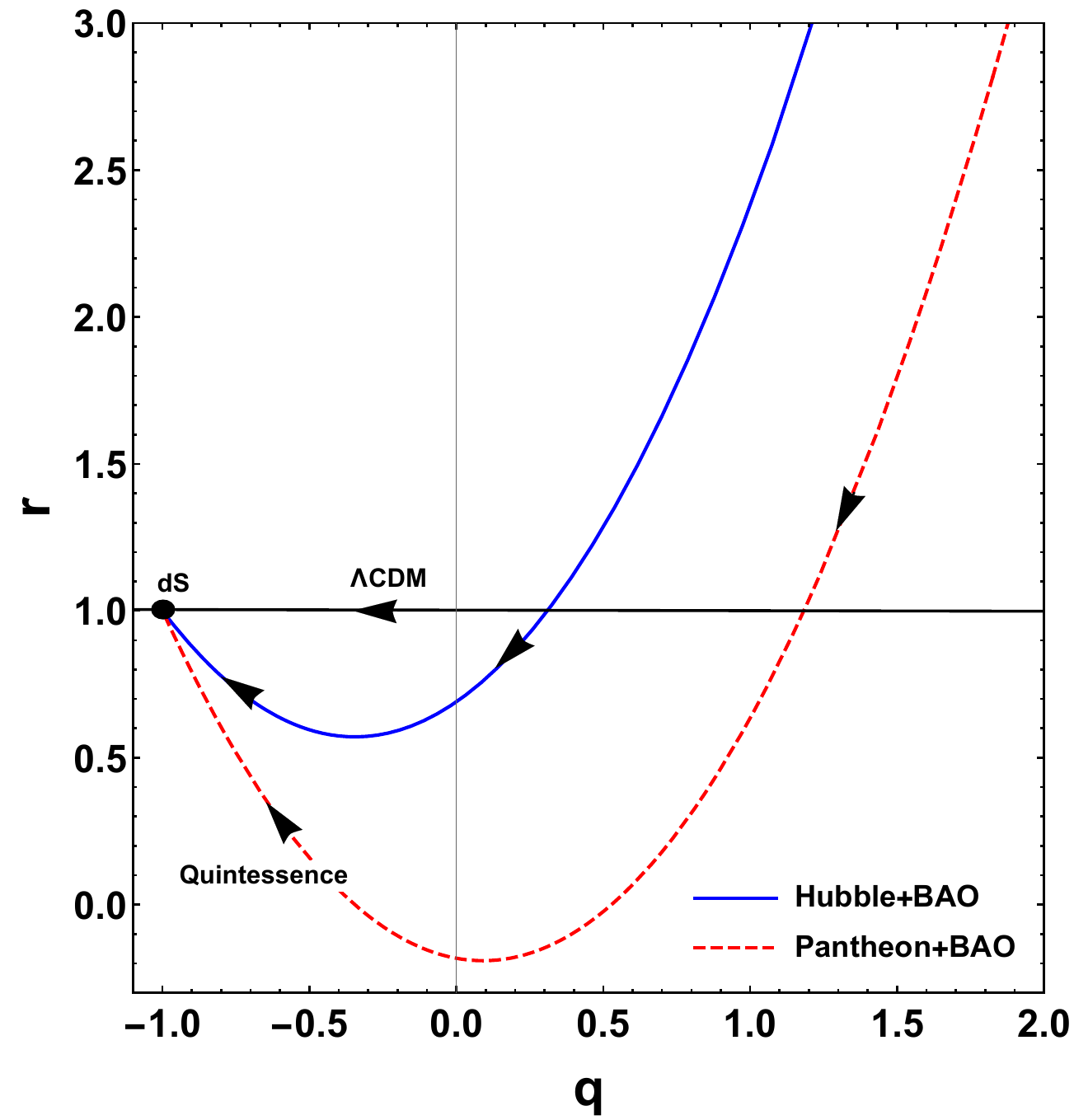}
\caption{The evolution trajectories of the given model in $q-r$ plane
for the free parameter values obtained by the Hubble
and Pantheon datasets together with BAO datasets.}
\label{r-q-BAOz}
\end{figure}

The fixed point $(s,r)=(0,1)$ in the $s-r$ diagram \eqref{r-sz} shows the
spatially flat $\Lambda $CDM model and $(q,r)=(-1,1)$ shows the de Sitter
point in Figure \eqref{r-qz}. We plot the $s-r$ and $q-r$ diagram for the values
of $\alpha $, $\xi _{0}$, $\xi _{1}$ and $\xi _{2}$ constrained by the
Hubble and the Pantheon datasets. Corresponding to the Hubble and the
Pantheon datasets, the present values of $(s,r)$ parameter are $%
(0.159,0.568) $ and $(0.065,0.748)$ respectively. Furthermore, we
have also plotted the $s-r$ and $q-r$ diagrams for the
other set of values of the model parameters $\alpha $, $\xi _{0}$%
, $\xi _{1}$ and $\xi _{2}$ as obtained by the
combined results of BAO datasets with the Hubble and the Pantheon datasets
and are shown in the following plot figures \eqref{r-s-BAOz} and \eqref{r-q-BAOz}. Corresponding to the combined results of BAO datasets with Hubble and the
Pantheon datasets, the present values of $(s,r)$ parameter are $%
(0.030,0.872) $ and $(0.096,0.621)$ respectively.

The statefinder diagnostic can differentiate the variety of dark energy
models like quintessence, the Chaplygin gas, braneworld models, etc., see the
references \cite{U.Alam,Gorini,Zim}. The departure of our bulk viscous model
from this fixed point establishes the distance of the given model from $%
\Lambda $CDM model. In present epoch the given model lie in Quintessence
region $(s>0,r<1)$. We can observe that the trajectories of $s-r$ diagram
will pass through the $\Lambda $CDM fixed point in the future. Thus the
statefinder diagnostic successfully shows that the given model is different
from other models of dark energy.

\section{Conclusions}\label{sec8z}

In this chapter, we analyzed the evolution of FLRW Universe dominated with
non-relativistic bulk viscous matter, where the time-dependent bulk
viscosity has the form $\xi =\xi _{0}+\xi _{1}H+\xi _{2}\left( \frac{\dot{H}%
}{H}+H\right) $. From the cosmic scale factor we found that in case of first
limiting conditions the deceleration parameter shows the transition from
deceleration to acceleration phase in past if \ $\bar{\xi}_{0}+\bar{\xi}_{1}>%
\bar{\alpha}$, at present if $\bar{\xi}_{0}+\bar{\xi}_{1}=\bar{\alpha}$ and
in the future if $\bar{\xi}_{0}+\bar{\xi}_{1}<\bar{\alpha}$. For second
limiting conditions transition occur in the past if $\bar{\xi}_{0}+\bar{\xi}%
_{1}<\bar{\alpha}$, at present if $\bar{\xi}_{0}+\bar{\xi}_{1}=\bar{\alpha}$
and in the future if $\bar{\xi}_{0}+\bar{\xi}_{1}>\bar{\alpha}$. While in
the absence of bulk viscosity i.e. $\xi _{0}=\xi _{1}=\xi _{2}=0$, the
deceleration parameter becomes $q=\frac{1}{2}$. Hence, to describe the
late-time acceleration of the expanding Universe without invoking any dark
energy component, the cosmic fluid with bulk viscosity is the most viable
candidate. The second limiting condition is not suitable for present
observational scenario, so we have considered the first limiting condition
for our analysis. Further, for constraining the model and bulk viscous
parameter we have used Hubble data and Pantheon datasets. Hence, from the
Hubble datasets, we have the best fit ranges for the model parameters are $%
\alpha =-1.03_{-0.55}^{+0.52}$, $\xi _{0}=1.54_{-0.79}^{+0.83}$, $\xi
_{1}=0.08_{-0.49}^{+0.49}$ and $\xi _{2}=0.66_{-0.83}^{+0.82}$ and from the
Pantheon datasets, we have $\alpha =-1.33_{-0.43}^{+0.45}$, $\xi
_{0}=0.10_{-0.12}^{+0.21}$, $\xi _{1}=1.81_{-0.87}^{+0.91}$ and $\xi
_{2}=2.08_{-0.96}^{+0.91}$. Furthermore, including the six data
points of the BAO datasets with the Hubble datasets and combine the results
of Hubble constrained values, we obtain the values of the model parameters
as, $\alpha =-1.06_{-0.82}^{+0.34}$, $\xi _{0}=2.25_{-1.7}^{+0.37}$%
, $\xi _{1}=0.08_{-0.96}^{+1.1}$ and $\xi
_{2}=0.7_{-1.1}^{+1.1}$. Similarly, including the six data points of
the BAO datasets with the Pantheon datasets and combine the results of
Pantheon constrained values, we obtain the values of the model parameters
as, $\alpha =-1.65_{-0.25}^{+0.85}$, $\xi _{0}=0.86_{-1.1}^{+0.29}$%
, $\xi _{1}=1.90_{-0.99}^{+0.97}$ and $\xi
_{2}=1.93_{-0.91}^{+1.2}$. For the above set of values of the model
parameters obtained are then used to plot the statefinder diagnostics.
 Finally, we conclude that the present bulk viscous model has been departed
from $\Lambda $CDM point, in the present scenario it lie in quintessence
region and it will again pass through the $\Lambda $CDM fixed point and
hence our model is different from other models of the dark energy. We
conclude that the bulk viscous theory can be considered as an alternate
theory to describe the late time acceleration of the Universe.


%% file: Chapters/Chapter3.tex

\chapter{Cosmic acceleration with bulk viscous matter in non-linear $f(Q)$ cosmology} 

\label{Chapter3} 

\lhead{Chapter 3. \emph{Cosmic acceleration with bulk viscous matter in non-linear $f(Q)$ cosmology}} 


\vspace{10 cm}
* The work, in this chapter, is covered by the following publication:

\textit{Bulk viscous fluid in symmetric teleparallel cosmology: theory versus experiment}, Universe \textbf{9(1)}, 12 (2023).
\clearpage
In this chapter, we consider a FLRW cosmological model dominated by bulk viscous cosmic fluid in $f(Q)$ gravity with the non-linear functional form $f(Q)=\alpha Q^n$, where $\alpha$ and $n$ are free  parameters of the model. We constrain our model with the Pantheon supernovae datasets of 1048 data points, Hubble datasets of 31 data points and BAO datasets consisting of six points. We present the evolution of deceleration parameter with redshift and it properly predicts a transition from decelerated to accelerated phases of the Universe expansion. Also, we present the evolution of density, bulk viscous pressure and the effective EoS parameter with redshift. We find that bulk viscosity in a cosmic fluid is a valid candidate to acquire the negative pressure to drive the cosmic expansion efficiently. We also examine the behavior of different energy conditions to test the viability of our cosmological $f(Q)$ model. Furthermore, the statefinder diagnostics are also investigated in order to distinguish among different dark energy models.

\section{Introduction}\label{sec1}

In this chapter we will work with the $f(Q)$ theory of gravity \cite{R51} with a non-linear functional form. The non-metricity formulation has been discussed earlier by Hehl and Ne'eman (see References \cite{R48,H2,H4}). The symmetric teleparallel gravity is proven in the so-called coincidence gauge by imposing that the connection is symmetric \cite{R50}. Weyl geometry is also observed to be a particular example of Weyl Cartan geometry in which torsion disappears. The non-metricity is interpreted as a massless spin 3-field in the case of symmetric connections \cite{N1,N2}. Also, it is noted in the literature that due to the appearance of non-metricity, the light cone structure is not preserved during parallel transport \cite{H5}. Further, fermions are an issue in TEGR because they couple to the axial contorsion of the Weitzenbock connection. This difficulty is eliminated in STEGR since Dirac fermions only couple to the completely antisymmetric component of the affine connection and is unaffected by any disformation piece. Although recently proposed, the $f(Q)$ gravity theory already presents some interesting and valuable applications in the literature \cite{Ferreira/2022,FDaa,FDab,FDac,FDad,FDae}. The first cosmological solutions in $f(Q)$ gravity appear in References \cite{R49,khyllep/2021}.

Here, we are going to consider $f(Q)$ cosmology in the presence of a viscous fluid. Basically, there are two viscosity coefficients, namely shear viscosity and bulk viscosity. Shear viscosity is related to velocity gradients in the fluid, and by considering the Universe as described by homogeneous and isotropic FLRW background, it can be omitted. Anyhow, by dropping the FLRW background assumption, several cosmological models with shear viscosity fluid have been constructed, as one can check, for instance, \cite{bali/1988,bali/1987,deng/1991,huang/1990}. On the other hand, bulk viscosity, which we are going to consider here, introduces damping associated with volumetric straining. To get in touch with bulk viscous fluid cosmological models, one can check References \cite{Davood/2019,Daa,Dab,Dac}. Moreover, some interesting applications of bulk viscous cosmology in black holes presented in \cite{bb1,bb2}.

Researchers examine dark energy reconstruction with numerous observations as data increases. The majority of studies has been concentrated on observable evidences from SN Ia, CMB and BAO, which are known to be helpful in constraining cosmological models. The Hubble parameter dataset shows the intricate structure of the expansion of the Universe. The ages of the most massive and slowly evolving galaxies offer direct measurements of the Hubble parameter $H(z)$ at various redshifts $z$, resulting in the development of a new form of standard cosmological probe \cite{Jimenez/2002}. In this chapter we include 31 measurements of Hubble expansion spanned using differential age method \cite{Sharov/2017} and BAO data having six measurements \cite{BAO1}, along with the Pantheon samples to constrain the cosmological model.

 This chapter is organized as follows. In Sec. \ref{sec3x} we describe the FLRW Universe dominated by bulk viscous non-relativistic matter and derive the expression for the Hubble parameter and the deceleration parameter. Further, in Sec. \ref{sec4x}, we analyze the observational data to find the model parameters value utilizing the Hubble dataset containing 31 points, BAO sample and the Pantheon dataset of 1048 samples. Moreover, we analyze the behavior of different evolutionary parameters like the Hubble, density, effective pressure, deceleration parameter and effective EoS (EoS) parameter. In Sec. \ref{sec5x}, we investigate the consistency of our bulk viscous fluid model by analyzing the different energy conditions. In Sec. \ref{sec6x}, we analyze the behavior of statefinder parameters on the values constrained by the observational data to differentiate between dark energy models. Lastly, we discuss our results in Sec. \ref{sec7x}.

\section{The cosmological model}\label{sec3x}

Firstly, to consider bulk viscosity in a fluid can be seen as an attempt to refine its  description, minimizing its ideal properties. This can be checked, for instance, in the stellar astrophysics realistic models in References  \cite{gusakov/2007,gusakov/2008,haensel/2002}. Under conditions of spatial homogeneity and isotropy (which refers to the cosmological principle, as one can check, for instance, Reference \cite{ryden/2003}), the bulk viscous pressure is the unique admissible dissipative phenomenon. In a gas dynamical model, the existence of an effective bulk pressure can be traced back to a non-standard self-interacting force on the particles of the gas \cite{colistete/2007}. The bulk viscosity contributes negatively to the total pressure, as one can check, for instance Reference \cite{meng/2009}.

Due to spatial isotropy, the bulk viscous pressure is the same in all spatial directions and hence proportional to the volume expansion $\theta=3H$.

The effective pressure of the cosmic fluid becomes \cite{brevik/2005,gron/1990,C.E.}
\begin{equation}
\bar{p}= p-\zeta\theta=p-3\zeta H,
\end{equation}
in which $p$ is the usual pressure and $ \zeta >0 $ is the bulk viscosity coefficient, which we will assume as a free parameter of the model.

The relation between normal pressure and matter-energy density follows $p=(\gamma-1)\rho$ \cite{J.Ren}, with $\gamma$ being a constant lying in the range $0\leq \gamma \leq 2$. Then the effective EoS for the bulk viscous fluid is given by the following
\begin{equation}\label{3ix}
\bar{p}= (\gamma -1)\rho - 3\zeta H.
\end{equation}

For our investigation of bulk viscous fluid cosmological model, we consider the following $f(Q)$ functional form,
\begin{equation}\label{3fx}
f(Q)= \alpha Q^n,  
\end{equation} 
with $\alpha\neq0 $ and constant $n$. This particular functional form for $f(Q)$ was motivated by a polinomial form applied, for instance, in Reference \cite{Sanjay}. In particular, for $f(Q)=-Q$ we retrieve the usual Friedmann equations \cite{Harko-2}, as expected, since as we have mentioned before, this particular choice for the functional form of the function $f(Q)$ is the STEGR limit of the theory. For the above choice of the $f(Q)$ function (equation (\ref{3fx})), we obtain the Friedmann like equations (\ref{3dz}) and(\ref{3ez}) as follows,

\begin{equation}\label{3gx}
\rho = \alpha6^n \left(\frac{1}{2}-n \right) H^{2n},
\end{equation}
\begin{equation}\label{3hx}
\dot{H}+\frac{3}{2n} H^2= \frac{6^{1-n}\bar{p}}{2\alpha n(2n-1)} H^{2(1-n)}.
\end{equation}  

From equations \eqref{3hx} and \eqref{3ix}, we have the following,
 
\begin{equation}\label{3jx}
\dot{H}+ \frac{3\gamma}{2n} H^2 = - \frac{6^{2-n}\zeta}{4\alpha n(2n-1)} H^{3-2n}. 
\end{equation} 

Now, we replace the term $d/dt$ by $d/dln a$ via the expression $d/dt=Hd/dln a$, such that equation \eqref{3jx} becomes
\begin{equation}\label{3kx}
\frac{dH}{dln a} + \frac{3\gamma}{2n}H = -\frac{6^{2-n}\zeta}{4\alpha n(2n-1) } H^{2(1-n)}.
\end{equation}

The integration of equation \eqref{3kx} yields the following solution,

\begin{equation}\label{3lx}
H(a) = \left\{( H_0 a^{-\frac{3\gamma}{2n}} )^{2n-1} + \frac{6^{1-n}\zeta}{\gamma \alpha (2n-1)^2}[ a^{-\frac{3\gamma(2n-1)}{2n}}-1 ] \right\}^{\frac{1}{2n-1}},
\end{equation}
 with $H_0$ being a constant of integration to be found below.

We obtain the Hubble parameter in terms of redshift by using relation $a(t)= 1/(1+z)$ \cite{ryden/2003} in equation \eqref{3mx}.  By making $z=0$ in \eqref{3mx} we find that $H(0)=H_0$. The deceleration parameter is defined as $q= -\ddot{a}a/\dot{a}^2 = -\ddot{a}/(H^2a) $. Henceforth from equation \eqref{3lx} we have,

\begin{eqnarray}\label{3mx}
H(z) &=& \left\{[ H_0 (1+z)^{\frac{3\gamma}{2n}} ]^{2n-1} + \frac{6^{1-n}\zeta}{\gamma \alpha (2n-1)^2} [ (1+z)^{\frac{3\gamma(2n-1)}{2n}}-1 ] \right\}^{\frac{1}{2n-1}},\\
\label{qqx}
q(z) &=&\frac{3}{2n}\left\{\frac{\zeta}{\alpha 6^{n-1}(2n-1) \left\{ [ H_0(1+z)^{\frac{3\gamma}{2n}} ]^{2n-1} + \frac{6^{1-n}\zeta}{\gamma\alpha (2n-1)^2} [ (1+z)^{\frac{3\gamma(2n-1)}{2n}} -1 ] \right\}} + \gamma\right\} -1. 
\end{eqnarray}

\section{Observational constraints}\label{sec4x}

To examine the observational features of our cosmological model, we use the most recent cosmic Hubble and Supernovae observations. We use 31 points of the Hubble datasets, 6 points of the BAO datasets and 1048 points from the Pantheon supernovae samples. We apply the Bayesian analysis and likelihood function along with the MCMC method in \texttt{emcee} python library \cite{R56}. 

\subsection{Hubble datasets}

We incorporate the set of 31 data points that are measured from the differential age approach \cite{sharov} to avoid extra correlation with BAO data. The mean values of the model parameters $\zeta$, $\alpha$, $\gamma$ and $n$ are calculated using the chi-square function as follows,

\begin{equation}
\chi _{H}^{2}(\zeta,\alpha,\gamma,n)=\sum\limits_{k=1}^{31}
\frac{[H_{th}(\zeta,\alpha,\gamma,n, z_{k})-H_{obs}(z_{k})]^{2}}{
\sigma _{H(z_{k})}^{2}}.  \label{4ax}
\end{equation}

\begin{figure}[H]
\centering
\includegraphics[scale=0.9]{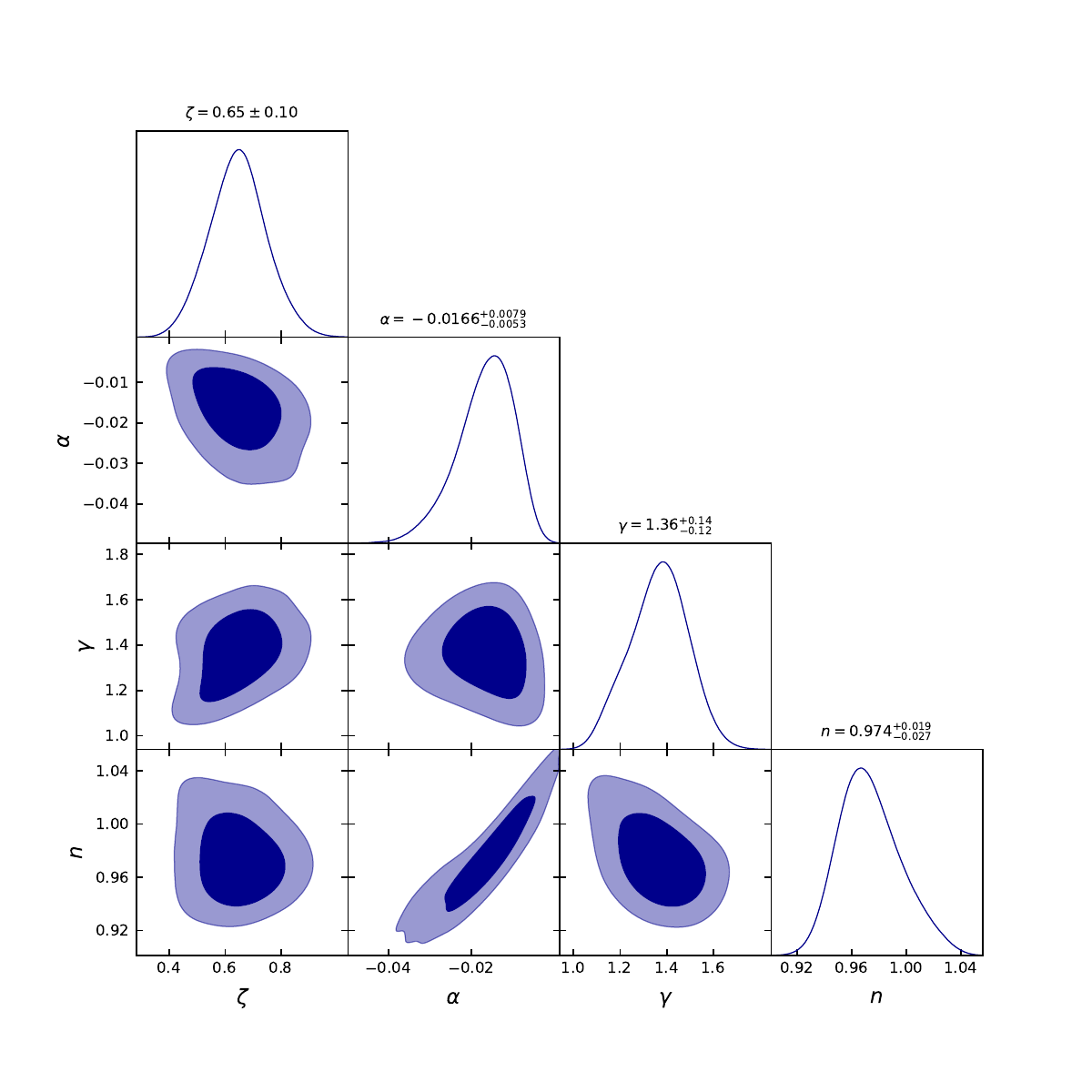}
\caption{The $1-\sigma$ and $2-\sigma$ likelihood contours for the model parameters using the Hubble datasets.}
\label{ContourHubx}
\end{figure}

Here, $H_{th}$ is the Hubble parameter value predicted by the model, $H_{obs}$ represents its observed value and the standard error is given by $\sigma _{H(z_{k})}$. From the Hubble dataset, we obtain values for model parameters $\zeta$, $\alpha$, $\gamma$, $n$ as the $1-\sigma$ and $2-\sigma$ contour plots in Figure \eqref{ContourHubx}. 

\begin{figure}[H]
\includegraphics[scale=0.6]{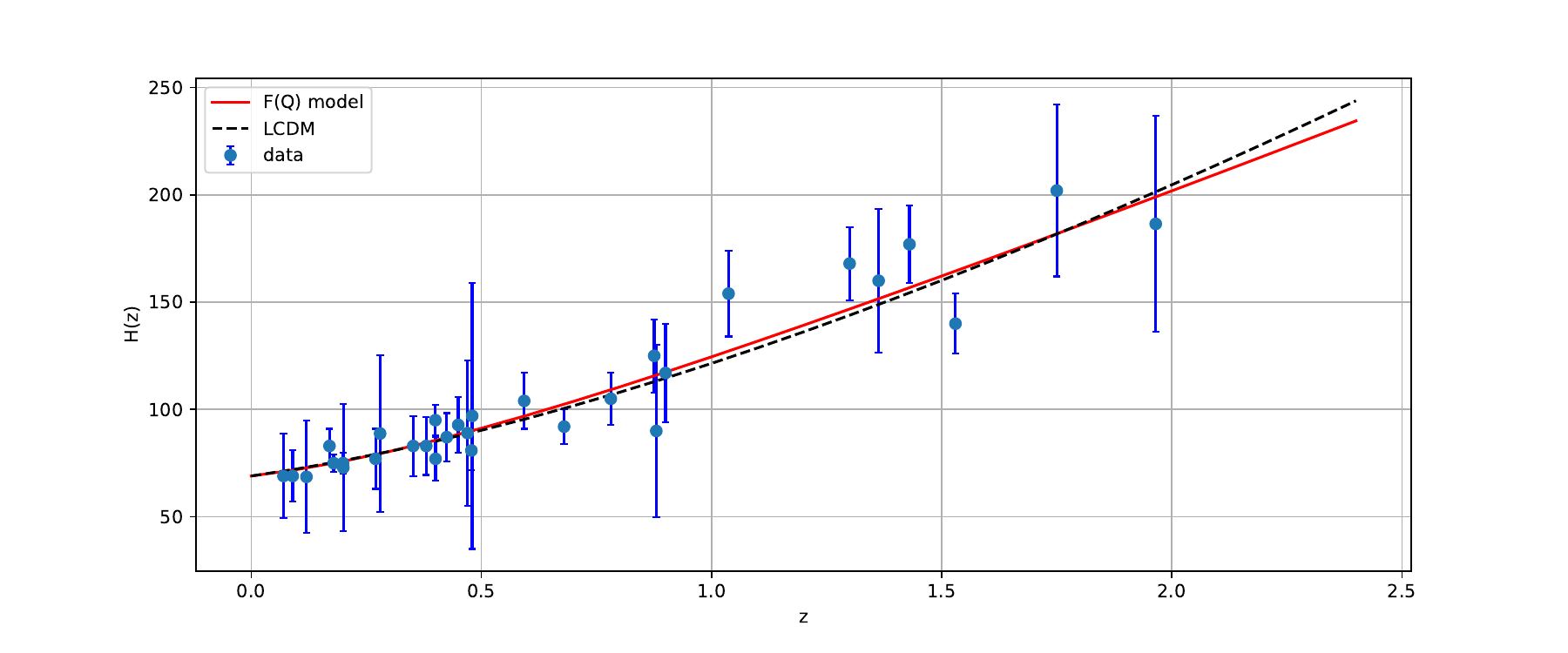}
\caption{The error bar plot of $H$ versus $z$ for the considered $f(Q)$ model. The solid red line is the curve for $f(Q)$ model whereas the black dotted line represents the $\Lambda$CDM model. The blue dots depict the 31 points of the Hubble data.} \label{ErrorHubblex} 
\end{figure}
	
The obtained suitable values of the free parameters are $\zeta= 0.65^{+0.10}_{-0.10} $, $\alpha=-0.0166^{+0.0079}_{-0.0053}$, $\gamma=1.36^{+0.14}_{-0.12}$ and $n=0.974^{+0.019}_{-0.027}$. Figure \eqref{ErrorHubblex} shows the error bar plot of the assumed model and $\Lambda$CDM or  standard cosmological model, with  cosmological constant density parameter $\Omega_{\Lambda_{0}}=0.7$, matter density parameter $\Omega_{m_{0}}=0.3$ and $H_{0}= 69$ km/s/Mpc. 

\subsection{BAO datasets}
\begin{figure}[H]
\centering
\includegraphics[scale=0.9]{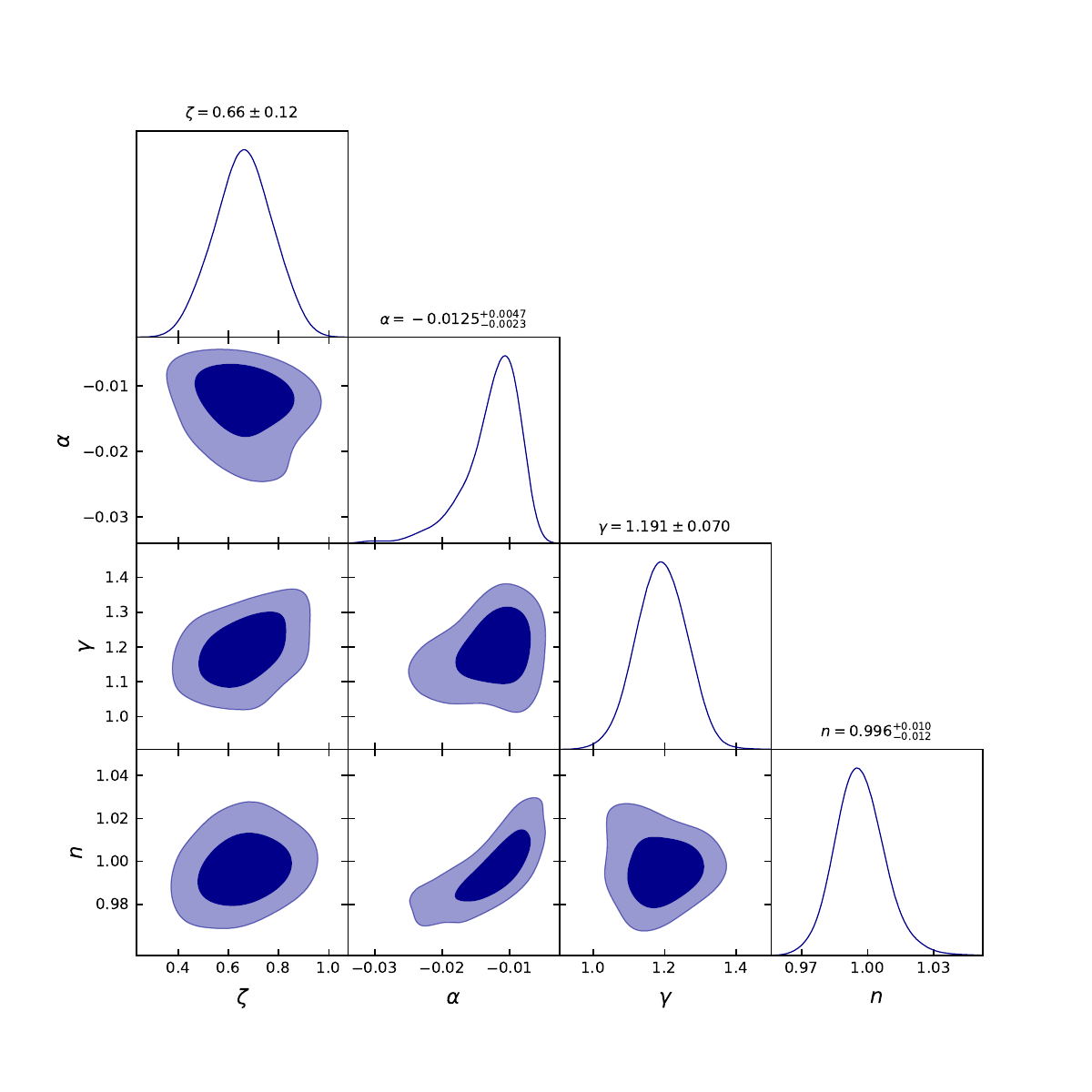}
\caption{The $1-\sigma$ and $2-\sigma$ likelihood contours for the model parameters using the BAO datasets.}
\label{ContourBAOx}
\end{figure}

The BAO distance dataset, which includes the 6dFGS, SDSS and WiggleZ surveys, comprise BAO measurements at six different redshifts in Table \eqref{Table-2z}. The chi-square function for the BAO datasets is taken to be \cite{BAO6} 
 
\begin{equation}\label{4ex}
\chi _{BAO}^{2}=X^{T}C^{-1}X\,,  
\end{equation}
where the details of the vector $X$ and the covariance matrix $C$ are already discussed in the previous chapter. The values that fit observations are  $\zeta=0.66^{+0.12}_{-0.12}$, $\alpha=-0.0125^{+0.0047}_{-0.0023}$, $\gamma=1.191^{+0.070}_{-0.070}$ and $n=0.996^{+0.010}_{-0.012}$ in Fig.\ref{ContourBAOx}.

\subsection{Pantheon datasets} 

Scolnic et al. \cite{R18} put together the Pantheon samples consisting of 1048 type Ia supernovae in the redshift range $0.01 < z < 2.3$. The PanSTARSS1 Medium Deep Survey, SDSS, SNLS and numerous low-z and HST samples contribute to it. The empirical relation used to calculate the distance modulus of SN Ia from the observation of light curves is given by $\mu= m_{B}^{*}+\alpha X_{1}-\beta C-M_{B} + \Delta_{M}+ \Delta_{B}$, where $X_{1}$ and $C$ denote the stretch and color correction parameters, respectively \cite{R18,Mukherjee/2021}, $m_{B}^*$ represents the observed apparent magnitude and $M_{B}$ is the absolute magnitude in the B-band for SN Ia. The parameters $\alpha$ and $\beta$ are the two nuisance parameters describing the luminosity stretch and luminosity color relations, respectively. Further, the distance correction factor is $\Delta_{M}$ and $\Delta_{B}$ is a distance correction based on predicted biases from simulations.

The nuisance parameters in the Tripp formula \cite{Tripp/1998} were reconstructed using a novel technique called BEAMS with Bias Corrections \cite{Kessler/2017, Fotios/2021} and the observed distance modulus was reduced to the difference between the corrected apparent magnitude $m_{B}$ and the absolute magnitude $M_{B}$, which is $\mu= m_{B}-M_{B}$. We shall avoid marginalizing the over nuisance parameters $\alpha$ and $\beta$ but marginalize over the Pantheon data for $M_{B}$. Hence, we ignore the values of $\alpha$ and $\beta$ for the present investigation of the model.

\begin{figure}[H]
\centering
\includegraphics[scale=0.9]{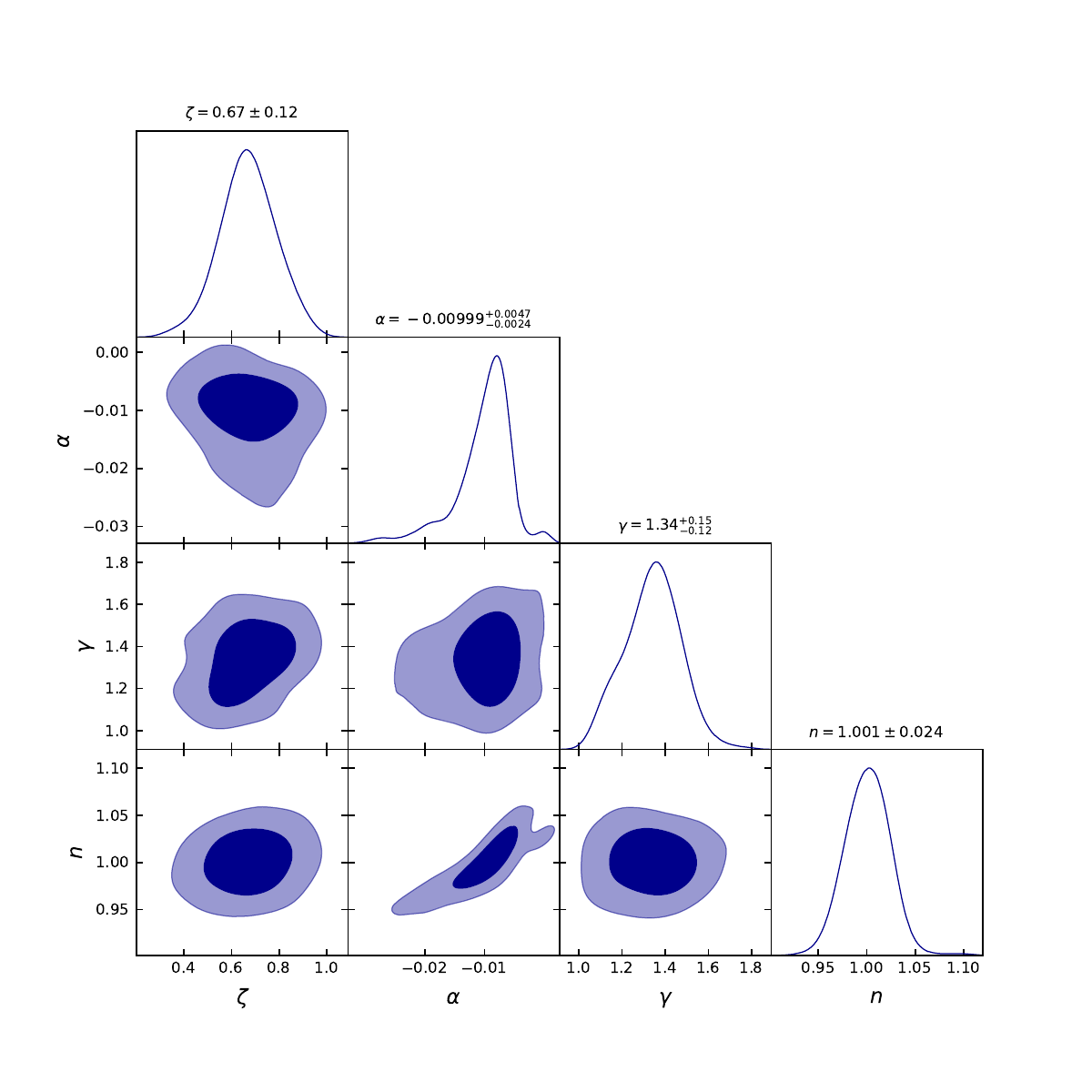}
\caption{The $1-\sigma$ and $2-\sigma$ likelihood contours for the model parameters using the Pantheon datasets.}
\label{ContourPanx}
\end{figure}

The $\chi^{2}$ function for type Ia supernovae is given as follows,

\begin{equation}\label{4ix}
\chi^{2}_{SN}(\zeta, \alpha, \gamma, n)= \sum _{k=1}^{1048} \dfrac{\left[ \mu_{obs}(z_{k})-\mu_{th}(\zeta, \alpha, \gamma, n, z_{k})\right] ^{2}}{\sigma^{2}(z_{k})},
\end{equation}
where $\mu_{th}$ is the theoretical value of distance modulus,  $\mu_{obs}$ is the observed value whereas $\sigma^{2}(z_{k})$ is the standard error. Using the Pantheon supernovae datasets, we obtain the values for model parameters $\zeta$, $\alpha$, $\gamma$ and $n$ as the $1-\sigma$ and $2-\sigma$ contour plots in Figure \eqref{ContourPanx}. The values that fit the model are $\zeta=0.67^{+0.12}_{-0.12} $, $\alpha=-0.00999^{+0.0047}_{-0.0024}$, $\gamma=1.34^{+0.15}_{-0.12}$ and $n=1.001^{+0.024}_{-0.024}$. 

\subsection{Cosmological parameters}

The evolution trajectories of the Hubble function, deceleration parameter, energy density, pressure with bulk viscosity and the effective EoS parameter for the redshift range $-1 < z <8$ are presented below, in order to test the late time cosmic expansion history and the future of expanding Universe \cite{Sunny}. In order to do so we use the set of values constrained by Hubble, BAO and the Pantheon datasets for the model parameters.

\begin{figure}[H]
\centering
\includegraphics[scale=0.51]{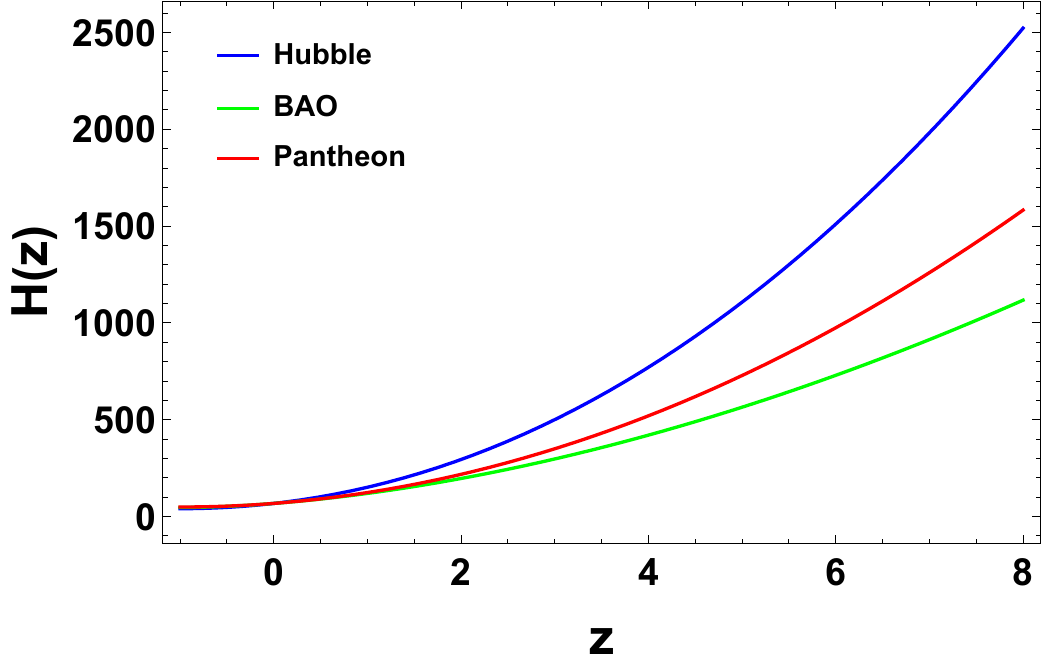}
\caption{Profile of the Hubble parameter for the given model with the parameter values estimated by the Hubble, BAO and the Pantheon data point sets.}
\label{hx}
\end{figure}

\begin{figure}[H]
\centering
\includegraphics[scale=0.5]{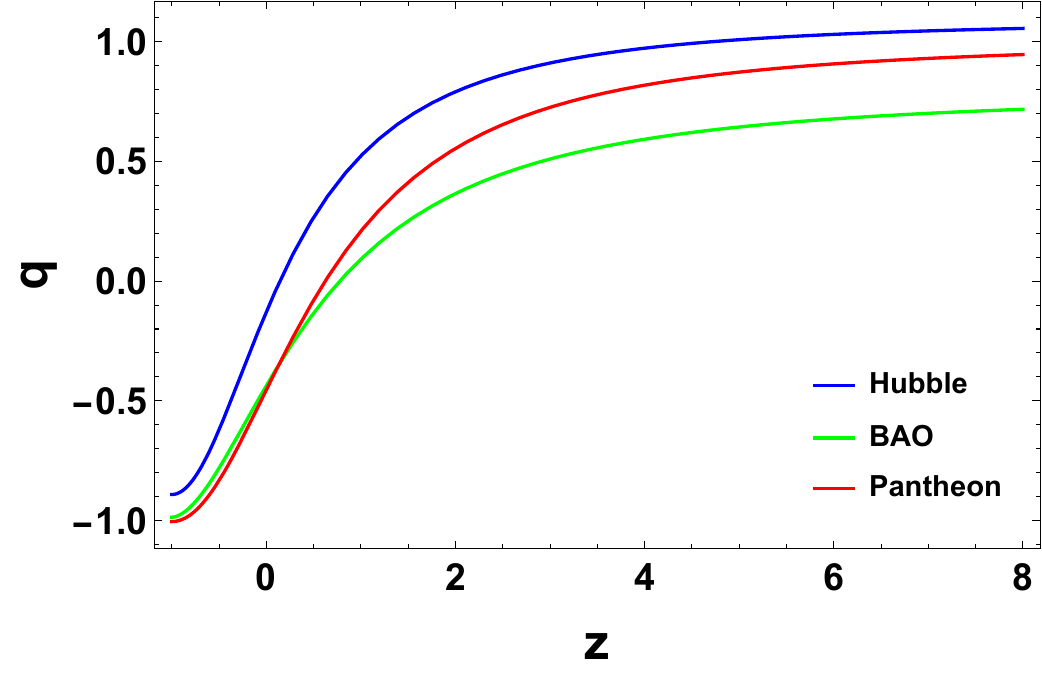}
\caption{Profile of the deceleration parameter for the given model with the parameter values estimated by the Hubble, BAO and the Pantheon data point sets.}
\label{qx}
\end{figure}

\begin{figure}[H]
\centering
\includegraphics[scale=0.52]{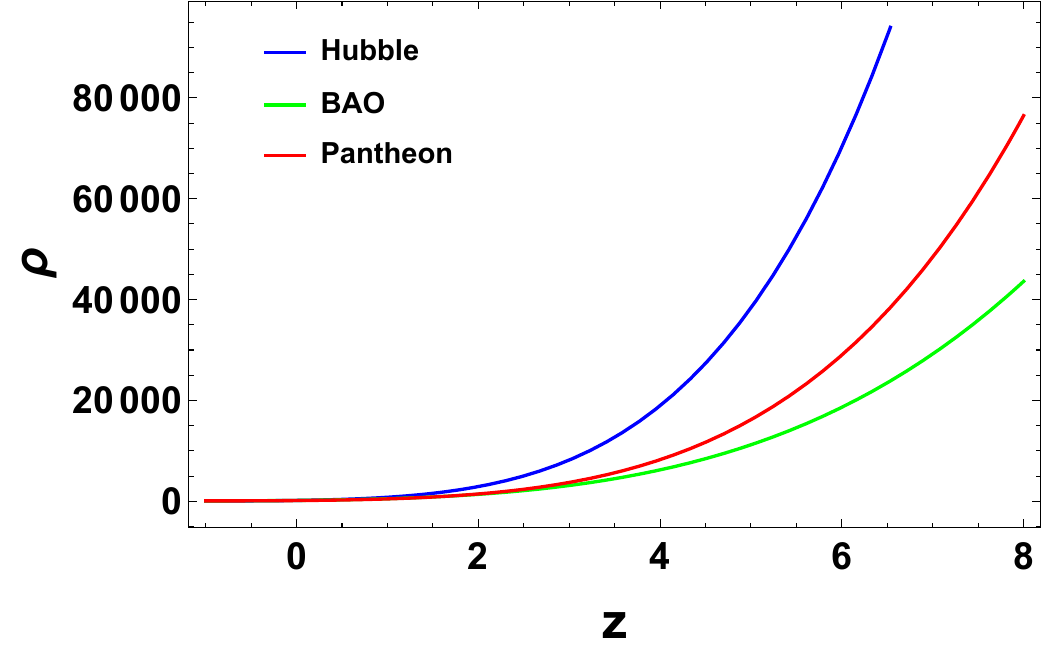}
\caption{Profile of the density parameter for the given model with the parameter values estimated by the Hubble, BAO and the Pantheon data point sets.}
\label{rhox}
\end{figure}

\begin{figure}[H]
\centering
\includegraphics[scale=0.41]{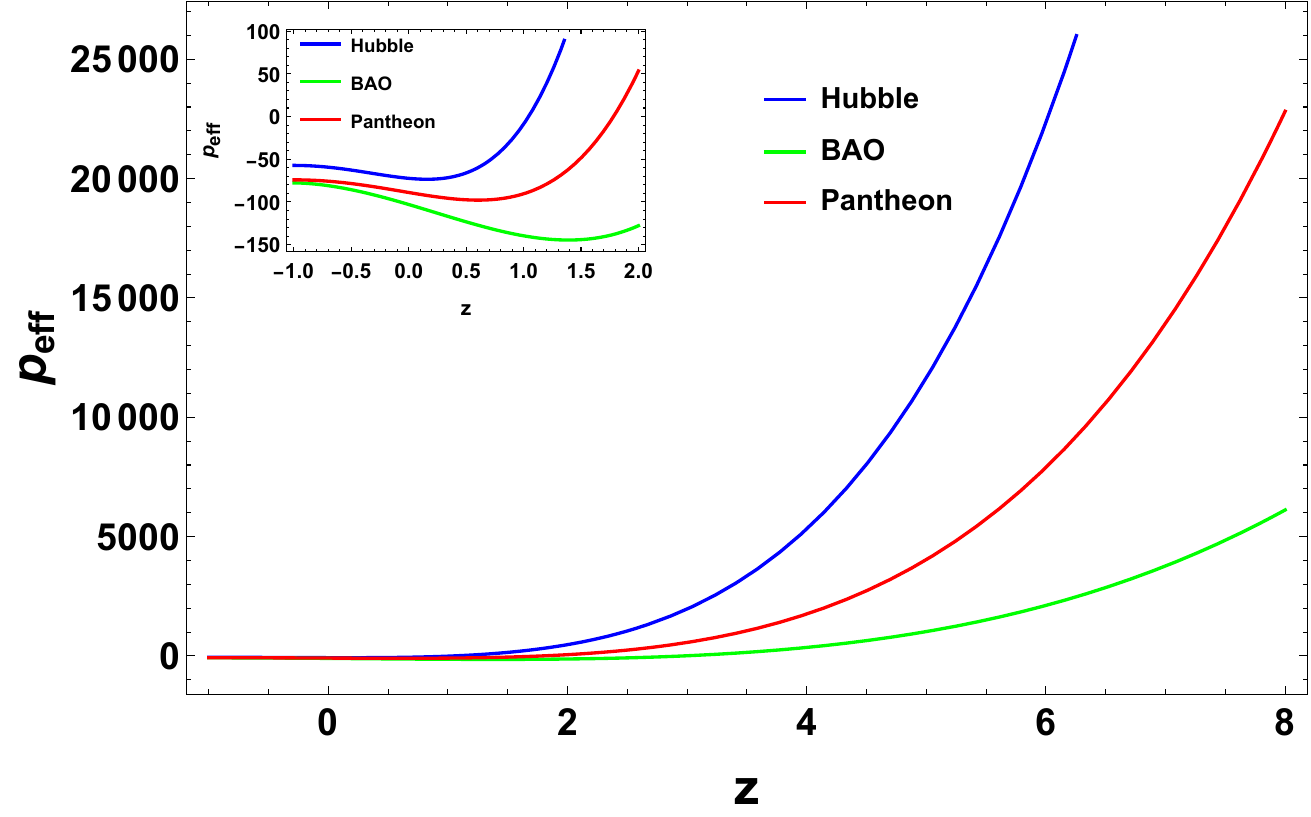}
\caption{Profile of the pressure for the given model with the parameter values estimated by the Hubble, BAO and the Pantheon data point sets.}
\label{px}
\end{figure}

\begin{figure}[H]
\centering
\includegraphics[scale=0.5]{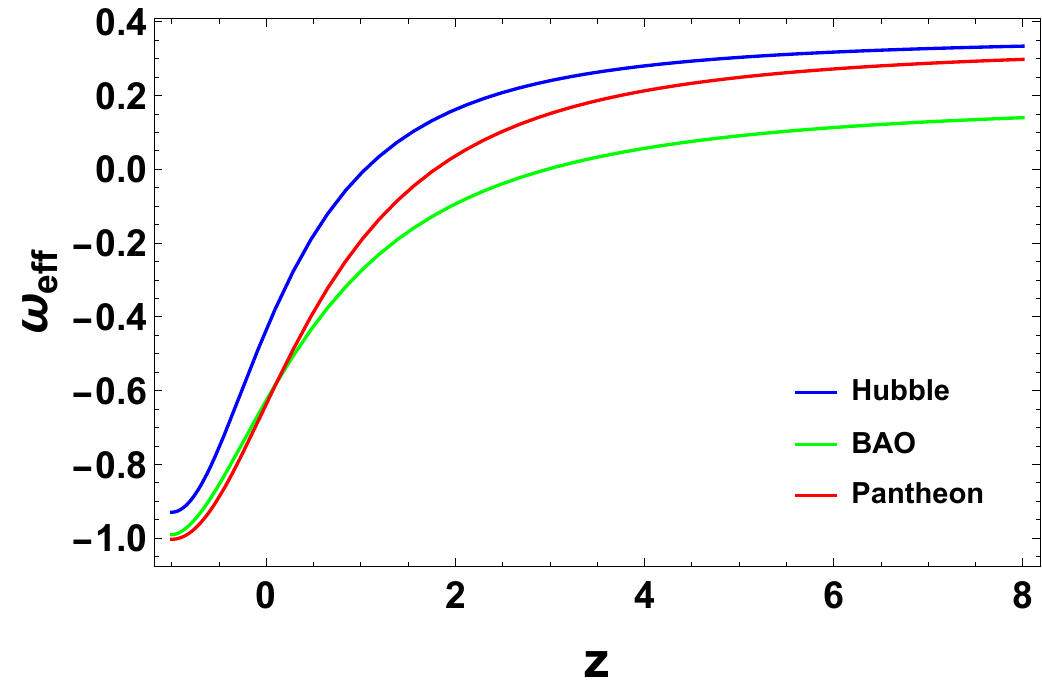}
\caption{Profile of the EoS parameter for the given model with the parameter values estimated by the Hubble, BAO and the Pantheon data point sets.}
\label{wx}
\end{figure}

From figure \eqref{qx}, it is clear that the deceleration parameter shows the transition from a decelerated ($q>0$) to an accelerated ($q<0$) phase of the Universe expansion for the estimated values of free parameters. The transition redshift is $z_t \thickapprox 0.142$,   $ z_t \thickapprox 0.776 $ and $ z_t \thickapprox 0.622 $ corresponding to the Hubble, BAO and the Pantheon datasets, respectively. The present value of the deceleration parameter is, respectively,  $q_0 = -0.127$ , $q_0= -0.436$ and  $q_0= -0.454$. 

From figures \eqref{hx} and \eqref{rhox} it is clear that the Hubble and density parameter shows the positive behavior for the estimated values of free parameters, which is expected.

Figure \eqref{px} indicates that the bulk viscous cosmic fluid exhibits, for lower redshifts, the negative pressure that make bulk viscosity to be a viable candidate to drive the cosmic acceleration. Furthermore, the effective EoS parameter presented in figure \eqref{wx} indicates that the cosmic viscous fluid behaves like quintessence dark energy. The present values of EoS parameter with respect to the Hubble, BAO and the Pantheon samples are  $\omega_{0}= -0.433$ , $\omega_0= -0.625$, and  $\omega_0= -0.635$.

\section{Energy conditions}\label{sec5x}

In the present section we are going to construct the energy conditions for the solutions of the present model. The energy conditions are relations applied to the matter energy-momentum tensor with the purpose of satisfying positive energy. The energy conditions are derived from the Raychaudhuri equation and are written as \cite{EC} \\

\begin{itemize}
\item \textbf{Null energy condition (NEC):} $\rho_{eff}+p_{eff}\geq 0$;  
\item \textbf{Weak energy condition (WEC):} $\rho_{eff} \geq 0$ and  $\rho_{eff}+p_{eff}\geq 0$; 
\item \textbf{Dominant energy condition (DEC):} $\rho_{eff} \pm p_{eff}\geq 0$; 
\item \textbf{Strong energy condition (SEC):} $\rho_{eff}+ 3p_{eff}\geq 0$,
\end{itemize}
 with $\rho_{eff}$ being the effective energy density.

\begin{figure}[H]
\centering
\includegraphics[scale=0.52]{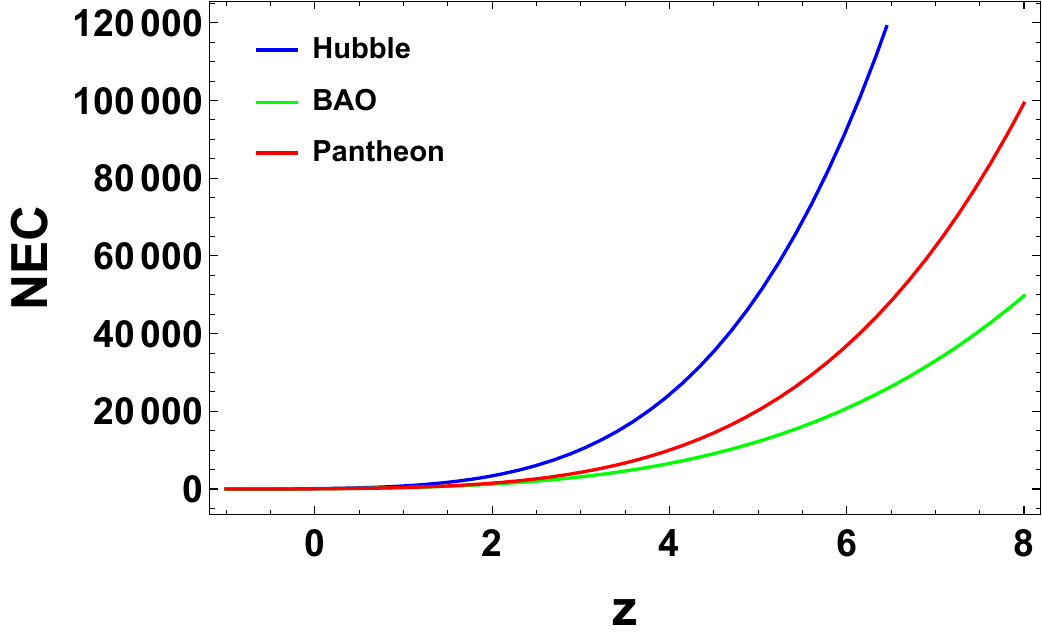}
\caption{Profile of the null energy condition for the given model with the parameter values estimated by the Hubble, BAO and the Pantheon data point sets.}
\label{necx}
\end{figure}

\begin{figure}[H]
\centering
\includegraphics[scale=0.51]{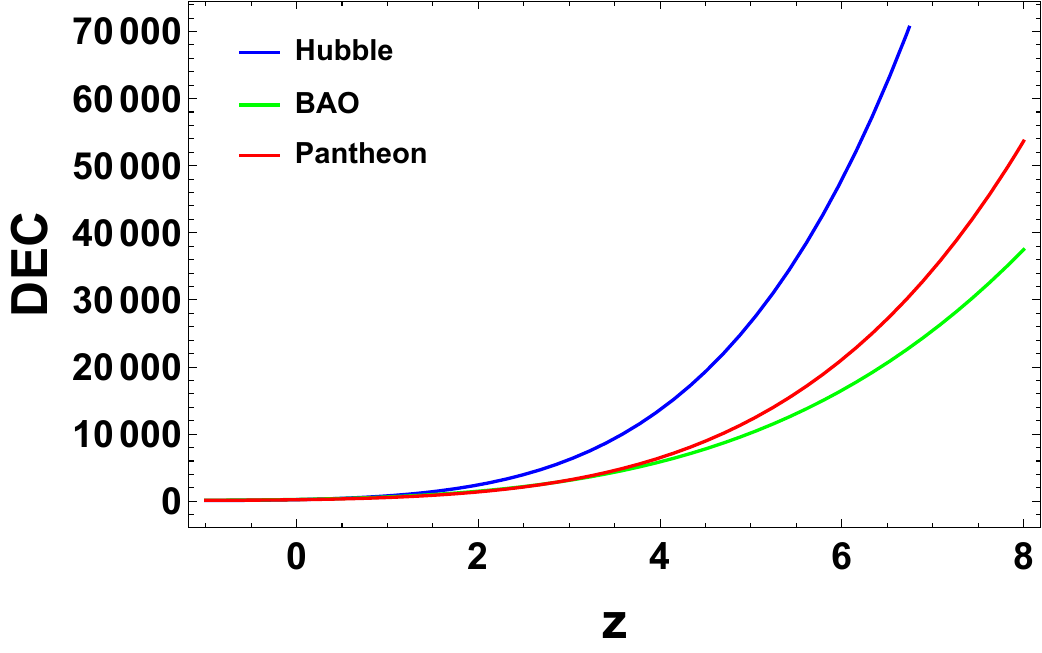}
\caption{Profile of the dominant energy condition for the given model with the parameter values estimated by the Hubble, BAO and the Pantheon data point sets.}
\label{decx}
\end{figure}

\begin{figure}[H]
\centering
\includegraphics[scale=0.45]{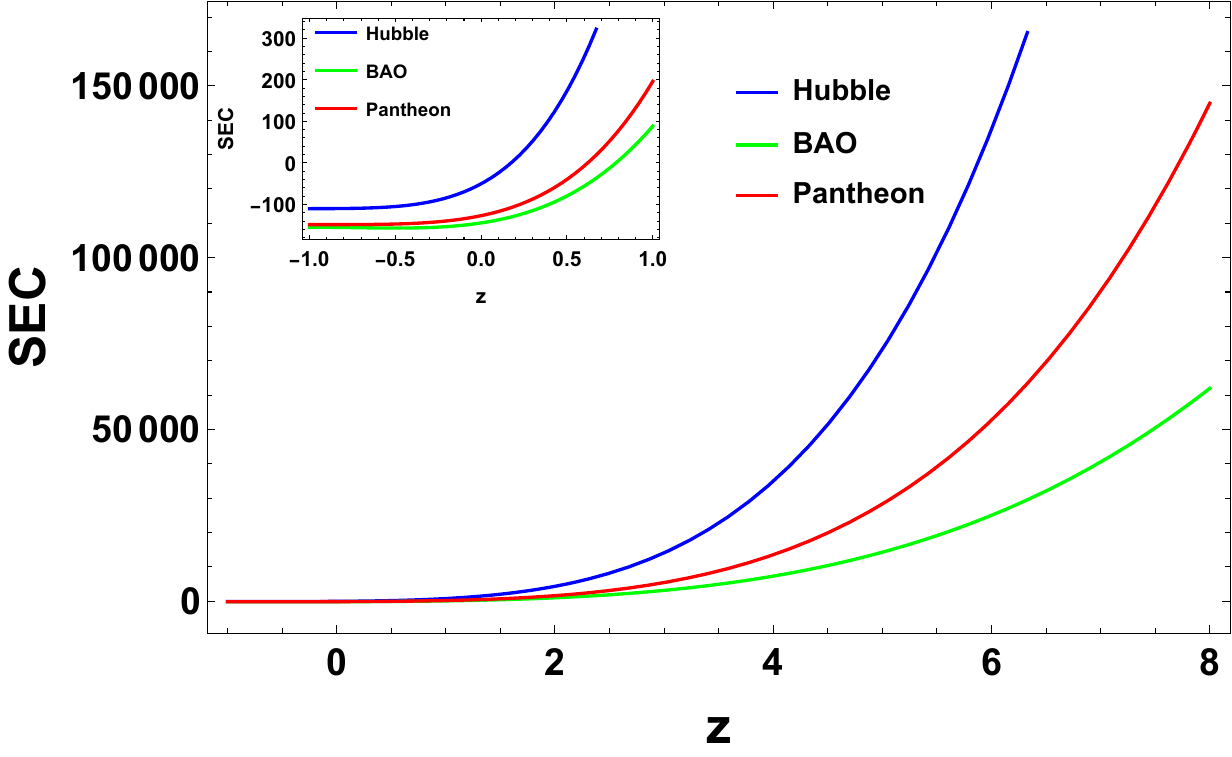}
\caption{Profile of the strong energy condition for the given model with the parameter values estimated by the Hubble, BAO and the Pantheon data point sets.}
\label{secx}
\end{figure}

In figures \eqref{necx} and \eqref{decx} it is evident that the NEC and DEC exhibit positive behavior for the estimated values of free parameters. As WEC is the combination of energy density and NEC, we conclude that NEC, DEC and WEC are all satisfied in the entire domain of redshift. Figure \eqref{secx} indicates that the SEC exhibits, for lower redshifts, the negative behavior that is related to cosmic acceleration \cite{MandalC}. This is also reflected in the deceleration parameter behavior in figure \eqref{qx}.

\section{Statefinder analysis}\label{sec6x}

For different values of the statefinder pair $(r,s)$, it represents the following dark energy models:
\begin{itemize}
    \item ${r=1,s=0}$ represents $\Lambda$CDM model.
    \item ${r>1,s<0}$ represents Chaplygin gas model.
    \item ${r<1,s>0}$ represents quintessence model.
\end{itemize}

\begin{figure}[H]
\centering
\includegraphics[scale=0.47]{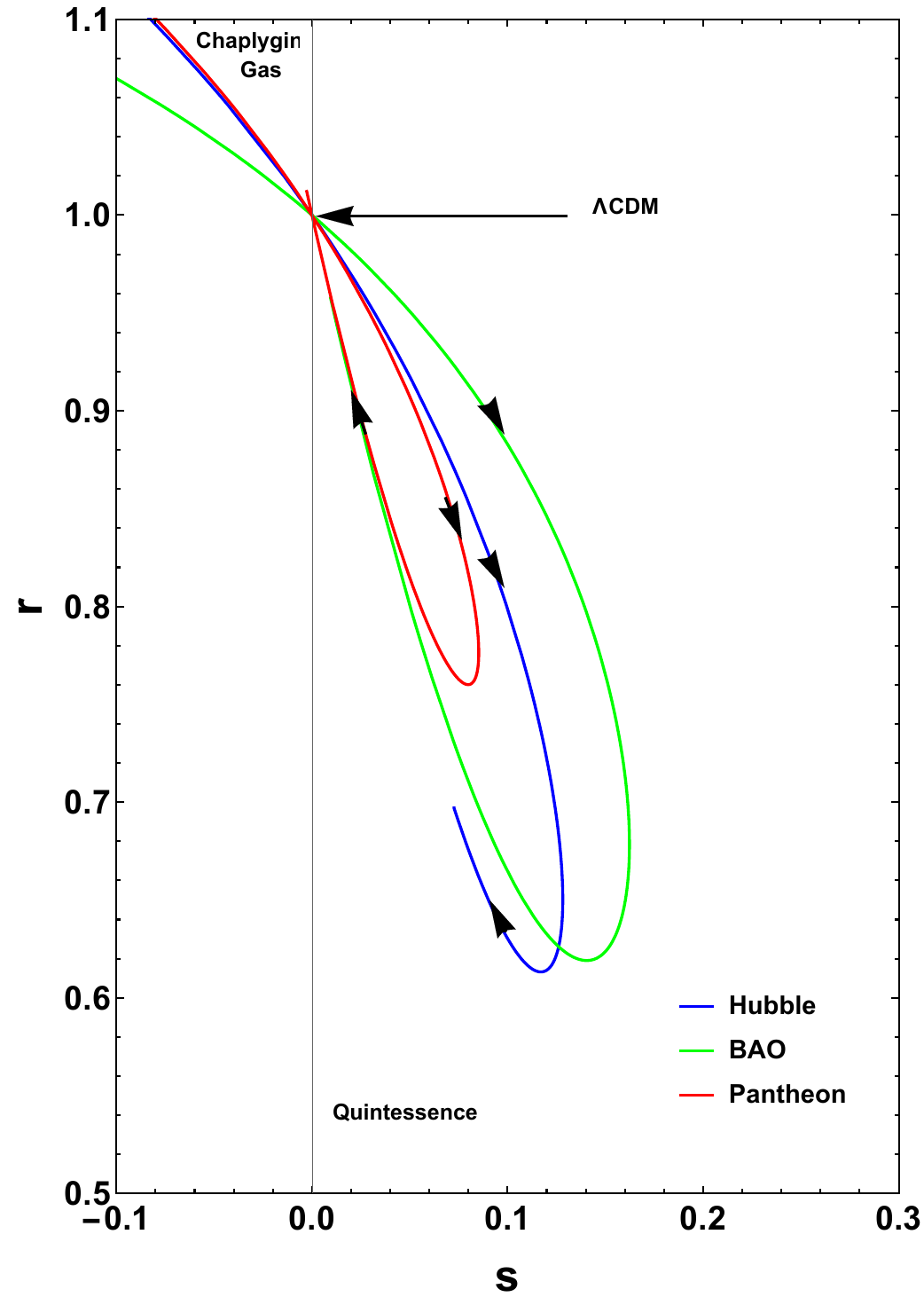}
\caption{Plot of the trajectories in the $r-s$ plane for the given cosmological model with the parameter values estimated by the Hubble, BAO and the Pantheon datasets.}
\label{rsx}
\end{figure}

\begin{figure}[H]
\centering
\includegraphics[scale=0.48]{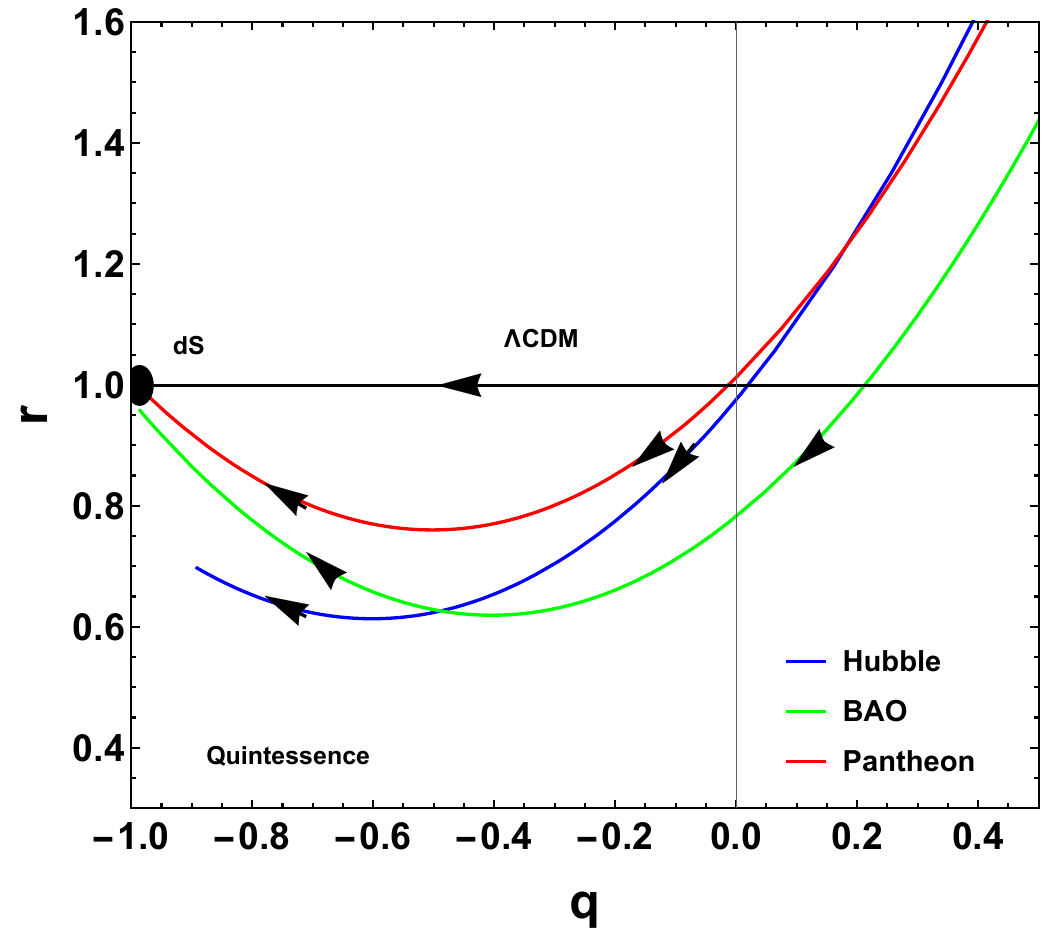}
\caption{Plot of the trajectories in the $q-r$ plane for the given cosmological model with the parameter values estimated by the Hubble, BAO, and the Pantheon datasets.}
\label{qrx}
\end{figure}

In figures \eqref{rsx} and \eqref{qrx}, we plot the $s-r$ and $q-r$ diagrams for our cosmological model by taking the parameter values estimated by the Hubble, BAO and the Pantheon datasets. 

Figures \eqref{rsx} and \eqref{qrx} show that our bulk viscous model lies in the quintessence region. Also, the evolutionary trajectories of our model departure from the $\Lambda$CDM point. The present values of the statefinder parameters corresponding to the parameter values estimated by the Hubble, BAO and the Pantheon samples are 
$ r_0= 0.837$ and $s_0= 0.086$, $r_0= 0.62$ and $s_0= 0.135$,  $r_0= 0.762$ and $s_0=  0.083$ respectively.

\section{Discussions and conclusions}\label{sec7x}

Cosmology has been on the agenda mainly for two reasons: dark energy and dark matter. While dark energy has been deeply discussed throughout the paper, dark matter is predicted within $\Lambda$CDM model as a sort of matter that does not interact electromagnetically, so that it cannot be seen, but its gravitational effects otherwise can well be detected. Still we have not yet detected or even associated dark matter to a particle of standard model or beyond \cite{cdmsii_collaboration/2010,akerib/2014,essig/2012}. Modified (or alternative) theories of gravity have also been used to describe dark matter effects \cite{bohmer/2008,mannheim/2012}. In these cases, dark matter is simply an effect of modification of gravity.

In the present chapter, as an attempt to describe dark energy we have assumed the symmetric teleparallel gravity as the underlying gravity theory. Our $f(Q)$ cosmological model was based on a spatially homogeneous and isotropic flat metric and an energy-momentum tensor describing a bulk viscous fluid. We let the $f(Q)$ function be a power of $n$ as $f(Q)\sim Q^n$, with $n$ a free parameter.

In Section III we started testing our cosmological solutions. We started confronting the Hubble parameter \eqref{3mx} with 31 data points that are measured from differential age approach. In figure \eqref{ContourHubx} we have obtained suitable values for the free parameters. We have then plotted figure \eqref{ErrorHubblex}, in which our $H(z)$ is confronted with cosmological data and compared with $\Lambda$CDM prediction. We can see the $f(Q)$ model describes observations with good agreement and specially for higher values of redshift it is clear that it provides a better fit when compared to $\Lambda$CDM model. Further, in figures \eqref{ContourBAOx} and \eqref{ContourPanx}, we have obtained values for the free parameters.

From figures \eqref{hx} and \eqref{rhox} we found that the Hubble and density parameter shows the expected positive behavior for the estimated values of free parameters.
Figure \eqref{px} indicates that the bulk viscous cosmic fluid exhibit the negative pressure that make bulk viscosity to be a viable candidate to drive the cosmic acceleration. This is also reflected in the deceleration parameter behavior in figure \eqref{qx}, which shows a transition from decelerated to accelerated phases of the Universe expansion. Furthermore, the effective EoS parameter presented in figure \eqref{wx} indicates that the cosmic viscous fluid behaves like quintessence dark energy.

In Section IV, we investigated the consistency of our model by analyzing the different energy conditions. We found that NEC, DEC, and WEC all are satisfied in the entire domain of redshift (presented in figures \eqref{necx} and \eqref{decx}) while the SEC, presented in figure \eqref{secx}, is violated for lower redshifts that implies the cosmic acceleration and satisfied for higher redshifts that implies a decelerated phase of the Universe.

In Section V, figures \eqref{rsx} and \eqref{qrx} show that the evolutionary trajectories of our model is departed from $\Lambda$CDM fixed point {${r=1,s=0}$}. In the present epoch they lie in the quintessence region {${r<1,s>0}$}. The present model is, therefore, a good alternative to explain the Universe dynamics, particularly with no necessity of invoking the cosmological constant.

%% file: Chapters/Chapter4.tex

\chapter{Cosmological constraints on $f(Q)$ gravity models in the coincident gauge formalism} 

\label{Chapter4} 

\lhead{Chapter 4. \emph{Cosmological constraints on $f(Q)$ gravity models in the coincident gauge formalism}} 

\vspace{8 cm}
* The work, in this chapter, is covered by the following publications: \\
 
\textit{Complete dark energy scenario in $f(Q)$ gravity}, Physics of the dark Universe \textbf{36}, 100996 (2022).

\textit{Accelerating expansion of the Universe in modified symmetric teleparallel gravity}, Physics of the dark Universe \textbf{36}, 101053 (2022).

\clearpage

In this chapter, we assume a $f(Q)$ model that contains a linear and a non-linear form of non-metricity scalar, particularly $f(Q)=\alpha Q + \beta Q^n$, where $\alpha$, $\beta$, and $n$ are free model parameters. Then we find the values of our model parameters that would be in agreement with the observed value of cosmographic parameters. We analyze the behavior of different cosmological parameters like deceleration parameter, density, and the EoS parameter with the energy conditions for our cosmological model. We found that for higher positive values of $n$ specifically $n \geq 1$, dark energy fluid part evolving due to non-metricity behaves like quintessence type dark energy while for higher negative values of $n$ specifically $n \leq -1$ it follows phantom scenario. Further for $n=0$, our cosmological $f(Q)$ model behaves like $\Lambda$CDM model. Moreover, we find the exact solution and parameter constraints for the case $f(Q)=\alpha Q + \beta$. Thus, we conclude that the geometrical generalization of GR can be a viable candidate for the description of origin of the dark energy.

\section{Introduction}\label{sec1v}

The cosmological constant $\Lambda$ in GR is the most successful description of the dark energy so far and it is represented by $\omega=-1$. Another widely analyzed  DE model is the model with quintessence dark energy which is characterized by $-1<\omega<-\frac{1}{3}$ \cite{RP,M.T.}. In quintessence scenario, dark energy density decreases with cosmic time \cite{LX}. Further, the least theoretically understood DE is the phantom energy that is characterized by $\omega<-1$. Nowadays researchers attracted towards the phantom energy case due to its unusual charateristics. In phantom scenario, dark energy density increases with time that results a finite-time future singularity. For the classification of singularities, see the references \cite{KI,DB,DB-2}. To bypass the undetected dark energy, modified gravity theories were proposed in the literature as an alternative to GR. In the present work, we investigate different dark energy scenarios within the $f(Q)$ gravity background, where $Q$ is the non-metricity scalar. Of course, $f(T)$ theory is more restrictive than $f(Q)$ as the corresponding connection has to satisfy more conditions, in the sense, more independent equations involved in $Q_{abc}=0$ than $T_{abc}=0$ in $f(Q)$ theory. The presence of anti-symmetric part in the field equations of the $f(T)$ theory and the local Lorentz invariance problem hinder the Bianchi identity to hold automatically whereas the field equations in $f(Q)$ theory is purely symmetric and free of this issue. Several investigations have been done in the context of $f(Q)$ gravity with different aspects such as covariant formulation \cite{DZ}, spherically symmetric configuration \cite{RHL}, and signature of $f(Q)$ gravity \cite{Noemi/2021}. The growth index of matter perturbations have been studied in the context of $f(Q)$ \cite{khyllep/2021}. Dimakis et al. studied quantum cosmology for a $f(Q)$ polynomial model \cite{ND}. The geodesic deviation equation in $f(Q)$ gravity has been studied and some fundamental results were obtained in \cite{GDV}. F. K. Anagnostopoulos has presented an interesting study in which they provide evidence that non-metricity $f(Q)$ gravity could challange $\Lambda$CDM model \cite{FKA}.

This chapter is organized as follows: In Sec \ref{sec3v}, flat FLRW Universe in $f(Q)$ gravity along with a dark energy component is presented. In Sec \ref{sec4v}, we assume a cosmological $f(Q)$ model and derive the expressions for density, EoS, and the deceleration parameter. Then we estimate the model parameter values that would be in agreement with the observations. Further in Secs. \ref{sec5v},\ref{sec6v}, and \ref{sec7v} we discuss the different behaviors of our cosmological model depending upon the power $n$ in the $f(Q)$ function. Moreover, in Sec \ref{secaddv}, we find the exact solution and parameter constraints for the case $f(Q)=\alpha Q + \beta$. Finally, we present our outcomes in Sec \ref{sec8v}. 

\section{Flat FLRW Universe in $f(Q)$ cosmology}\label{sec3v}

The Friedmann equations (\ref{3dz})-(\ref{3ez}) governing the dynamics of the Universe filled with perfect fluid for the functional form $-Q+f(Q)$, can be written as follows,

\begin{equation}\label{3d4v}
f+Q-2Qf_Q = 2\rho
\end{equation}
and
\begin{equation}\label{3e4v}
\dot{H}=\frac{\rho+p}{2 \left(-1+f_Q+2Qf_{QQ} \right)} \text{.} 
\end{equation}  

One can retrieve the Friedmann equations of GR (i.e. STEGR limit) for the function $-Q \:\: \ \ \text{i.e.} \:\: f(Q)=0$. These equations \eqref{3d4v} and \eqref{3e4v} can be interpreted as STGER cosmology with an additional component arising due to non-metricity of space-time which behaves like dark energy fluid part. These dark energy fluid components evolving due to non-metricity are defined by, 

\begin{equation}\label{3i4v}
\rho_{DE}=-\frac{f}{2}+Qf_Q
\end{equation}
and 
\begin{equation}\label{3j4v}
p_{DE}= -\rho_{DE} - 2\dot{H} \left(f_Q+2Qf_{QQ} \right) \text{.} 
\end{equation}

Now, the dimensionless density parameter corresponding to the dark energy fluid part is defined as,

\begin{equation}\label{3k4v}
\Omega_{DE}= \frac{\rho_{DE}}{3H^2} \text{.} 
\end{equation}

In addition, the EoS parameter that relates the dark energy density and its pressure is given by,

\begin{equation}\label{3l4v}
\omega_{DE}= \frac{p_{DE}}{\rho_{DE}} = -1 + 4\dot{H} \left( \frac{f_Q+2Qf_{QQ}}{f-2Qf_Q} \right) \text{.} 
\end{equation}

Using \eqref{3d4v} and \eqref{3e4v}, we get,

\begin{equation}\label{3m4v}
\omega_{DE} = -1 + \left(1+\omega\right) \frac{\left(f+Q-2Qf_Q\right) \left(f_Q+2Qf_{QQ} \right)}{\left(-1+f_Q+2Qf_{QQ} \right) \left(f-2Qf_Q \right)} \text{.} 
\end{equation}

Thus, the effective Friedmann equations along with a dark energy fluid part arising due to non-metricity reads,

\begin{equation}\label{3n4v}
H^2=\frac{1}{3} \left[ \rho+\rho_{DE} \right]
\end{equation}
and
\begin{equation}\label{3o4v}
\dot{H}=-\frac{1}{2} \left[ \rho+p+\rho_{DE}+p_{DE} \right] \text{.} 
\end{equation}

Moreover, these dark energy fluid part satisfy the standard continuity equation,

\begin{equation}\label{3p4v}
\dot{\rho}_{DE} + 3H \left( \rho_{DE} + p_{DE} \right) = 0 \text{.} 
\end{equation}

\section{Cosmological $f(Q)$ model}\label{sec4v}

For our analysis, we consider the following $f(Q)$ model which is a combination of a linear and a non-linear term of non-metricity scalar, 
\begin{equation}\label{4a4v}
f(Q)= \alpha Q + \beta Q^n \text{.} 
\end{equation}
 
Here $\alpha$, $\beta$ and $n\neq1$ are arbitrary parameters.

This particular functional form was motivated by a polynomial form applied, for instance, in Reference \cite{Harko-2}. Then for this specific $f(Q)$ cosmological model, we obtained a first-order differential equation for a Universe consisting of non-relativistic pressureless matter,

\begin{equation}\label{4b4v}
\dot{H} \left[ \alpha-1+ \beta n (2n-1)6^{n-1} H^{2n-2}   \right] + \frac{3}{2} H^2  \left[ \alpha-1- \beta (2n-1)6^{n-1} H^{2n-2} \right]=0 \text{.} 
\end{equation}

By using equations \eqref{3i4v} and \eqref{4a4v}, we obtain,

\begin{equation}\label{4c4v}
\rho_{DE}= 3\alpha H^2 + \beta \frac{(2n-1)}{2} 6^n H^{2n} \text{.} 
\end{equation}

Again, by using equations \eqref{3k4v} and \eqref{3m4v}, we obtain,

\begin{equation}\label{4d4v}
\Omega_{DE}= \frac{\rho_{DE}}{3H^2} = \alpha + \beta (2n-1) 6^{n-1} H^{2n-2}
\end{equation}
and
\begin{equation}\label{4e4v}
\omega_{DE}= -1+ \frac{\left( \alpha-1+ \beta (2n-1) 6^{n-1} H^{2n-2} \right) \left( \alpha+ \beta n (2n-1) 6^{n-1} H^{2n-2}  \right)}{\left( \alpha-1+ \beta n (2n-1) 6^{n-1} H^{2n-2}\right)\left( \alpha+ \beta (2n-1) 6^{n-1} H^{2n-2}\right)} \text{.} 
\end{equation}

Thus, the effective EoS parameter for our model is, 

\begin{equation}\label{4f4v}
w_{eff}= \frac{p_{eff}}{\rho_{eff}}=\frac{p_{DE}}{\rho + \rho_{DE}} \text{,} 
\end{equation}

where, $p_{eff}$ and $\rho_{eff}$ correspond to the total pressure and energy density of the Universe. Then we have,

\begin{equation}\label{4g4v}
w_{eff} = - \left( \alpha+ \beta (2n-1) 6^{n-1} H^{2n-2}\right) +  \frac{\left( \alpha-1+ \beta (2n-1) 6^{n-1} H^{2n-2} \right) \left( \alpha+ \beta n (2n-1) 6^{n-1} H^{2n-2}  \right)}{\left( \alpha-1+ \beta n (2n-1) 6^{n-1} H^{2n-2}\right)} \text{.} 
\end{equation}

Another key component that characterize the expansion phase of the Universe is deceleration parameter which is defined as \cite{Muj},

\begin{equation}\label{4h4v}
q= \frac{1}{2} \left( 1+3\Omega_{DE} \omega_{DE} \right) \text{.} 
\end{equation}

Using equations \eqref{4d4v} and \eqref{4e4v}, we get,

\begin{equation}\label{4i4v}
q=\frac{1}{2} + \frac{3}{2} \biggl\{ - \left( \alpha+ \beta (2n-1) 6^{n-1} H^{2n-2} \right) +  \frac{\left( \alpha-1+ \beta (2n-1) 6^{n-1} H^{2n-2} \right) \left( \alpha+ \beta n (2n-1) 6^{n-1} H^{2n-2}  \right)}{\left( \alpha-1+ \beta n (2n-1) 6^{n-1} H^{2n-2}\right)}  \biggr\} \text{.} 
\end{equation}

Now, we rewrite the Friedmann equations \eqref{3d4v} and \eqref{3e4v} with the help of dimensionless matter density parameter as,

\begin{equation}\label{4j4v}
H^2=\frac{1}{12(-1+f_Q)} \left[ -Q \left( \Omega +1 \right) + f \right]
\end{equation}
and
\begin{equation}\label{4k4v}
 \dot{H}= \frac{1}{4(-1+f_Q)} \left[ \Omega Q-4f_{QQ} \dot{Q} H \right] \text{,} 
\end{equation}

where $\Omega=\frac{\rho}{3H^2}$.

Our aim is to estimate the value of parameters of a given $f(Q)$ model that would be in agreement with the recently observed values of the cosmographic parameters. By using equations \eqref{4j4v} and \eqref{4k4v}, we obtain the values of model parameters $\alpha$ and $\beta$ in terms of present value cosmographic parameters and the power of non-metricity term $n$, as,

\begin{equation}\label{4l4v}
\alpha = 1+\frac{\Omega_0}{(1-n)} \left[ n- \frac{3}{2 \left(1+q_0 \right)} \right]
\end{equation}
and
\begin{equation}\label{4m4v}
\beta=\frac{\Omega_0}{(2n-1)(1-n)6^{n-1} H_0^{2n-2}} \left[  \frac{3}{2 \left(1+q_0 \right)}-1 \right] \text{,} 
\end{equation}

where $n\neq \frac{1}{2} \: $ and $\: n\neq 1$.

Now, we solve the differential equation \eqref{4b4v} by the numerical algorithm with initial conditions as $H_0=67.9 \: km/s/Mpc$, $q_0=-0.55$, $\Omega_0=0.303$ with present time $t_0=13.8$ Gyrs \cite{R2}. We have obtained the different behaviors of solution of our model depending upon the choice of power $n$ of the non-metricity term.

\section{Quintessence like behavior of $f(Q)$ gravity model}\label{sec5v}

\subsection{Cosmological parameters}

\begin{figure}[H]
\centering
{\includegraphics[scale=0.5]{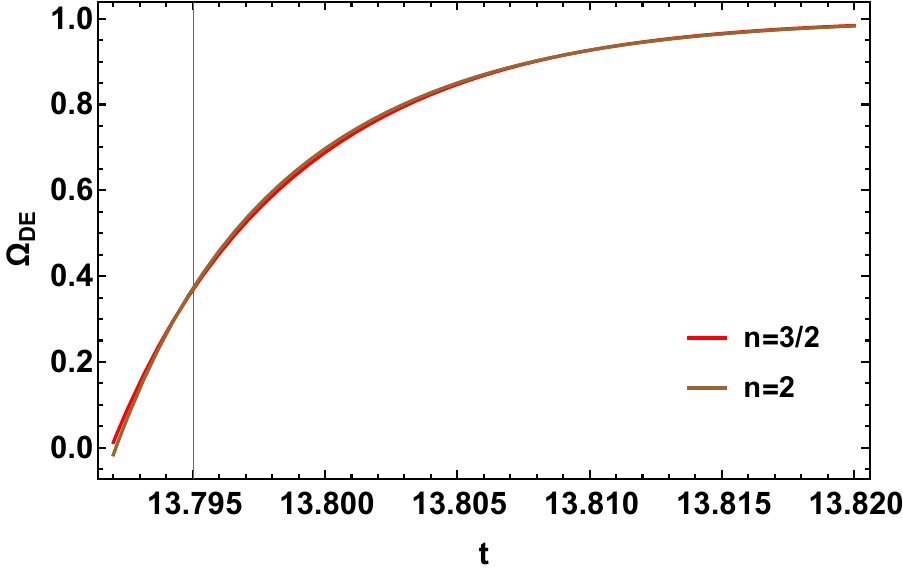}}
{\includegraphics[scale=0.505]{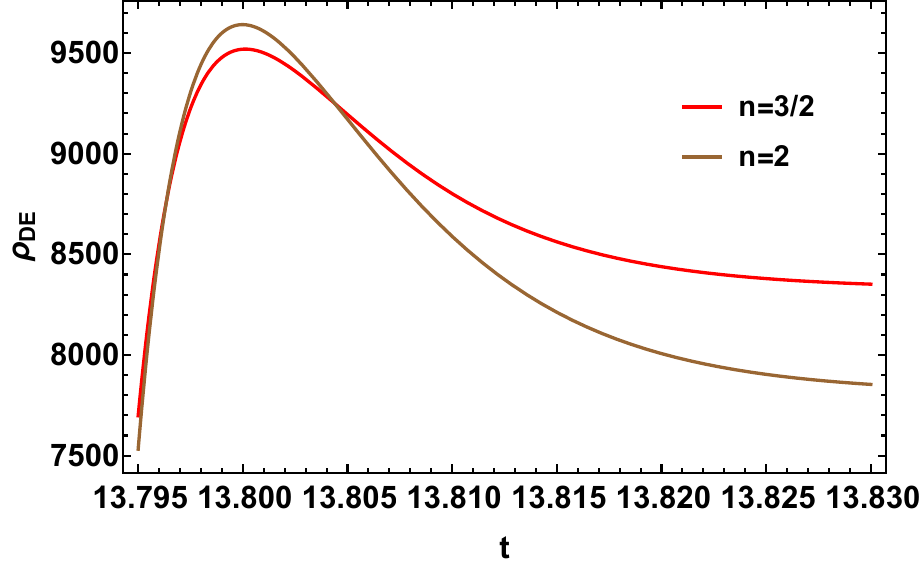}}
\caption{Profile of the dimensionless density parameter and the energy density for the dark energy component vs cosmic time t.}\label{f1v}
\end{figure}

\begin{figure}[H]
\centering
{\includegraphics[scale=0.45]{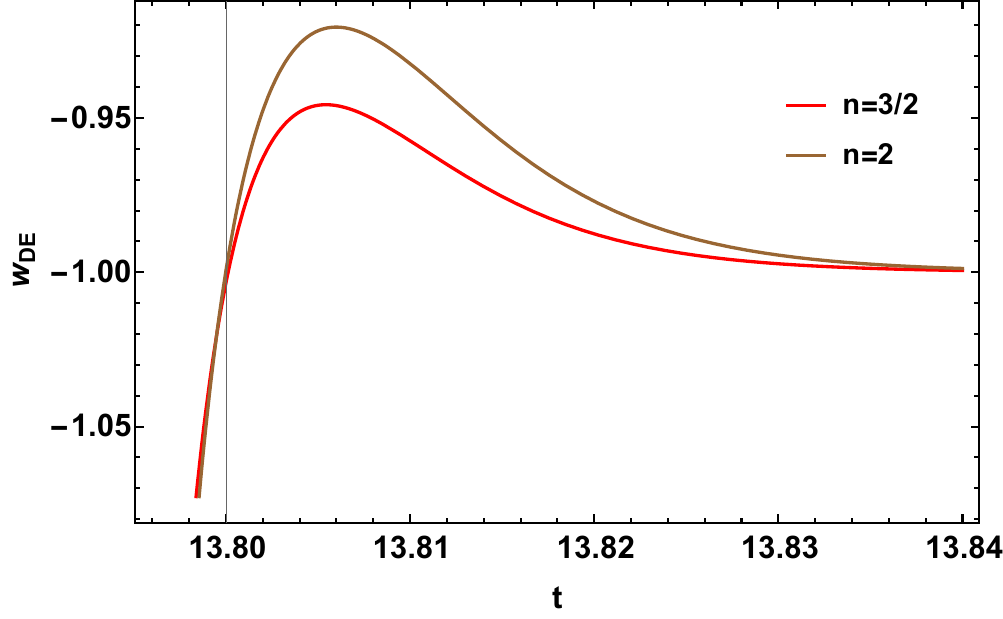}}
{\includegraphics[scale=0.43]{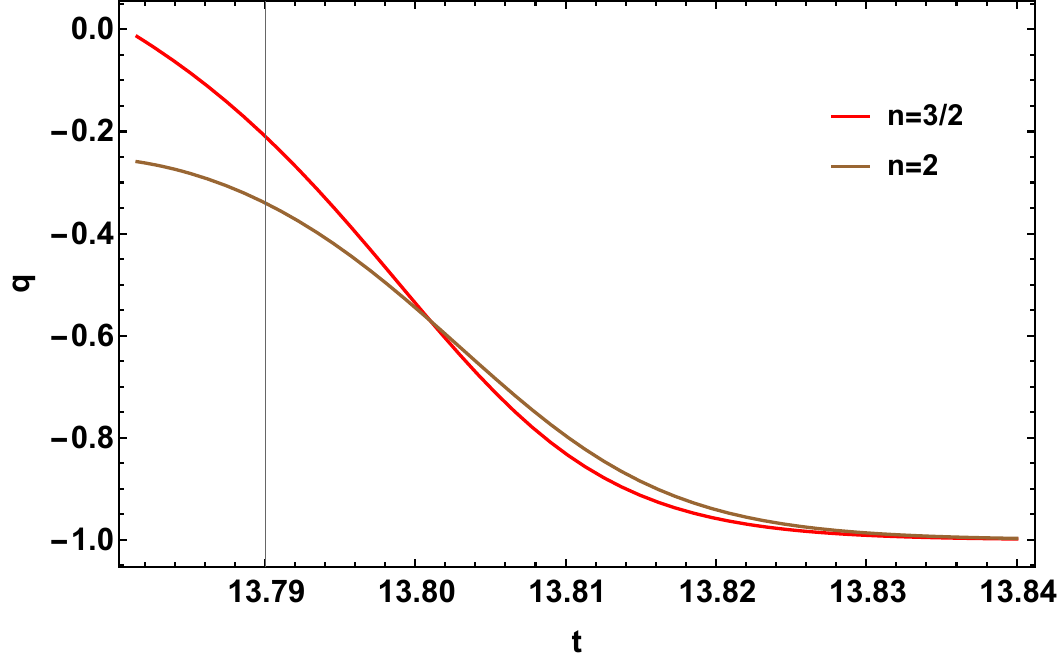}}
\caption{Profile of the EoS parameter for the dark energy component and the deceleration parameter vs cosmic time t. }\label{f3v}
\end{figure}

\begin{figure}[H]
\centering
{\includegraphics[scale=0.438]{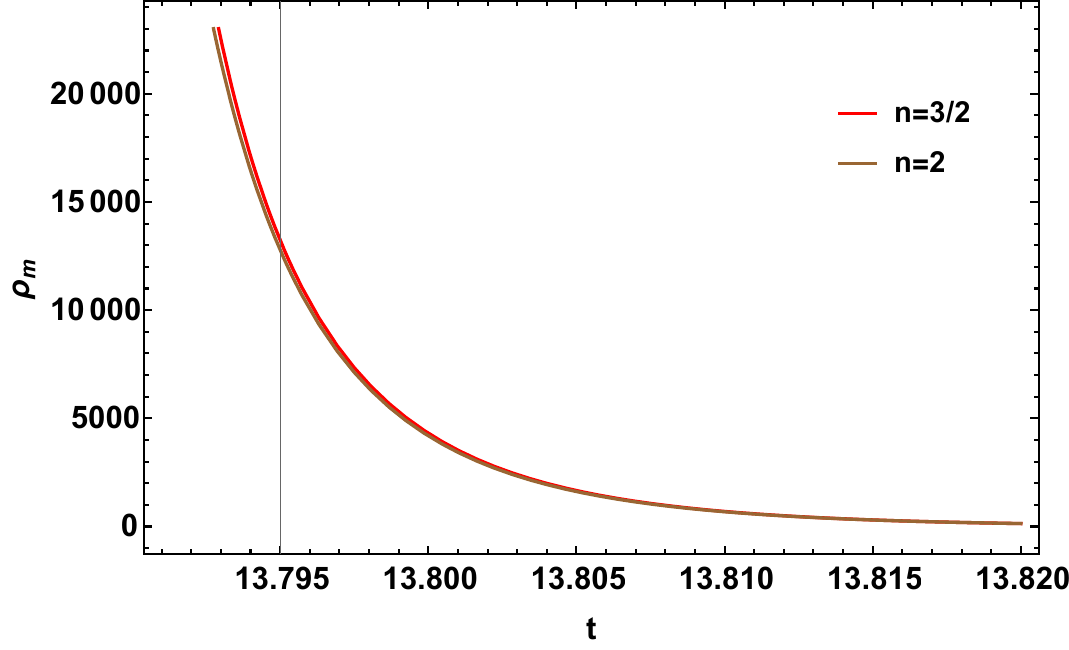}}
{\includegraphics[scale=0.44]{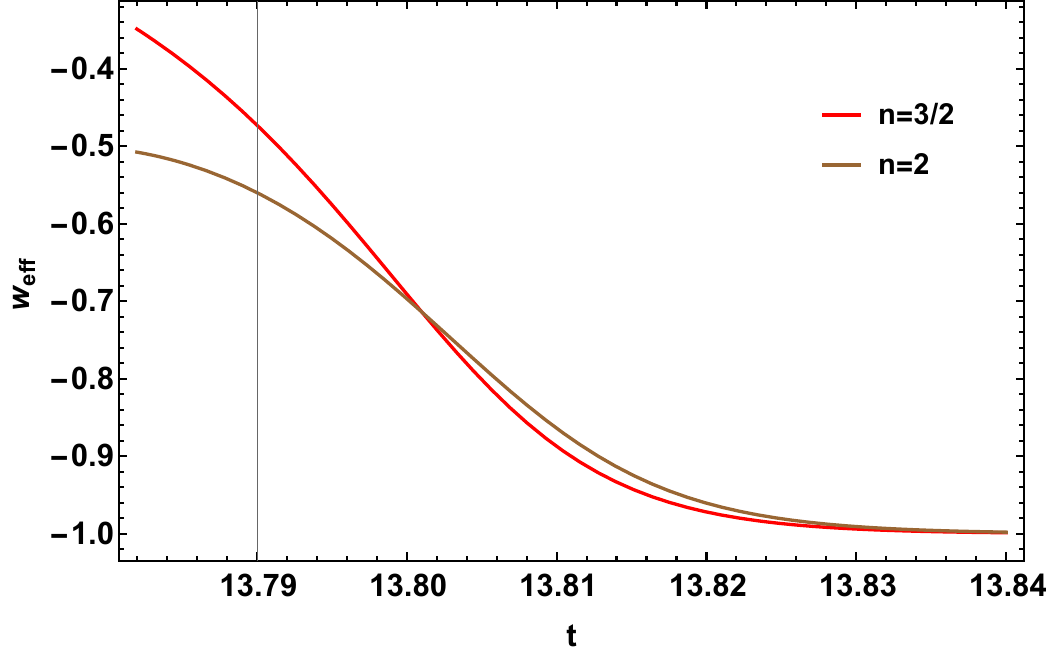}}
\caption{Profile of the matter-energy density and the effective EoS parameter vs cosmic time t.}\label{f5v}
\end{figure}

From figures \eqref{f1v} and \eqref{f5v} (left panel), we found that the both matter-energy density and dark energy density of the Universe decrease with cosmic time and matter energy density falls off to be zero while dark energy density becomes constant in the far future. Also, from figures \eqref{f3v} (left panel) and \eqref{f5v} (right panel) it is clear that the EoS parameter for both the dark energy and effective fluid converges to the $\Lambda$CDM EoS and the dark energy evolving due to non-metricity shows quintessence like behavior \cite{KF,KC}. Further, figure \eqref{f3v} (right panel) indicates a transition from decelerating to accelerating phase of the Universe in the recent past. 

\subsection{Energy conditions}

The energy conditions have  a fundamental role to describe the geodesics of the Universe and it can be derived from the Raychaudhuri equation \cite{EC}. The ECs are defined for the perfect fluid type effective matter content in $f(Q)$ gravity as $\rho_{eff} \geq 0$, $\rho_{eff}+p_{eff}\geq 0$, $\rho_{eff} \pm p_{eff}\geq 0$, and $\rho_{eff}+ 3p_{eff}\geq 0$.

\begin{figure}[H]
\centering
{\includegraphics[scale=0.42]{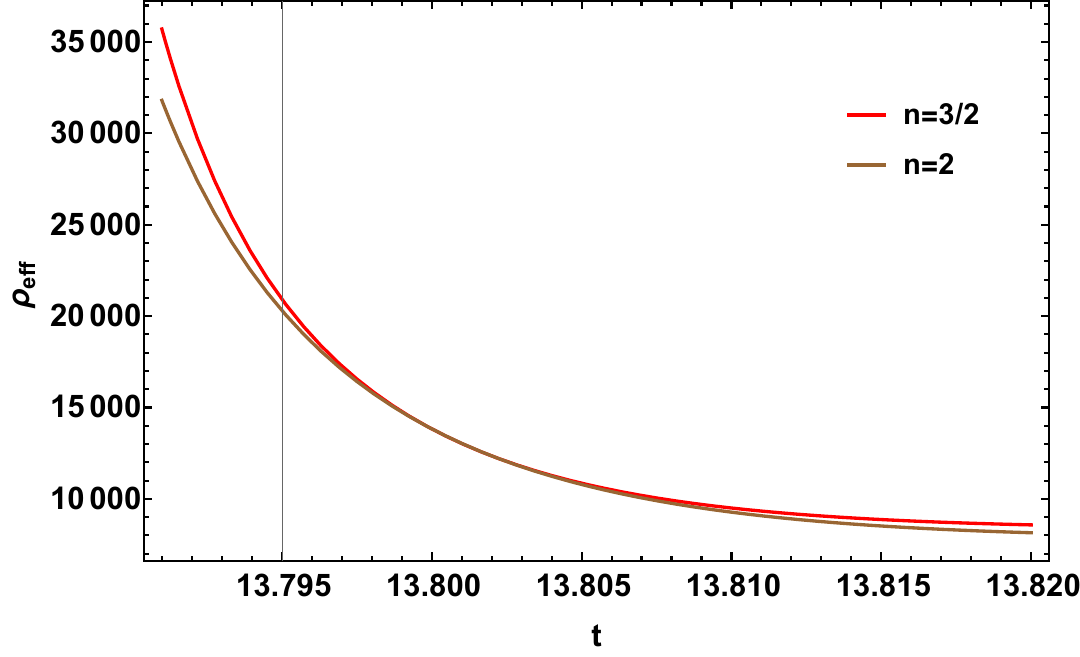}}
{\includegraphics[scale=0.43]{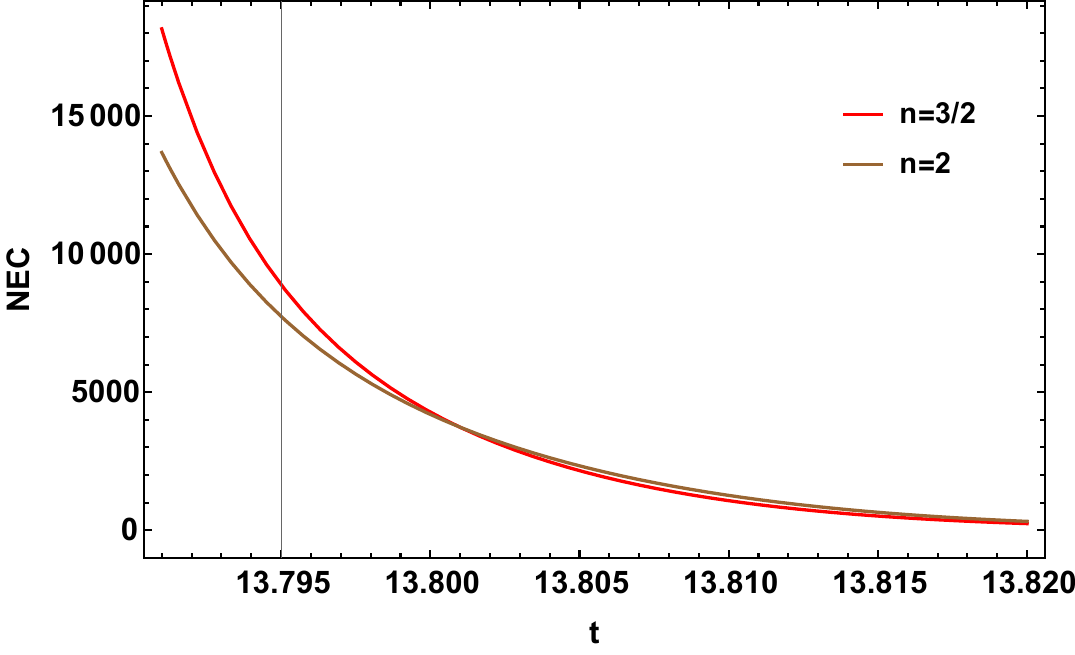}}
\caption{Profile of the effective energy density and the NEC vs cosmic time t.}\label{f7v}
\end{figure}

\begin{figure}[H]
\centering
{\includegraphics[scale=0.42]{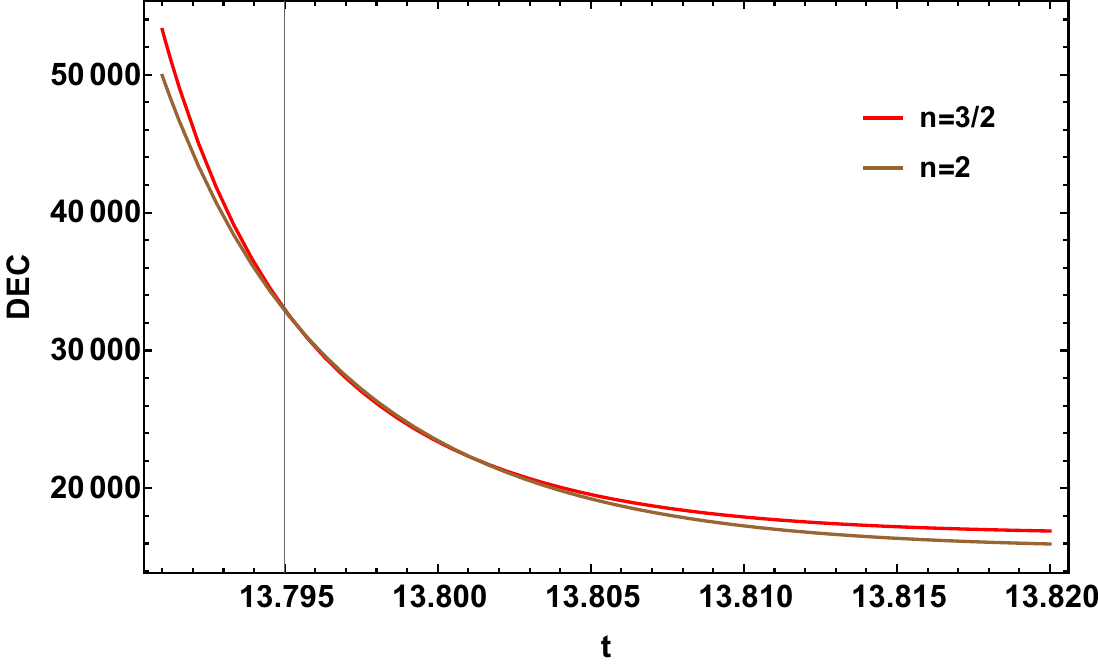}}
{\includegraphics[scale=0.46]{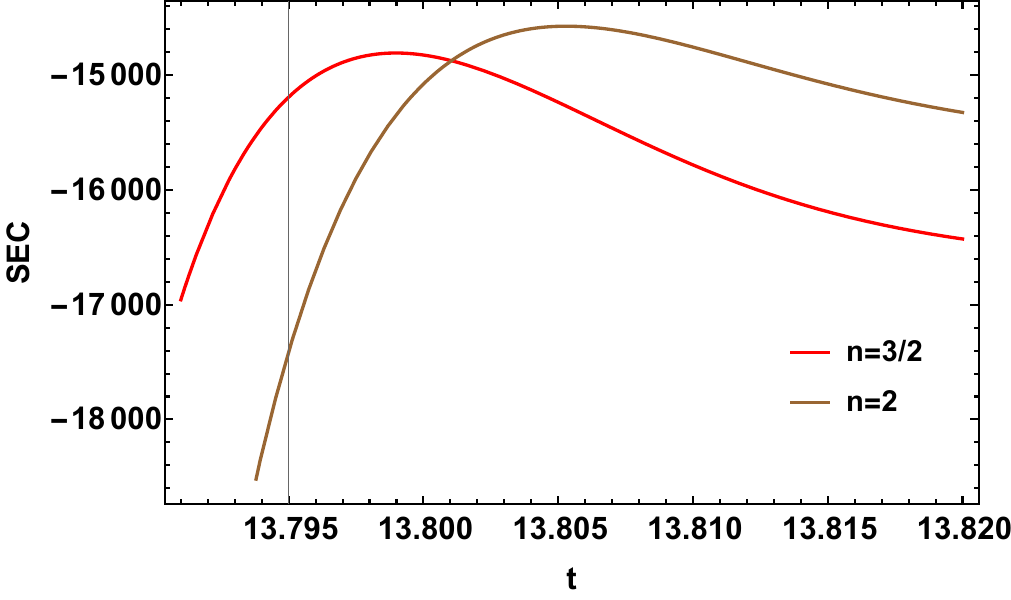}}
\caption{Profile of the DEC and the SEC vs cosmic time t.}\label{f9v}
\end{figure}

From figure \eqref{f7v} (left panel) it is clear that the effective energy density show positive behavior. Moreover from figures \eqref{f7v} (right panel) and \eqref{f9v} we found that NEC, WEC, and DEC are satisfied while the SEC is violated. Violation of SEC depicts the acceleration phase of the Universe. 

\section{Phantom like behavior of $f(Q)$ gravity model}\label{sec6v}

\subsection{Cosmological parameters}

\begin{figure}[H]
\centering
{\includegraphics[scale=0.46]{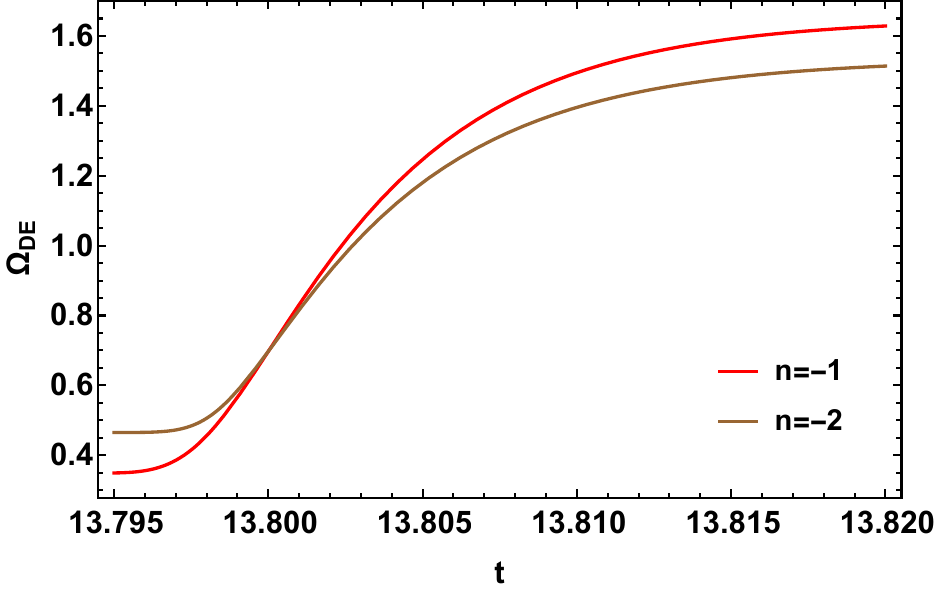}}
{\includegraphics[scale=0.45]{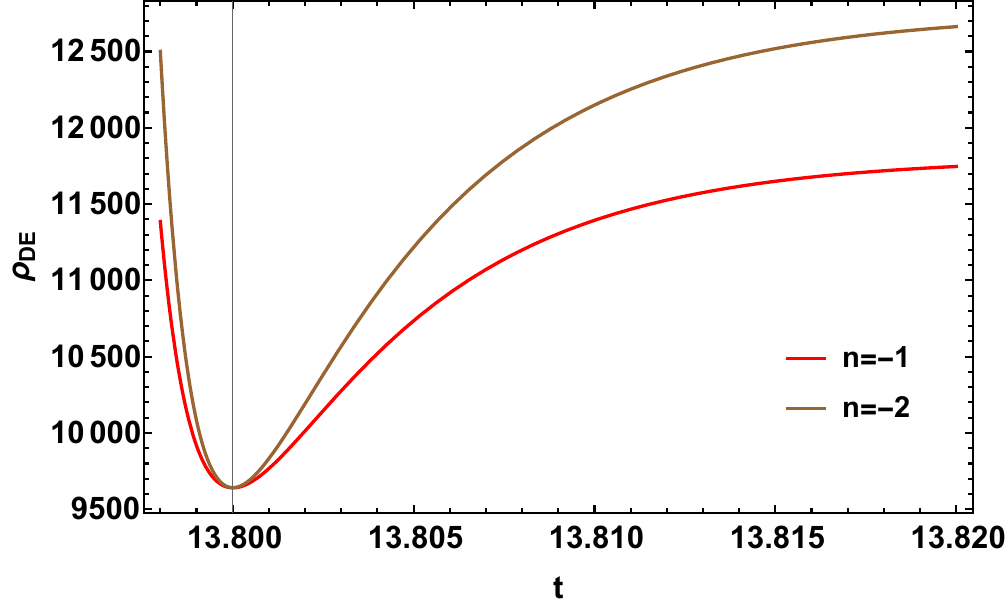}}
\caption{Profile of the dimensionless density parameter and the energy density for the dark energy component vs cosmic time t.}\label{f11v}
\end{figure}

\begin{figure}[H]
\centering
{\includegraphics[scale=0.46]{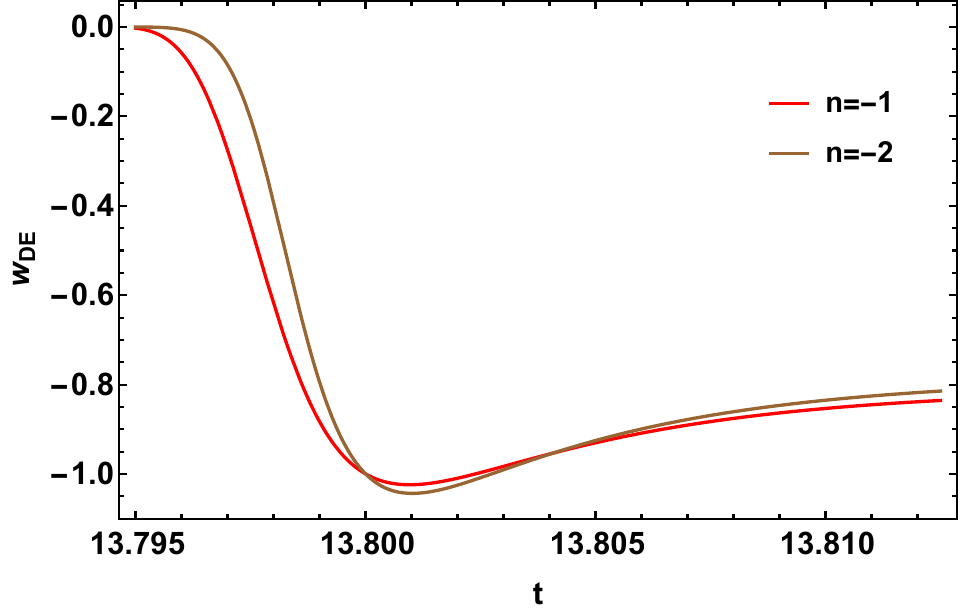}}
{\includegraphics[scale=0.45]{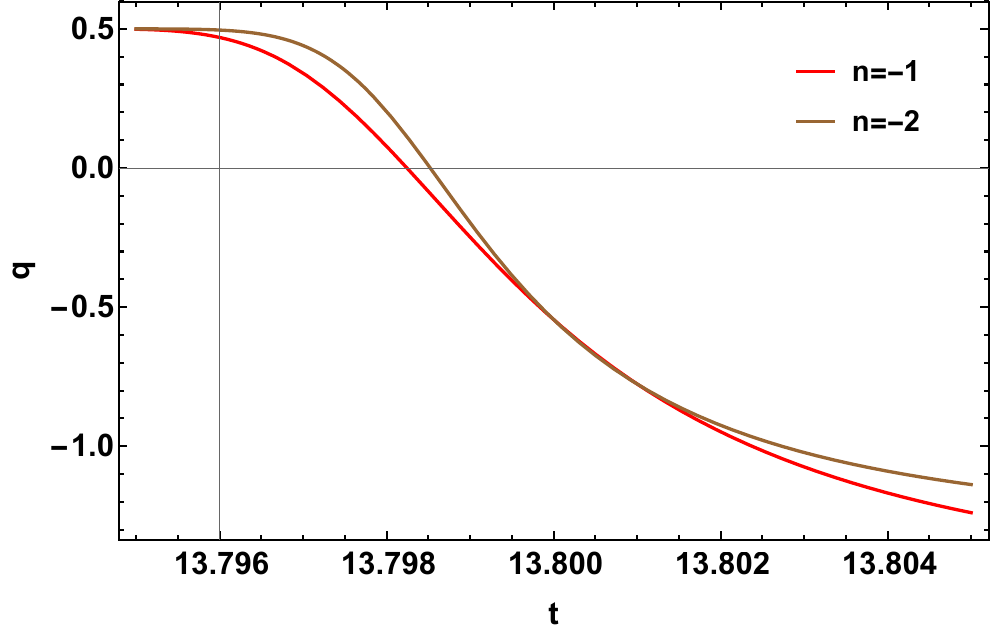}}
\caption{Profile of the EoS parameter for the dark energy component and the deceleration parameter vs cosmic time t. }\label{f13v}
\end{figure}

\begin{figure}[H]
\centering
{\includegraphics[scale=0.455]{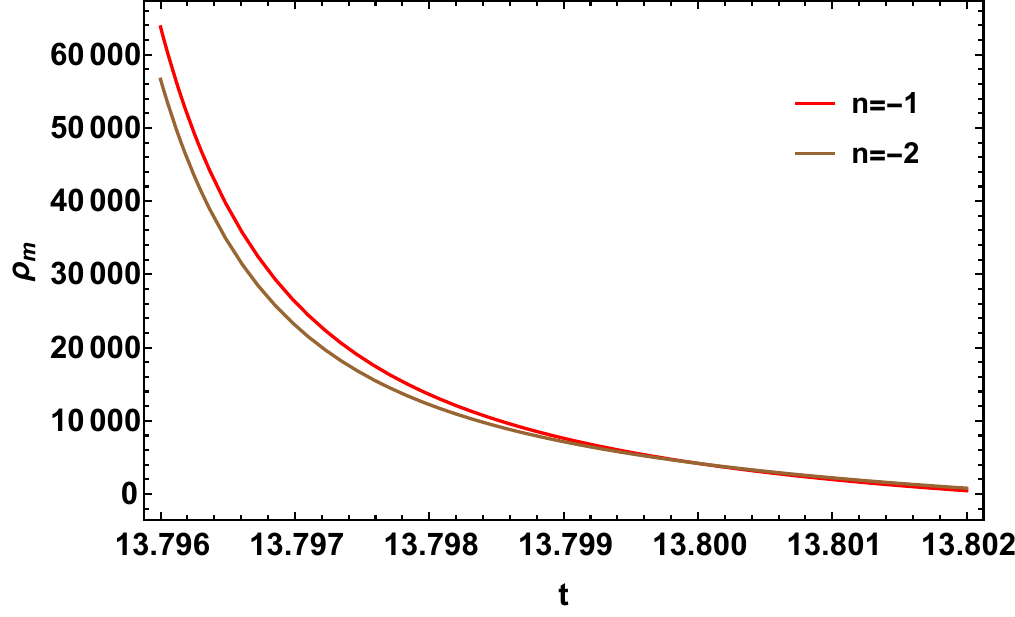}}
{\includegraphics[scale=0.45]{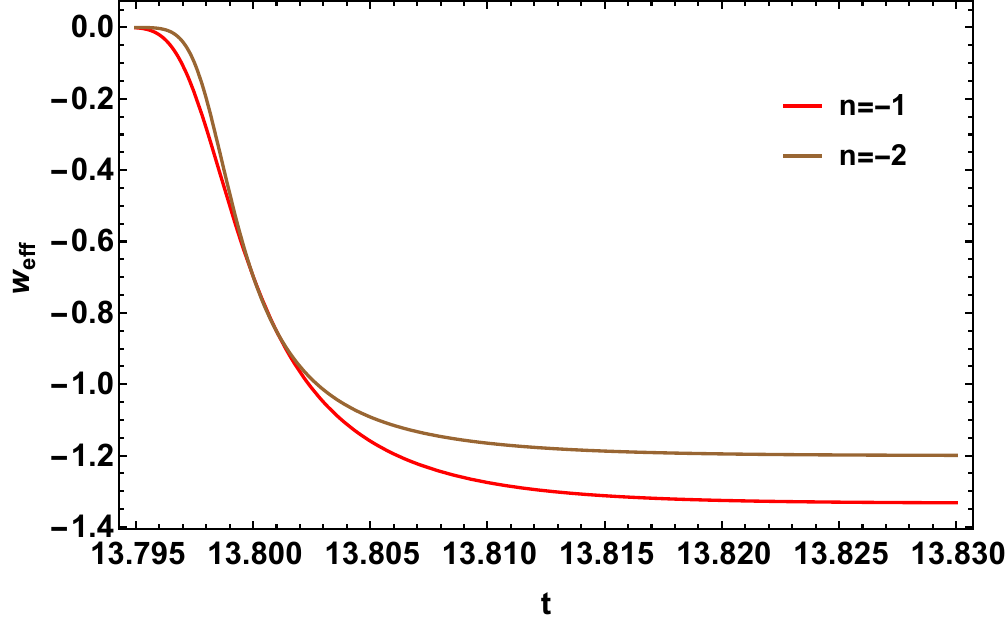}}
\caption{Profile of the matter-energy density and the effective EoS parameter vs cosmic time t.}\label{f15v}
\end{figure}

From figures \eqref{f11v} and \eqref{f15v} (left panel), we found that the matter-energy density of the Universe decrease with cosmic time while dark energy density increases with time. Moreover, the matter energy density falls off to be zero while dark energy density becomes constant in the far future.  Also, from figures \eqref{f13v} (left panel) and \eqref{f15v} (right panel) we found that the EoS parameter for the dark energy fluid shows a transiting behavior from phantom to non-phantom phase while the effective EoS parameters crosses the $\Lambda$CDM line and then converges in phantom region. Hence, the dark energy evolving due to non-metricity shows phantom like behavior. Further, figure \eqref{f13v} (right panel) indicates a transition from decelerating to accelerating phase of the Universe in the recent past.

\subsection{Energy conditions}

\begin{figure}[H]
\centering
{\includegraphics[scale=0.43]{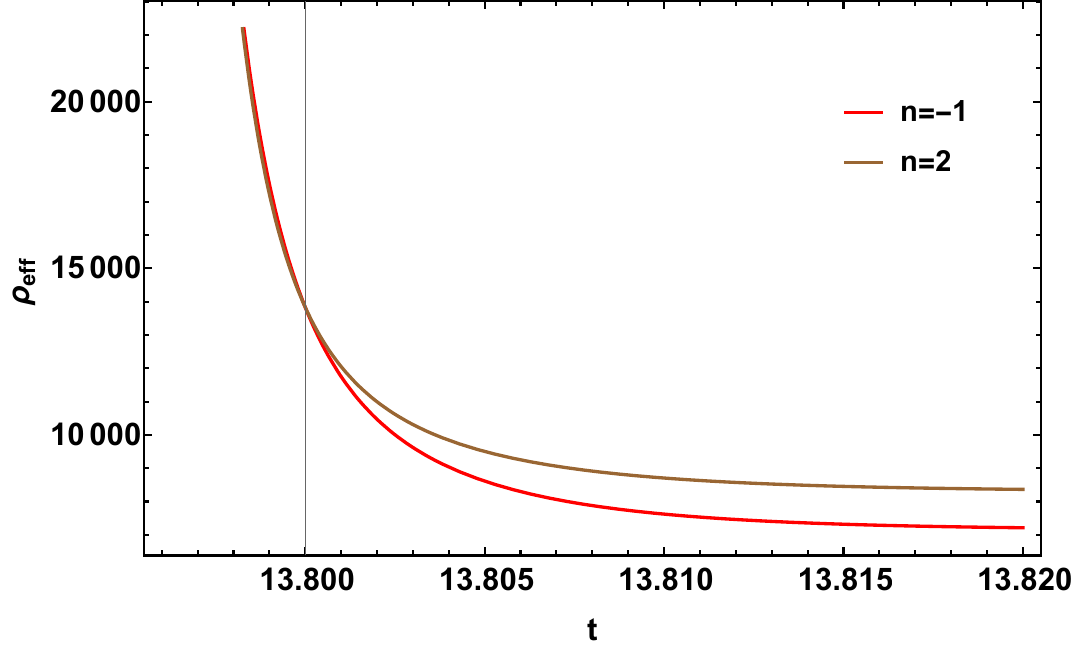}}
{\includegraphics[scale=0.45]{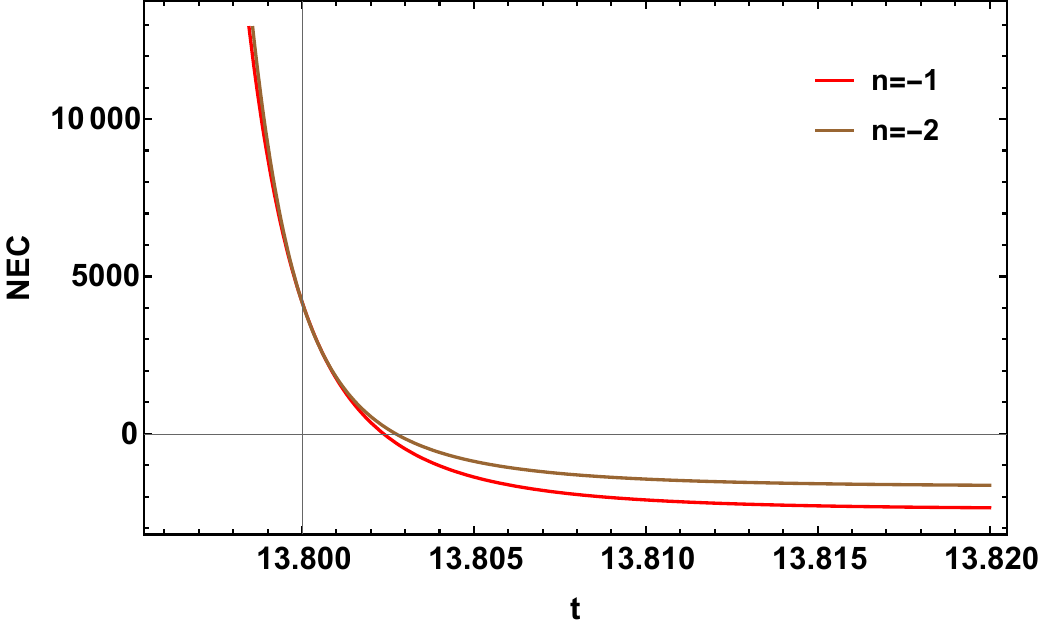}}
\caption{Profile of the effective energy density and the NEC vs cosmic time t.}\label{f17v}
\end{figure}

\begin{figure}[H]
\centering
{\includegraphics[scale=0.45]{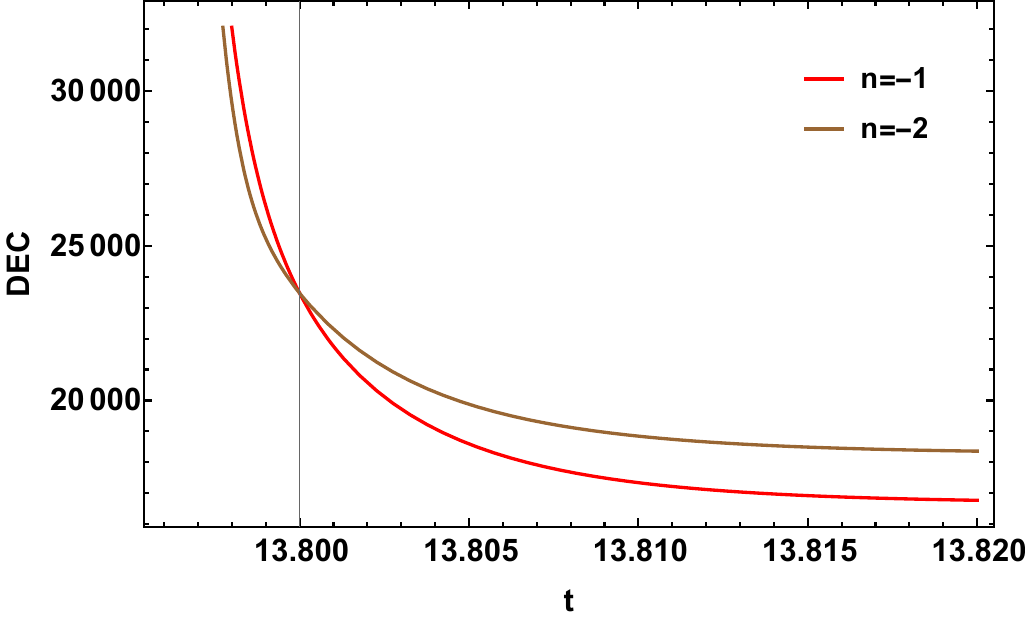}}
{\includegraphics[scale=0.45]{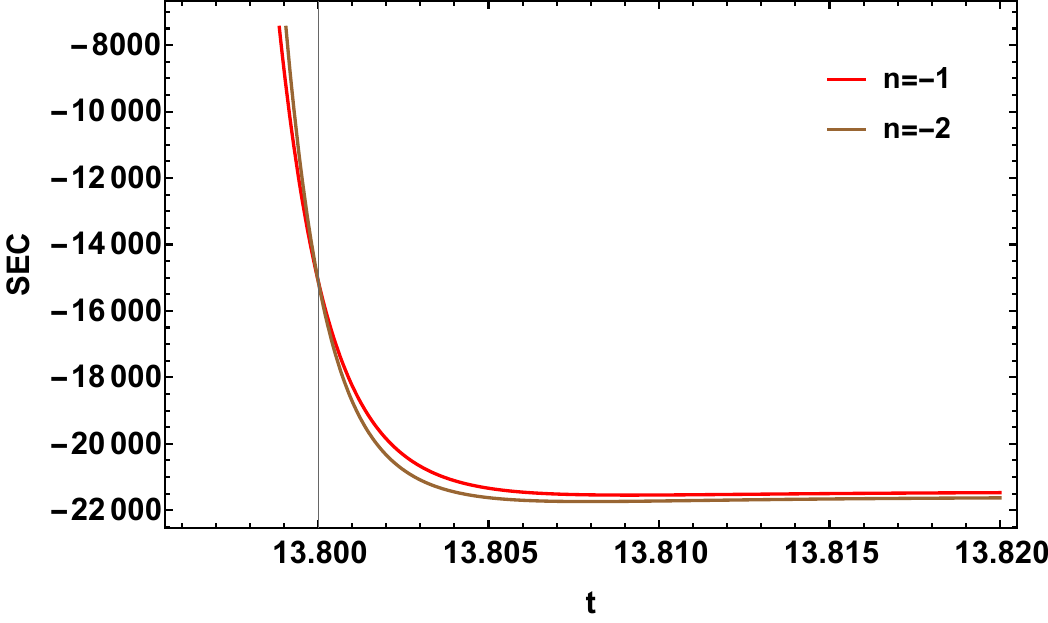}}
\caption{Profile of the DEC and the SEC vs cosmic time t.}\label{f19v}
\end{figure}

From figure \eqref{f17v} (left panel) it is clear that the effective energy density show positive behavior. Moreover from figures \eqref{f17v} (right panel) and \eqref{f19v} we found that NEC and SEC are violated. By the definition of energy conditions violation of NEC implies the violation of WEC and DEC. Thus all the ECs are violated in this case. Violation of SEC depicts the acceleration phase of the Universe while the violation of NEC depicts the existence of exotic matter. Thus the violation of ECs also shows that the dark energy evolving due to non-metricity behaves like phantom energy.

\section{$\Lambda$CDM like behavior of $f(Q)$ gravity model}\label{sec7v}

Consider the case $\alpha=0 \:$, $\: n=0 \:$ and $\: \beta < 0 \:$. In this case we have $\rho_{DE}= -\frac{\beta}{2} = constant \:$, $\: p_{DE}= \frac{\beta}{2} = constant \:$ and $\: \omega_{DE}=-1 $.
Also, the density parameter and effective EoS parameter becomes 

\begin{equation}
\Omega_{DE}= - \frac{\beta}{6H^2}
\end{equation}
and
\begin{equation}
w_{eff} = \frac{\beta}{6H^2}
\end{equation}
and
\begin{equation}
q=\frac{1}{2} + \frac{\beta}{4H^2} \text{.} 
\end{equation}

\begin{figure}[H]
\centering
{\includegraphics[scale=0.44]{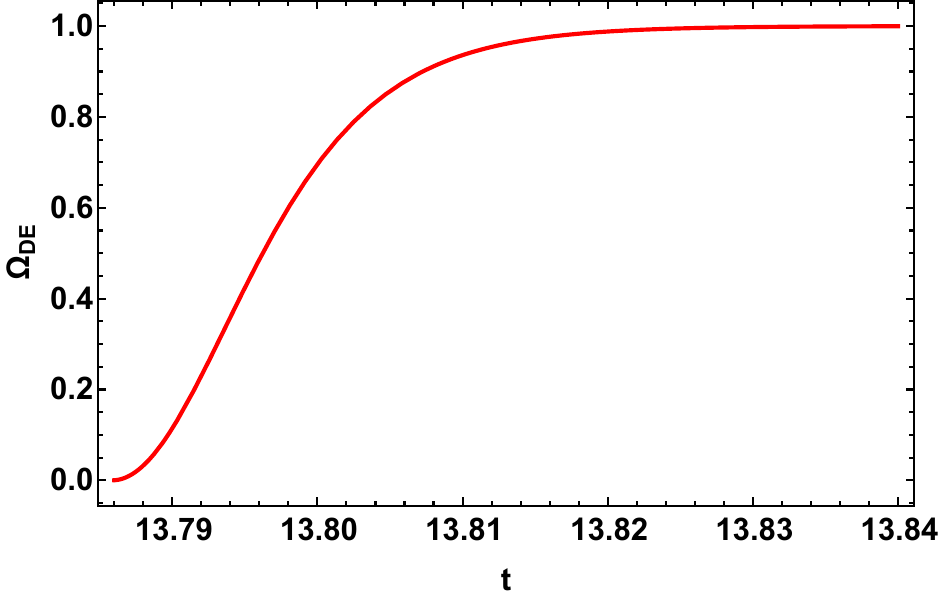}}
{\includegraphics[scale=0.44]{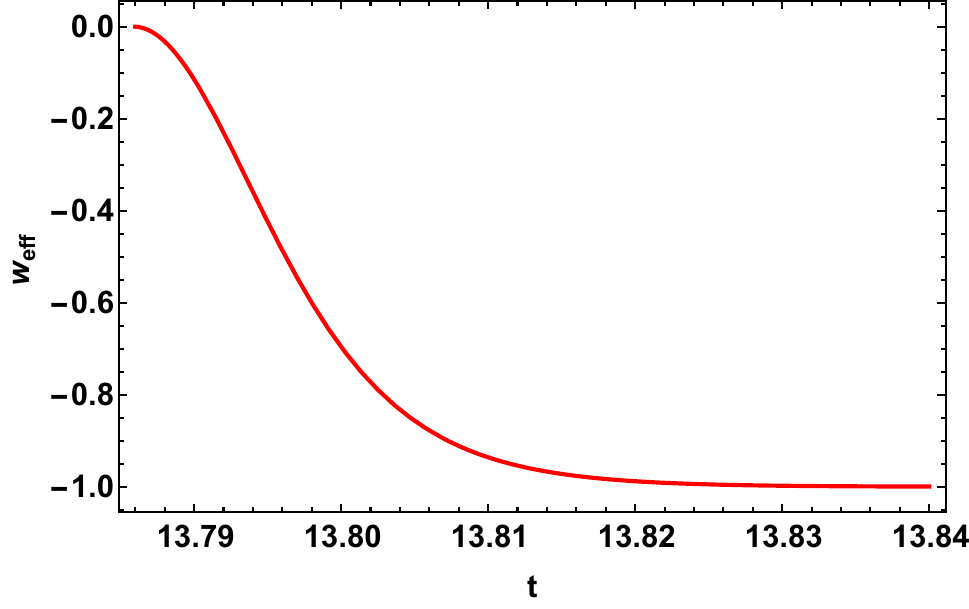}}
{\includegraphics[scale=0.45]{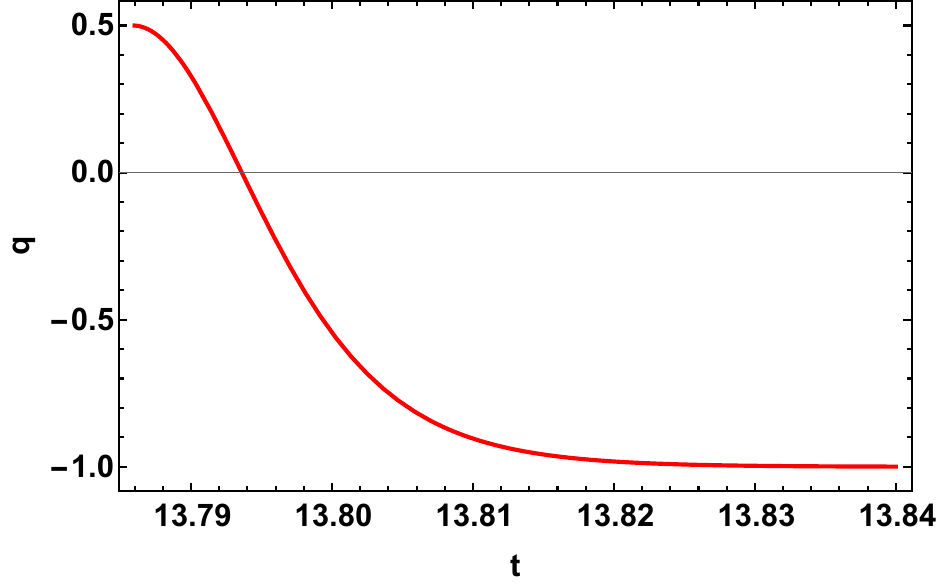}}
\caption{Profile of the dimensionless density parameter for the dark energy component, the effective EoS parameter, and the deceleration parameter vs cosmic time t.}\label{f21v}
\end{figure}

In this case, we have obtained constant negative pressure with dark energy EoS parameter follows the $\Lambda$CDM EoS. The effective EoS parameter in the figure \eqref{f21v} (right panel) converges to the $\Lambda$CDM line and the deceleration parameter in the figure \eqref{f21v} (below) shows a transition from deceleration phase to acceleration phase in the recent past. Hence, this cosmological $f(Q)$ model mimics the standard $\Lambda$CDM model.

\section{Exact solution and parameter constraints for the case $f(Q)=\alpha Q + \beta$}\label{secaddv}

On putting the $n=0$ i.e. for the case $f(Q)=\alpha Q + \beta$, the equation \eqref{4b4v} for the non-relativistic pressureless matter becomes,
\begin{equation}\label{4ba1v}
\dot{H} \left[ \alpha-1 \right] + \frac{3}{2} H^2  \left[ \alpha-1+ \frac{\beta}{6H^2} \right]=0 \text{.} 
\end{equation}

\begin{figure}[H]
\centering
\includegraphics[scale=0.85]{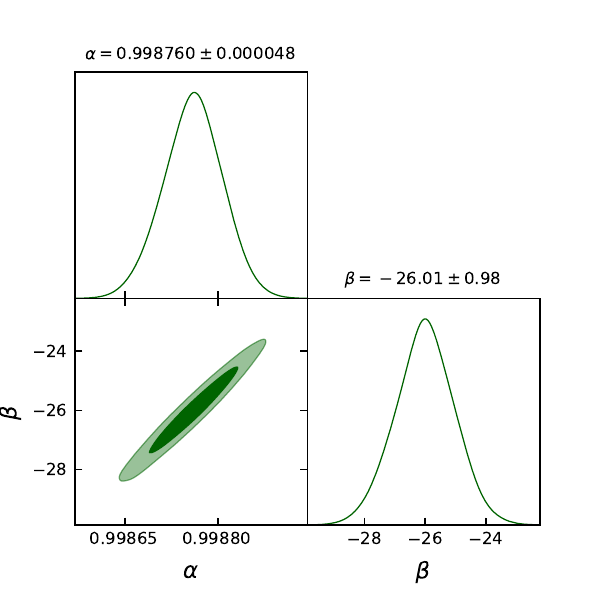}
\caption{The $1-\sigma$ and $2-\sigma$ likelihood contours for the model parameters using the combination CC+BAO+Pantheon datasets.}\label{f3a1v}
\end{figure}

Now by solving equation \eqref{4ba1v}, we obtained the analytical expression for the Hubble parameter in terms of redshift as
\begin{equation}\label{4ja2v}
H(z)= \bigl\{ H_0^2(1+z)^3+ \frac{\beta}{6(\alpha-1)} \left[ 1-(1+z)^3 \right] \bigr\}^{\frac{1}{2}} \text{,} 
\end{equation}
where $H_0=67.9\pm 0.5$ km/s/Mpc \cite{R2} is the present Hubble parameter value. To constrain the free parameters of the above function, we use the Hubble datasets of 31 points using DA method (Cosmic Chronometers), BAO datasets consisting of 6 points, and Pantheon Supernovae observations. We have calculated the best fit ranges for parameters $\alpha$ and $\beta$ by minimizing the chi-square function for the combination CC+BAO+Pantheon. The obtained values for the arbitrary parameters are $\alpha= 0.998760 \pm 0.000048$ and $\beta=-26.01 \pm 0.98$. The $1-\sigma$ and $2-\sigma$ likelihood contour for the model parameters using the combine CC+BAO+Pantheon datasets is presented in figure \eqref{f3a1v}.

\subsection{Cosmological parameters}

The evolution of different cosmological parameters of our cosmological model such as density, deceleration, and the EoS parameters for the estimated values of the free parameters are presented below.
\begin{figure}[H]
\centering
{\includegraphics[scale=0.49]{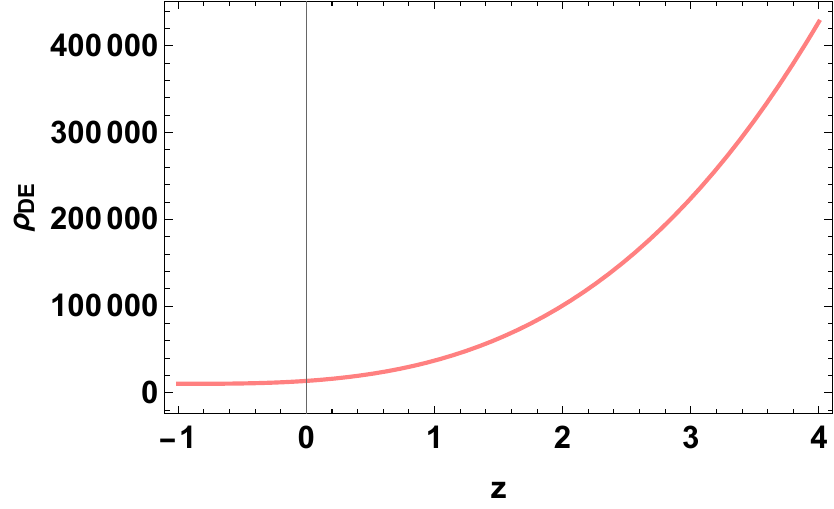}}
{\includegraphics[scale=0.47]{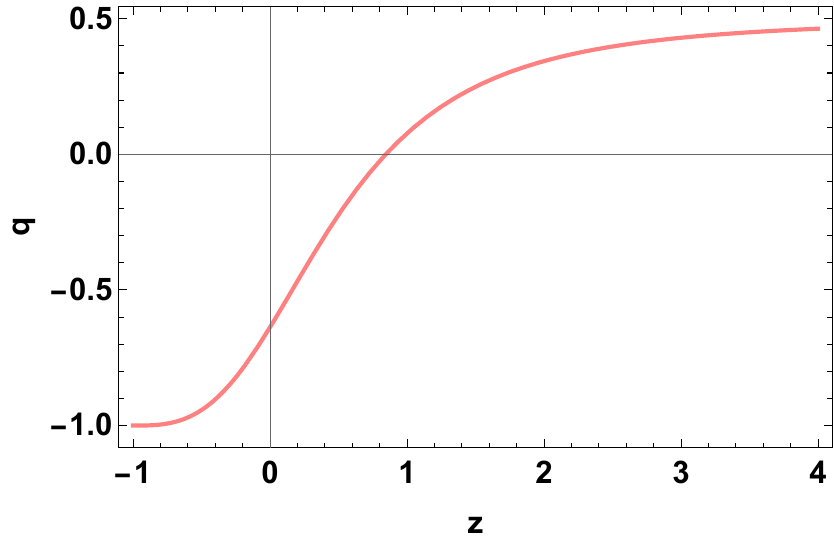}}
\caption{Profile of the density of the dark energy component and the deceleration parameter vs redshift z .}\label{f4a2v}
\end{figure}

\begin{figure}[H]
\centering
{\includegraphics[scale=0.47]{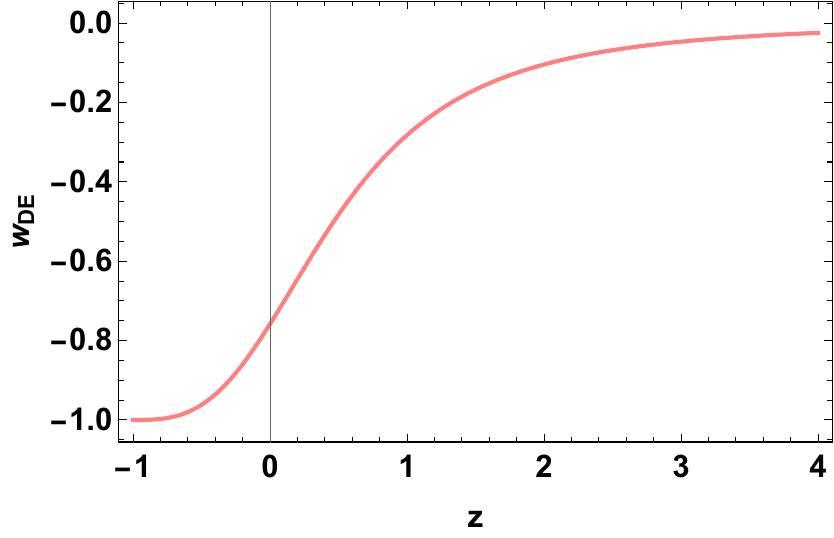}}
{\includegraphics[scale=0.47]{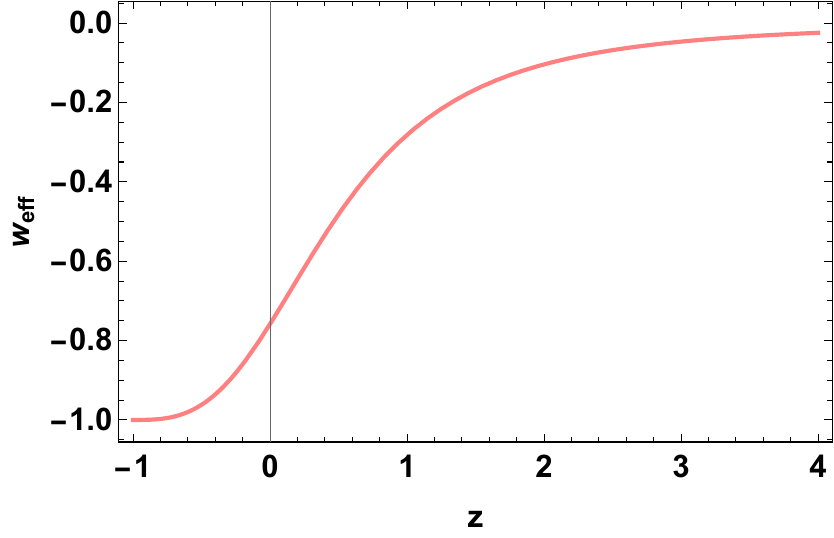}}
\caption{Profile of the EoS parameter for the dark energy component and the effective EoS parameter vs redshift z .}\label{f6a4v}
\end{figure}

Form figure \eqref{f4a2v} (left panel) it is clear that the energy density of the dark energy component of the Universe decreases with cosmic time and it falls off to be zero in the far future. Hence our model represents a decaying dark energy model. Figure \eqref{f4a2v} (right panel) show that a transition from decelerating to accelerating phase has been experienced by the Universe in the recent past with transition redshift $z_t=0.844^{+0.0027}_{-0.0025}$. Further from figure \eqref{f6a4v}, we observe that the EoS parameter for the dark energy fluid evolving due to non-metricity and effective EoS parameter shows quintessence like behavior. The present values of the deceleration parameter and EoS parameter for the dark energy fluid part are $q_0=-0.63^{+0.0011}_{-0.0012}$ and $w_0=-0.75^{+0.0007}_{-0.0008}$.

\section{Conclusions}\label{sec8v}

In this article, we attempted to describe the evolution of dark energy from the geometry of spacetime. We considered a $f(Q)$ model which contains a linear and a non-linear form of non-metricity scalar that is $f(Q)=\alpha Q + \beta Q^n$, where $\alpha$, $\beta$, and $n$ are free parameters. Then we derived the expressions for density, deceleration, and the EoS parameters for our cosmological model. We found that for higher positive values of $n$ specifically $n \geq 1$, dark energy fluid part evolving due to non-metricity behaves like quintessence type dark energy while for higher negative values of $n$ specifically $n \leq -1$ it follows phantom scenario. For $n=0$, our cosmological $f(Q)$ model mimics the $\Lambda$CDM model of GR. Furthermore, we have analyzed the behavior of different energy conditions. We obtained that, except SEC all the energy conditions are satisfied for the case $n \geq 1$ while all the energy conditions are violated for the case $n \leq -1$. Therefore, depending upon the choice of $n$ we found that one can obtain quintessence, phantom, and the $\Lambda$CDM like behavior without invoking any dark energy component or exotic fluid in the matter part. 

In addition, we found the exact solution for the case $f(Q)=\alpha Q + \beta $, where $\alpha$ and $\beta$ are free parameters.  Further, to constrain the model parameters we used CC datasets consisting 31 data points, 6 points of the BAO datasets, and 1048 points from the Pantheon supernovae samples. We have calculated the suitable values of free parameters $\alpha$ and $\beta$ for the combine CC+BAO+Pantheon datasets. The obtained free parameter values are $\alpha= 0.998760 \pm 0.000048$ and $\beta=-26.01 \pm 0.98$.  Further, we have studied the evolution of different cosmological parameters corresponding to these best fit values of the model parameters. The evolution trajectory of the deceleration parameter shows that our Universe had experienced a transition from deceleration to acceleration phase in the recent past with the transition redshift $z_t=0.844^{+0.0027}_{-0.0025}$. The obtained present value of the deceleration parameter is $q_0=-0.63^{+0.0011}_{-0.0012}$. Further, the present value of EoS parameter i.e. $w_0=-0.75^{+0.0007}_{-0.0008}$ for the dark energy fluid part indicating a quintessence type behavior of our model. Thus, we conclude that the geometrical generalization of GR can be a viable candidate for the description of origin of the dark energy.

%% file: Chapters/Chapter5.tex

\chapter{Cosmological constraints on $f(Q)$ gravity models in the non-coincident formalism} 

\label{Chapter5} 

\lhead{Chapter 5. \emph{Cosmological constraints on $f(Q)$ gravity models in the non-coincident formalism}} 
\vspace{10 cm}
* The work presented in this chapter is covered by the following two publications: \\
 
\textit{Cosmological constraints on $f(Q)$ gravity models in the non-coincident formalism}, Journal of High Energy Astrophysics \textbf{43}, 258-267 (2024).

\clearpage
The chapter investigates cosmological applications of $f(Q)$ theories in a non-coincident formalism. We explore a new $f(Q)$ theory dynamics utilizing a non-vanishing affine connection involving a non-constant function $\gamma(t)=-a^{-1}\dot{H}$, resulting in Friedmann equations that are entirely distinct from those of $f(T)$ theory. In addition, we propose a new parameterization of the Hubble function that can consistently depicts the present deceleration parameter value, transition redshift, and the late time de-Sitter limit. We evaluate the predictions of the assumed Hubble function by imposing constraints on the free parameters utilizing Bayesian statistical analysis to estimate the posterior probability by employing the CC, Pantheon+SH0ES, and the BAO samples. Moreover, we conduct the AIC and BIC statistical evaluations to determine the reliability of MCMC analysis. Further, we consider some well-known corrections to the STEGR case such as an exponentital $f(Q)$ correction, logarithmic $f(Q)$ correction, and a power-law $f(Q)$ correction and then we find the constraints on the parameters of these models via energy conditions. Finally, to test the physical plausibility of the assumed $f(Q)$ models we conduct the thermodynamical stability analysis via the sound speed parameter.

\section{Introduction}\label{sec1m}

Upon a thorough review of the literature on cosmological applications of 
$f(Q)$ theories, we observed a prevalent trend of utilizing only the vanishing affine connection in the spatially flat FLRW spacetime. However, in this scenario, the Friedmann equations are identical to those of the 
$f(T)$ theory \cite{R49}. As a result, this specific gauge limits researchers to the outcomes already achieved in $f(T)$ theory, undermining the significance of $f(Q)$ theory as a novel modified gravity theory. A new framework for $f(Q)$ theory has been introduced in \cite{ab2}, based on a different gauge equivalence class of non-vanishing affine connections, incorporating a function
$\gamma$ in the spatially flat FLRW background.
In this chapter, we explore the observational constraints on several well known class of $f(Q)$ gravity models in this novel non-coincident formalism, along with a new parameterization of the Hubble function that can adequately describes the different cosmological epochs such as late time acceleration with recent transtion from deceleration epoch to acceleration epoch with observational compatibility.

In section \ref{sec3m}, we present Friedmann like equations for the $f(Q)$ gravity formalism in a non-coincident gauge setting. We also propose a new parameterization scheme of the Hubble function. In section \ref{sec4m}, we carry out a statistical analysis to evaluate the predictions of the proposed Hubble function using recent observational data. Further in section \ref{sec5m}, we test the viability of different $f(Q)$ gravity models utilizing the thermodynamical stability and produce the parameter constraints on these models with the help of energy conditions. In section \ref{sec6m}, we conclude our outcomes of the investigation.

\section{Equations of motion}\label{sec3m}
The FLRW metric for  spatially flat homogeneous and isotropic case, expressed in spherical coordinates, is given by,

\begin{eqnarray}\label{metricm}
ds^2= & -dt^2+a(t)^{2}\left( dr^2+r^{2} d\theta^2 +r^{2} \sin^{2}\theta d\phi^2 \right) \text{.} 
\end{eqnarray}

Several significant publications have recently emerged on the modified 
$f(Q)$ gravity theory and its cosmological implications, in the references \cite{2,2aa,2ab,2ac}. Additionally, the dynamical system analysis of the $f(Q)$ theory at both background and perturbative levels for a spatially flat FLRW spacetime has been conducted in \cite{16}. These analyses, however, have predominantly utilized the coincident gauge choice. In this gauge, the line element is expressed in cartesian coordinates with $\mathring{\Gamma}^{\lambda}{}_{\mu \nu} = 0$. This simplification reduces the covariant derivative to a partial derivative, facilitating easier calculations.

In the present discussion, we examine a significant class of affine connections with $\mathring{\Gamma} \neq 0$. The non-vanishing components of the affine connection $\mathring{\Gamma}$ corresponding to the metric (\ref{metricm}) are as follows,

\begin{eqnarray}
\nonumber
  &&\hspace{0cm}   \mathring{\Gamma}^{t}{ }_{t t}=\gamma+\frac{\dot{\gamma}}{\gamma}, \quad \mathring{\Gamma}^{r}{ }_{t r}=\gamma, \quad \mathring{\Gamma}^{r}{ }_{\theta \theta}=-r \text{,} \\ \nonumber
&&\hspace{0cm} \mathring{\Gamma}^{r}{ }_{\phi \phi}=-r \sin ^{2} \theta, \quad \mathring{\Gamma}^{\theta}{ }_{t \theta}=\gamma \text{,} 
 \\ \nonumber
&&\hspace{0cm} \mathring{\Gamma}^{\theta}{ }_{r \theta}=\frac{1}{r}, \quad \mathring{\Gamma}^{\theta}{ }_{\phi \phi}=-\cos \theta \sin \theta, \quad \mathring{\Gamma}^{\phi}{ }_{t \phi}=\gamma \text{,}  \\
 &&\hspace{0cm} \mathring{\Gamma}^{\phi}{ }_{r \phi}=\frac{1}{r}, \quad \mathring{\Gamma}^{\phi}{ }_{\theta \phi}=\cot \theta \text{.} 
\end{eqnarray}

We can then calculate the required tensors in the general setting,

\begin{eqnarray}
&&\hspace{0cm}Q_{t t t}=2 \big(\gamma+\frac{\dot{\gamma}}{\gamma}\big), \quad Q_{t r}{ }^r=Q_{t \theta}{ }^\theta=Q_{t \phi}{ }^\phi=2\left(H-\gamma\right), ~~~~\\
&&\hspace{0cm} \quad Q^r{ }_{r t}=Q^\theta{ }_{\theta t}=Q^\phi{ }_{\phi t}=-\gamma, \\
&&\hspace{0cm} \quad L_{t t t}=-\big(\gamma+\frac{\dot{\gamma}}{\gamma}\big), \quad L_{t r}{ }^r=L_{t \theta}{ }^\theta=L_{t \phi}{ }^\phi=H,\\&&\hspace{0cm} \quad L_{r t}^r=L^\theta{ }_{\theta t}=L^\phi{ }_{\phi t}=\gamma-H, \\
&&\hspace{0cm}\quad P_{t t t}=\frac{-3\gamma}{4}, \quad P_{t r}{ }^r=P_{t \theta}{ }^\theta=P_{t \phi}{ }^\phi=\frac{1}{4}\left(4 H-3 \gamma\right),\\
&&\hspace{0cm}\quad P_{r t}^r=P^\theta{ }_{\theta t}=p^\phi{ }_{\phi t}=\frac{1}{4}\left(2\gamma+\frac{\dot{\gamma}}{\gamma}-H\right) .
\end{eqnarray}

In the above expression, $\gamma$ represents the nonvanishing function of time (t). An overdot denotes a time derivative. This function generates nonzero components in the nonmetricity tensor, offering new insights into the dynamics of $f(Q)$ in a spatially flat FLRW background, distinctly separate from the dynamics of $f(T)$ theory. The non-metricity scalar $Q$ can be calculated as follows,
\begin{eqnarray}
    Q=-6 H^{2}+9 \gamma H+3 \dot{\gamma} \text{.} 
\end{eqnarray}
The equations analogous to the Friedmann equations, derived from the field equation (\ref{e106}), are as follows

\begin{eqnarray}\label{f1}
  &&\hspace{0.5cm}  \kappa \rho=\frac{1}{2} f+\left(3 H^{2}-\frac{Q}{2}\right) f_Q+\frac{3}{2} \dot{Q} \gamma f_{QQ},\\ \label{f2}
&&\hspace{0.0cm}\kappa p=-\frac{1}{2} f+\left(-2 \dot{H}-3 H^{2}+\frac{Q}{2}\right) f_Q+\frac{\dot{Q}}{2}(-4 H+3 \gamma) f_{QQ}.~~~~~~~
\end{eqnarray}
In particular, for the choice $\gamma=0$, one can retrieve the usual Friedmann-like equation of the $f(Q)$ gravity in a \textit{coincident gauge} formalism. 

For further analysis, we propose a new parameterization Hubble function as follows,
\begin{equation}\label{NH}
H(z) = H_0 (z+1)^n + \beta \left[1-(z+1)^n\right].   
\end{equation}
Here, $H_0$, $\beta$, and $n$ are free parameters. Several parameterizations have been explored for the EoS parameter, including the CPL (Chevallier-Polarski-Linder), BA (Barboza-Alcaniz), and LC (Low Correlation) models, as well as the deceleration parameter \cite{2,p1}. Beyond the parameterization of the deceleration parameter, numerous other schemes for parameterizing different cosmological parameters have been extensively discussed in the literature \cite{p2,p3}. These schemes aim to address issues in cosmological investigations, such as the initial singularity problem, the persistent accelerating expansion problem, the horizon problem, and the Hubble tension. However, several previous parameterization schemes \cite{sib1,sib2,sib3} leads to a discrepancy from the standard $\Lambda$CDM and hence cannot consistently describe the present deceleration parameter value, transition redshift, and the late time de-Sitter limit as obtained in standard $\Lambda$CDM model. This motivated us to consider a new parameterization function that can describe the aforementioned cosmological epochs with observational compatibility.

\section{Data and methodology}\label{sec4m}

In this section, we carry out a statistical analysis to evaluate how the predictions of the modified gravity models align with cosmological observational data. Our goal is to impose constraints on the free parameters. We use a dataset from Cosmic Chronometers, containing 31 measurements, and the Pantheon+SH0ES sample, which includes 1701 data points, as well as BAO data with 6 data points for our analysis.

\subsection{Pantheon+SH0ES}
\justifying
The Pantheon+SH0ES samples span a wide range of redshifts, from 0.001 to 2.3, surpassing earlier collections of SN Ia by incorporating the latest observational data. SN Ia, known for their consistent brightness, serve as dependable standard candles for measuring relative distances using the distance modulus technique. Over the past two decades, several compilations of Type Ia supernova data have emerged, including Union \cite{R15}, Union2 \cite{R16}, Union2.1 \cite{R17}, JLA \cite{MBAA}, Pantheon \cite{R18}, and the most recent addition, Pantheon+SH0ES \cite{PSSS}. The corresponding $\chi^2$ function is given as,
\begin{equation}\label{4b}
\chi^2_{SN}= D^T C^{-1}_{SN} D.
\end{equation}
$C_{SN}$ \cite{PSSS} denotes the covariance matrix related to the Pantheon+SH0ES samples, encompassing both statistical and systematic uncertainties. Additionally, the vector $D$ is defined as $D=m_{Bi}-M-\mu^{th}(z_i)$, where $m_{Bi}$ and $M$ denote the apparent magnitude and absolute magnitude, respectively. Furthermore, $\mu^{th}(z_i)$ represents the distance modulus of the assumed theoretical model, which can be expressed as,
\begin{equation}\label{4c}
\mu^{th}(z_i)= 5log_{10} \left[ \frac{D_{L}(z_i)}{1 Mpc}  \right]+25 \text{.} 
\end{equation}

Here, $D_{L}(z)$ represents the luminosity distance in the proposed theoretical model, and it can be formulated as:
\begin{equation}\label{4d}
D_{L}(z)= c(1+z) \int_{0}^{z} \frac{ dx}{H(x,\theta)} \text{.} 
\end{equation}
Here, $\theta$ denotes the parameter space of our constructed model.

The Pantheon+SH0ES compilation resolves the degeneracy between the parameters $H_0$ and $M$ differently from the Pantheon datasets by redefining the vector $D$ as follows,
\begin{equation}\label{4e}
\bar{D} = \begin{cases}
     m_{Bi}-M-\mu_i^{Ceph} & i \in \text{Cepheid hosts} \\
     m_{Bi}-M-\mu^{th}(z_i) & \text{otherwise}
    \end{cases}  \text{,} 
\end{equation}
where $\mu_i^{Ceph}$ is independently estimated using Cepheid calibrators. Consequently, the equation \eqref{4b} becomes $\chi^2_{SN} = \bar{D}^T C^{-1}_{SN} \bar{D}$.

\subsection{CC+SN Ia+BAO}
\begin{figure}[H]
\includegraphics[scale=0.85]{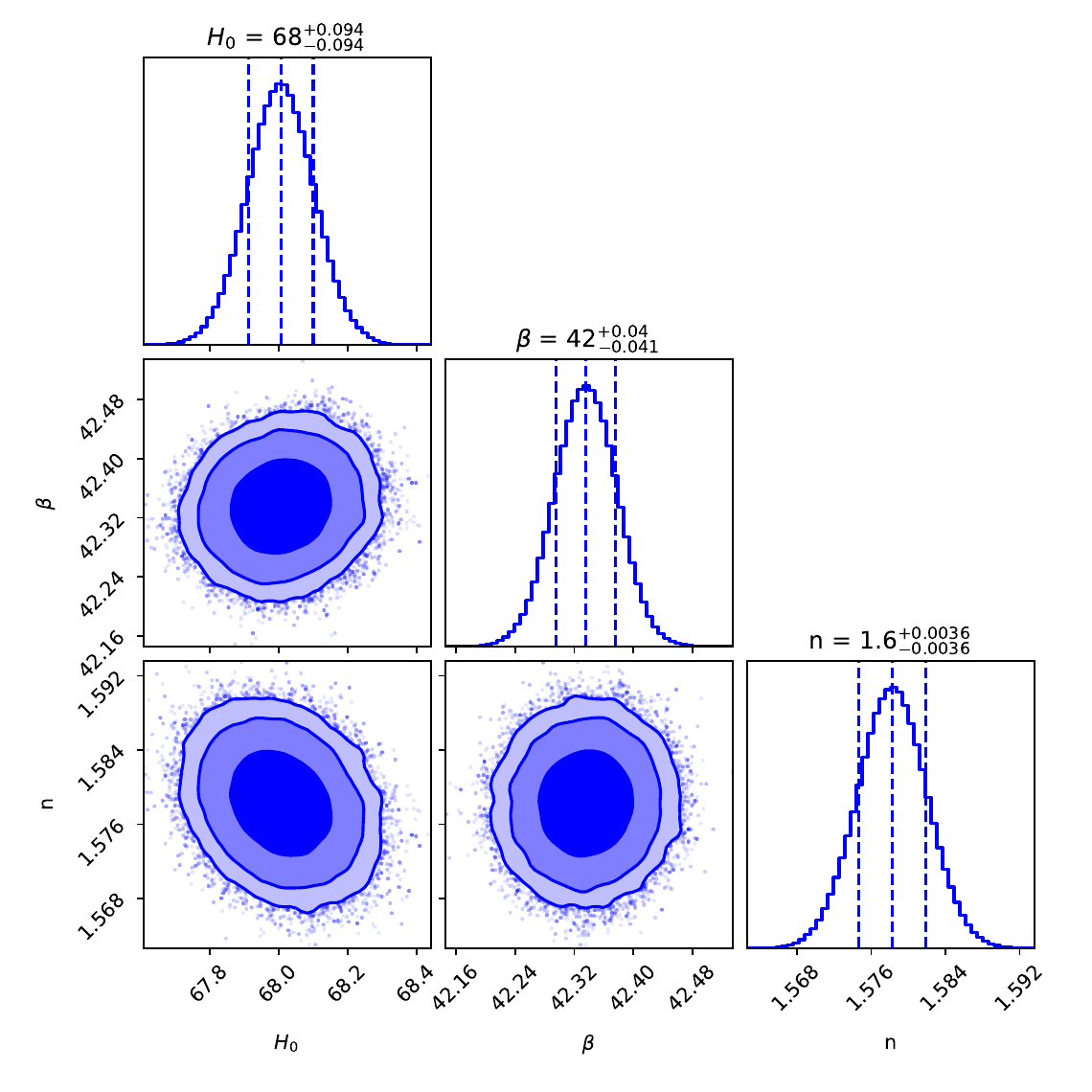}
\caption{The contour plot for the free parameter space $(H_0, \beta, n)$ corresponding to the given Hubble function within the $1\sigma-3\sigma$ confidence interval using CC+Pantheon+SH0ES+BAO samples.}\label{fig1} 
\end{figure}

\justifying
We determine the constraints on the free parameter space for the combined CC+Pantheon+SH0ES+BAO samples using the Gaussian priors within the range of $H_0 \in [50,100]$, $\beta \in [30, 60]$, and $n \in [0,5]$. To find the best fit values for the parameters, we minimize the total $\chi^2_{total}$ function, defined as follows,
\begin{equation}\label{4f}
\chi^2_{total}= \chi^2_{CC}+\chi^2_{SN} +\chi^2_{BAO} \text{.} 
\end{equation}
The contour plot describing the correlation between the free parameters within the $1\sigma-3\sigma$ confidence interval is presented in figure \eqref{fig1}.

We obtained the free parameter constraints as $H_0=68 \pm 0.094 \: km/s/Mpc$, $\beta=42^{+0.04}_{-0.041}$, and $n=1.6 \pm 0.0036$ within $68 \%$ confidence limit. In addition, we obtained the minimum value of the $\chi^2_{total}$ as $\chi^2_{min}=1648.988$.

\subsection{Model comparison with the $\Lambda$CDM}
\justifying
To determine the reliability of the MCMC analysis, it is essential to conduct a statistical evaluation using the Akaike Information Criterion (AIC) and Bayesian Information Criterion (BIC). The formulation of AIC could be written as \cite{chi},
\begin{eqnarray}
    AIC=\chi^2_{min}+2d \text{.} 
\end{eqnarray}

Here, $d$ denotes the number of parameters in the specified model. When comparing the model with the established $\Lambda$CDM model, we are establishing $\Delta \text{AIC}=|\text{AIC}_{\text{Model}}-\text{AIC}_{ \Lambda\text{CDM}}|$. A value in $\Delta$AIC of less than 2 indicates strong support for the assumed theoretical model. A value in $\Delta$AIC between 4 and 7 suggests moderate support. If the value of $\Delta$AIC exceeds 10, there is no evidence to support the assumed model. The second criterion, BIC, is formulated in the following manner,
\begin{eqnarray}
    \text{BIC}=\chi^2_{\text{min}}+d\times\ln(N) \text{.} 
\end{eqnarray}
In this context, $N$ denotes the number of data samples utilized in the MCMC analysis. Likewise, $\Delta\text{BIC}<2$ indicates strong evidence supporting the assumed theoretical model best, while a decrease in BIC between 2 and 6 suggests moderate support. By utilizing the aforementioned $\chi^2_{\text{min}}$ value, we obtained $\Delta\text{AIC}=1.408$ and $\Delta\text{BIC}=6.869$ where we have determined $\text{AIC}_{\text{Model}}=1654.988$, $\text{AIC}_{ \Lambda\text{CDM}}=1653.58$, $\text{BIC}_{\text{Model}}=1671.369$ and $\text{BIC}_{ \Lambda\text{CDM}}=1664.50$. Thus, it is evident from the $\Delta\text{AIC}$ and $\Delta\text{BIC}$ value that there is strong evidence in favor of our proposed Hubble function over the standard $\Lambda$CDM.

\subsection{Evolutionary profile of cosmological parameters}
\justifying
In this section, we investigate the behavior of some cosmological evolutionary parameters, for instance, the deceleration parameter $q$, the jerk parameter $j$, and the snap parameter $s$, that plays a vital role to explore the expansion phase of the Universe.

\begin{figure}[H]
\centering
{\includegraphics[width=6.3cm, height=4.5cm]{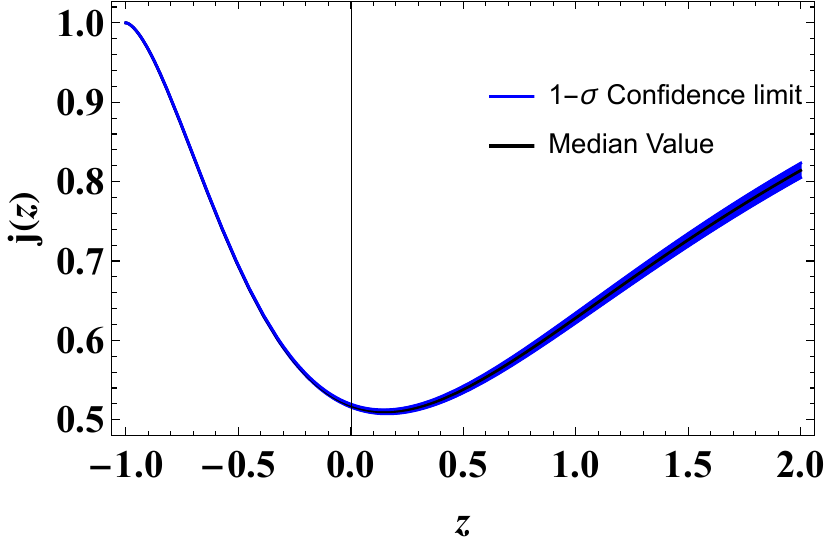}}
{\includegraphics[width=6.2cm, height=4.5cm]{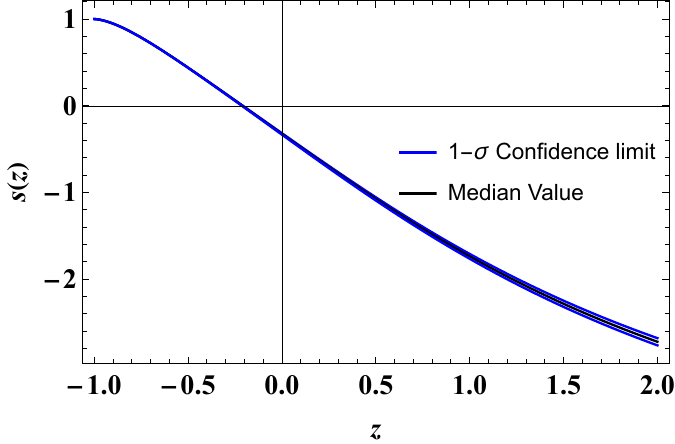}}
{\includegraphics[width=7.3cm, height=4.9cm]{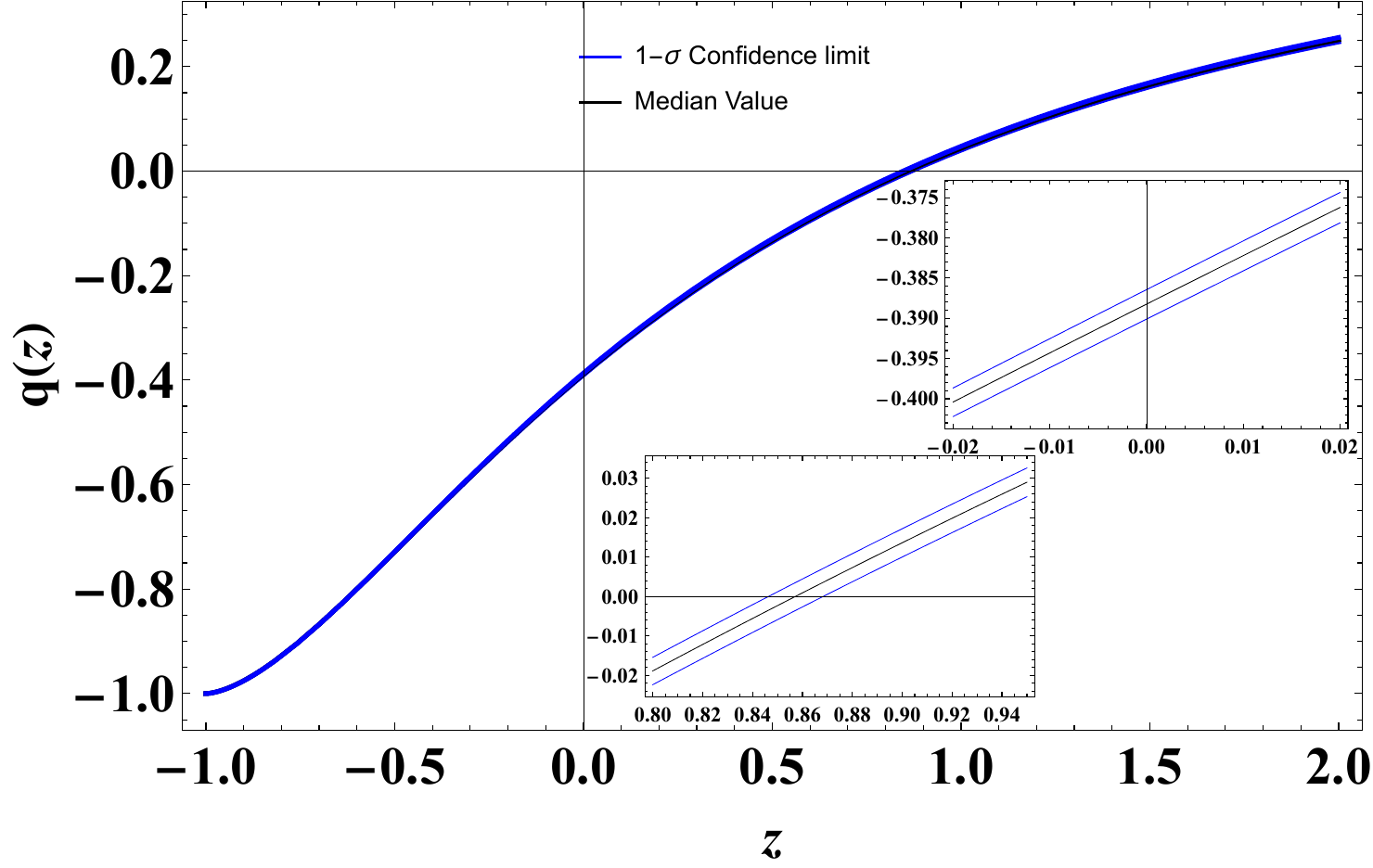}}
\caption{Profile of the jerk, snap, and the deceleration parameter vs redshift.}\label{fig2} 
\end{figure}     

It is well known that the cosmographic parameters can be obtained as the coefficient of the Taylor series expansion of the scale factor with the present time as the center. The coefficient of terms $(t-t_0)^2$, $(t-t_0)^3$, and $(t-t_0)^4$ are defined as the deceleration parameter $q$, the jerk parameter $j$, and the snap parameter $s$, whereas the coefficient of the term $(t-t_0)$ known as the Hubble parameter. The explicit expression can be expressed as $q=-\frac{\ddot{a}}{aH^2}$, $j=\frac{\dddot{a}}{aH^3}$, and $s=\frac{\ddddot{a}}{aH^4}$ \cite{CSM}. The evolutionary trajectories of these parameters corresponding to the proposed Hubble function have been presented in figure \eqref{fig2}. From figure \eqref{fig2} it is evident that the proposed function predicts the de-Sitter type accelerated expansion phase at the late times via a transition epoch from the decelerated epoch to accelerated epoch in the recent past with the transition redshift $z_t=0.857 \pm 0.011$. The present value of all these parameters are $q_0=-0.388 \pm 0.002$, $j_0=0.517 \pm 0.002$, and $s_0=-0.325^{+0.008}_{-0.009}$ within $68 \%$ confidence limit. Note that $j(z) \equiv 1$, $s(z) \equiv 1$ in case of the $\Lambda$CDM model. The obtained present deceleration parameter value with transition redshift is quite consistent with the cosmological observations \cite{R2}.

\section{Cosmological $f(Q)$ models}\label{sec5m}
\justifying
The choice of the function $f(Q)=Q$ presents an equivalent formulation to GR, and hence in the case, the theory is known as STEGR. The correction to the STEGR case provides an interesting class of solutions applicable either in early or late-time cosmology \cite{R49}. We consider some well-known corrections to the STEGR case that have been previously investigated in the coincident gauge formalism as follows,\\

\textbf{Model-I: $f(Q)=Q+\eta e^{\alpha Q}$} \cite{Sanjay} 

\textbf{Model-II: $f(Q)=Q+\eta log{\alpha Q}$} \cite{KV}

\textbf{Model-III: $f(Q)=Q+\eta Q^\alpha$}  \cite{Sanjay}

\textbf{Model-IV: $f(Q)=Q+\eta Q^{-1}$ }  \cite{R49}

\textbf{Model-V: $f(Q)=Q+\eta Q^2$ } \cite{R49}\\

For our further investigation, we have considered a non-constant function $\gamma(t)=-a^{-1}\dot{H}$, where $\dot{H}$ represents the time derivative of $H$. Note that, the choice $\gamma(t)=0$ reduces to coincident gauge formalism, whereas the case of constant $\gamma$ is the trivial one i.e. not much of the physical interest. Therefore we have assumed a non-constant choice of $\gamma$ function, however, one can construct the different choice of $\gamma(t)$ as done in \cite{GA}.

\subsection{Energy conditions}
Energy conditions are crucial for comprehending the Universe's geodesics. These conditions can be derived from the established Raychaudhuri equations \cite{EC}. These conditions involve the contraction of timelike or null vector fields with the Einstein tensor and the energy-momentum tensor. These conditions help to define what is considered physically reasonable or not clearly. The mathematical expression for the SEC is $\rho+3p\geq 0$, whereas for the NEC is $\rho+p\geq 0$. 
\begin{figure}[H]
\includegraphics[width=8.6cm, height=5.5cm]{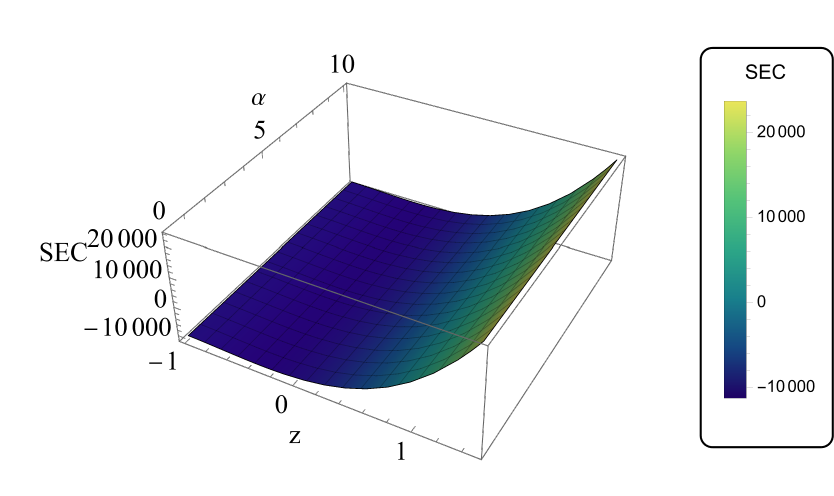}   
\includegraphics[width=8.8cm, height=5.5cm]{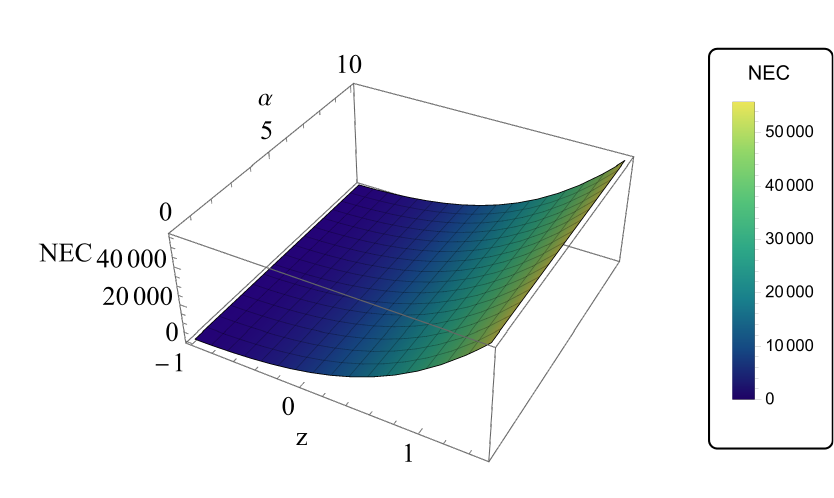}
\caption{Profile of the SEC and the NEC vs redshift, corresponding to the Model-I with varying $\alpha$ and fixed $\eta=1$.}\label{M1_p} 
\end{figure}    

\begin{figure}[H]
\includegraphics[width=8.6cm, height=5.5cm]{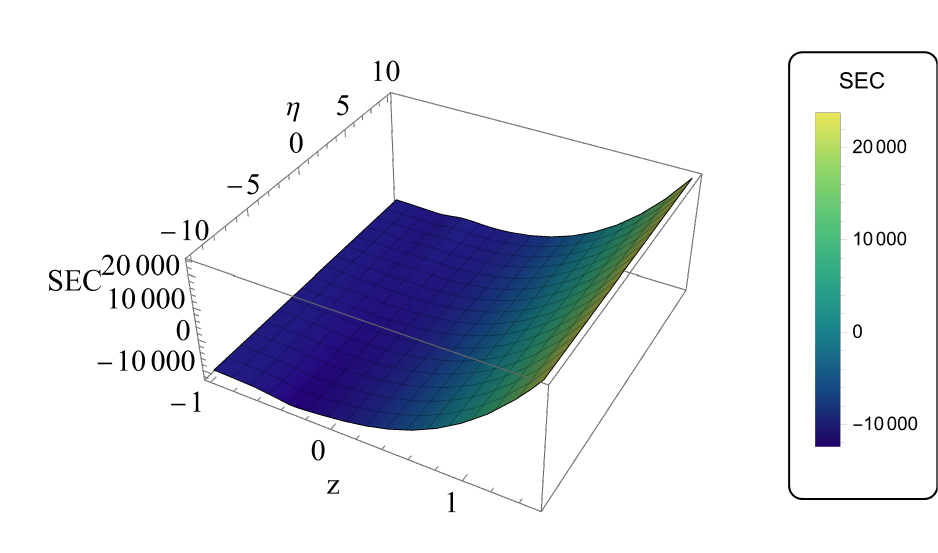}      
\includegraphics[width=8.8cm, height=5.5cm]{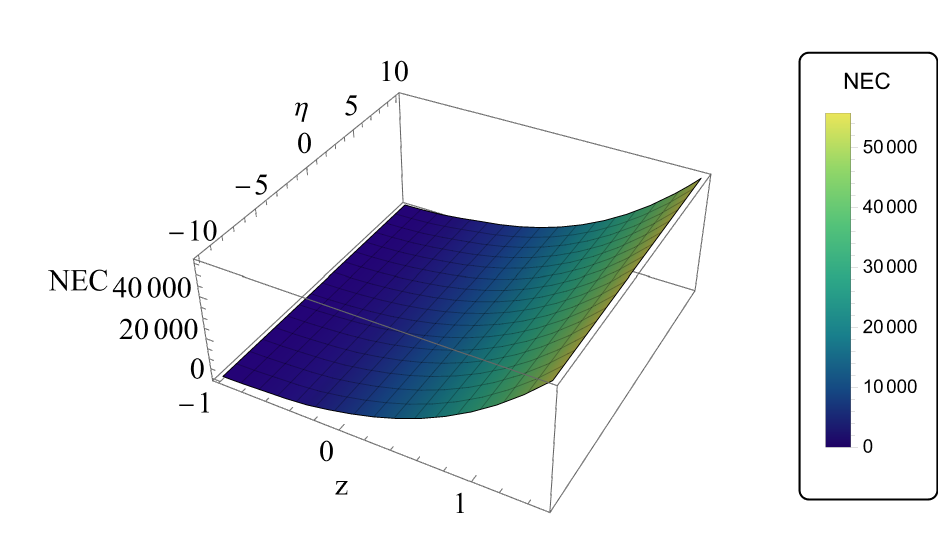}    
\caption{Profile of the SEC and the NEC vs redshift, corresponding to the Model-I with varying $\eta$ and fixed $\alpha=1$.}
\label{M1_c} 
\end{figure}        

\textbf{Model-I:} The Model-I is an exponential correction to the STEGR case. From figures \eqref{M1_p} and \eqref{M1_c}, it is clear that, in the entire range of redshift, the NEC is satisfied for both cases i.e. $\alpha \in [0,10]$ with $\eta=1$ and $\eta \in [-10,10]$ with $\alpha=1$. Further, the SEC is violated corresponding to both case in the recent past that favors the accelerating nature of expansion along with transition epoch in the recent past.\\

\begin{figure}[H]
\includegraphics[width=8.6cm, height=5.5cm]{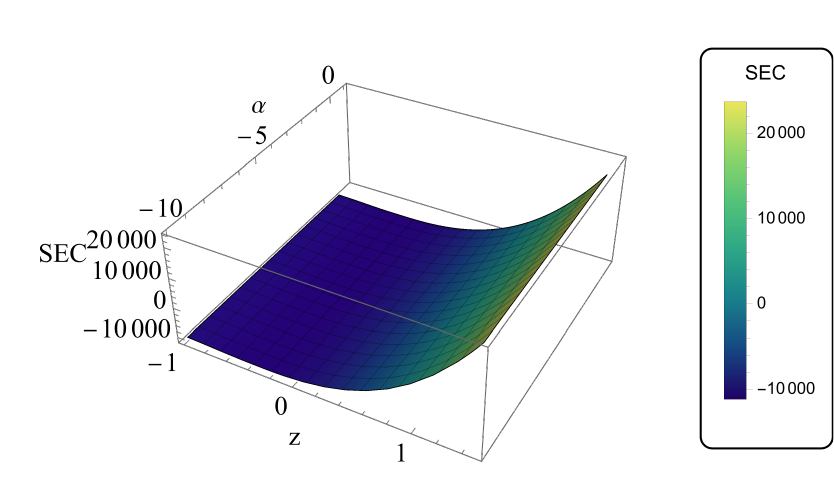}   
\includegraphics[width=8.8cm, height=5.5cm]{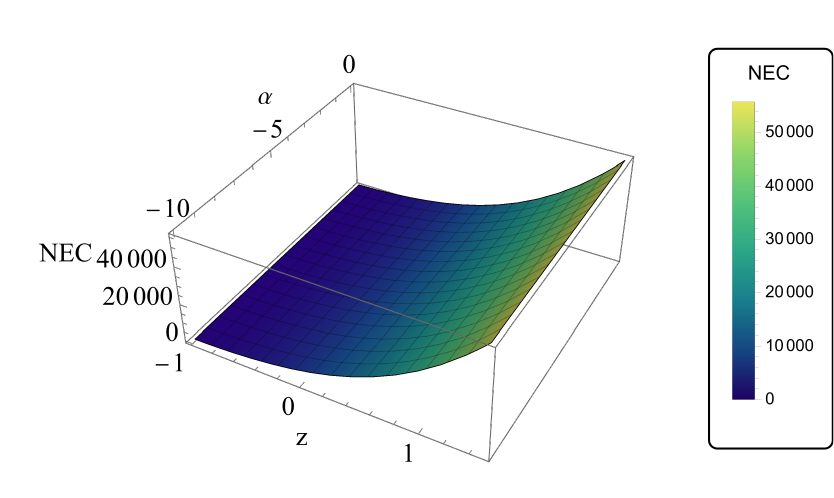}    
\caption{Profile of the SEC and the NEC vs redshift, corresponding to the Model-II with varying $\alpha$ and fixed $\eta=1$. }\label{M2_p} 
\end{figure}

\begin{figure}[H]
\includegraphics[width=8.6cm, height=5.5cm]{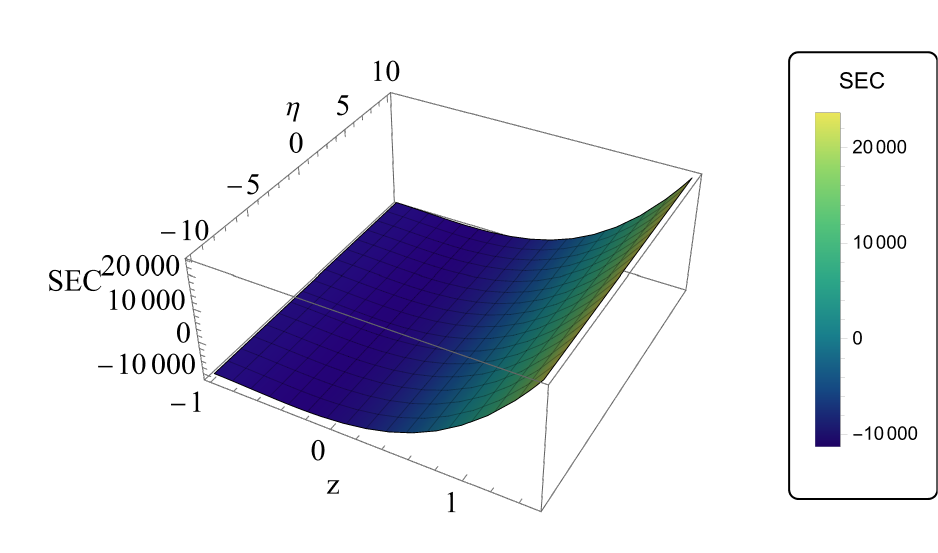}  
\includegraphics[width=8.8cm, height=5.5cm]{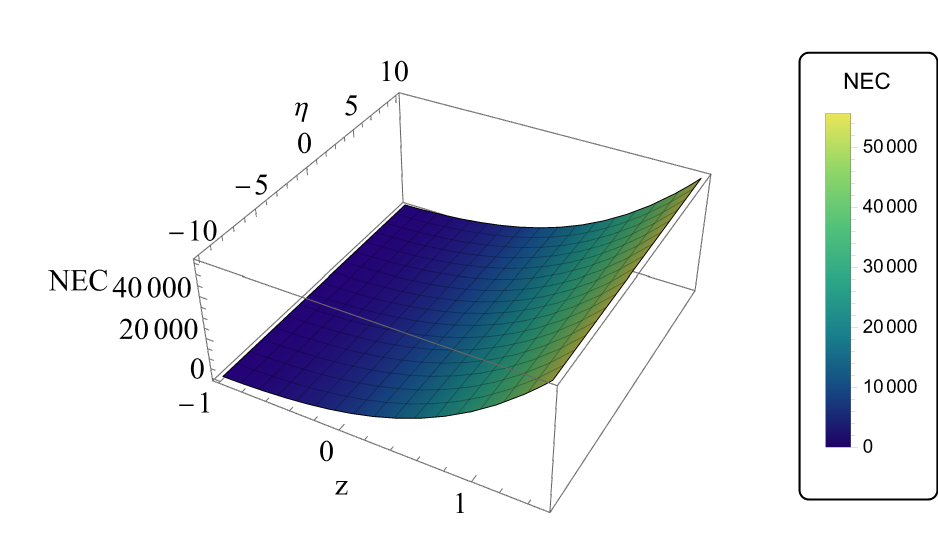}    
\caption{Profile of the SEC and the NEC vs redshift, corresponding to the Model-II with varying $\eta$ and fixed $\alpha=-1$.}\label{M2_c} 
\end{figure}

\textbf{Model-II:} The Model-II is an logarithmic correction to the STEGR case. From figures \eqref{M2_p} and \eqref{M2_c}, it is clear that, in the entire range of redshift, the NEC is satisfied for both cases i.e. $\alpha \in [-10,0]$ with $\eta=1$ and $\eta \in [-10,10]$ with $\alpha=-1$. Further, the SEC is violated corresponding to both case in the recent past that favors the accelerating nature of expansion along with transition epoch in the recent past.\\

\begin{figure}[H]
\includegraphics[width=8.6cm, height=5.5cm]{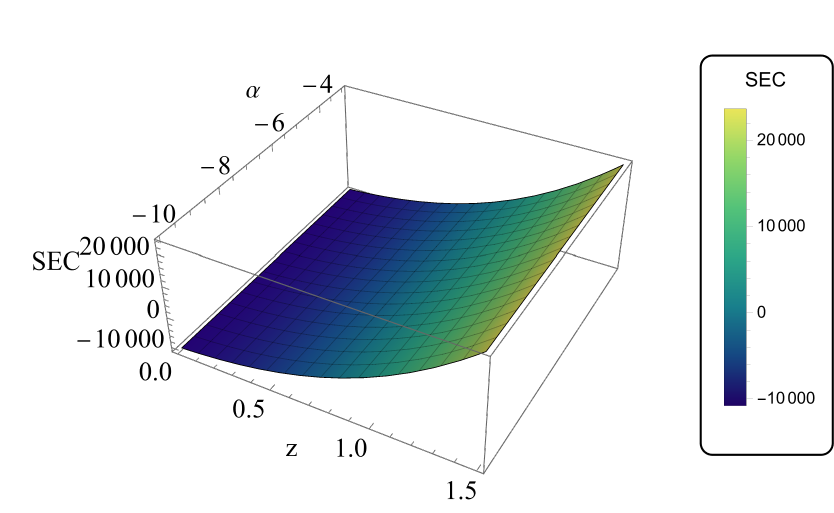}  
\includegraphics[width=8.8cm, height=5.5cm]{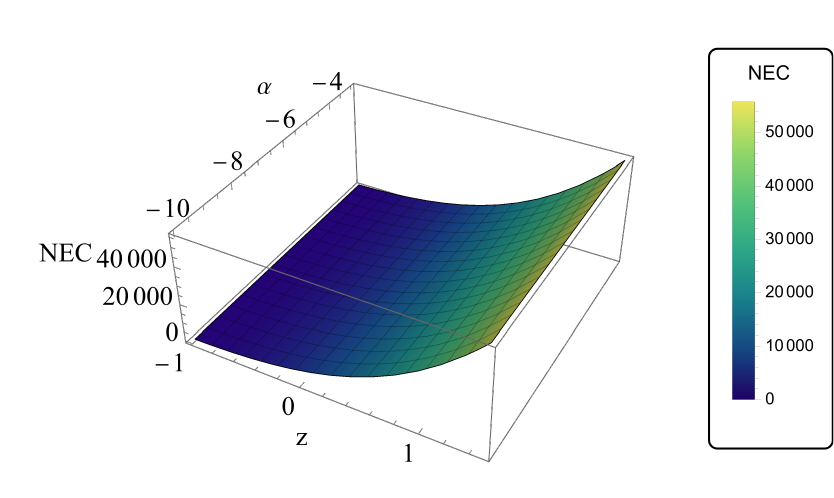} 
\caption{Profile of the SEC and the NEC vs redshift, corresponding to the Model-III with varying $\alpha$ and fixed $\eta=1$.}\label{M3_p} 
\end{figure}

\textbf{Model-III:} The Model-III is a power-law correction to the STEGR case. From figure \eqref{M3_p}, it is evident that, in the entire range of redshift, the NEC is satisfied for the parameter value $\alpha \in [-10,-4]$ with $\eta=1$. Further, the SEC is violated in the recent past that favors the accelerating nature of expansion along with transition epoch in the recent past. It is interesting to note that the same power-law correction model considered in \cite{Sanjay} in the \textit{coincident gauge connection} favors the positive values of $\alpha$ with the negative value of coefficient $\eta$. \\

\begin{figure}[H]
\includegraphics[width=8.6cm, height=5.5cm]{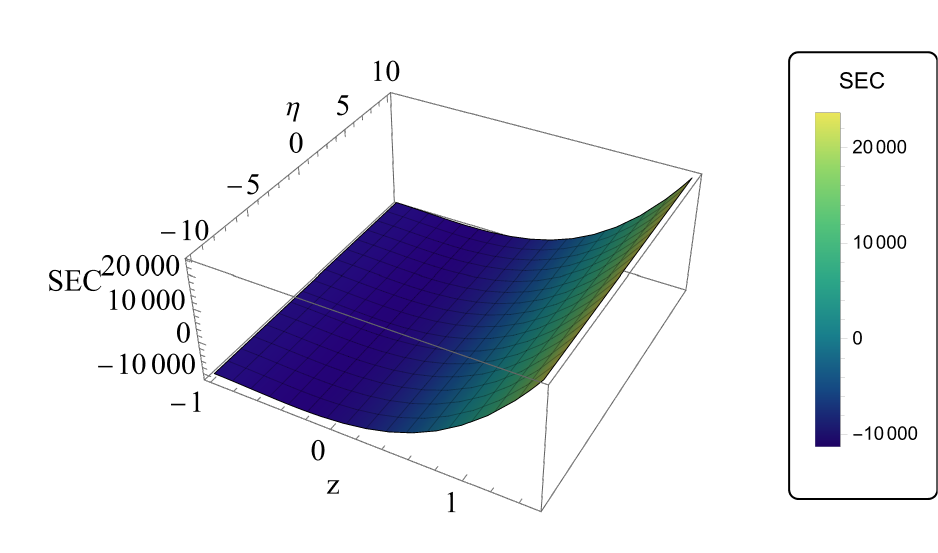} 
\includegraphics[width=8.8cm, height=5.5cm]{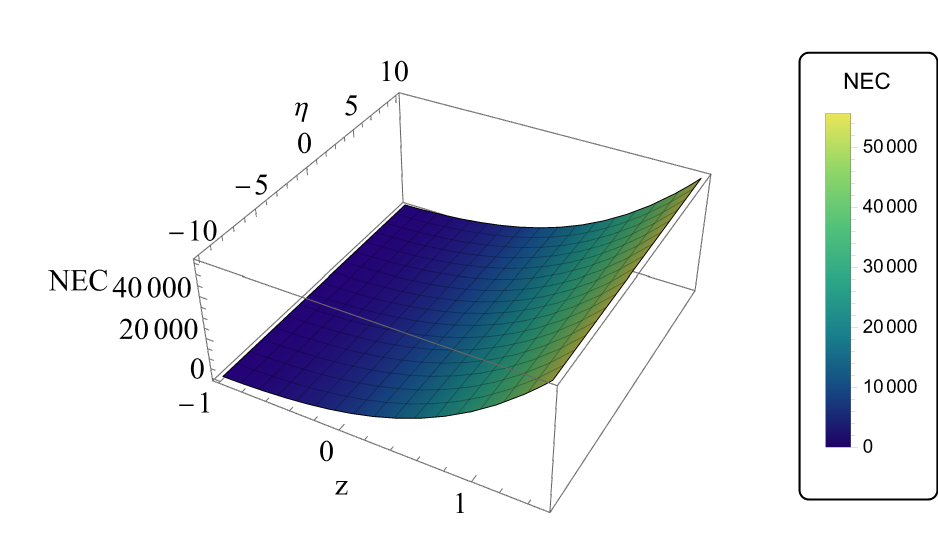} 
\caption{Profile of the SEC and the NEC vs redshift, corresponding to the Model-IV with varying $\eta$ and fixed $\alpha=-1$.}
\label{M4_p} 
\end{figure}

\textbf{Model-IV:} The Model-IV is a negative power-law correction to the STEGR case. A negative power correction to the STEGR case provides a correction to the late-time cosmology, where they can give rise to the dark energy \cite{R49}. From figure \eqref{M4_p}, it is evident, in the entire range of redshift, the NEC is satisfied for the parameter value $\eta \in [-10,10]$ with $\alpha=-1$. Further, the SEC is violated in the recent past that favors the accelerating nature of expansion along with transition epoch in the recent past. \\

\begin{figure}[H]
\includegraphics[width=6.3cm, height=5.5cm]{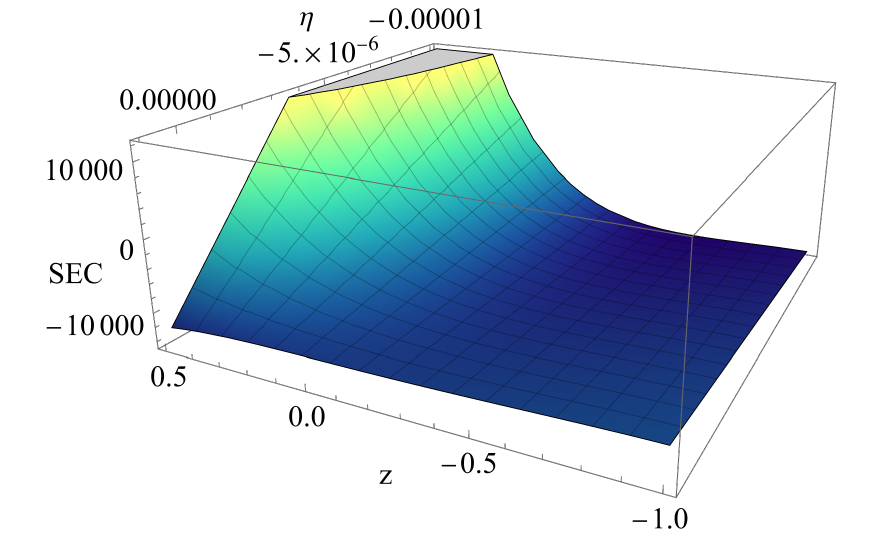} 
\includegraphics[width=1.4cm, height=5.0cm]{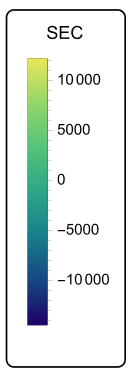} 
\includegraphics[width=6.3cm, height=5.5cm]{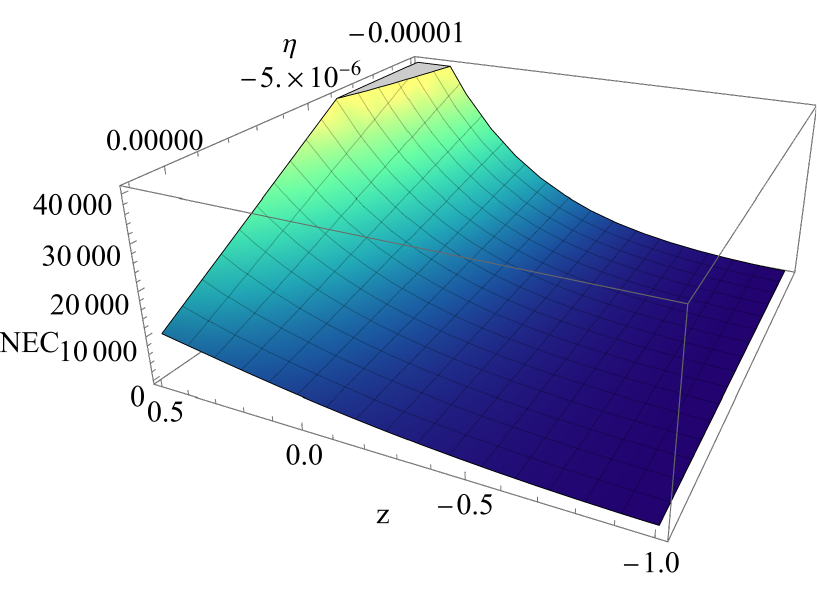} 
\includegraphics[width=1.4cm, height=5.0cm]{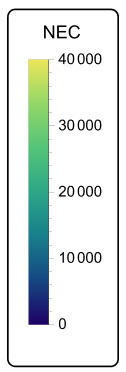}
\caption{Profile of the SEC and the NEC vs redshift, corresponding to the Model-V with varying $\eta$ and fixed $\alpha=2$.}
\label{M5_p} 
\end{figure} 

\textbf{Model-V:} The Model-V is a positive power-law correction to the STEGR case. A positive power correction to the STEGR case provides a correction to the early epochs of the Universe with potential applications to inflationary solutions \cite{R49}. From figure \eqref{M5_p}, it is evident that the NEC is satisfied for the parameter value $\alpha=2$ and for a tiny value of $\eta$ nearly close to $0$. Further, the SEC is violated in the recent past that favors the accelerating nature of expansion along with transition epoch in the recent past. 

\subsection{Thermodynamical stability}
\justifying
A rigorous criterion for validating a cosmological model is the square of the speed of sound $v_s^2$. For a model to be physically plausible, the square of the speed of sound must be less than the speed of light $c$. This relationship establishes the stability requirement for cosmological models. Therefore, if the condition  $0<v_s^2<c$ is satisfied, the model is considered physically plausible. These constraints enhance the suitability of this type of modeling, and various models featuring variable sound speed have been documented in the literature \cite{sos1,sos2}.
For this analysis, we consider the Universe as an adiabatic system, implying no heat or mass exchange with its surroundings and thus maintaining zero entropy perturbation. Under these assumptions, the primary focus shifts to how pressure varies with energy density, which leads us to define the sound speed parameter by the following expression,
\begin{eqnarray}
v_s^2=\frac{dp}{d\rho} \text{.} 
\end{eqnarray}

\begin{figure}[H]
\includegraphics[width=6.5cm, height=4.8cm]{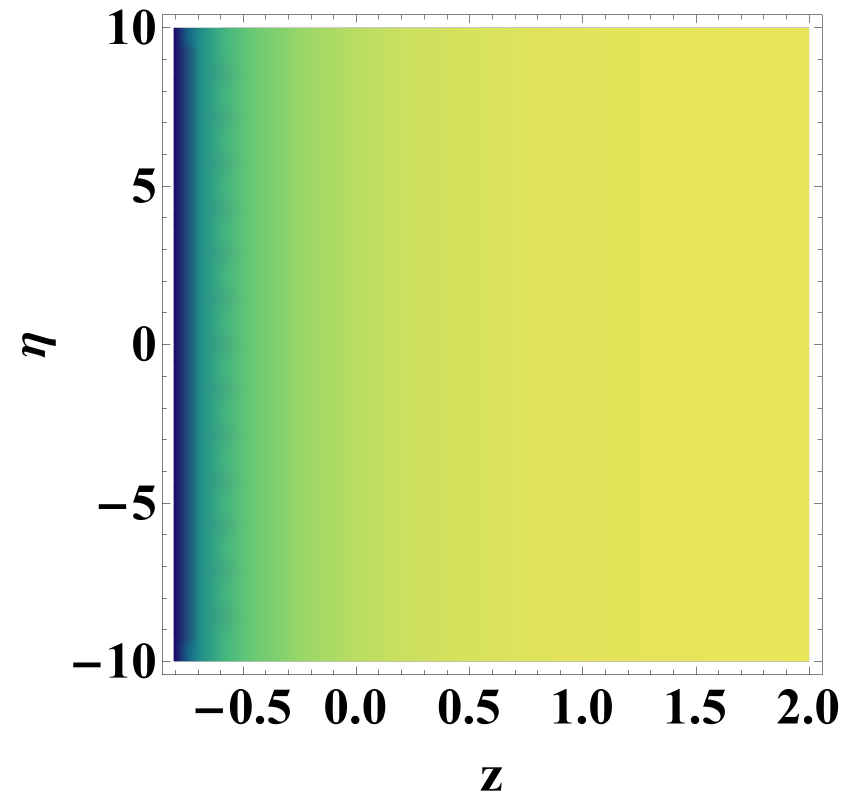}
\includegraphics[width=0.9cm, height=4.8cm]{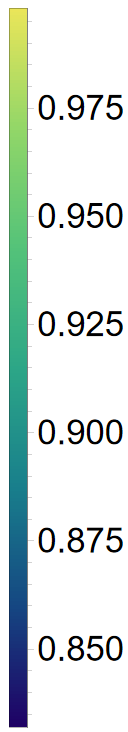}
\includegraphics[width=6.5cm, height=4.8cm]{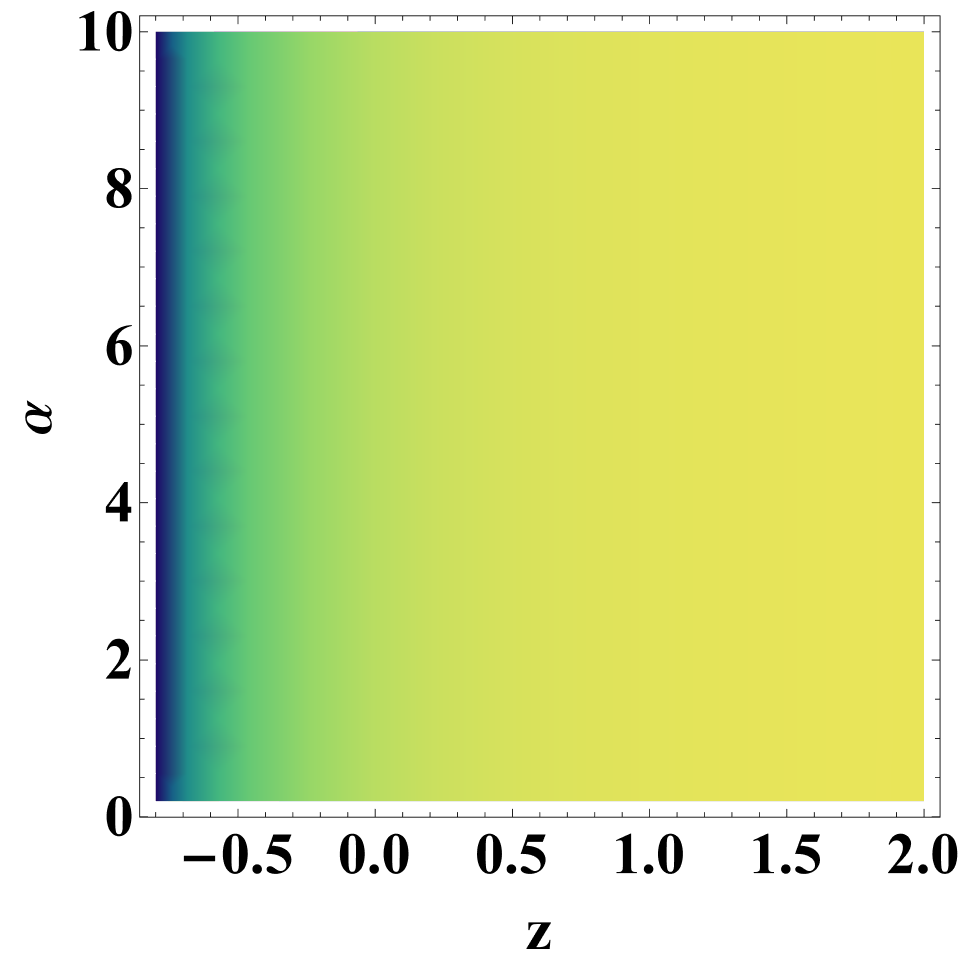}
\includegraphics[width=0.9cm, height=4.8cm]{Figures/ch5/m3stab2bar2.png}
\caption{Profile of the sound speed parameter vs z (redshift) for the Model-I corresponding to the case varying $\eta$ with $\alpha=1$ (left panel) and the varying $\alpha$ with $\eta=1$ (right panel).}\label{stab1} 
\end{figure}
From figure \eqref{stab1}, one can observe that the sound speed parameter for the Model-I lies between $0$ and $1$ corresponding to both cases i.e. $\eta \in [-10,10]$ with $\alpha=1$ (left panel) and $\alpha \in [0,10]$ with $\eta=1$ (right panel). Thus the assumed exponential $f(Q)$ correction model show stable behavior in the same range obtained from the energy conditions.

\begin{figure}[H]
\includegraphics[width=6.5cm, height=4.8cm]{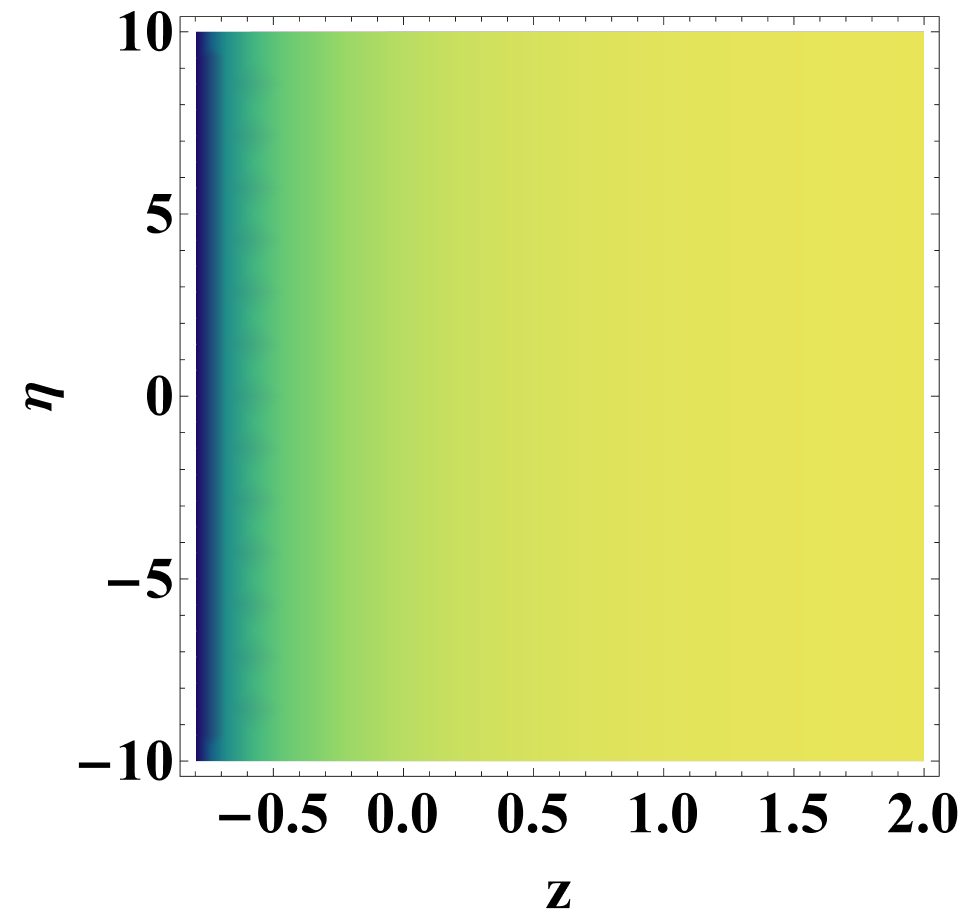}
\includegraphics[width=1.0cm, height=4.8cm]{Figures/ch5/m3stab2bar2.png}
\includegraphics[width=6.5cm, height=4.8cm]{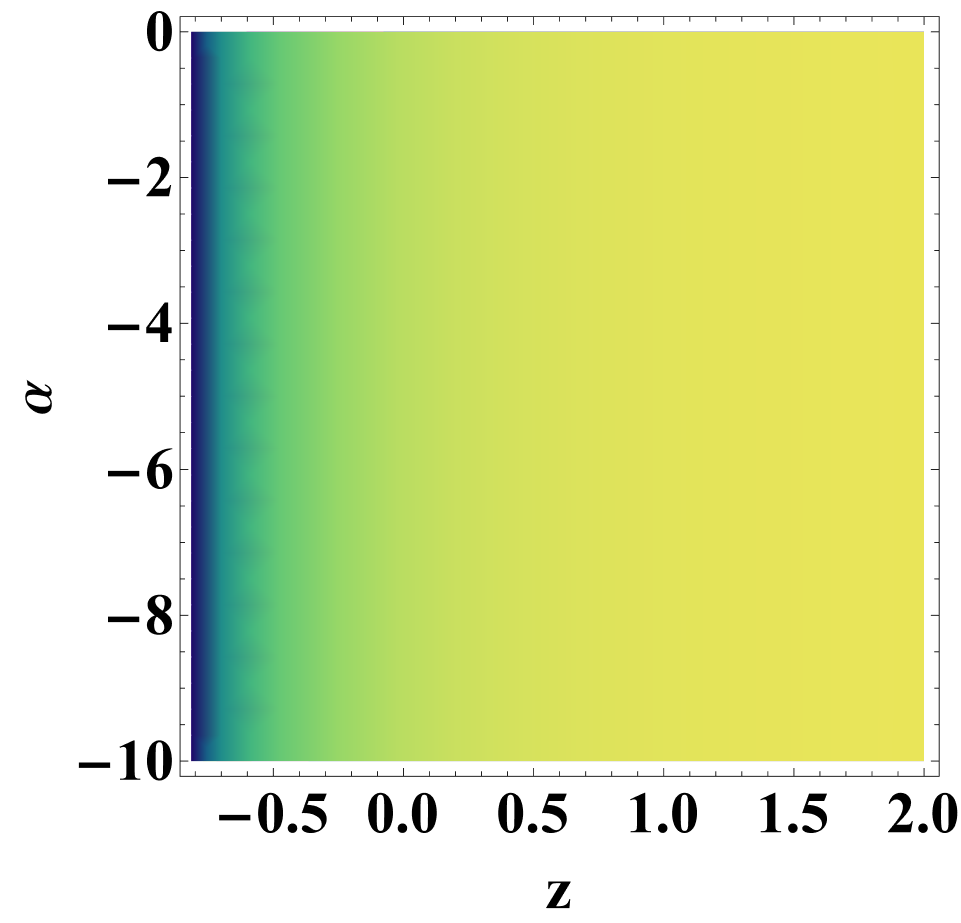}
\includegraphics[width=1.0cm, height=4.8cm]{Figures/ch5/m3stab2bar2.png}
\caption{Profile of the sound speed parameter vs z (redshift) for the Model-II corresponding to the case varying $\eta$ with $\alpha=-1$ (left panel) and the varying $\alpha$ with $\eta=1$ (right panel).}\label{stab2} 
\end{figure}
From figure \eqref{stab2}, it is evident that the sound speed parameter for the Model-II lies between $0$ and $1$ corresponding to both cases i.e. $\eta \in [-10,10]$ with $\alpha=-1$ (left panel) and $\alpha \in [-10,0]$ with $\eta=1$ (right panel). Thus the assumed logarithmic $f(Q)$ correction model show stable behavior in the same range obtained from the energy conditions.

\begin{figure}[H]
\includegraphics[width=4.1cm, height=4.1cm]{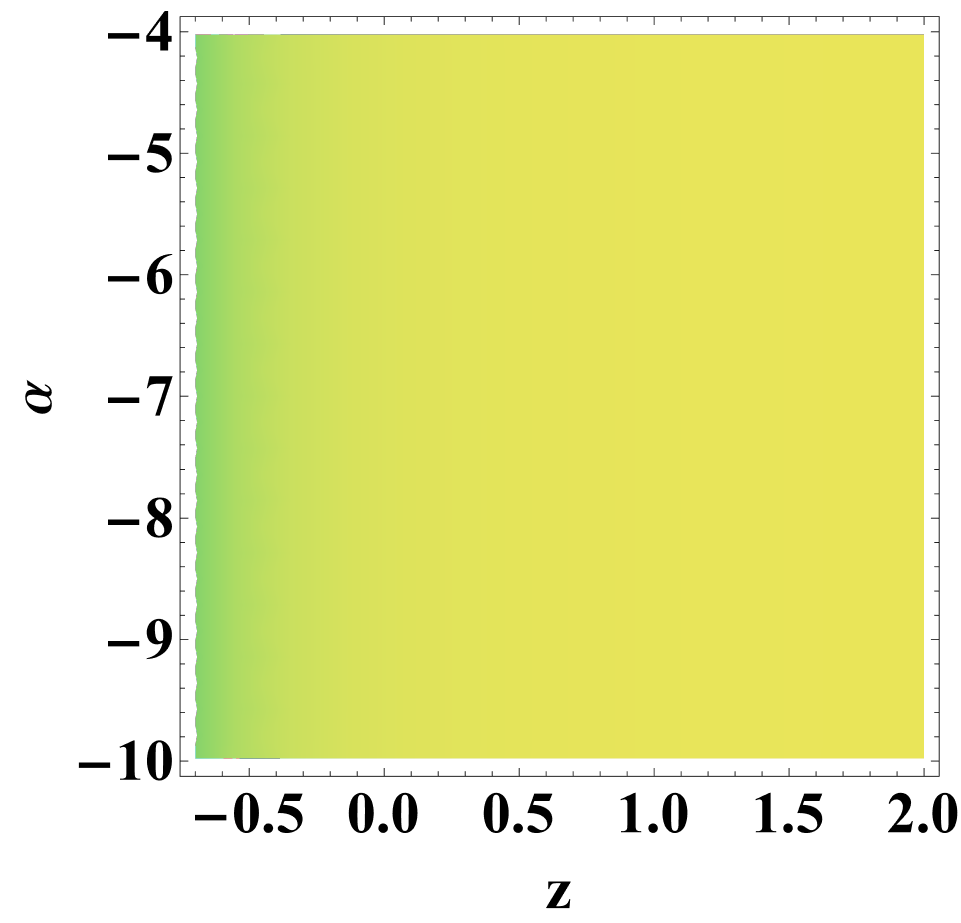}
\includegraphics[width=0.68cm, height=4.1cm]{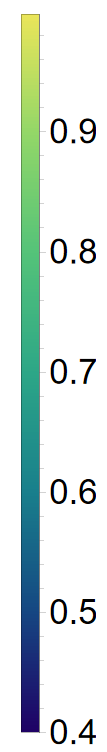}
\includegraphics[width=4.1cm, height=4.1cm]{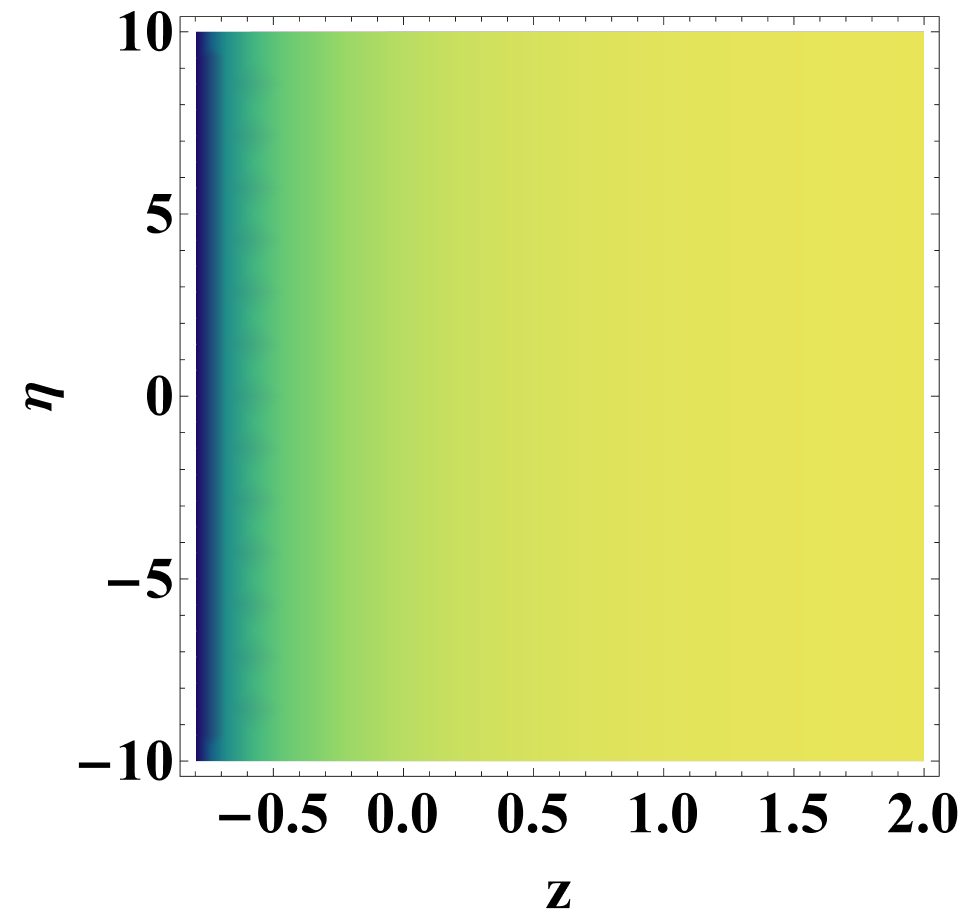}
\includegraphics[width=0.67cm, height=4.1cm]{Figures/ch5/m3stab2bar2.png}
\includegraphics[width=4.7cm, height=4.1cm]{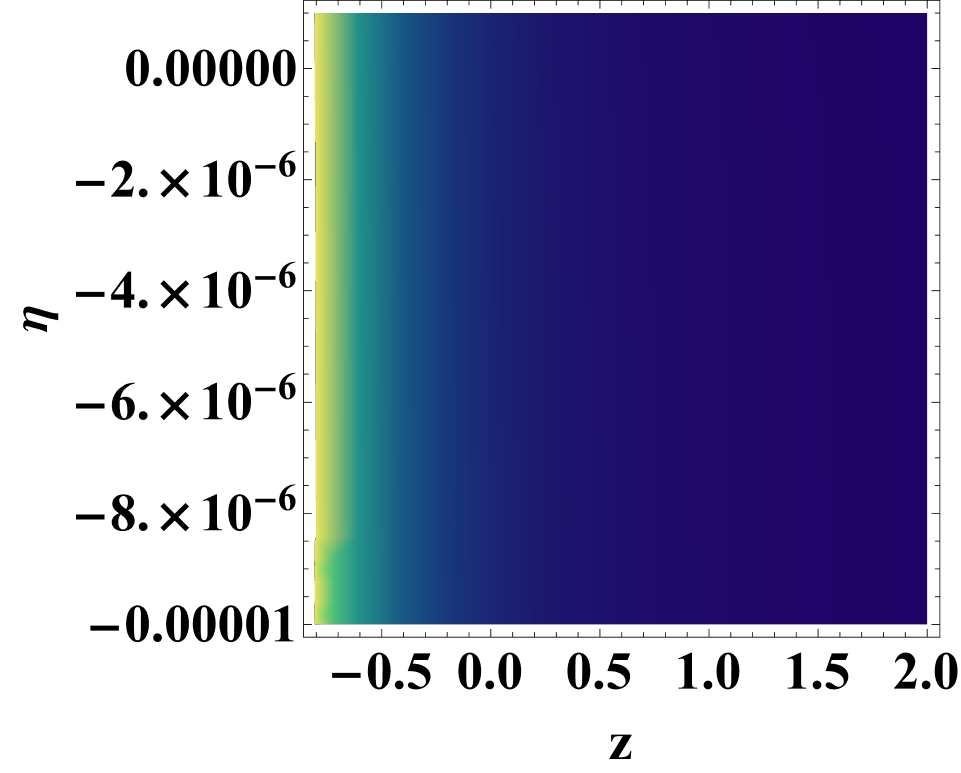}
\includegraphics[width=0.67cm, height=4.1cm]{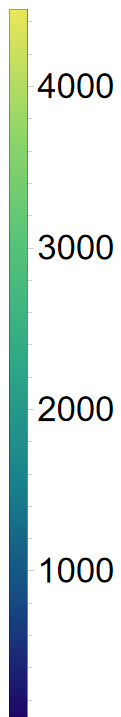}
\caption{Profile of the sound speed parameter vs z (redshift) for the Model-III with varying $\alpha$ and $\eta=1$ (left panel), Model-IV with varying $\eta$ and $\alpha=-1$ (middle panel), and the Model-V with varying $\eta$ and $\alpha=2$ (right panel).}\label{stab3} 
\end{figure}
From figure \eqref{stab3} (left panel), we find that the sound speed parameter for the Model-III lies between $0$ and $1$ corresponding to the parameter choice $\alpha \in [-10,-4]$ with $\eta=1$ (left panel). Thus the considered power-law $f(Q)$ correction model show stable behavior in the same range obtained from the energy conditions. Further, we have assumed two particular cases of the power-law model with both negative and positive power. Both assumed model have significantly utilized in the case of coincident gauge choice and found to be physically viable. The Model-IV is a negative power-law correction that can provides a correction to the late-time cosmology, where they can give rise to the dark energy, whereas the Model-V is a positive power-law correction that can provides a correction to the early epochs of the Universe with potential applications to inflationary solutions \cite{R49}. In this investigation, we observe that the sound speed parameter for the Model-IV lies between $0$ and $1$ corresponding to the parameter choice $\eta \in [-10,10]$ with $\alpha=-1$ (see figure \eqref{stab3} middle panel), whereas for the Model-V, it does not lies between $0$ and $1$ corresponding to the parameter choice obtained from energy condition constraints (see figure \eqref{stab3} right panel).  Thus the considered negative power-law $f(Q)$ correction model show stable behavior, whereas the positive power-law $f(Q)$ correction model show unstable behavior.

\section{Conclusions}\label{sec6m}

In this chapter, we have utilized a new $f(Q)$ theory dynamics utilizing a non-vanishing affine connection involving a function $\gamma$, resulting in Friedmann equations that are entirely distinct from those of $f(T)$ theory. The basic mathematical construction of $f(Q)$ gravity from a non-vanishing affine connection involves an arbitrary function of time $\gamma(t)$. As this function is arbitrary, one can investigate the different class of this function, as done in \cite{GA}. We have considered a non-constant function $\gamma(t)=-a^{-1}\dot{H}$, where $\dot{H}$ represents the time derivative of $H$. Note that, the choice $\gamma(t)=0$ reduces to coincident gauge formalism, whereas the case of constant $\gamma$ is the trivial one i.e. not much of the physical interest. Therefore we have assumed a non-constant choice of $\gamma$ function, however, one can construct the different choice of $\gamma(t)$ as done in \cite{GA}. In addition, we have proposed a new parameterization of the Hubble function in the equation (\ref{NH}). As several previous parameterization schemes \cite{sib1,sib2,sib3} leads to a discrepancy from the standard $\Lambda$CDM and hence cannot consistently describe the present deceleration parameter value, transition redshift, and the late time de-Sitter limit as obtained in standard $\Lambda$CDM model. This motivated us to consider a new parameterization function that can describe the aforementioned cosmological epochs with observational compatibility.

We carried out a statistical analysis to evaluate the predictions of the assumed Hubble function parameterization by imposing constraints on the free parameters. We utilized Bayesian statistical analysis to estimate the posterior probability by employing the likelihood function and the MCMC sampling technique by invoking the CC, Pantheon+SH0ES, and the BAO data samples for our analysis. We obtained the free parameter constraints as $H_0=68 \pm 0.094 \: km/s/Mpc$, $\beta=42^{+0.04}_{-0.041}$, and $n=1.6 \pm 0.0036$ within $68 \%$ confidence limit. The corresponding contour plot describing the correlation between the free parameters within the $1\sigma-3\sigma$ confidence interval is presented in figure \eqref{fig1}. To determine the reliability of MCMC analysis we have conducted the AIC and BIC statistical evaluations. We have determined $\Delta AIC=1.408$ and $\Delta BIC= 6.869$ which supports a shred of strong evidence in favor of our proposed Hubble function over the standard $\Lambda \text{CDM}$.

In addition, we analyzed some cosmographic parameters like $q(z)$, $j(z)$ and $s(z)$. The evolutionary paths of these parameters, corresponding to the proposed Hubble function, are illustrated in figure \eqref{fig2}. As seen in figure \eqref{fig2}, the proposed function forecasts a de-Sitter type accelerated expansion phase at late times, following a transition from a decelerated epoch to an accelerated epoch in the recent past, with the transition redshift $z_t=0.857 \pm 0.011$. The current values of these parameters are $q_0=-0.388 \pm 0.002$, $j_0=0.517 \pm 0.002$, and $s_0=-0.325^{+0.008}_{-0.009}$ within $68 \%$ confidence limit. It is noteworthy that for the $\Lambda$CDM model, $j(z) \equiv 1$ and $s(z) \equiv 1$. The obtained present deceleration parameter value, along with the transition redshift, aligns well with cosmological observations \cite{R2}.

Further, we have considered some well-known corrections to the STEGR case such as an exponentital $f(Q)$ correction, logarithmic $f(Q)$ correction, and a power-law $f(Q)$ correction, that have been previously investigated in the coincident gauge formalism \cite{Sanjay,KV,R49}. We obtained constraints on free parameters of the considered $f(Q)$ correction models via energy conditions. For the Model-I (see figures \eqref{M1_p} and \eqref{M1_c}) the NEC is satisfied and the SEC is violated corresponding to both cases i.e. $\alpha \in [0,10]$ with $\eta=1$ and $\eta \in [-10,10]$ with $\alpha=1$. For the Model-II (see figures \eqref{M2_p} and \ref{M2_c}) the NEC is satisfied and the SEC is violated corresponding to both cases i.e. $\alpha \in [-10,0]$ with $\eta=1$ and $\eta \in [-10,10]$ with $\alpha=-1$. For the Model-III (see figure \eqref{M3_p}) the NEC is satisfied and the SEC is violated corresponding to the parameter value $\alpha \in [-10,-4]$ with $\eta=1$. It is interesting to note that the same power-law correction model (i.e. Model-III) considered in \cite{Sanjay} in the \textit{coincident gauge connection} favors the positive values of $\alpha$ with the negative value of coefficient $\eta$. For the Model-IV (see figure \eqref{M4_p}) the NEC is satisfied and the SEC is violated, corresponding to the parameter value $\eta \in [-10,10]$ with $\alpha=-1$. Lastly, for the Model-V (see figure \eqref{M5_p}) it is evident that the NEC is satisfied and the SEC is violated for the parameter value $\alpha=2$ and for a tiny value of $\eta$ nearly close to $0$.

Finally, to test the physical plausibility of the assumed $f(Q)$ models we conducted the thermodynamical stability analysis via the sound speed parameter. We found that the assumed exponential and logarithmic $f(Q)$ corrections in Model-I and Model-II respectively are thermodynamically stable in the parameter range obtained from the energy condition constraints (see figures \eqref{stab1} and \eqref{stab2}). Moreover, the considered power-law $f(Q)$ correction in Model-III show stable behavior in the same range obtained from the energy conditions (see figure \eqref{stab3} left panel). Further, we have assumed two particular cases of the power-law model with both negative and positive power. Both assumed models have been significantly utilized in the case of coincident gauge choice and found to be physically viable. The Model-IV is a negative power-law correction that can provides a correction to the late-time cosmology, where they can give rise to dark energy, whereas the Model-V is a positive power-law correction that can provides a correction to the early epochs of the Universe with potential applications to inflationary solutions \cite{R49}. In this investigation, we found that the considered negative power-law $f(Q)$ correction model shows stable behavior (see figure \eqref{stab3} middle panel), whereas the positive power-law $f(Q)$ correction model shows unstable behavior (see figure \eqref{stab3} right panel).

%% file: Chapters/Chapter6.tex
\chapter{$f(Q,\mathcal{T})$ gravity, its covariant formulation, energy conservation and phase-space analysis} 

\label{Chapter6} 

\lhead{Chapter 6. \emph{$f(Q,\mathcal{T})$ gravity, its covariant formulation, energy conservation and phase-space analysis}} 

\vspace{10 cm}
* The work, in this chapter, is covered by the following publications: \\
 
\textit{$f(Q,\mathcal{T})$ gravity, its covariant formulation, energy conservation and phase-space analysis}, The European Physical Journal C \textbf{83(3)}, 261 (2023).

\clearpage

In this chapter we analyze the matter-geometry coupled $f(Q,\mathcal{T})$ theory of gravity. We offer the fully covariant formulation of the theory, with which we construct the correct energy balance equation and employ it to conduct a dynamical system analysis in a spatially flat Friedmann-Lema\^{i}tre-Robertson-Walker spacetime. We consider three different functional forms of the $f(Q,\mathcal{T})$ function, specifically, $f(Q,\mathcal{T})=\alpha Q+ \beta \mathcal{T}$, $f(Q,\mathcal{T})=\alpha Q+ \beta \mathcal{T}^2$, and $f(Q,\mathcal{T})=Q+ \alpha Q^2+ \beta \mathcal{T}$. We attempt to investigate the physical capabilities of these models to describe various cosmological epochs. We calculate Friedmann-like equations in each case and introduce some phase space variables to simplify the equations in more concise forms. We observe that the linear model $f(Q,\mathcal{T})=\alpha Q+ \beta \mathcal{T}$ with $\beta=0$ is completely equivalent to the GR case without cosmological constant $\Lambda$. Further, we find that the model $f(Q,\mathcal{T})=\alpha Q+ \beta \mathcal{T}^2$ with $\beta \neq 0$ successfully depicts the observed transition from decelerated phase to an accelerated phase of the Universe. Lastly, we find that the model $f(Q,\mathcal{T})= Q+ \alpha Q^2+ \beta \mathcal{T}$ with $\alpha \neq 0$ represents an accelerated de-Sitter epoch for the constraints $\beta < -1$ or $ \beta \geq 0$.

\section{Introduction}\label{sec1}

Recently, a matter-geometry coupling in the form of $f(Q,\mathcal{T})$ theories were proposed \cite{Yixin} in which the Lagrangian was represented by the function of the non-metricity scalar $Q$, and the trace $T$ of the energy-momentum tensor. Harko et al. \cite{Harko} argued that this dependence can be caused by exotic imperfect fluids or quantum phenomena. Recently, several cosmological and astrophysical aspects of $f(Q,\mathcal{T})$ gravity have been tested, for instance, energy conditions \cite{Arora1}, baryogenesis \cite{BG}, cosmological inflation \cite{Maryam}, reconstruction of $f(Q,\mathcal{T})$ lagrangian \cite{Gadbail}, static spherically symmetric wormholes \cite{Moreshwar}, constraint on the effective EoS \cite{Arora2}, observational constraints on $f(Q,\mathcal{T})$ gravity models \cite{Arora3}, and cosmological perturbations \cite{Antonio}. However, except the introductory article, most other studies were primarily conducted to contact observational datasets and not much theoretical investigation was carried out in this gravity theory which is still at its infancy at best. On a closer look, it is noticed that a covariant formulation was eluded so far, and there is some missing terms in the energy balance equation. 

This motivated us to derive the covariant formulation and the energy balance equation of the $f(Q,\mathcal{T})$ gravity. In addition, we present the asymptotic behavior of some cosmological $f(Q,\mathcal{T})$ models by utilizing the dynamical system technique. This approach is quite efficient in describing the asymptotic behavior of non-linear modified gravity models. Several cosmological models of the modified gravity have been tested by utilizing the dynamical system technique \cite{Mirza,Rudra,Leon1,Leon2,Khyllep}. One can investigate the asymptotic behavior of the cosmological model by analyzing the nature of critical points obtained by solving an autonomous system of first-order ordinary differential equations. The most important feature of any cosmological model is to have late-time stable solutions that depict the late-time behavior of the model. 

The present chapter is organized as follows: In Sec. \ref{sec2}, we present the mathematical formulation of $f(Q,\mathcal{T})$ gravity. Then in Sec. \ref{sec3}, we derive the covariant formula for $f(Q,\mathcal{T})$ gravity. In Sec. \ref{sec4}, we derive the corrected version of the energy-balance equation, both from the original metric field equation as well as from the novel covariant form of it. Further in Sec. \ref{sec5}, we investigate the asymptotic behavior of some cosmological $f(Q,\mathcal{T})$ models with the help of dynamical system analysis. Finally in Sec. \ref{sec6}, we summarize the obtained results.


\section{The mathematical formulation}\label{sec2}
The action of $f(Q,\mathcal{T})$ gravity is given by \cite{Yixin},
\begin{equation*}
S = \int \left[\frac{1}{2\kappa}f(Q,T) + L_m \right] \sqrt{-g}\,d^4 x \text{,} 
\end{equation*}
where $Q = Q_{\lambda\mu\nu}P^{\lambda\mu\nu}$ is the non-metricity scalar, $L_m$ is the matter Lagrangian, and
$\mathcal{T}$ is the trace of the stress energy tensor $\mathcal{T}_{\mu\nu}$, which is defined as, 
\begin{align}
\mathcal{T}_{\mu\nu}=-\frac 2{\sqrt{-g}}\frac{\delta(\sqrt{-g} L_m)}{\delta g^{\mu\nu}}\,.
\end{align}
The variation of the action with respect to the metric, gives the metric field equation,
\begin{equation} \label{FE1}
\frac{2}{\sqrt{-g}} \partial_\lambda (\sqrt{-g}f_QP^\lambda{}_{\mu\nu}) -\frac{1}{2}f g_{\mu\nu}
+f_{\mathcal{T}}(\mathcal{T}_{\mu\nu}+\Theta_{\mu\nu})
 + f_Q(P_{\nu\rho\sigma} Q_\mu{}^{\rho\sigma} -2P_{\rho\sigma\mu}Q^{\rho\sigma}{}_\nu) = \kappa \mathcal{T}_{\mu\nu}\,,
\end{equation}
where $f_{(\cdot)}$ represents the partial derivative of $f$ with respect to $(\cdot)$ and 
\begin{align}
\Theta_{\mu\nu}=\frac{g_{\alpha\beta}\delta \mathcal{T}_{\alpha\beta}}{\delta g^{\mu\nu}}\,.
\end{align}
Furthermore, varying the action with respect to the connection, we obtain the following connection field equation,
\begin{align}\label{FE-connection}
\nabla_\mu\nabla_\nu(2\sqrt{-g}f_QP^{\mu\nu}{}_\lambda-\kappa H_\lambda{}^{\mu\nu})=0\,,
\end{align}
where 
\[
H_\lambda{}^{\mu\nu}=\frac{\sqrt{-g}}{2\kappa}f_\mathcal{T}\frac{\delta \mathcal{T}}{\delta \mathring{\Gamma}^\lambda{}_{\mu\nu}}
+\frac{\delta(\sqrt{-g} L_m)}{\delta\mathring{\Gamma}^\lambda{}_{\mu\nu}}\,,
\]
is the hypermomentum tensor density.

Noticing that the equation (\ref{FE1}) is valid only in the coincident gauge coordinate \cite{R51}, it is essential to express it in its covariant form in the following section which firstly is independent of the choice  of coordinate systems and secondly provides an relatively straightforward manner in identifying the effective energy density and pressure. 

\section{Covariant formulation of $f(Q,T)$ gravity}\label{sec3}
In the literature of cosmological application of symmetric teleparallel theory, we commonly observe the use of the so-called ``coincident gauge", that is, to identify a coordinate system whereby the connection vanishes 
and covariant derivatives reduce to partial derivatives. This is common in both isotropic spatially flat FLRW spacetime as well as in Bianchi type anisotropic spacetimes. As described in \cite{DZ, R46}, this sometime poses a major issue when we attempt to indulge into investigation of some other spacetimes using the same vanishing connection. Most of the time, the system is not consistent unless we force the non-metricity scalar $Q$ to be a constant, even worse, $Q=0$. To alleviate this problem, a fully-covariant formulation is very useful to employ non-vanishing connections into the game. In this section, we introduce the much-awaited covariant formulation of the $f(Q,\mathcal{T})$ theory.

Let us begin with earlier discussed curvature free and torsion free constraints providing 
\begin{equation}\label{mRicci}
R_{\mu\nu}+\nabla_\alpha L^\alpha{}_{\mu\nu}-\nabla_\nu\tilde L_\mu
+\tilde L_\alpha L^\alpha{}_{\mu\nu}-L_{\alpha\beta\nu}L^{\beta\alpha}{}_\mu=0\,,  
\end{equation}
\begin{equation}\label{mR}
 R+\nabla_\alpha(L^\alpha-\tilde L^\alpha)-Q=0\,.   
\end{equation}
As mentioned before, the coincident gauge is chosen in which  $\mathring{\Gamma}^\lambda{}_{\mu\nu}=0$ or $\Gamma^\lambda{}_{\mu\nu}=-L^\lambda{}_{\mu\nu}$.
Then we have   
\begin{align}\label{del:g}
\partial_\lambda\sqrt{-g}=-\sqrt{-g}\tilde L_\lambda\,.
\end{align}
Furthermore, we derive 
\begin{align}
\frac{2}{\sqrt{-g}}\partial_\lambda (\sqrt{-g}f_QP^\lambda{}_{\mu\nu})
+& f_Q(P_{\nu\rho\sigma} Q_\mu{}^{\rho\sigma} -2P_{\rho\sigma\mu}Q^{\rho\sigma}{}_\nu)
\notag\\
=&2(\mathring{\nabla}_\lambda f_Q)P^\lambda{}_{\mu\nu}
    +2f_Q(\nabla_\lambda P^\lambda{}_{\mu\nu}
    -L_{\alpha\beta\mu}P^{\beta\alpha}{}_\nu
    -L_{\alpha\beta\mu}P_\nu{}^{\beta\alpha}
    +L_{\nu\alpha\beta}P^{\alpha\beta}{}_\mu ) \notag\\
=&2(\mathring{\nabla}_\lambda f_Q)P^\lambda{}_{\mu\nu}
    +2f_Q(\nabla_\lambda P^\lambda{}_{\mu\nu}
    -\tilde L_\alpha L^\alpha{}_{\nu\mu}+L_{\alpha\beta\nu}L^{\beta\alpha}{}_\mu)\,.
    \label{I}
\end{align}
On the other hand, using (\ref{mRicci}) and (\ref{mR}),
we obtain
\begin{align}
2\nabla_\alpha P^\alpha{}_{\mu\nu}
=&-\nabla_\alpha L^\alpha{}_{\mu\nu}
    +\frac{\nabla_\alpha(L^\alpha-\tilde L^\alpha)}2g_{\mu\nu}
    +\frac{\nabla_\nu\tilde L_\mu+\nabla_\mu\tilde L^\nu}2 \notag\\
=& R_{\mu\nu} +\frac{Q- R}2g_{\mu\nu}
   +\tilde L_\alpha L^\alpha{}_{\mu\nu}-L_{\alpha\beta\nu}L^{\beta\alpha}{}_\mu\,.
   \label{II}
\end{align}
Combining (\ref{I})-(\ref{II}) we obtain
\begin{equation} \label{A}
\frac{2}{\sqrt{-g}} \partial_\lambda (\sqrt{-g}f_QP^\lambda{}_{\mu\nu}) 
 + f_Q(P_{\nu\rho\sigma} Q_\mu{}^{\rho\sigma} -2P_{\rho\sigma\mu}Q^{\rho\sigma}{}_\nu) =
 2(\mathring{\nabla}_\lambda f_Q)P^\lambda{}_{\mu\nu}+
f_Q \left( R_{\mu\nu} -\frac{ R-Q}2g_{\mu\nu}\right)\,.
\end{equation}
Finally, the metric field equation can be rewritten covariantly as,
\begin{equation} \label{FE2}
f_Q G_{\mu\nu} + \frac{1}{2}g_{\mu\nu}(Qf_Q - f)
+f_\mathcal{T}(\mathcal{T}_{\mu\nu}+\Theta_{\mu\nu})
 + 2(\mathring{\nabla}_\lambda f_Q) P^\lambda{}_{\mu\nu} 
 = \kappa \mathcal{T}_{\mu\nu} \text{,} 
\end{equation}
where $G_{\mu\nu}$ is the Einstein tensor corresponding to the Levi-Civita connection.
We define the effective stress energy tensor as
\begin{equation} \label{T^eff}
\kappa \mathcal{T}^{\text{eff}}_{\mu\nu} = \kappa \mathcal{T}_{\mu\nu}-f_\mathcal{T}(\mathcal{T}_{\mu\nu}+\Theta_{\mu\nu})-\frac{1}{2}g_{\mu\nu}(Qf_Q-f)
 -2(f_{QQ} \mathring{\nabla}_\lambda Q+f_{Q\mathcal{T}}\mathring{\nabla}_\lambda \mathcal{T}) P^\lambda{}_{\mu\nu} \,.
\end{equation}
In the present chapter, we consider a perfect fluid type spacetime, whose stress energy tensor takes the form 
\begin{align}
\mathcal{T}_{\mu\nu}=pg_{\mu\nu}+(p+\rho)u_\mu u_\nu \text{,} 
\end{align}
to which the matter Lagrangian can be taken as $ L_m=p$, where $\rho$, $p$ and $u^\mu$ denote 
the energy density, pressure and four velocity of the fluid respectively. It follows that 
\begin{align}
\Theta_{\mu\nu}=pg_{\mu\nu}-2\mathcal{T}_{\mu\nu} \text{.} 
\end{align}

\section{Energy conservation in $f(Q,\mathcal{T})$ gravity}\label{sec4}
The $f(Q,\mathcal{T})$ theory is not compatible with the energy conservation criterion, so an energy momentum balance equation was offered in \cite{Yixin}. Unfortunately, it seems to be not derived correctly as two crucial terms were noticed missing, particularly while applying the covariant derivative $\nabla_\mu$ to
\begin{align*}
\frac2{\sqrt{-g}}\mathring{\nabla}_\alpha(\sqrt{-g}f_QP^{\alpha\mu}{}_\nu) \qquad \text{  and  } \qquad f_\mathcal{T} \mathcal{T}^\mu{}_\nu.
\end{align*}
The corrected energy-momentum balance equation should read as 
\begin{align}\label{eqn:balance}
\nabla_\mu \mathcal{T}^\mu{}_\nu=\frac1{\kappa-f_\mathcal{T}}\left\{
\nabla_\mu(f_\mathcal{T}\Theta^\mu{}_\nu)+(\mathring{\nabla}_\mu f_\mathcal{T})\mathcal{T}^\mu{}_\nu-\frac12f_\mathcal{T}\mathring{\nabla}_\nu \mathcal{T}
+\frac\kappa{\sqrt{-g}}\mathring{\nabla}_\alpha\mathring{\nabla}_\mu H_\nu{}^{\alpha\mu}
\right\}\,.
\end{align}
In what follows, we briefly discuss the derivation of (\ref{eqn:balance}). Let us begin with
the field equation (\ref{FE1}) of type $(1,1)$ 
\begin{align}
 \label{FE1b}
 \kappa \mathcal{T}^\mu{}_\nu-f_\mathcal{T}(\Theta^\mu{}_\nu+\mathcal{T}^\mu{}_\nu)+\frac12f\delta^\mu{}_\nu
 =f_QP^{\mu\rho\sigma} Q_{\nu\rho\sigma}
 +\frac{2}{\sqrt{-g}}\mathring{\nabla}_\lambda (\sqrt{-g}f_QP^{\lambda\mu}{}_\nu) \,. 
\end{align}
The divergence of the preceding equation gives 
\begin{align}\label{eqn:divT-b}
(\kappa-f_\mathcal{T})\nabla_\mu \mathcal{T}^\mu{}_\nu-(\mathring{\nabla}_\mu f_\mathcal{T})\mathcal{T}^\mu{}_\nu
-\nabla_\nu(f_\mathcal{T}\Theta^\mu{}_\nu)+\frac12\mathring{\nabla}_\nu f
&
=(\mathring{\nabla}_\lambda f_Q P^{\mu\rho\sigma} Q_{\nu\rho\sigma})
 +\nabla_\mu\left\{\frac{2}{\sqrt{-g}}\mathring{\nabla}_\lambda (\sqrt{-g}f_QP^{\lambda\mu}{}_\nu) \right\}\,.
\end{align}
To simplify the above equation, we explicitly expand the two terms in the right hand side as 
\begin{align}
 \nabla_\mu(f_QP^{\mu\rho\sigma} Q_{\nu\rho\sigma})
 =&Q_{\nu\rho\sigma}(\mathring{\nabla}_\mu-\tilde L_\mu)(f_QP^{\mu\rho\sigma})
 +f_QP^{\mu\rho\sigma}(\mathring{\nabla}_\mu Q_{\nu\rho\sigma}
 +L^\beta{}_{\nu\mu}Q_{\beta\rho\sigma}) \nonumber\\
 =&-2L_{\sigma\nu\rho}(\mathring{\nabla}_\mu-\tilde L_\mu)(f_QP^{\mu\rho\sigma})
 +f_QP^{\mu\rho\sigma}(\nabla_\nu Q_{\mu\rho\sigma}
 -2L^\beta{}_{\rho\nu}Q_{\mu\beta\sigma})\,,
\end{align}
\begin{align}
\nabla_\mu\left\{
\frac{2}{\sqrt{-g}}\mathring{\nabla}_\lambda (\sqrt{-g}f_QP^{\lambda\mu}{}_\nu) 
\right\}
=&(\mathring{\nabla}_\mu-\tilde L_\mu)\left\{
\frac{2}{\sqrt{-g}}\mathring{\nabla}_\lambda (\sqrt{-g}f_QP^{\lambda\mu}{}_\nu) 
\right\}+
\frac{2L^\beta{}_{\nu\mu}}{\sqrt{-g}}\mathring{\nabla}_\lambda (\sqrt{-g}f_QP^{\lambda\mu}{}_\beta) \nonumber\\
=&\frac{2}{\sqrt{-g}}\mathring{\nabla}_\mu\mathring{\nabla}_\lambda (\sqrt{-g}f_QP^{\lambda\mu}{}_\nu) 
+
2L^\beta{}_{\nu\rho}
(\mathring{\nabla}_\mu-\tilde L\mu)(f_QP^{\mu\rho}{}_\beta) 
\,,
\end{align}
where we have used the relation (\ref{del:g}).
In addition, we have the relations
\begin{align}\label{ad1}
2L^\beta{}_{\nu\rho}(\mathring{\nabla}_\mu-\tilde L\mu)(f_QP^{\mu\rho}{}_\beta) 
=L_{\sigma\nu\rho}(\mathring{\nabla}_\mu-\tilde L\mu)(f_QP^{\mu\rho\sigma}) 
 +L^\beta{}_{\nu\rho}(f_QP^{\mu\rho}{}_\beta)Q_{\mu\sigma\beta} \text{,} 
\end{align}
\begin{align}\label{ad2}
\mathring{\nabla}_\nu Q=\nabla_\nu(P^{\mu\rho\sigma}Q_{\mu\rho\sigma})
=2P^{\mu\rho\sigma}\nabla_\nu Q_{\mu\rho\sigma}\,.
\end{align}
Finally, (\ref{eqn:balance}) can be obtained after 
substituting (\ref{FE-connection}) and the above relations into (\ref{eqn:divT-b}).

Now that we offer the fully covarinat formulation of the field equation (\ref{FE2}), we can directly use the Bianchi identity to derive another equivalent form of the energy momentum balance equation by taking divergence of (\ref{FE2})
\begin{align}\label{eqn:divT-c}
(\kappa-f_\mathcal{T})\nabla_\mu \mathcal{T}^\mu{}_\nu
-&(\mathring{\nabla}_\mu f_\mathcal{T})\mathcal{T}^\mu{}_\nu
-\nabla_\nu(f_\mathcal{T}\Theta^\mu{}_\nu)+\frac{f_\mathcal{T}}2\mathring{\nabla}_\nu Q \nonumber\\
=&(\mathring{\nabla}_\lambda f_Q)\left(G^\lambda{}_\nu+\frac Q2\delta^\lambda{}_\nu+2\nabla_\mu P^{\lambda\mu}{}_\nu\right)
 +2(\nabla_\mu\nabla_\lambda f_Q)P^{\lambda\mu}{}_\nu\,.
\end{align}
Note that (\ref{eqn:balance}) and (\ref{eqn:divT-c}) of the energy-balance in $f(Q,\mathcal{T})$ theory are identical once the affine connection field equation (\ref{FE-connection}) are taken into account.

\section{Cosmological applications}\label{sec5}

In this section we explore some cosmological applications of the previously obtained covariant formulation and energy balance equation. For this purpose we consider the spatially flat homogeneous and isotropic FLRW spacetime with the line element given in Cartesian coordinates by
\begin{align}\label{metric}
    ds^2=-dt^2+a^2(t)\left[dx^2+dy^2+dz^2\right].
\end{align}
For simplicity, we utilize the usual vanishing affine connection in this coincident gauge choice to obtain the non-metricity scalar as
\begin{align}
    Q=-6H^2.
\end{align}
Using equations \eqref{FE2} and \eqref{metric}, we obtain the following Friedmann-like equations,
\begin{align}
(\kappa +f_\mathcal{T})\rho+f_\mathcal{T} p &= \frac{f}{2} + 6H^2 f_Q,  \label{rho-temp}\\
\kappa p &= -\frac{f}{2} -6H^2 f_Q - \frac{\partial}{\partial t}(2Hf_Q)\,.
    \label{p}
\end{align}
When particularly $f_\mathcal{T}=-\kappa$, we obtain  the equation
\begin{align}
\frac{\partial}{\partial t}(2Hf_Q)=0\,,
\end{align}
whose solution is given as  
\begin{align}
f(Q,\mathcal{T})=\alpha\sqrt{-Q}-\kappa \mathcal{T}\,,
\end{align}
where $\alpha$ is a constant. 
As this solution does not include GR, we conclude that this is not an adequate model. In what follows, we consider only $f_\mathcal{T} \neq-\kappa$. Using (\ref{rho-temp})-(\ref{p}), we obtain
\begin{align}
\kappa\rho &=\frac f2+6H^2f_Q+\frac{f_\mathcal{T}}{\kappa+f_\mathcal{T}} \frac{\partial}{\partial t}(2Hf_Q)\,.
        \label{rho}
\end{align}
From the corrected energy balance equation (\ref{eqn:divT-c}) derived in the last section, we can write the continuity relation as  
\begin{align}\label{gen}
-\dot\rho-3H(\rho+p)
=-\frac1\kappa\frac{\partial}{\partial t}\left(\frac f2+6H^2f_Q\right)
-\frac1{\kappa+f_\mathcal{T}}\left(\frac{\dot f_\mathcal{T}}{\kappa+f_\mathcal{T}}-3H\right)\frac{\partial}{\partial t}(2Hf_Q)
  -\frac1\kappa\frac{f_\mathcal{T}}{\kappa+f_\mathcal{T}}\frac{\partial^2}{\partial t^2}(2Hf_Q).
\end{align}
In particular, for pressureless dust era ($p=0$) it reduces to
\begin{align}
\dot\rho+3H\rho
=\frac1{\kappa+f_\mathcal{T}}
 \left\{\frac{\dot f}2+12\dot HHf_Q+6H^2\dot{f_Q}
+\left(3H-\frac{\dot f_\mathcal{T}}{\kappa+f_\mathcal{T}}\right)\left(\frac f2+6H^2f_Q\right)
\right\}\,.\label{dust}
\end{align}

On the other hand, we can derive the effective energy density and pressure equations using (\ref{T^eff})
 \begin{align}
\kappa \rho^{\text{eff}}=&(\kappa +f_\mathcal{T})\rho+f_\mathcal{T} p- \frac{f}{2} - 3H^2 f_Q\,,  \\
 \kappa p^{\text{eff}} 
=&\kappa p +\frac{f}{2} + 3H^2f_Q+2H\dot{f}_Q 
\,.
\end{align}

Now that we pull the necessary things together, in the following subsections we are going to examine the dynamical behavior of some well known cosmological $f(Q,\mathcal{T})$ models by incorporating the phase space approach.

\subsection{Linear $f(Q,\mathcal{T})$ model}
We consider the following linear $f(Q,T)$ model
\begin{equation}\label{a0}
f(Q,\mathcal{T})= \alpha Q+ \beta \mathcal{T} \text{.} 
\end{equation}
Here, $\alpha$ and $\beta$ are free model parameters. The cosmological implications of the considered model have been investigated in \cite{Yixin}. We are going to investigate the asymptotic behavior of the model. 
The Friedmann equations \eqref{p} and \eqref{rho} corresponding to our linear $f(Q,\mathcal{T})$ model, for the dust case, becomes
\begin{equation}\label{a1}
    (2\kappa+3\beta) \rho = 6 \alpha H^2
\end{equation}
and
\begin{equation}\label{a2}
    2 \dot{H} + 3 H^2 = \frac{\beta}{2\alpha} \rho\,,
\end{equation}
and the equation \eqref{dust} becomes,
\begin{equation}\label{a3}
 \dot{\rho}  + \frac{6(\kappa+\beta)}{(2\kappa+3\beta)} H \rho = 0 \text{.} 
\end{equation}
We define the following phase space variables 
\begin{equation}
    x=\frac{ (2\kappa+3\beta) \rho}{6 \alpha H^2}  \:\: \text{and} \:\: y= \frac{1}{\frac{H_0}{H}+1} \text{.} 
\end{equation}\label{a4}
Then we have constraints $x=1$ and $0 \leq y \leq 1$.
Now we define $N=dln(a)$, then corresponding to our linear $f(Q,T)$ model we obtained the following autonomous system by using \eqref{a1}-\eqref{a3},
\begin{equation}\label{a5}
    x'= \frac{dx}{dN} = \frac{3\beta}{(2\kappa+3\beta)} x (1-x)
\end{equation}
and
\begin{equation}\label{a6}
    y'= \frac{dy}{dN} = \frac{3}{2} \big[ \frac{\beta x }{(2\kappa+3\beta)} -1 \big] y (1-y) \text{.} 
\end{equation} 

Now by the definition of deceleration and EoS parameter, we have for $x=1$
\begin{equation}\label{a8}
    q=-1+ \frac{3(\kappa+\beta)}{(2\kappa+3\beta)} \: \text{and} \: \omega = -1+ \frac{2(\kappa+\beta)}{(2\kappa+3\beta)} \text{.} 
\end{equation}

The nature of the critical points obtained by solving the above autonomous equations with $\kappa=1$, are presented below in the Table \eqref{Table-1}.

\begin{table}[H]
\begin{center}\caption{Table shows the critical points and their behavior corresponding to the model $f(Q,\mathcal{T})=\alpha Q+ \beta \mathcal{T}$.}
\begin{tabular}{|c|c|c|c|c|c|}
\hline
 Critical Points $(x_c,y_c)$ & Eigenvalues $\lambda_1$ and $\lambda_2$ & Nature of critical point  & $q$ & $\omega$ \\
\hline 
 $P(1,1)$ & $ -\frac{3 \beta }{(2+3 \beta) } \:\: \text{and} \:\: \frac{3 (1+\beta )}{(2+3 \beta) }$ & Stable for $-1<\beta <-\frac{2}{3}$ & $\frac{1}{2+3\beta}$ & $-\frac{\beta}{2+3\beta}$ \\
$Q(1,0)$ & $-\frac{3 (1+\beta )}{(2+3 \beta )} \:\: \text{and} \:\: -\frac{3 \beta }{(2+3 \beta) }$ & Stable for $\beta < -1$ or $ \beta \geq 0$ & $\frac{1}{2+3\beta}$ & $-\frac{\beta}{2+3\beta}$ \\
\hline
\end{tabular}\label{Table-1}
\end{center}
\end{table}

The asymptotic behaviour of our linear $f(Q,\mathcal{T})$ model corresponding to the case $\beta=0$  with $\kappa=1$, presented below in figure \eqref{f1}.

\begin{figure}[H]
\begin{center}
\includegraphics[scale=0.585]{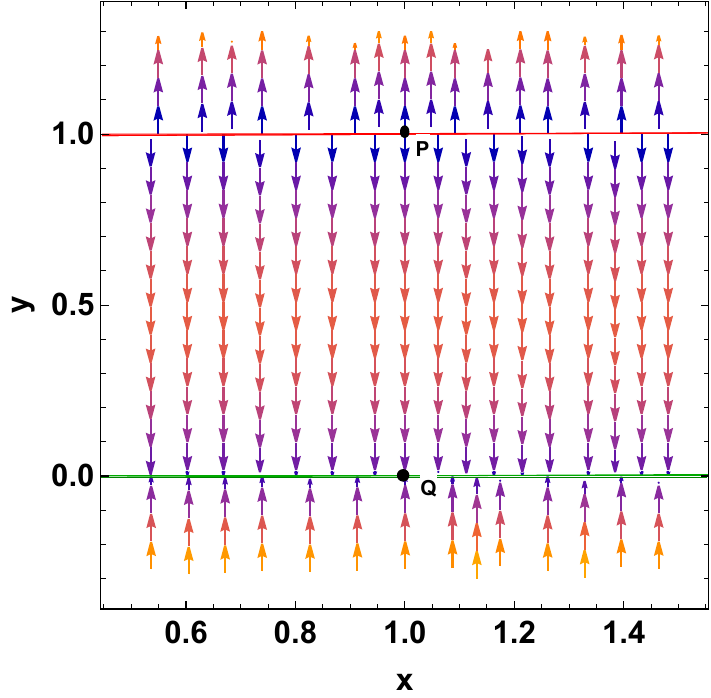}
\caption{Phase plots corresponding to the case $\beta=0$ with $\kappa=1$.}\label{f1}
\end{center}
\end{figure}

For the case $\beta=0$, the obtained critical points are P(1,1) and Q(1,0) with corresponding eigenvalues $0, \frac{3}{2}$ and $0, -\frac{3}{2}$ respectively. From figure \eqref{f1}, it is evident that the trajectories are emerging from the unstable past attractor P(1,1) and converging to future attractor Q(1,0). Hence, point P(1,1) is a source, whereas point Q(1,0) is a sink. From equation \eqref{a8}, we obtained $q=\frac{1}{2}$ and $\omega=0$ for both P and Q. Therefore, we can conclude that our model $f(Q,\mathcal{T})=\alpha Q+ \beta \mathcal{T}$ with $\beta=0$ remains lies in the matter dominated epoch and cannot describe the accelerated era as well as the initial singularity. Hence, it is completely equivalent to the GR case without cosmological constant $\Lambda$. 

\justify From the equation \eqref{a8}, one can notice that the linear model $f(Q,\mathcal{T})=\alpha Q + \beta \mathcal{T}$ cannot describes the transition from decelerated to accelerated era for any parameter values.

\subsection{Non-linear $f(Q,\mathcal{T})$ models}
\subsubsection{Model-I}

We consider the following non-linear $f(Q,\mathcal{T})$ model
\begin{equation}\label{b0}
f(Q,\mathcal{T})= \alpha Q+ \beta \mathcal{T}^2 \text{.} 
\end{equation}
Here, $\alpha$ and $\beta \neq 0$ are free model parameters. The cosmological implications of the considered model have been investigated in \cite{Yixin}. We are going to investigate the asymptotic behavior of the model. 
The Friedmann equations \eqref{p} and \eqref{rho} corresponding to our non-linear $f(Q,\mathcal{T})$ model, for the dust case, becomes
\begin{equation}\label{b1}
    (2\kappa-5\beta \rho ) \rho = 6 \alpha H^2 
\end{equation}
and
\begin{equation}\label{b2}
    2 \dot{H} + 3 H^2  = -\frac{\beta}{2 \alpha} \rho^2 \text{.} 
\end{equation}
and the equation \eqref{dust} becomes
\begin{equation}\label{b3}
 \dot{\rho}  + \frac{3(\kappa-2\beta \rho)}{(\kappa-5\beta \rho)}  H \rho = 0 \text{.} 
\end{equation}

We define the following phase space variables 
\begin{equation}\label{b4}
    x=\frac{\kappa \rho}{3 \alpha H^2},  \:\: y= \frac{1}{\frac{H_0}{H}+1} \:\:  \text{and} \:\: z= -\frac{5 \beta \rho^2}{6 \alpha H^2} \text{.} 
\end{equation}\label{a4}
Then we have constraints $x+z=1$ and $0 \leq y \leq 1 $.
Now corresponding to our non-linear $f(Q,\mathcal{T})$ model we obtained the following autonomous system with respect to the variable $N=dln(a)$, by using \eqref{b1}-\eqref{b3} and the constrain $x+z=1$,
\begin{equation}\label{b5}
    x'= \frac{dx}{dN} = \frac{3x (x-1)(x+4)}{5(x-2)} 
\end{equation}
and
\begin{equation}\label{b6}
    y'= \frac{dy}{dN} = \frac{3}{10} y (y-1) (x+4) \text{.} 
\end{equation} 

Now by the definition of deceleration and EoS parameter, we have 
\begin{equation}\label{b8}
    q= \frac{1}{10} (3x+2) \: \text{and} \: \omega = \frac{1}{5} (x-1) \text{.} 
\end{equation}

The nature of the critical points obtained by solving the above autonomous equations with $\kappa=1$, are presented below in the Table \eqref{Table-2}.

\begin{table}[H]
\begin{center}\caption{Table shows the critical points and their behavior corresponding to the model $f(Q,\mathcal{T})=\alpha Q+ \beta \mathcal{T}^2$.}
\begin{tabular}{|c|c|c|c|c|}
\hline
 Critical Points $(x_c,y_c)$ & Eigenvalues $\lambda_1$ and $\lambda_2$ & Nature of critical point  & $q$ & $\omega$ \\
\hline 
 $A(0,0)$ & $-\frac{6}{5} \:\: \text{and} \:\: \frac{6}{5}$ & Saddle & $\frac{1}{5}$ & $-\frac{1}{5}$ \\
 \hline 
 $B(1,1)$ & $-3 \:\: \text{and} \:\: \frac{3}{2}$ & Saddle & $\frac{1}{2}$ & $0$ \\
\hline
 $C(1,0)$  & $-3 \:\: \text{and} \:\: -\frac{3}{2}$ & Stable & $\frac{1}{2}$ & $0$ \\
\hline
 $D(0,1)$  & $\frac{6}{5} \:\: \text{and} \:\: \frac{6}{5}$ & Unstable & $\frac{1}{5}$ & $-\frac{1}{5}$ \\
 \hline
 $E(-4,y)$  & $-2 \:\: \text{and} \:\: 0$ & Stable & $-1$ & $-1$ \\
 \hline 
\end{tabular}\label{Table-2}
\end{center}
\end{table}

The asymptotic behaviour of our non-linear $f(Q,\mathcal{T})$ model with $\kappa=1$, presented below in figure \eqref{f2}.

\begin{figure}[H]
\begin{center}
\includegraphics[scale=0.6]{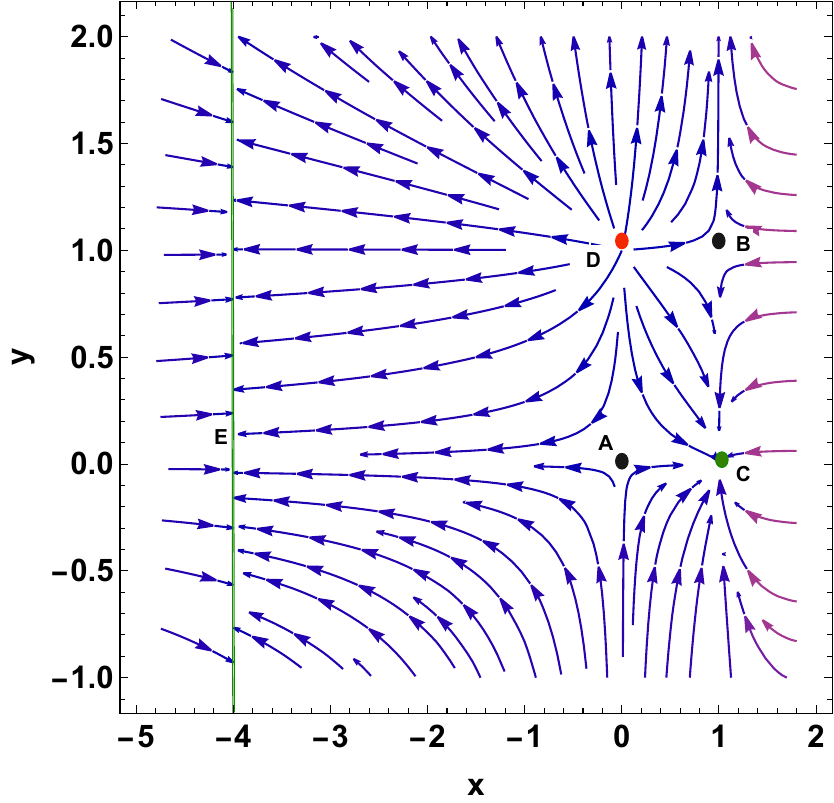}
\caption{Phase plot corresponding to the model $f(Q,T)=\alpha Q + \beta T^2 $ with $\kappa=1$. }\label{f2}
\end{center}
\end{figure}

\justify From figure \eqref{f2}, it is evident that the critical points A(0,0) and B(1,1) are saddle points, whereas the critical point C(1,0) indicates a stable matter dominated epoch. The critical point D(0,1) is a source while the critical point E(-4,y), for any y, is a sink representing a future attractor. The trajectories emerging from the D(0,1), indicating a decelerated epoch, are then converging to the point E(-4,y) representing an accelerated epoch. Thus we conclude that our non-linear f(Q,T) model efficiently describes the observed transition from decelerated phase to an accelerated phase of the Universe. The considered model $f(Q,\mathcal{T})=\alpha Q+ \beta \mathcal{T}^2$ with $\beta \neq 0$ behaves like standard $\Lambda$CDM model, and hence this model may represent an alternative to the $\Lambda$CDM.

\subsubsection{Model-II}
We consider the following non-linear $f(Q,\mathcal{T})$ model
\begin{equation}\label{c0}
f(Q,T)= Q+ \alpha Q^2+ \beta \mathcal{T} \text{.} 
\end{equation}
Here, $\alpha \neq 0$ and $\beta$ are free model parameters. The astrophysical implications of the considered model have been investigated in \cite{Moreshwar}. We are going to examine the cosmological behavior of the model. 
The Friedmann equations \eqref{p} and \eqref{rho} corresponding to this non-linear $f(Q,\mathcal{T})$ model, for the dust case, becomes
\begin{equation}\label{c1}
    (2\kappa+3\beta ) \rho = 6 H^2 (1-18\alpha H^2)
\end{equation}
and
\begin{equation}\label{c2}
    2 \dot{H} (1-36\alpha H^2) + 3 H^2 (1-18\alpha H^2) = \frac{\beta}{2} \rho \text{,} 
\end{equation}
and the equation \eqref{dust} becomes
\begin{equation}\label{c3}
 \dot{\rho}  + \frac{6(\kappa+\beta)}{(2\kappa+3\beta)} H \rho = 0 \text{.} 
\end{equation}
We define the following phase space variables, 
\begin{equation}
    x=\frac{ (2\kappa+3\beta) \rho}{6H^2 (1-18\alpha H^2)}  \:\: \text{and} \:\: y= \frac{1}{\frac{H_0}{H}+1} \text{.} 
\end{equation}\label{c4}
Then we have constraints $x=1$ and $0 \leq y \leq 1 $.
Now we define $N=dln(a)$, then corresponding to our non-linear $f(Q,\mathcal{T})$ model we obtained the following autonomous system by using \eqref{c1}-\eqref{c3},
\begin{equation}\label{c5}
    x'= \frac{dx}{dN} = \frac{3\beta}{(2\kappa+3\beta)} x (1-x)
\end{equation}
and
\begin{equation}\label{c6}
    y'= \frac{dy}{dN} = \frac{3(\kappa+\beta)}{(2\kappa+3\beta)}  \frac{[(1-y)^2-\bar{\alpha}y^2] }{[(1-y)^2-2 \bar{\alpha}y^2]} y (y-1) \text{.} 
\end{equation} 
Here, $\bar{\alpha}=18\alpha H_0^2$.

Again, by the definition of deceleration and EoS parameter, we have 
\begin{equation}\label{c7}
    q= -1 + \frac{3 (\kappa+\beta ) [(1-y)^2-\bar{\alpha}  y^2]}{(2\kappa+3 \beta ) [(1-y)^2-2 \bar{\alpha}  y^2]} \text{,} 
\end{equation}
\begin{equation}\label{c8}
    \omega= -1 + \frac{2 (\kappa+\beta ) [(1-y)^2-\bar{\alpha}  y^2]}{(2\kappa+3 \beta ) [(1-y)^2-2 \bar{\alpha}  y^2]} \text{.} 
\end{equation}

The nature of the critical points obtained by solving the above autonomous equations with $\kappa=1$, are presented below in the Table \eqref{Table-3}.

\begin{table}[H]
\begin{center}\caption{Table shows the critical points and their behavior corresponding to the model $f(Q,\mathcal{T})=Q+ \alpha Q^2+ \beta \mathcal{T}$.}
\begin{tabular}{|c|c|c|c|c|c|}
\hline
 Critical Points $(x_c,y_c)$ & Eigenvalues $\lambda_1$ and $\lambda_2$ & Nature of critical point  & $q$ & $\omega$ \\ 
\hline 
 $A(1,0)$ & $-\frac{3(1+\beta)}{(2+3\beta)} \:\: \text{and} \:\: -\frac{3\beta}{(2+3\beta)}$ & Stable for $\beta < -1$ or $ \beta \geq 0$ & $\frac{1}{2+3\beta}$ & $-\frac{\beta}{2+3\beta}$ \\
  $B(1,1)$ & $-\frac{3\beta}{(2+3\beta)} \:\: \text{and} \:\: \frac{3(1+\beta)}{2(2+3\beta)}$ & Stable for $-1 < \beta < -\frac{2}{3}$  & $-\frac{1+3\beta}{2(2+3\beta)}$ & $-\frac{1+2\beta}{(2+3\beta)}$ \\
$C(1,\frac{1}{1+\sqrt{\bar{\alpha}}}), \bar{\alpha}>0$  & $-\frac{3\beta}{(2+3\beta)} \:\: \text{and} \:\: -\frac{6(1+\beta)}{(2+3\beta)}$ &  Stable for $\beta < -1$ or $ \beta \geq 0$ & $-1$ & $-1$ \\
\hline
\end{tabular}\label{Table-3}
\end{center}
\end{table}

The asymptotic behaviour of our non-linear $f(Q,\mathcal{T})$ model corresponding to the case $\beta=0$ with $\kappa=1$ and $\bar{\alpha}=1$, presented below in figure \eqref{f3}.

\begin{figure}[H]
\begin{center}
\includegraphics[scale=0.585]{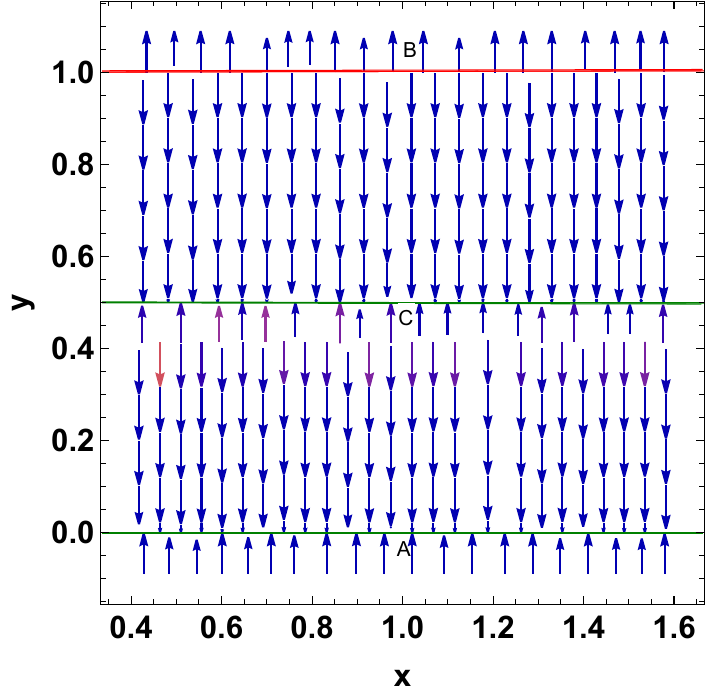}
\caption{Phase plots corresponding to the case $\beta=0$ with $\kappa=1$ and $\bar{\alpha}=1$. }\label{f3}
\end{center}
\end{figure}

For the case $\beta=0$ with $\kappa=1$ and $\bar{\alpha}=1$, the obtained critical points are A(1,0), B(1,1), and C(1,0.5). From Table \eqref{Table-3} and figure \eqref{f3}, it is evident that the critical points A(1,0) and C(1,0.5) with corresponding eigenvalues $-\frac{3}{2}, 0$ and $0,-3$ are stable and representing the matter dominated decelerated epoch and accelerated de-Sitter epoch respectively, whereas the point B(1,1) with eigenvalues $\frac{3}{4}, 0$ is unstable.   

\section{Conclusions}\label{sec6}
In the present chapter, we have examined the matter-geometry coupling in the form of the $f(Q,\mathcal{T})$ theory which shows promising development and rising research interest in the cosmological sector. We have derived an equivalent covariant formulation of the theory which yields an explicit comparison with the GR. The non-conservation of the energy-momentum tensor in this theory evokes a generally non-zero covariant vector which has been obtained from the divergence of this covariant formula of the field equation, and by using the Bianchi identity. In account of this result, we have computed the modified continuity relation, specific to this theory. The covariant formulation of the theory has also helped us to efficiently express the effective pressure and energy terms. 

Further, we have considered three different functional forms of the $f(Q,\mathcal{T})$ function, specifically, $f(Q,\mathcal{T})=\alpha Q+ \beta \mathcal{T}$, $f(Q,\mathcal{T})=\alpha Q+ \beta \mathcal{T}^2$, and $f(Q,\mathcal{T})=Q+ \alpha Q^2+ \beta \mathcal{T}$. These $f(Q,\mathcal{T})$ functions frequently appear in the literature. We have made an attempt to investigate the physical capabilities of the studied models to describe various cosmological epochs. We have incorporated the dynamical system technique to investigate the asymptotic behavior of the considered $f(Q,T)$ models. The obtained outcomes for the assumed linear and two non-linear $f(Q,\mathcal{T})$ models have been presented in Tables \eqref{Table-1}-\eqref{Table-3}, and the corresponding phase plots in figures \eqref{f1}-\eqref{f3}. Naturally, we have observed that the linear model $f(Q,\mathcal{T})=\alpha Q+ \beta \mathcal{T}$ with $\beta=0$ is completely equivalent to the GR case without cosmological constant $\Lambda$. Further, we have found that the model $f(Q,\mathcal{T})=\alpha Q+ \beta \mathcal{T}^2$ with $\beta \neq 0$ successfully describes the observed transition from decelerated phase to an accelerated phase of the Universe and hence the considered non-linear model I behaves like standard $\Lambda$CDM model. Moreover, our obtained results agrees with the cosmological implications of the same model found in \cite{Yixin}. Lastly, we found that the non-linear model II $f(Q,\mathcal{T})= Q+ \alpha Q^2+ \beta \mathcal{T}$ with $\alpha \neq 0$ represents an accelerated de-Sitter epoch for the constraint $\beta < -1$ or $ \beta \geq 0$. Moreover, a stable matter dominated epoch with an accelerated de-Sitter type epoch can be obtained in the absence of energy momentum scalar term $\mathcal{T}$ i.e., for the case $\beta=0$. From the basis of our findings we can conclude that both considered non-linear models can efficiently predicts the de-Sitter type expansion of the Universe and may represent a viable geometric alternative to dark energy.\\
 

%% file: Chapters/Conclusion.tex

\chapter{Concluding remarks and future perspectives} 

\label{Chapter7} 

\lhead{Chapter 7. \emph{Concluding remarks and future perspectives}} 

 \clearpage
 
In this thesis, we have discussed different cosmological accelerating models that can adequately describe the evolution of the late-time phase utilizing the non-Reimannian spacetime, particularly the modified symmetric teleparallel gravity based on non-metricity. The summary of the discussion, investigation, and the corresponding outcomes of each chapter are discussed below.

\section{Concluding remarks}

In Chapter-\ref{Chapter1}, we began our introductory phase with the description of the observed Universe and its key ingredients such as cosmological scale, cosmic distance ladder, galaxy rotation curves, etc. Then, we briefly discussed the fundamentals of relativity from Newtonian to Einstein's GR, which describes gravity as the geometric theory of curved spacetime. Further, we have discussed several cosmological solutions to GR. Moreover, we highlighted the pros and cons of the standard model of cosmology. Lastly, we have introduced the fundamentals of some important non-Riemannain spacetime geometry, such as teleparallel and symmetric teleparallel gravity and its extensions. 
 
In Chapter-\ref{Chapter2}, we analyzed the evolution of the FLRW Universe with non-relativistic bulk viscous matter, where the time-dependent bulk viscosity has the form $\xi =\xi _{0}+\xi _{1}H+\xi _{2}\left( \frac{\dot{H}}{H}+H\right) $ whose components is proportional to the velocity and acceleration of the expanding Universe, in the framework of linear $f(Q)$ gravity, particularly, $f(Q)=\alpha Q $. We obtained two sets of limiting conditions on the bulk viscous parameters $\xi _{0},$ $\xi _{1},$ $\xi _{2}$ and model parameter $\alpha$, out of which one condition favors the present scenario of cosmic acceleration with a phase transition and corresponds to the Universe with a Big Bang origin. We calculated the analytical solution of the model, and then we found the free parameters value by invoking the $H(z)$ datasets having $57$ points, Pantheon datasets having $1048$ data samples along with the BAO datasets having six data points. The obtained constraints show that our model has good compatibility with observations. We conclude that the bulk viscous theory can be considered as a good candidate to invoke negative pressure that contributes to describing the late time acceleration of the Universe.

In Chapter-\ref{Chapter3}, as an attempt to describe dark energy, we have assumed the non-linear $f(Q)$ gravity model, particularly the power-law functional $f(Q)=\alpha Q^n$, along with the energy-momentum tensor describing a bulk viscous fluid. We found the analytical solution of the model, and then we started testing our cosmological solutions. We started confronting the Hubble parameter with 31 data points that are measured from a differential age approach, which is also known as cosmic chronometers. In addition, we have considered Pantheon and BAO samples to constrain the obtained cosmological solution. Further, we investigated the consistency of our constrained solution by analyzing the different energy conditions. We found that NEC, DEC, and WEC are all satisfied in the entire domain of redshift, while the SEC is violated for lower redshifts, which implies the cosmic acceleration, and satisfied for higher redshifts, which implies a decelerated phase of the Universe. Lastly, we employed the statefinder diagnostic test, which shows that the evolutionary trajectories of our model are departed from $\Lambda$CDM. In the present epoch, they lie in the quintessence region, and hence the present model is, therefore, a good alternative to explain the Universe's late-time dynamics, particularly with no necessity of invoking the cosmological constant. 

In Chapter-\ref{Chapter4}, we explored the cosmological constraints on $f(Q)$ gravity models in the coincident gauge formalism. We considered a $f(Q)$ model that contains a linear and a non-linear form of non-metricity scalar, particularly $f(Q)=\alpha Q + \beta Q^n$, where $\alpha$, $\beta$, and $n$ are free model parameters. As the analytical solution to this class of model does not exist in the case of a perfect fluid, we analyzed the solution behavior numerically through NDSolve in Mathematica. We found that for higher positive values of $n$ specifically $n \geq 1$, dark energy fluid part evolving due to non-metricity behaves like quintessence type dark energy, while for higher negative values of $n$ specifically $n \leq -1$, it follows phantom scenario. For $n=0$, our cosmological $f(Q)$ model mimics the $\Lambda$CDM model of GR. Furthermore, we obtained that, except SEC, all the energy conditions are satisfied for the case $n \geq 1$ while all the energy conditions are violated for the case $n \leq -1$. Therefore, depending upon the choice of $n$ we found that one can obtain quintessence, phantom, and the $\Lambda$CDM like behavior without invoking any dark energy component or exotic fluid in the matter part. In addition, we found the exact solution for the particular case $f(Q)=\alpha Q + \beta $, where $\alpha$ and $\beta$ are free parameters. We have calculated the best fit ranges of the model parameters $\alpha$ and $\beta$ for the combined CC+BAO+Pantheon datasets and discussed the corresponding evolutionary parameters. Thus, we conclude that the geometrical generalization of GR can be a viable candidate for the description of the origin of the dark energy. 

The Chapter-\ref{Chapter5} investigates cosmological constraints on $f(Q)$ gravity models in a non-coincident formalism. We studied a new $f(Q)$ theory dynamics utilizing a non-vanishing affine connection involving a non-constant function $\gamma(t)=-a^{-1}\dot{H}$. In addition, we proposed a new parameterization of the Hubble function that consistently depicted the present deceleration parameter value, transition redshift, and the late time de-Sitter limit. We evaluated the predictions of the assumed Hubble function by imposing constraints on the free parameters utilizing Bayesian statistical analysis to estimate the posterior probability by employing the CC, Pantheon+SH0ES, and the BAO samples. Moreover, we conducted the AIC and BIC statistical evaluations to determine the reliability of MCMC analysis. We have determined $\Delta AIC=1.408$ and $\Delta BIC= 6.869$ which supports a shred of strong evidence in favor of our proposed Hubble function over the standard $\Lambda \text{CDM}$. Further, we considered some well-known corrections to the STEGR case such as an exponentital $f(Q)$ correction, particularly, $f(Q)=Q+\eta e^{\alpha Q}$, logarithmic $f(Q)$ correction $f(Q)=Q+\eta log{\alpha Q}$, and a power-law $f(Q)$ correction $f(Q)=Q+\eta Q^\alpha$, and then we find the constraints on the parameters of these models via energy conditions. Finally, to test the physical plausibility of the assumed $f(Q)$ models we conducted the thermodynamical stability analysis via the sound speed parameter.

In Chapter-\ref{Chapter6}, we presented the fully covariant formulation of the $f(Q,\mathcal{T})$ gravity, utilizing the corrected version of the energy balance equation. We employed this proposed formulation to conduct a dynamical system analysis in a spatially flat Friedmann-Lema\^{i}tre-Robertson-Walker spacetime. We considered three different functional forms of the $f(Q,\mathcal{T})$ function, specifically, $f(Q,\mathcal{T})=\alpha Q+ \beta \mathcal{T}$, $f(Q,\mathcal{T})=\alpha Q+ \beta \mathcal{T}^2$, and $f(Q,\mathcal{T})=Q+ \alpha Q^2+ \beta \mathcal{T}$. We attempted to investigate the physical capabilities of these models to describe various cosmological epochs. We calculated the Friedmann-like equations in each case and introduced some phase space variables to simplify the equations in more concise forms. We observed that the linear model $f(Q,\mathcal{T})=\alpha Q+ \beta \mathcal{T}$ with $\beta=0$ is completely equivalent to the GR case without cosmological constant $\Lambda$. Further, we found that the model $f(Q,\mathcal{T})=\alpha Q+ \beta \mathcal{T}^2$ with $\beta \neq 0$ successfully depicted the observed transition from decelerated phase to an accelerated phase of the Universe. Lastly, we found that the model $f(Q,\mathcal{T})= Q+ \alpha Q^2+ \beta \mathcal{T}$ with $\alpha \neq 0$ represents an accelerated de-Sitter epoch for the constraints $\beta < -1$ or $ \beta \geq 0$. 

\section{Future perspectives}

The present investigation is completely focused on the physical capabilities of the non-metricity based modified symmetric teleparallel gravity to reconstruct the dark energy scenario. Lots of investigation along with the present thesis have shown that this modified theory can be efficient in describing the accelerating phase via bypassing the need of $\Lambda$, as well as in resolving $H_0$ tension. From a future perspective, it would be interesting to study the early Universe behavior of this modified symmetric teleparallel gravity, particularly the inflationary scenario, Big Bang nucleosynthesis constraints, as well as the formation of large-scale structures.